\documentclass[11pt]{article}

\usepackage[T1]{fontenc}
\usepackage{lmodern, microtype}
\usepackage{fnpct}
\usepackage{caption}
\usepackage[left=1.25in, right=1.25in, top=1.25in, bottom=1.25in]{geometry}
\usepackage[onehalfspacing]{setspace}
\usepackage[small]{titlesec}

\makeatletter
\newcommand*{\addFileDependency}[1]{
  \typeout{(#1)}
  \@addtofilelist{#1}
  \IfFileExists{#1}{}{\typeout{No file #1.}}
}

\usepackage{xr-hyper}
\usepackage{hyperref}

\newcommand*{\myexternaldocument}[1]{%
  \externaldocument{#1}%
  \addFileDependency{#1.tex}%
  \addFileDependency{#1.aux}%
}
\myexternaldocument{OnlineAppendix}

 \usepackage{multicol}

\usepackage{xfrac,epigraph,enumerate,comment,amssymb,amsmath,multirow,booktabs}
\usepackage{amsthm}
\usepackage{fancyhdr}
\usepackage{bbm}
\usepackage{soul}
\usepackage{xcolor,paralist}
\usepackage{hyperref}
\usepackage[section]{placeins}
\hypersetup{
  colorlinks=true,
  linkcolor=blue,
  citecolor=blue
}
\usepackage{natbib}
\setcitestyle{authoryear,open={(},close={)}}

\theoremstyle{definition}

\usepackage{graphicx}

\usepackage{pstricks, enumerate, pst-node, pst-text, pst-plot}

\usepackage{changepage}

\usepackage{xparse}

\DeclareDocumentCommand\Pr{ m g }{\ensuremath{
    {   \IfNoValueTF {#2}
      {\mathbb{P}\left[{#1}\right]}
      {\mathbb{P}\left[{#1}\middle\vert{#2}\right]}
    }
}}
\DeclareDocumentCommand\E{ m g }{\ensuremath{
    {   \IfNoValueTF {#2}
      {\mathbb{E}\left[{#1}\right]}
      {\mathbb{E}\left[{#1}\middle\vert{#2}\right]}%
    }
  }}

\title{\bfseries\Large Learning Through Imitation: An Experiment%
\thanks{This work was presented at Princeton University, MiddExLab virtual seminar, Chicago Harris, UCSB, NYU, University of Michigan, Osaka University, Texas A\&M, Stanford University, CREST, University of Hamburg, University of Maryland, Caltech, Essex University, NYU Abu Dhabi, University of Southampton, Kansas University, Shanghai Jiao Tong University, and GAMES 2020. We thank the participants of these seminars and conferences and Carlo Cusumano for their helpful comments.}
}

\author{
\begin{tabular}{cc}
Marina Agranov\thanks{Caltech and NBER. Email: magranov@hss.caltech.edu. Experiments were funded by the grant received from the Ronald and Maxine Linde Institute of Economic and Management Sciences at Caltech.}
&
Gabriel Lopez-Moctezuma\thanks{Caltech. Email: glmoctezuma@caltech.edu.}
\\[1ex]
Philipp Strack\thanks{Yale University. Email: philipp.strack@yale.edu.}
&
Omer Tamuz\thanks{Caltech. Email: tamuz@caltech.edu. Omer Tamuz was supported by a grant from the Simons Foundation (\#419427), a Sloan fellowship, a BSF award (\#2018397), and a National Science Foundation CAREER award (DMS-1944153).}
\end{tabular}
}

\date{\today}

\begin{document}

\maketitle

\begin{abstract}
We compare how well agents aggregate information in two repeated social learning environments. In the first setting agents have access to a public data set. In the second they have access to the same data, and also to the past actions of others. Despite the fact that actions contain no additional payoff-relevant information, and despite potential herd behavior, free riding and information overload issues, observing and imitating the actions of others leads agents to take the optimal action more often in the second setting. We also investigate the effect of group size, as well as a setting in which agents observe private data and others' actions.  
\end{abstract}



\clearpage 
\section{Introduction}
A large literature in economics and  finance has shown that imitation in social learning leads to inefficiencies: agents who imitate do not reveal their private information, giving rise to information cascades and herd behavior. This point has been made both theoretically and experimentally in a variety of settings, including those with rational agents \citep{banerjee1992simple, bikhchandani1992theory, anderson1997information, celen2004distinguishing, harel2020rational} and those that consider behavioral biases and heuristics \citep{enke2019, eyster2015experiment}. 

We study a social learning setting in which all information is \emph{public}, and ask whether agents perform better or worse when given the opportunity to imitate the actions of others. Although these actions do not contain additional payoff-relevant information beyond the public raw data, and despite their potential to create herd behavior, free riding, and information overload, we find that agents perform better when allowed to observe others' actions. 

In our experiment, participants play the following game: An urn is chosen to be either majority green or majority red, each with probability one half. In the former case the urn contains six green balls and four red balls, whereas in the latter case the numbers are reversed. A signal is a random draw (with replacement) of a ball from the urn. A game consists of twenty periods. In each period, each member of a group of eight participants observes some information and then has to guess the majority color of the urn.  Participants are rewarded for guessing correctly.

In each period, each participant observes a signal. Furthermore, depending on the treatment, participants observe some additional information. In the {\sc signals} treatment, participants observe others' signals. In the {\sc all} treatment, they likewise observe all the others' signals, and also all the others' past actions. Thus, in both treatments all signals are public, but actions are public only in the latter.

The comparison between the {\sc signals} and {\sc all} treatments sheds light on the usefulness of observing other people's decisions, in addition to the raw public data. Rational agents are expected to perform identically in these two treatments, since both offer agents exactly the same information about the state; in both cases all agents can see all the signals available to society. Some behavioral heuristics suggest that agents should do better when observing only signals, as observing the actions of others can---through information overload and correlation neglect---lead to wrong conclusions, potentially through herd behavior and groupthink. Information overload has been known to interfere with information processing and may lead to sub optimal decisions \citep{caplin2011, scheibehenne2010}. Since the {\sc all} treatment features twice as many pieces of information as the {\sc signals} treatment, these  phenomena suggest that performance in the {\sc all} treatment should be worse. Furthermore, the literature has documented a tendency of people to neglect correlation among dependent pieces of information \citep{enke2019}. In the {\sc signals} treatment, correlation neglect plays no role because the signals of other group members are independent conditional on the state of nature. However, in the {\sc all} treatment, correlation neglect might be detrimental since others' actions are not independent and regarding them as independent will hinder learning. Finally, pure imitation by all agents is clearly inefficient, as it leaves no room for the signals to inform the actions.



Nevertheless, we find that agents perform better in the {\sc all} treatment than in the {\sc signals} treatment: they are better off when they are privy to others' actions, in addition to the information that those actions are based on. A possible explanation for this finding is that observing the actions of others allows subjects to benefit from the aggregation of information done by other subjects. We observe that subjects rely on the actions of others more heavily when signals are ``weak'', by which we mean relatively uninformative (i.e., roughly the same number of red and green balls have been drawn) and it is thus harder to determine which color was drawn more. Individual level analysis shows that both high-IQ and low-IQ subjects condition on the actions of others to a similar extent, and do so almost exclusively when signals are weak. Both groups benefit from imitating others.


Having documented the benefits of imitation in a repeated social learning setting with \emph{public} information, we complement our findings by studying a similar setting with \emph{private} information. We focus on the effects of information structures and group size and conduct two additional treatments. In the {\sc no info} treatment each participant observes her own signals only, and does not see others' signals or actions. In the {\sc actions} treatment, subjects again observe  their own signals and not the signals of others, but do observe the past actions of others.

The comparison between the {\sc no info} and {\sc actions} treatments reveals to what extent people are capable of extracting the information contained in their peers' private signals by observing their actions, in a repeated setting. Theory predicts that such inference is difficult and has limited benefits \citep{harel2020rational}, which leads to the natural question of how well people perform in practice in such settings. We find that participants perform significantly better in the {\sc actions} treatment as compared with the {\sc no info} treatment, despite the complexity of the information extraction problem. 

Finally, we compare how changing the group size affects results, by repeating these experiments with groups of only four agents. While in the {\sc all} treatment larger groups perform better, in the {\sc actions} treatment performance is identical across group sizes. This result qualitatively matches a theoretical result due to \cite{harel2020rational}, who show that myopic Bayesian agents, in the long-run, do not perform significantly better in larger groups.

We conclude the analysis by presenting a simple behavioral model that
captures the main features of our experiment. In this model, agents
noisily respond to a weighted sum of two stimuli: signals and others’
actions. Estimating this model on our experimental data reproduces all
of our main findings. A natural direction for future research is a
formal analysis of the model, including proofs of the conjectures we
put forward.

We proceed as follows. Related literature is surveyed in Section~\ref{sec:Literature}. Section~\ref{sec:Design} contains the detailed description of our experiment and Section~\ref{sec:Theory} outlines theoretical predictions for each of the treatments. Section~\ref{sec:Results} reports the experimental results: Section~\ref{sec:RedundantInfo} focuses on the setting with public information ({\sc all} versus {\sc signals} treatments) and Section~\ref{sec:InfoGroupSize}  considers settings with private information (the {\sc actions} treatment) and documents group size effects. Section~\ref{sec:Model} presents the behavioral model and simulation results. Finally, we offer some discussion of implications of our experimental findings in Section~\ref{sec:Discussion}.

\section{Related Literature}\label{sec:Literature}

Our paper contributes to a large literature on social learning. Most of the theoretical literature focuses on sequential settings, in which exogenously ordered agents act once after observing some statistics about their predecessors' actions and their own private signal (see for example \citealp{banerjee1992simple,bikhchandani1992theory, smith2000pathological}, or a recent survey by \citealp{bikhchandani2021information}).  

The experimental literature on sequential social learning is vast, and a complete survey is beyond the scope of this paper. The first experimental paper that documents informational cascades is \cite{anderson1997information}. Since then, a fruitful literature in experimental economics has studied the determinants and the limitations of herding behavior in sequential settings. \cite{HungPlott} report the robustness of informational cascades under different rewards systems. \cite{celen2004distinguishing} elicit subjects' beliefs in order to identify the source of imitation behavior. \cite{KublerWeiszacker} allow subjects to purchase a private signal for a small fee. \cite{CiprianiGuarino} embed this game in a financial market setting. \cite*{GoereeEtAl} look at longer-horizon games. \cite{ZiegelmeyerEtAl} study the fragility of information cascades patterns using groups of differently informed agents. \cite{Weizsacker} conducts a meta-study of sequential social learning experiments and finds that subjects follow their own private information more frequently than is empirically optimal.\footnote{\cite*{ZiegelmeyerEtAl2013} enlarge the data and use a modified methodology compared with \cite{Weizsacker}. They find that subjects tend to over-weigh their private signals but do so to a lesser degree than previously found. \cite{DeFilippisEtAl} identify that such over-weighting occurs only when private signals are in conflict with social information. See also \cite{AngrisaniEtAl} who study social learning in a continuous action space experiment and disentangle different theories that deliver over-weighting of private information. \cite{eyster2014extensive} investigate agents who naively believe that the actions of others solely represent their private signals.} \cite*{DuffyEtAl2019} allow subjects to choose between private and public information prior to guessing the state, while \cite*{DuffyEtAl2020} in addition vary the persistence of the state across periods.

In comparison to this work, the repeated social learning game we study is quite different from the sequential move games surveyed above as, in our case, the same group of agents interacts repeatedly. This setting allows us to study how the information and the actions of others impact the beliefs and actions in an arguably more realistic setting. Accordingly, our paper is related to a new and exciting experimental literature that compares social learning outcomes under different network structures \citep*{MuellerFrankNeri, GrimmMengel, ChandrasekharEtAl,DasarathaHe, choi2012, agranovetalNetworks}. These papers vary the connections between agents in a group, i.e., who can observe whose actions, and ask how well subjects aggregate dispersed private information through repeated observations of neighbors' actions. Contrary to this literature, we fix the network structure to be complete, and vary the type of information our agents observe.\footnote{Moreover, in our game, participants get private signals in each game round, while in the papers above private signals are distributed only once at the very beginning of the game. 
} 

Two recent laboratory games deserve special attention. The first one is \cite*{eyster2015experiment} which compares social learning outcomes in a standard sequential move game and a cleverly constructed four-at-a-time move game. The sequential move game features participants who move one at a time after observing their predecessors' actions. The four-at-a-time game has four participants move in each period after observing predecessors from previous periods. Theoretically, Bayesian inference prescribes subjects to anti-imitate their predecessors in the latter but not the former game. The experimental data shows that subjects rarely anti-imitate in both games, which is at odds with the rational model. The four-at-a-time move game is interesting as it is one of the first environments in which, due to the incorrect processing of social information by participants, social learning is harmful on average, i.e., participants would do much better if they just ignored their predecessors actions and focused on their own private information instead. 

The second paper is \cite{EvdokimovGarfagini} who focus on groups with two players and vary whether both players, only one, or no one observes others' actions in a repeated social learning game. Similarly to our game, each agent gets a conditionally independent private signal in each period and reports her best guess about the state. However, contrary to our main results, the authors find no significant differences in the average quality of guesses across informational treatments. This difference between our papers indicates that group size plays an important role when aggregating social information.

From a theory perspective, our {\sc no info}, {\sc signals} and {\sc all} treatments are trivial to analyze, as they are equivalent to a single agent problem for a Bayesian decision maker. In contrast, the {\sc actions} treatment is difficult to analyze within a rational framework, because higher order beliefs play a significant role. We elaborate on this in Section~\ref{sec:Theory}. \cite{vives1993fast} studies the speed of learning in a similar setting, but with a continuum of agents and continuous signals.  \cite{harel2020rational} study a setting that is identical to ours, under the assumption that agents are Bayesian and myopic. They focus on the long-term rate of learning, and show that even if many more agents are added to the group, the speed of learning remains bounded. \cite{huang2021learning} extend this result to a network setting and to forward-looking agents.

\section{Experimental Design}\label{sec:Design}

Subjects are presented with the following scenario. There are two possible urns: a ``red'' urn and a ``green'' urn. The red urn contains six red and four green balls, and the green urn contains six green and four red balls. The color of the urn represents the state of nature (the superior policy, the best candidate for a job, the tastier croissant, etc.), which subjects are trying to learn by interacting repeatedly throughout the game.  The structure and a sample of instructions are presented in sections \ref{app_payments} and \ref{app_instructions} of the Online Appendix, respectively.

In all experiments, subjects play 10 games. At the start of each game, subjects are randomized into a group of eight subjects and one of the urns is chosen by a toss of a fair coin. This urn then remains unchanged for the duration of the game, and is used by all the participants of the game. Each game consists of 20 rounds. In each round, a subject guesses the color of the urn. We  refer to these guesses as her actions. She then receives an independent draw (with replacement) from the chosen urn. The color of the drawn ball matches the color of the urn with probability 60\%, which is therefore the precision of one's private signal; we purposefully chose an environment with low precision signals because in this environment the signals and actions of others are particularly valuable.    

Depending on the treatment, subjects may observe additional information at the end of each round, which may help them choose their actions for the follow-up rounds. At the end of a session, a random round of a random game is selected uniformly. Subjects are rewarded for guessing correctly  in that round: the correct action earns \$20, while the incorrect one earns \$5.\footnote{Paying for a single round chosen at random (as opposed to paying for all rounds) eliminates hedging incentives, which might cause subjects to switch between guesses. Moreover, the standard “isolation” assumption---that participants treat each game as independent---coupled with myopic optimization assumption---that participants report their best guess in each round of a game---imply that incentive compatibility is preserved \citep{AzrieliEtAl}. In an alternative design, payment is based on the accuracy of last-round guess alone. In principle, such a payment design could induce a wide variety of behavior. For example, in the actions treatment (where agents observe others’ actions but not their signals) agents could in principle spend all periods but the last playing actions equal to their signals, thus revealing their signals to each other. Alternatively, in any treatment, given the cognitive cost of guessing well, agents could just (say) always choose red until the last round, where they could make a good guess.  } 

Our main variation between treatments is the  information available to subjects at the time they choose their actions (in addition to their own signals). We consider four information structures:
\begin{enumerate}
    \item[(1)] The {\sc no info} treatment, in which subjects observe only their private signals in every round of a game.  
    \item[(2)] The {\sc actions} treatment, in which in addition to observing their own private signals, subjects also get to see the actions chosen by their group members in all previous rounds of the game. 
    \item[(3)] The {\sc signals} treatment, in which subjects observe both their own private signals and the private signals of their group members in all previous rounds of the game. 
    \item[(4)] The {\sc all} treatment, in which subjects observe their own private signals, their group members' private signals and the actions chosen by their group members in all previous rounds of the game.
\end{enumerate}

We conduct two additional treatments with the same information structure as the {\sc actions} and the {\sc all} treatments, but with  smaller groups of four members each. We call these treatments {\sc actions4} and {\sc all4}. These treatments allow us to explore how fast small and large groups learn depending on the information available to their members. Note that when subjects take their first action  in round 1, they have no additional information about the state except for the prior. This sequencing of events within a round allows for a clean comparison between information treatments as we describe below, where all the information is delivered at the end of the round before the next-round actions are taken. 

Throughout the game, subjects have access to a table that keeps track of all signals and actions they observed in the past rounds of the game. This table presents information in an intuitive and visual way. This feature of the design ensures that our results are not affected by the  memory subjects may have about the events that transpired during the game (see section \ref{app_instructions} in the Online Appendix for screenshots).

At the end of the experiment, we elicit subjects' strategies with a series of open-ended questions as well as their beliefs about the fraction of correct actions of the other participants in the last round of the game in various treatments. Subjects are paid for the accuracy of their prediction in one randomly selected belief question. Subjects also report their gender and major, and complete a series of control tasks including risk attitudes elicitation using two investment tasks \citep{GneezyPotters}, IQ questions (ICAR, Condon and Revelle, 2014), and overconfidence.\footnote{See sections \ref{app_strat}, \ref{app_beliefs} and \ref{app_controls} in the Online Appendix for details on the strategies, beliefs and other controls, respectively.} In the analysis of experimental data, we classify subjects into \emph{low-IQ} and \emph{high-IQ} based on their answers to six IQ questions. We use their self-described strategies and beliefs to measure how useful low-IQ and high-IQ subjects think observing different types of information is, and whether it relates to their behavior in the experiment. 

\begin{table}[h!]
    \centering 
    \caption{Experimental Design}\label{tab:design}
     
{\footnotesize  \begin{tabular}{l|c|cc|cc|c|c}
    \hline \hline
        & Group  & \multicolumn{4}{c|}{Information} & &   \\ 
       &  size &\multicolumn{2}{c|}{signals}  &\multicolumn{2}{c|}{actions} && \\ \hline
       Treatment &  & own  & others & own & others& \# of sessions &\# of subjects  \\
         \hline 
       {\sc no info} & 1 & yes  & no & yes & no  & 4 sessions & 80 subjects \\ 
        {\sc actions} &8 & yes  & no& yes &  yes & 8 sessions & 136 subjects\\ 
       {\sc signals} & 8 & yes & yes & yes &no  & 3 sessions & 82 subjects \\ 
        {\sc all} & 8 & yes & yes& yes & yes & 8 sessions & 152 subjects \\
        {\sc actions4} &4 & yes  & no& yes &  yes & 4 sessions & 76 subjects\\ 
       {\sc all4} &4 & yes  & yes& yes &  yes & 4 sessions & 80 subjects\\ 
       \hline
        Total & & & & & & 31 sessions & 606 subjects \\
         \hline \hline
    \end{tabular} }
\end{table}

Table \ref{tab:design} summarizes the experimental design and number
of participants. Experimental sessions were conducted at two
locations: University of California in San Diego (UCSD) and the Ohio
State University (OSU).\footnote{Because of the closure of physical labs
  due to COVID-19, our data collection process at UCSD was interrupted
  and we had to finish collecting the data online at OSU using the
  same subject pool of students who normally participate in laboratory
  experiments. {\sc no info}, {\sc actions}, and 4 sessions of {\sc
    all} treatments were conducted in the experimental laboratory at
  UCSD, while the remaining 4 sessions of {\sc all} treatment and {\sc
    signals} treatment were conducted online at OSU. In  section
  \ref{sec:AppOnlineSessions} of the Online Appendix, we compare data
  collected at the physical lab at UCSD and in the online lab at OSU
  for the {\sc all} treatment. We find that aggregate outcomes and
  individual level behavior of subjects in these two types of sessions
  are very similar to each other and statistically indistinguishable. \label{foot:design}} The experiment was programmed and conducted with the oTree software \citep{chen2016otree} and was pre-registered in the AEA RCT registry (AEARCTR-0003315). Overall, 606 subjects participated in 31 sessions, and no subject participated in more than one session. The experiment lasted about 90 minutes. Subjects earned on average \$25.7, including a \$7 participation fee.

\section{Theoretical Preliminaries}\label{sec:Theory}

We model each game played by subjects in the experiment as follows. Denote by $\omega \in \{R,G\}$ the state of nature, which is chosen uniformly at random and stays fixed throughout each game. Denote by $N$ the set of players. At the beginning of each round $t \in \{1,\ldots,20\}$, each player $i$ chooses an action $a_{i,t} \in \{R,G\}$. She then observes a random signal $s_{i,t} \in \{R,G\}$ such that conditioned on the state $\omega$, the probability that the signal equals the state (i.e., $s_{i,t} = \omega$) is 60\%. Conditional on the state, the signals are independent, both between time periods and players. Depending on the treatment, the players observe additional information; we denote by $I_{i,t}$ all that player $i$ has observed before choosing her action $a_{i,t}$ at the beginning of round $t$. We use the terms action and guess interchangeably throughout the text.  

Each player plays some number of games. At the conclusion of player $i$'s session one of these games is chosen uniformly at random, and in that game a round $t$ is chosen uniformly at random. Player $i$ receives a payoff  equal to \$20 if she chose the correct action $a_{i,t}=\omega$, and to \$5 otherwise. We assume that players are Bayesian and that they have a strictly increasing utility for money. Under this assumption,  behavior depends on the information players observe, but \emph{not} on their preferences over risks.\footnote{As there are only two outcomes, the possible payoff distributions are ranked in first order stochastic dominance. Therefore, the player will choose the same strategy under any preference that is monotone in first order stochastic dominance (e.g., cumulative prospect theory, etc.), as long as she forms beliefs using Bayes' rule.}

\paragraph{ The {\sc no info} and {\sc signals} treatments.} In these treatments the information $I_{i,t}$ available to player $i$ in round $t$ consists of a collection of signals that she observed up until this round. Specifically, in the {\sc no info} treatment this collection consists of her own signals only, $I_{i,t} = (s_{i,\tau})_{\tau < t}$, while in the {\sc signals} treatment this collection contains signals of all group members, $I_{i,t} = (s_{j,\tau})_{j \in N, \tau < t}$. A standard calculation shows that the action $a_{i,t}$ is optimal if it is equal to any color that appears in $I_{i,t} $ at least as often as the other. In other words, the optimal action $a_{i,t}$ is equal to the majority color observed in the past signals, and can be both in case of a tie. Note the distinction between the correct action, i.e., the one that matches the state, and the optimal action, which is the one that matches the majority of the observed signals. 

Given that subjects observe eight times more signals in the {\sc signals} treatment than in the {\sc no info} treatment, theory predicts that subjects should be better at guessing the state in the former case in every round of the game (except the first, where they have no information). In particular, given our parameters, in the second round of the game, subjects should guess the state correctly 60\% of the time in the {\sc no info} treatment and 71\% of the time in the {\sc signals} treatment. In the last round of the game, subjects are expected to guess the state 81\% of the time in the {\sc no info} treatment and 99\% of the time in the {\sc signals} treatment. In Table~\ref{tab:simulation} we provide the predicted probability of choosing correctly in each round.

\paragraph{The {\sc all} Treatment.} In the {\sc all} treatment, subject $i$ in round $t$ observes all the signals and actions of her group members up until round  $t$, so that  $I_{i,t} = (s_{j,\tau},a_{j,\tau})_{j \in N, \tau < t}$. Since the past actions $(a_{j,\tau})$ can contain no further information than the signals $(s_{j,\tau})$, in this case too optimal behavior implies that $a_{i,t}$ is equal to any color that appears in $(s_{j,\tau})_{j \in N, \tau < t}$ at least as often as the other. In other words, for Bayesian subjects we expect identical results in the {\sc all} and {\sc signals} treatments, since subjects should ignore others' actions and base their decisions on the observed signals.
 
\begin{table}[t!]
\begin{center}
  \caption{Probabilities of Correct Actions}
  \label{tab:simulation}\medskip
{\scriptsize \begin{tabular}{r|ccccc}
    \toprule 
    $t$  & {\sc no info} &  {\sc actions} & {\sc signals}/{\sc all} & {\sc actions4} &  {\sc all4}\\
    \hline
      1&     0.50&  0.50&   0.50 & 0.50 & 0.50\\
      2&     0.60&  0.60&   0.71 & 0.60 & 0.65 \\
      3&     0.60&  0.73&  0.79 & 0.68  & 0.71 \\
      4&   0.65&  0.77&  0.84 & 0.71 & 0.75 \\
      5&   0.65&  0.78&  0.87 & 0.73 & 0.79\\ 
      6 & 0.68 & 0.80 & 0.90 & 0.75 & 0.81 \\
      8 & 0.71 & 0.83 & 0.93 & 0.77 & 0.86 \\
      10 & 0.73 & 0.84 & 0.96 & 0.79 & 0.89 \\
      12 & 0.75 & 0.86 & 0.97 & 0.81 & 0.91  \\
      14 & 0.77 & 0.87 & 0.98 & 0.83 & 0.93  \\
      16 & 0.79 & 0.88 & 0.99  & 0.84 & 0.94 \\
      18 & 0.80 & 0.89 & $>$0.99 & 0.85 & 0.95 \\
      20 & 0.81 & 0.90 & $>$0.99& 0.86 & 0.96 \\
         \bottomrule
\end{tabular}}
\vspace{2mm}
\end{center}
\par
{ \footnotesize \underline{Notes:} Predicted probability of correct actions for parameters used in the experiment. In the {\sc actions} treatments we assume common knowledge of rationality and that players act myopically.}
\end{table}

\paragraph{The {\sc actions} Treatment.} The situation is more complicated in the {\sc actions} treatment. Here, the information available to subject $i$ in round $t$ consists of her private signals and of the actions of the other subjects in all previous rounds, i.e., $I_{i,t} = (s_{i,\tau}, a_{j,\tau})_{j \in N, \tau < t}$. To the best of our knowledge, this case is analytically intractable.
Nevertheless, a simulation of equilibrium behavior is possible under the assumption that players act myopically, i.e. do not change their action to manipulate the future behavior of others.\footnote{This assumption is commonly made in the literature, see e.g., \cite{parikh1990communication,gale2003bayesian,mossel2014asymptotic, harel2020rational} and seems plausible given the complexity of the environment.} 
Formally, in this simulation we assume that in each period each player maximizes the probability of making a correct choice, i.e.
\begin{equation}\label{eq:myopicdecisions}
    a_{i,t} = \begin{cases} R \text{ if } p_{i,t} > 0.5\\
    G \text{ if } p_{i,t} < 0.5
    \end{cases}    
\end{equation}
where $p_{i,t}$ denotes the probability player $i$ assigns to state $R$ at the beginning of period $t$. When $p_{i,t}=0.5$, the player is indifferent and we assume that she randomizes uniformly over the two actions. Each player computes this probability using Bayes' rule taking into account the actions of others. We assume common knowledge of rationality, so each player knows that the others are making their decisions according to \eqref{eq:myopicdecisions}. We thus assume that each player does not only know the signal generating process, but also the strategies of all other players, which is arguably a strong rationality requirement. In Table~\ref{tab:simulation} we report the probabilities of correct actions under these assumptions. We provide these simulation results as a benchmark for what Bayesian agents can hope to achieve in this setting; our results do not hinge on a comparison of the subjects' behavior to this benchmark.

\section{Results}\label{sec:Results}

We present results from our experiments in the following order. First, we focus on the public data setting, and compare the {\sc signals} and {\sc all} treatments (Section \ref{sec:RedundantInfo}). Second, we study the private data setting (the {\sc actions} treatment) and investigate group size effects (Section \ref{sec:InfoGroupSize}). As we go through the analysis, we summarize the main empirical findings as observations.  

\paragraph{Data Analysis Approach.} Our main analysis considers all
ten games in all sessions.\footnote{We see moderate learning across
  games within a session. Section \ref{app_learn_games} in the Online
  Appendix shows aggregate results subsetting our data for early
  (first 5) and late (last 5) games.} We investigate the effect of the
treatment on the fraction of correct actions and on the consensus
rates using regression analysis. To compare performance across
treatments, we estimate a linear regression of the indicator of a
correct action by a given participant $i$ in a round $t$ of game $g$ on a
round-specific treatment effect.\footnote{Specifically, let $v_{it}^g$ equal one if participant $i$ guessed optimally in round $t$ of game $g$, and zero otherwise. Let $\text{treat}_i \in \{\text{SIGNALS,ALL}\}$ denote the treatment assigned to participant $i$. We estimate the regression
\[
v_{it}^g = \alpha + \beta\,\text{treat}_i + \gamma_t + \delta_t\,\text{treat}_i + \varepsilon_{it}^g,
\]
where $\gamma_t$ are round fixed effects and $\delta_t$ capture
round-specific treatment effects, normalized so that $\delta_1 =
0$. Under this normalization, $\beta$ measures the treatment effect in
round 1, and $\delta_t$ captures the differential treatment effect in
round $t$ relative to round 1. Identification comes from
within-participant variation across rounds and across games.} To compare consensus rates across treatments, we regress the
relative size of the majority in a given round $t$ on a round-specific treatment
effect.\footnote{Specifically, let $c_t^g$ denote the share of
  participants in the majority subgroup in round $t$ of game $g$, and
  let $\text{treat}^g \in \{\text{SIGNALS,ALL}\}$ denote the
  treatment assigned to participants in game $g$. We estimate the regression
\[
c_t^g = \alpha + \beta\,\text{treat}^g + \gamma_t + \delta_t\,\text{treat}^g + \varepsilon_t^g,
\]
where $\gamma_t$ are round fixed effects and $\delta_t$ capture
round-specific treatment effects, normalized so that $\delta_1 =
0$. Under this normalization, $\beta$ measures the treatment effect in
round 1, and $\delta_t$ captures the differential treatment effect in
round $t$ relative to round 1. Identification comes from within-game
variation across rounds.}\footnote{All standard errors are clustered
at the session level to account for the inter-dependencies of
observations that come from re-matching subjects within a session. Appendix \ref{app_aggregate} reports the average treatment effects across rounds as well as treatment effects in early (rounds 2 to 10) and  late rounds (rounds 11 to 20).}

Throughout our analysis, we classify the collection of agents' signals
$(s_{i,\tau})_{\tau <t}$ according to their strength, as given by the difference in the number
of signals of one color relative to the other. We define five categories of signal strength: \emph{Very Strong Green}, \emph{Strong Green}, \emph{Weak}, \emph{Strong Red}, and \emph{Very Strong Red}. These correspond, respectively, to the difference between the number of red and green signals being less than $-26$, in $[-26,-15]$, in $[-14,14]$, in $[15,26]$, and above $26$. The boundaries between these categories correspond to the 
10\textsuperscript{th}, 25\textsuperscript{th}, 75\textsuperscript{th}
and 90\textsuperscript{th} percentile of the distribution at the end
of the game. For the small group treatments we adjust the intervals to
match the same percentiles.\footnote{In section
  \ref{app_learn_others} of the Online Appendix we show that the
  qualitative results reported in this section are robust to using
  different cutoffs of these categories.}

In some discussions, we only distinguish between ``weak'' and ``strong'' signals by pooling together the \emph{Strong} and \emph{Very Strong} categories for each color. An alternative approach of defining the signal strength based on the \textit{fraction of signals} instead of the difference in the number of signals of each color yields similar results. We discuss these two approaches in detail in Section \ref{sec:RedundantInfo}.

\subsection{The Public Data Setting and the Effect of Redundant Information}\label{sec:RedundantInfo}
\begin{figure}[h!]
\begin{center}
\begin{tabular}{cc}
\scriptsize{Panel (a): Correct Actions, by round} & \scriptsize{Panel (b): Consensus, by round}\\ 
\includegraphics[scale=0.35]{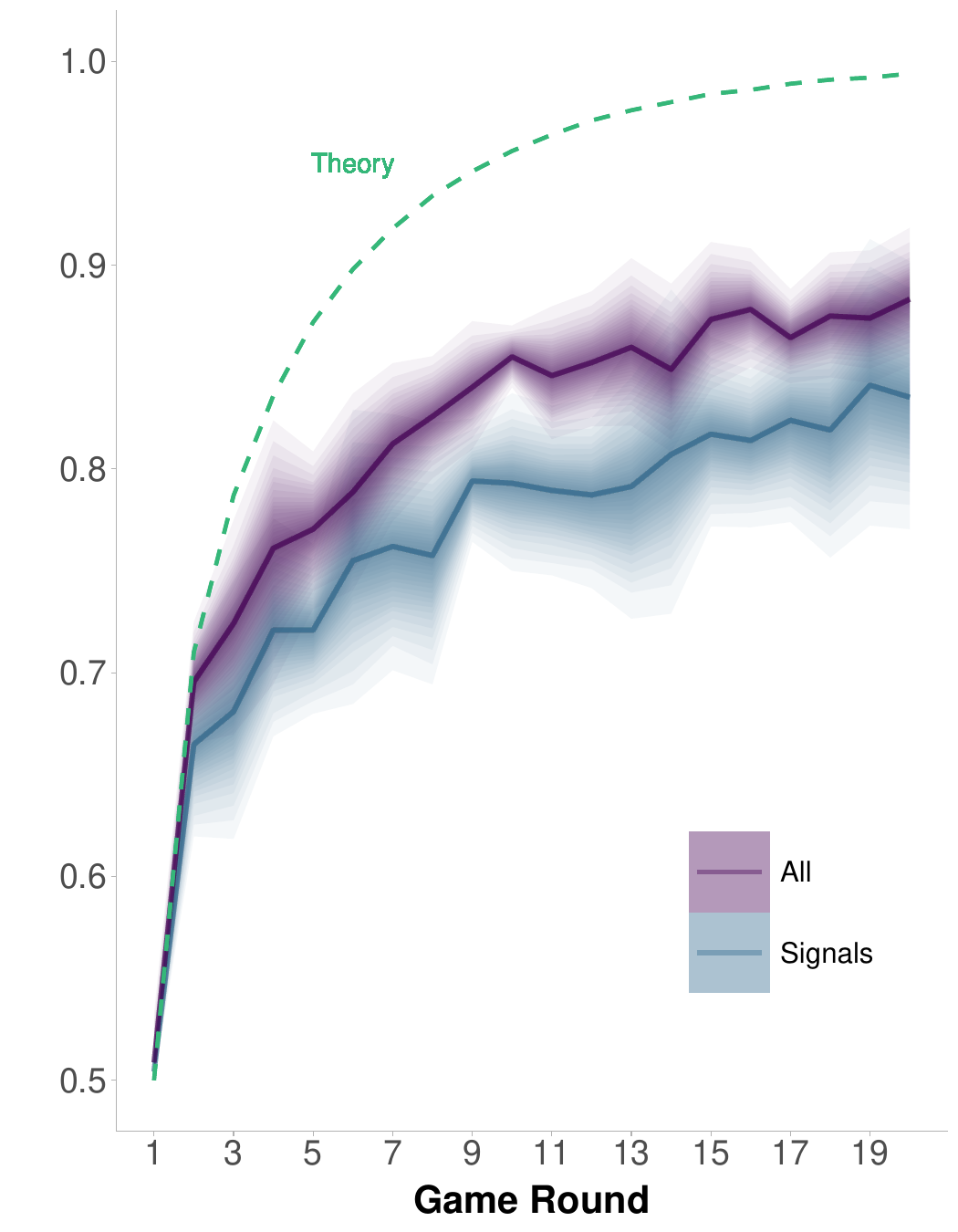}  & \includegraphics[scale=0.35]{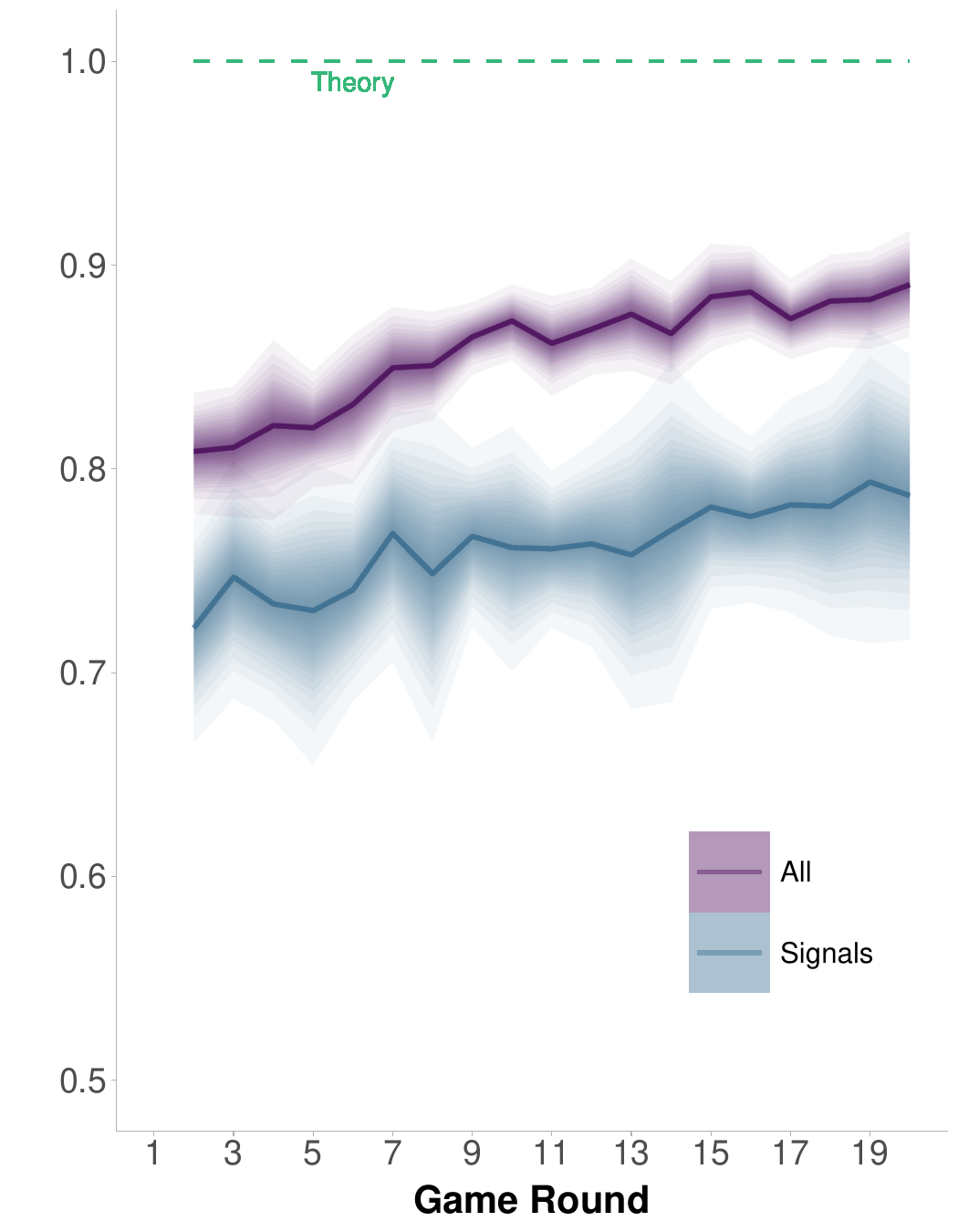}  
\end{tabular}
\caption{Aggregate statistics in the {\sc all} and {\sc signals}  treatments}\label{fig:Correct} 
\end{center}
{\footnotesize \underline{Notes:} Panel (a) presents the average frequency of correct actions in each round, averaged across games. Panel (b) depicts the evolution of consensus in each round, i.e., the relative size of the majority, averaged across games. For panel (b) we exclude cases with an equal number of green and red signals. Shaded regions represent $95\%$ confidence intervals from 50\% (darkest) to 95\% (faintest) probability levels. Confidence intervals are constructed with a variance-covariance matrix clustered by session.}
\end{figure}

In this section we compare the {\sc all} and {\sc signals} treatments.
Panel (a) in Figure \ref{fig:Correct} presents our main outcome of interest: how often subjects guess the state correctly in each round of the game, in comparison to the Bayesian benchmark.

In both treatments, the first action that subjects make is a coin toss. This is expected since this action is made before any signals are observed, so the probability of guessing correctly is one half. As the game progresses and more information arrives, we observe an upward trend in the likelihood of choosing the correct action.

The main finding of this paper is that subjects perform better in the
{\sc all} treatment, as compared to the {\sc signals} treatment. This
happens despite the aforementioned fact that the only difference
between the two treatments is the observability of other players'
actions, which are informationally redundant. In Online Appendix \ref{app:hetero}, we show that the performance
advantage of the {\sc all} treatment over the {\sc signals} treatment
remains robust after conditioning on the distribution of prediction
quality across participants. In particular, participants throughout
the distribution perform better in the {\sc all} treatment than in the
{\sc signals} treatment.

Across rounds, the
percentage of correct actions is on average $5\%$ larger in the {\sc
  all} treatment, as compared to the {\sc signals}
treatment.\footnote{See panel (a) of Figure
  \ref{fig_app_corr_diff_signals_all} in Appendix A and columns (1)
  and (2) in Table \ref{table_app_signals_all} for point estimates and
  standard errors.}  In the next section, we explore in depth why this
is the case. We note that both treatments significantly underperform
relative to optimal behavior, which means that in both treatments at
least some subjects deviate from the Bayesian
benchmark.\footnote{Online Appendix \ref{app:random} analyzes the extent to which
these deviations are driven by participants who behave as if random,
ignoring all relevant information in their signals.}

Panel (b) of Figure \ref{fig:Correct} measures the polarization of opinions, presenting  the evolution of consensus rates. The consensus rate is defined as the relative size of the majority subgroup in a given round, based on members' actions. This rate varies between one half and one, where a consensus rate of one half indicates a maximally polarized group with half of the members choosing each of the two actions, while a consensus rate of one indicates the case where all members choose the same action in a particular round. 

Overall, the consensus rates are increasing in both treatments as the
game progresses, showing that group members' opinions tend to align
the longer they interact with each other. This is to be expected,
since actions become more correlated with the state, and hence with
each other, as subjects gather more information. Nevertheless,
consensus rates are significantly less than one, which would be the
theoretical prediction for Bayesian agents.\footnote{Note that Panel
  (b) excludes cases in which the number of red and green signals is
  the same, for which theory does not provide a unique prediction on
  the size of the majority.} 
  
  More interestingly, the consensus rates
are significantly higher in the {\sc all} treatment than in the {\sc
  signals} treatment: subjects agree more when they see others'
actions and raw data, than when they only see raw data. On average,
consensus rates are $10 \%$ higher in the {\sc all} treatment than in
the {\sc signals} treatment.\footnote{Additional evidence is presented
  in panel (b) of Figure \ref{fig_app_corr_diff_signals_all} and
  average point estimates and standard errors are in columns (3) and
  (4) in Table \ref{table_app_signals_all} in Appendix A.} This is
consistent with subjects being influenced by the actions of others. In Online Appendix \ref{app:rmse}, we complement these results by
quantifying the improvement in performance from using the redundant
information in others’ actions beyond the use of signals. Specifically,
we compute the root mean squared error (RMSE) between the Bayesian
benchmark probabilities of correct actions (as reported in Table
\ref{tab:simulation}) and model-predicted probabilities under two
specifications: one in which participants’ behavior depends only on
signals, and another in which it depends on both signals and others’
actions. We find that incorporating others’ actions significantly
reduces RMSE, with a magnitude equivalent to approximately one-third
of the performance improvement attributable to signals.

Finally, focusing on cases with strict majorities, we note that the probability that the majority is correct is extremely high and exceeds 90\% on average in both the {\sc all} and {\sc signals} treatments. Furthermore, it is 4\% higher in the {\sc all} treatment, as compared to the {\sc signals} treatment.\footnote{The evidence is presented in Figure \ref{fig_app_corr_maj_signals_all} and columns (5) and (6) of Table \ref{table_app_signals_all} in Appendix A.}  

\medskip

\textit{\textbf{Observation 1:} People learn faster and develop more unified opinions when they observe each other's signals and actions, compared to observing signals only.}

\medskip

\paragraph{What Drives These Aggregate Results?} 
 
To what extent do subjects' actions follow the information contained in the signals they observe? We start with the simplest statistic, i.e., the second round behavior, which, when performed optimally, entails reporting the color of the majority of the eight signals observed in the first round of the game. In both treatments, a large fraction of around 84\% of our subjects choose optimally in the second round.

\begin{figure}[h!]
    \begin{center}
    \begin{tabular}{cc}
    \scriptsize{Panel (a): Observed Fraction of Red Actions} & \scriptsize{Panel (b): Estimated Probability of Optimal Action}\\
    \includegraphics[scale=0.35]{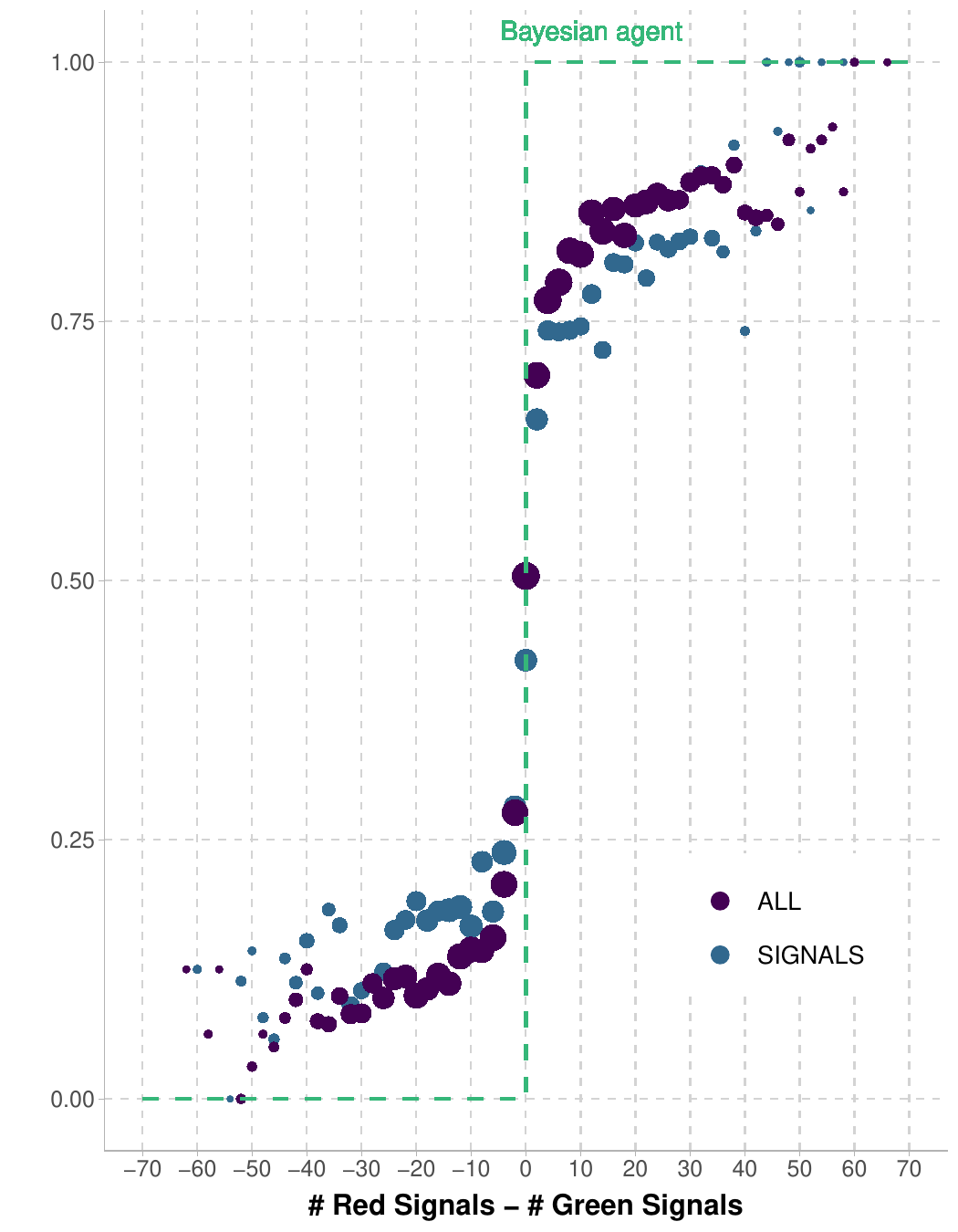} &   \includegraphics[scale=0.35]{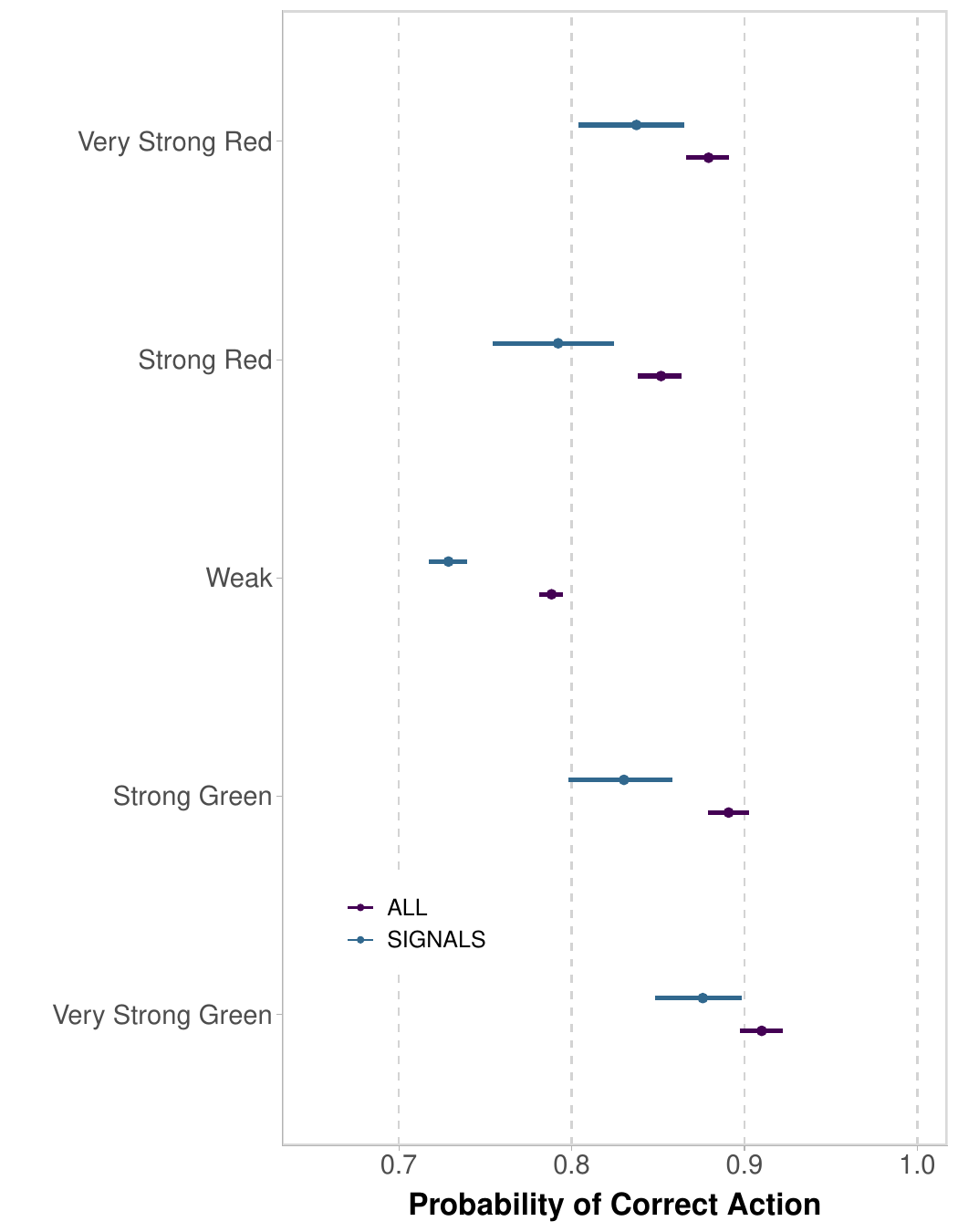} 
    \end{tabular}
    \vspace{2mm}\caption{Learning from signals in the {\sc signals} and {\sc all} treatments}\label{fig:learning_signals_all}
    \end{center}
    
    {\footnotesize \underline{Notes:} Panel (a) depicts the fraction
      of red actions as a function of the difference between the
      number of red and green signals. The size of the dot corresponds
      to the number of observations in each bin. Panel (b) shows the
      estimated probability of an optimal action (i.e., reporting the
      color of the majority of signals) as a function of the treatment
      and signal strength. We estimate a Bayesian logistic regression
      of  the probability of taking the red action on treatment  that
      varies by signal strength and controlling for session random effects. We present the estimated posterior median and $95\%$ confidence intervals. The categories in Panel (b) for signal strength are as defined in  Section~\ref{sec:Results}.} 
\end{figure} 

As more and more signals are observed, a Bayesian subject would keep a tally of the signals and report the majority color. 
Panel (a) in Figure \ref{fig:learning_signals_all} shows the frequency
with which subjects chose the color red as aggregated over the
observed difference between the total number of red and green
signals. Panel (b) shows the estimated probability of an optimal
action, i.e., reporting the color of the majority of signals, as a
function of signal strength.\footnote{To recover the probabilities in panel (b) of
  Figure \ref{fig:learning_signals_all}, we estimate a Bayesian
  logistic regression pooling the data from the {\sc signals} and {\sc
    all} treatments, where the probability of a red action,
  $Pr(a_{it}=R)$, is modeled as a function of treatment and signal
  strength, accounting for session-level random effects. Specifically,
  \begin{equation}
    \label{eq:prob_bet_strength}
 \text{logit}\big(Pr(a_{it}=R)\big)
=
\beta_{\text{strength}_j}\,\text{treat}_i
+
\gamma_{\text{session}(i)},   
  \end{equation}
where $\text{treat}_i \in \{\text{SIGNALS,ALL}\}$ denotes the treatment assigned to participant $i$, $\beta_{\text{strength}_j} \sim \mathcal{N}(0,\sigma_{\text{strength}})$ is a random slope that varies across signal-strength categories $j$, and $\gamma_{\text{session}} \sim \mathcal{N}(0,\sigma_{\gamma})$ is a session-level random intercept. As a robustness check, Table \ref{app_table_red_bet} in the Online Appendix reports similar estimates using linear probability models with standard errors clustered at the session level. We obtain similar results when additionally accounting for game, round, and participant effects.}

This figure suggests several insights. First, while panel (a) would
look like a step function for a Bayesian (the green line),  in the
data we see a gradual increase in the frequency of choosing red. Thus,
subjects respond to signals, and do so in the correct direction, but
not perfectly. Second, the mistakes are much more frequent when
signals are weak, i.e., when the majority signal color has occurred
only slightly more than the other color. Because the expected utility
is a function of the difference between the number of red and green
signals, this is consistent with a Luce model of behavior, in which
choice probabilities are related to the difference in expected payoffs
\citep{luce1958probabilistic}. It is also consistent with standard
models of perception where close-by states are harder to
distinguish. Third, subjects systematically perform better in the {\sc
  all} treatment across different signal strengths. In particular, when
signals are weak subjects are $6\%$ more likely to choose correctly in
the {\sc all} treatment compared to the {\sc signals} treatment
($73\%$ for the  {\sc signals} versus $79\%$ for the {\sc all}
treatment).

\begin{figure}[h!]
    \begin{center}
    \includegraphics[scale=0.5]{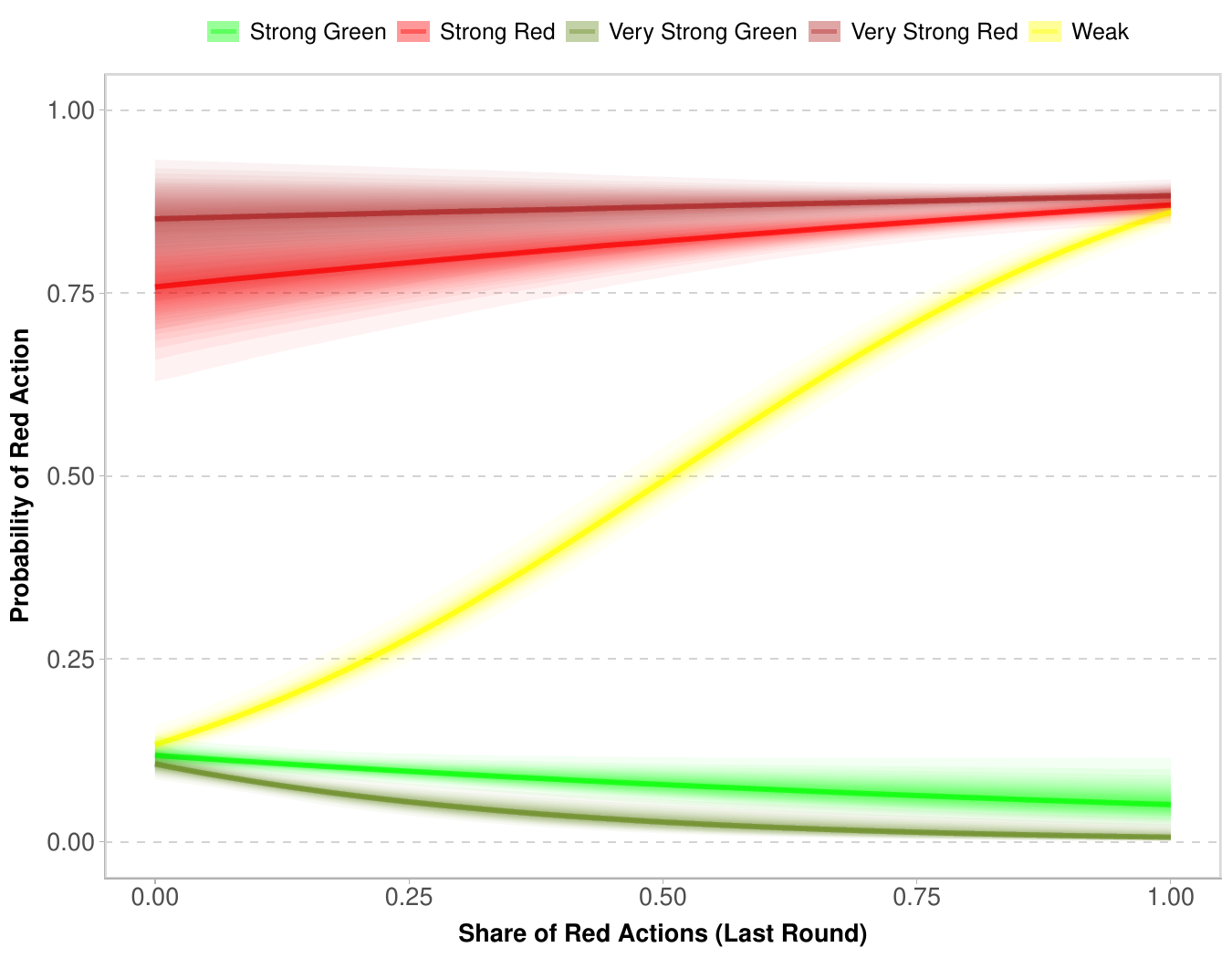}
    \caption{Learning from others' actions in the {\sc all} treatment}
    \label{fig:learning_from_guesses} 
    \end{center}
    
    {\footnotesize \underline{Notes:} This figure depicts the probability of choosing red as a function of the share of red actions of other group members. The estimates are obtained from a Bayesian logistic regression of subjects' actions on the share of others' actions in the previous round conditional on signal strength and session random effects. Shaded regions represent $95$\% confidence intervals from 50\% (darkest) to 95\% (faintest) probability levels.}
\end{figure}

Figure~\ref{fig:learning_from_guesses} provides a closer look at
behavior in the {\sc all} treatment. To quantify the informational
value of observing others’ actions, beyond the raw data, we estimate a
Bayesian logistic regression in which a subject’s action is regressed
on the share of others’ actions in the previous round. The effect of
others’ actions is allowed to vary across signal-strength categories,
as previously defined.\footnote{Specifically, we estimate
\begin{equation}
  \label{eq:prob_bet_strength_actions}
\text{logit}\big(Pr(a_{it}=R)\big)
=
\beta_{\text{strength}_j}\,\text{share}_{a_{k \neq i,t-1}}
+
\gamma_{\text{session}(i)},
\end{equation}
where $\text{share}_{a_{k \neq i,t-1}}$ is the share of others' actions in
round $t-1$. $\beta_{\text{strength}_j} \sim
\mathcal{N}(0,\sigma_{\text{strength}})$ is a random slope for signal-strength category $j$, and $\gamma_{\text{session}}
\sim \mathcal{N}(0,\sigma_{\gamma})$ is a session-level random
intercept.} We find that when signals are strong or very strong, a subject's actions are barely influenced by those of others: regardless of what others do, subjects tend to follow the majority of signals. This can be seen by mostly flat lines in the bottom and top of Figure \ref{fig:learning_from_guesses}. In contrast, when the number of signals of each color is comparable, subjects rely heavily on the actions of others, as is indicated by the steep middle line, which shows a strong responsiveness of subjects to others' actions. 

An alternative approach, in which one classifies signal strength by
the  \textit{fraction} of signals instead of the difference in the
\textit{number} of signals of each color yields similar results. This
alternative approach is inspired by prior experiments, which found
that some people rely on sample proportion over sample size in
probability judgement tasks \citep[see, e.g.,][]{Benjamin}.\footnote{One may ask which of the two models provides a better fit to
the data, i.e., whether participants behave as if they condition their
actions on the fraction or on the number of signals of each color. Our
data slightly favors the former specification, although the difference
is minimal: the average fraction of correctly predicted guesses is
\(72\%\) versus \(71\%\). Figure
\ref{app_fig_pred_diff_share} in the Online Appendix replicates Figure
\ref{fig:learning_from_guesses} using the fraction of signals to
measure signal strength and reports the in-sample fit, in terms of the
fraction of correctly predicted guesses, for the two alternative
approaches.}
\medskip

\textit{\textbf{Observation 2:} In both the {\sc signals} and {\sc all} treatments, subjects use signals to choose their actions, but do so less well than in the Bayesian optimum. In particular, subjects tend to make more mistakes when the tally of signals is close, i.e., when signals are not very informative. In the {\sc all} treatment, subjects partially correct for these mistakes by relying on others' actions when signals are weak, which accelerates learning as compared to the {\sc signals} treatment.}

\paragraph{Individual Level Analysis.} Our results so far show that agents condition on the social signal, i.e., on the actions of others, which improves performance in the {\sc all} treatment, as compared to the {\sc signals} treatment. 
In this section we calculate a number of statistics describing a
subject's behavior in our game and explore its joint distribution in
the population. We analyze subjects' behavior by their level of IQ,
given our auxiliary IQ measure.\footnote{Online Appendix \ref{app:balance} examines covariate balance
across treatments. Using a paired matched sample to address any
imbalances, we replicate the main result that participants perform
better in the {\sc all} treatment than in the {\sc signals}
treatment. Relative to the full sample, performance in the matched
sample is higher in both treatments, with outcomes approaching the
Bayesian benchmark, in particular in the {\sc all} treatment.}

We classify subjects into low-IQ and high-IQ based on their answers to six IQ questions: three matrix reasoning questions, which are similar to Raven's Progressive Matrices, and three 3-D rotation questions \citep{ChapmanEtAl}.\footnote{The questions are drawn from the International Cognitive Ability Resource, a public domain intelligence measure (ICAR; \cite{CondonRevelle}). In each of the first three questions, participants are asked to determine which of the options completes a graphic pattern. In the remaining questions, participants are asked to identify which of the presented drawings of a cube were compatible with another drawing of a cube. In general, these questions have been shown to capture a variation of fluid intelligence in the general population and is becoming one of the popular measures of IQ on par with CRT tests and general knowledge tests.} The low-IQ subjects are those who answered at most half of the IQ questions correctly and the high-IQ subjects are the remaining group.\footnote{The average number of correct answers is 3.7 in the {\sc all} treatment and 3.4 in the {\sc signals} treatment. This classification delivers roughly similar proportions of low-IQ subjects across treatments: 47\% in the {\sc all} treatment and 40\% in the {\sc signals} treatment. } Despite its obvious coarseness, our IQ measure correlates with subjects' performance in the main game: high-IQ subjects are approximately $9\%$ more likely  to guess the state correctly than the low-IQ ones in both the {\sc all} and the {\sc signals} treatments ($p<0.01$).\footnote{Tables \ref{app:table_corr_iq} and \ref{app:table_corr_naive} in the Online Appendix show the estimates and clustered standard errors of a linear probability model of correct actions on the 6-question IQ measure and the low-IQ/high-IQ indicator, respectively. The relationship between accuracy and IQ is robust to controlling for game and round effects as well as for other subject characteristics.} 

\begin{figure}[h!] 
    \begin{center}   
      \begin{tabular}{cc}
        \scriptsize{Panel (a): Response to Signals ({\sc signals} Treatment)} & \scriptsize{Response to Signals ({\sc all} Treatment)}\\ 
  \includegraphics[scale=0.3]{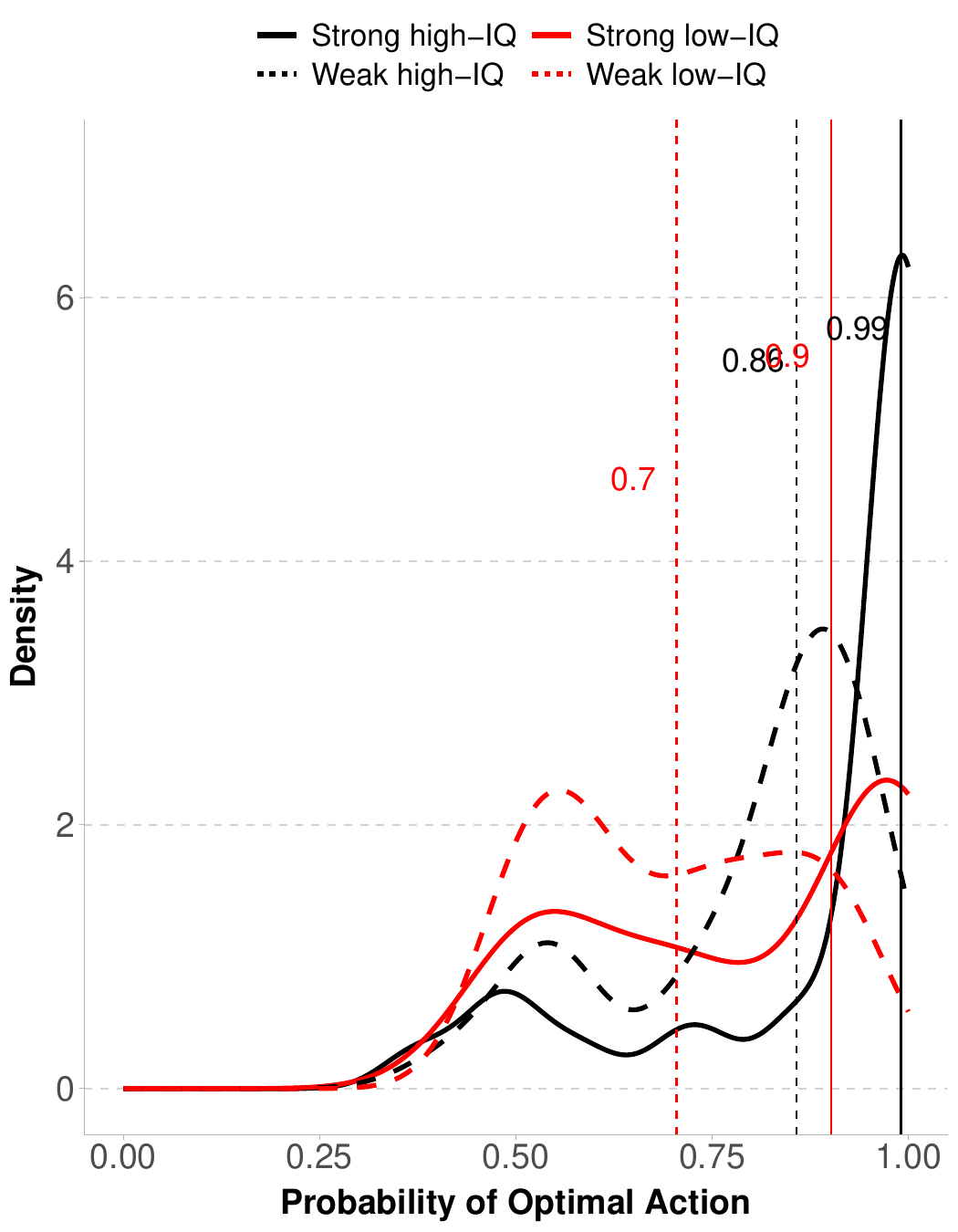} &
                                                                 \includegraphics[scale=0.3]{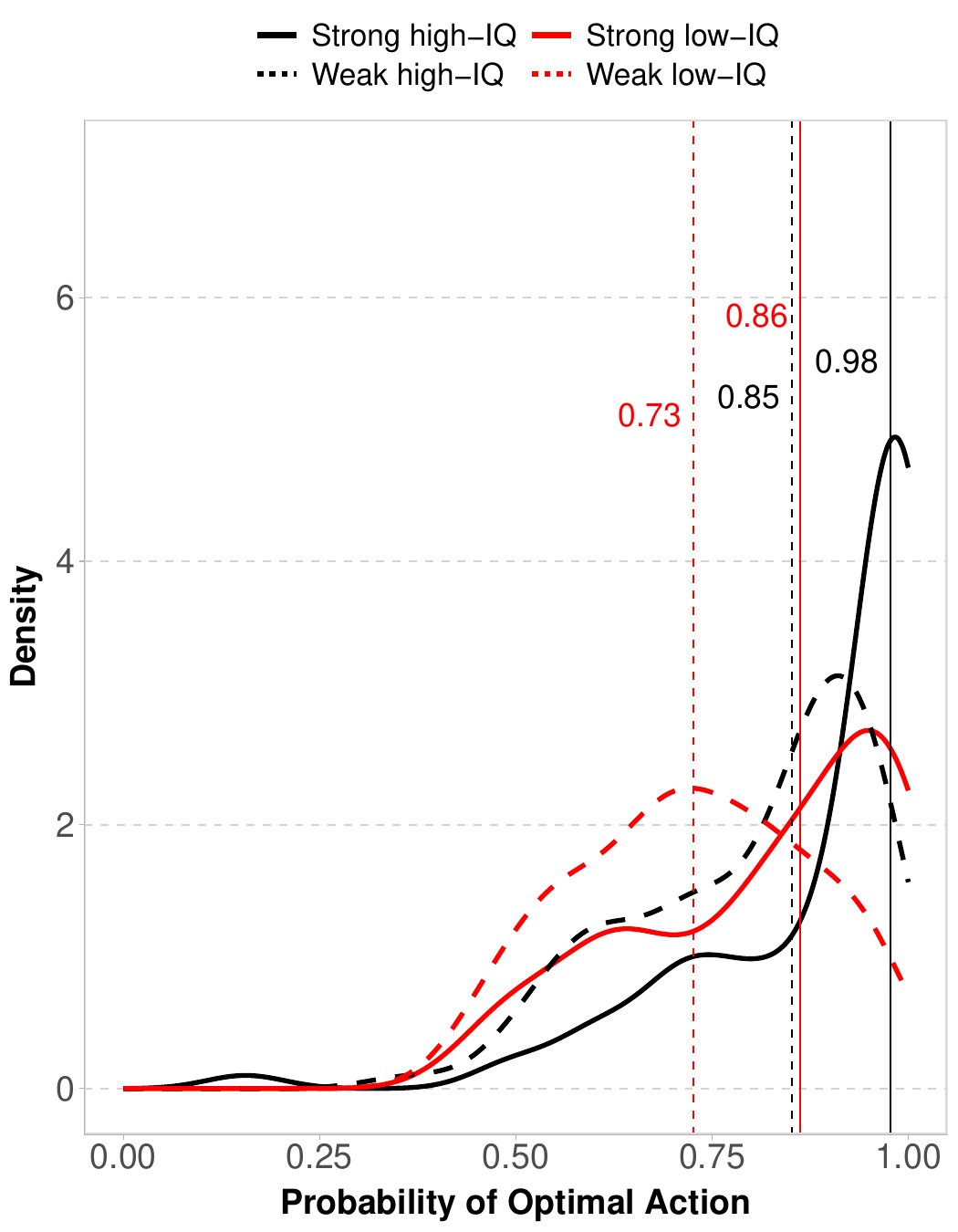}\\
    \multicolumn{2}{c}{\scriptsize{Panel (c): Response to Others'
        Actions ({\sc all} Treatment)}}\\       
    \multicolumn{2}{c}{\includegraphics[scale=0.3]{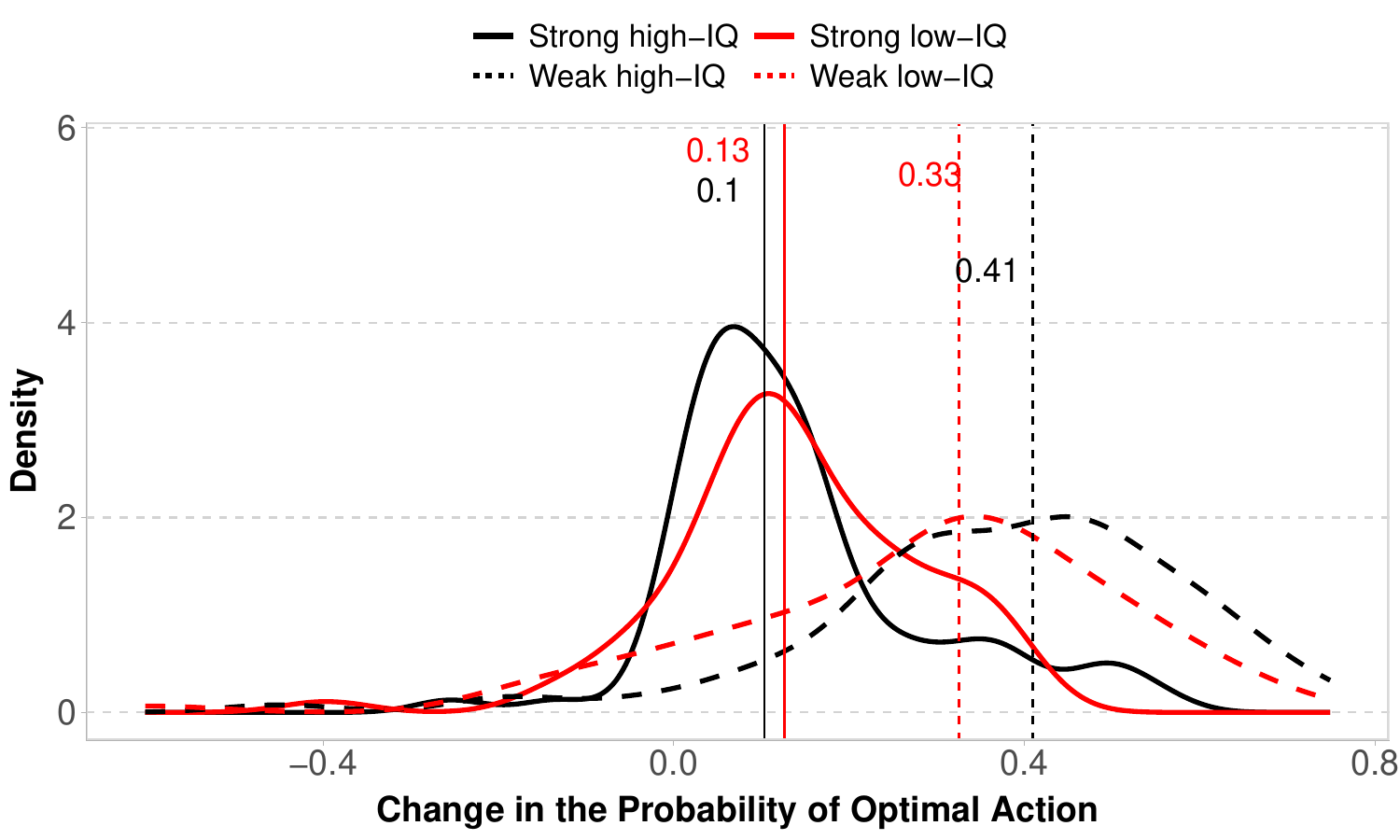}}
        \end{tabular}
        \end{center}
    \caption{Responsiveness to signals and actions, individual level data}
    \label{fig:RespSignals}
    \vspace{2mm}
{\footnotesize \underline{Notes:} 
Kernel distributions of participants' responsiveness to weak and
strong signals are presented in the top part of the graph (Panel (a)
for {\sc signals} and Panel (b) for {\sc all}
treatments). Responsiveness to signals is calculated based on
regressions (\ref{eq:prob_bet_strength_ind}) and
(\ref{eq:prob_bet_strength_actions_ind}) for the {\sc signals} and {\sc all}
treatments, respectively. Responsiveness is given by the probability
that a participant's action is optimal. For the {\sc all} treatment,
we also set the actions of others group members in the previous round
to be uninformative, i.e., split equally between green and red. Kernel
distributions of participants' responsiveness to others' actions are
presented in panel (c) for the {\sc all} treatment. Responsiveness to others' actions is measured by the change in the probability of choosing the action of the majority of signals when all versus none of the other group members choose the majority of signals in the last round. The vertical lines and the numbers next to them are median responsiveness for each group.}
\end{figure}

Next, we investigate whether there are systematic differences in how
low-IQ and high-IQ subjects process the information available to
them. We start with the {\sc signals} treatment and calculate two
statistics for each subject: (i) \textit{responsiveness to weak
  signals}: the probability that they choose optimally  (i.e., with
the majority of signals, as a Bayesian would) when signals are weak
and (ii) \textit{responsiveness to strong signals}: the probability
that they choose optimally when signals are strong. This probability
is computed via a Bayesian logistic regression of the subject's action
on our measure of signal strength. To compute responsiveness at the
individual level, we exploit the variation across rounds and games for
a given subject, which allows us to recover measures of responsiveness
at the individual level via individual-specific random
effects.\footnote{Specifically, we estimate and individual-level
  specification of equation (\ref{eq:prob_bet_strength}) for the {\sc
    signals} treatment:
  \begin{equation}
    \label{eq:prob_bet_strength_ind}
\text{logit}\big(Pr(a_{it}=R)\big)
=
\alpha_{i} +\beta_{i,\text{strength}_{j}}
+
\gamma_{\text{session}(i)},
 \end{equation}
where $\alpha_{i}$ is an individual-specific random intercept,
$\beta_{i,\text{strength}_j}$ is an individual-specific random slope that varies
across signal-strength categories $j$, with $\begin{pmatrix}
\alpha_{i} \\
\beta_{i,\text{strength}_j}
\end{pmatrix}
\sim \mathcal{N}(0, \Sigma)$ and $\gamma_{\text{session}}
\sim \mathcal{N}(0,\sigma_{\gamma})$ is a session-level random
intercept. For a fully Bayesian participant $i$ we expect to see 100\%
responsiveness for both weak and strong signals.} The top part of
Figure \ref{fig:RespSignals} presents the kernel distributions of an
individual  responsiveness to signals in the {\sc signals} and {\sc
  all} treatments. The responsiveness is calculated as the posterior
median of the probability of an optimal action by the participant. For
the {\sc all} treatment we interact it with the social signal measured
by the actions of other group members in the previous round as in
equation (\ref{eq:prob_bet_strength_actions}).\footnote{Specifically, the logit regression takes the form
\begin{equation}
  \label{eq:prob_bet_strength_actions_ind}
\text{logit}\big(Pr(a_{it}=R)\big)
=\alpha_i + \beta_{i,\text{strength}_j}\,\text{share}_{a_{k \neq i,t-1}}
+
\gamma_{\text{session}(i)},
\end{equation}
where $\text{share}_{a_{k \neq i,t-1}}$ is the share of others' actions in
round $t-1$. $\beta_{i,\text{strength}_j}$ is an individual-specific random slope that varies
across signal-strength categories $j$, with $\begin{pmatrix}
\alpha_{i} \\
\beta_{i,\text{strength}_j}
\end{pmatrix}
\sim \mathcal{N}(0, \Sigma)$ and $\gamma_{\text{session}}
\sim \mathcal{N}(0,\sigma_{\gamma})$ is a session-level random
intercept.} To isolate the responsiveness to raw signals, we set the number of others who take each action to be equally split between red and green.

Two patterns are apparent from Figure~\ref{fig:RespSignals}. First, in both treatments, the high-IQ participants are on average much more responsive to raw signals than the low-IQ ones, both when signals are weak and when they are strong. 
In fact, high-IQ subjects are very close to the Bayesian benchmark when it comes to strong signals, but some of them make a significant number of mistakes with weak signals. These mistakes are, however, less frequent than the mistakes of the low-IQ subjects, which explains why low-IQ subjects perform worse than the high-IQ ones in the {\sc signals} treatment. 

Second, there is substantial heterogeneity in individual responsiveness to raw signals among participants in both treatments. For instance, about a quarter of high-IQ subjects in the {\sc all} treatment have estimated responsiveness of above 90\% for both weak and strong signals, compared to only 7\% of low-IQ subjects with similar estimates. This means that the social signal, measured by the average action of all group members in the {\sc all} treatment, is informative since it contains a sizable fraction of correct actions.

Who uses this social signal,  and does it affect one's performance?
Panel (c) of Figure~\ref{fig:RespSignals} sheds light on this question. For both low-IQ and high-IQ participants, we compute individual responsiveness to social signals as the difference in the  probability of choosing optimally when all others agree with the majority of signals minus this probability when all others disagree with the majority of signals.\footnote{For example, if a participant chooses the correct action with
probability 0.69 when all observed actions from the previous round
align with the majority of signals, and with probability 0.42 when
they disagree, then the implied responsiveness to social signals is
27 percentage points.} As shown in the figure, both groups of participants rarely condition
on others’ actions when signals are strong, but do so to a substantial
degree when signals are weak. On average, high-IQ participants rely on
others’ actions to a greater extent than low-IQ participants
(\(p=0.004\)).

Importantly, because the social signal is informative, conditioning on
it improves performance for all participants, not only low-IQ
individuals. Among low-IQ participants, those whose responsiveness to
others’ actions under weak signals is above the group median guess the
state correctly 86\% of the time, compared to 70\% for those below the
median (\(p<0.001\)). The same pattern holds for high-IQ participants,
with success rates of 91\% versus 81\% for those above and below the
median, respectively (\(p=0.001\)).

In other words, a substantial fraction of participants effectively
aggregate information, making the social signal informative and worth
conditioning on.

\medskip

\textit{\textbf{Observation 3:} The high-IQ subjects process the raw signals and perform well, regardless of whether social information is available. The low-IQ subjects do not process well the information contained in the signals, and hence perform badly when social information is not available. When social information is available, it is used by  subjects in both groups and boosts the performance of those who use it.}
 
\bigskip

\subsection{The Private Data Setting and the Effect of Group Size}\label{sec:InfoGroupSize}

We now turn our attention to the  {\sc actions} treatment, in which people observe each other's actions but not others' signals. We ask whether they can parse useful information from these choices. This task is complex since actions of group members are inherently correlated and making such inferences requires forming beliefs about how others act. Even under heroic assumptions about common knowledge of rationality among group members, theory is only able to characterize long run outcomes such as the speed of learning, but not how individuals should act in the short term  \citep{harel2020rational}. Our experimental data is particularly useful in these circumstances as it provides the first step at documenting how people actually behave in this complex situation. 

\begin{figure}[h!]
\begin{center}
\caption{Frequency of correct actions, by information structure and group size}\label{fig:CorrectGuesses} 
\begin{tabular}{cc}
\scriptsize{Panel (a): Information Structures} & \scriptsize{Panel (b): Group Size}\\
\includegraphics[scale=0.32]{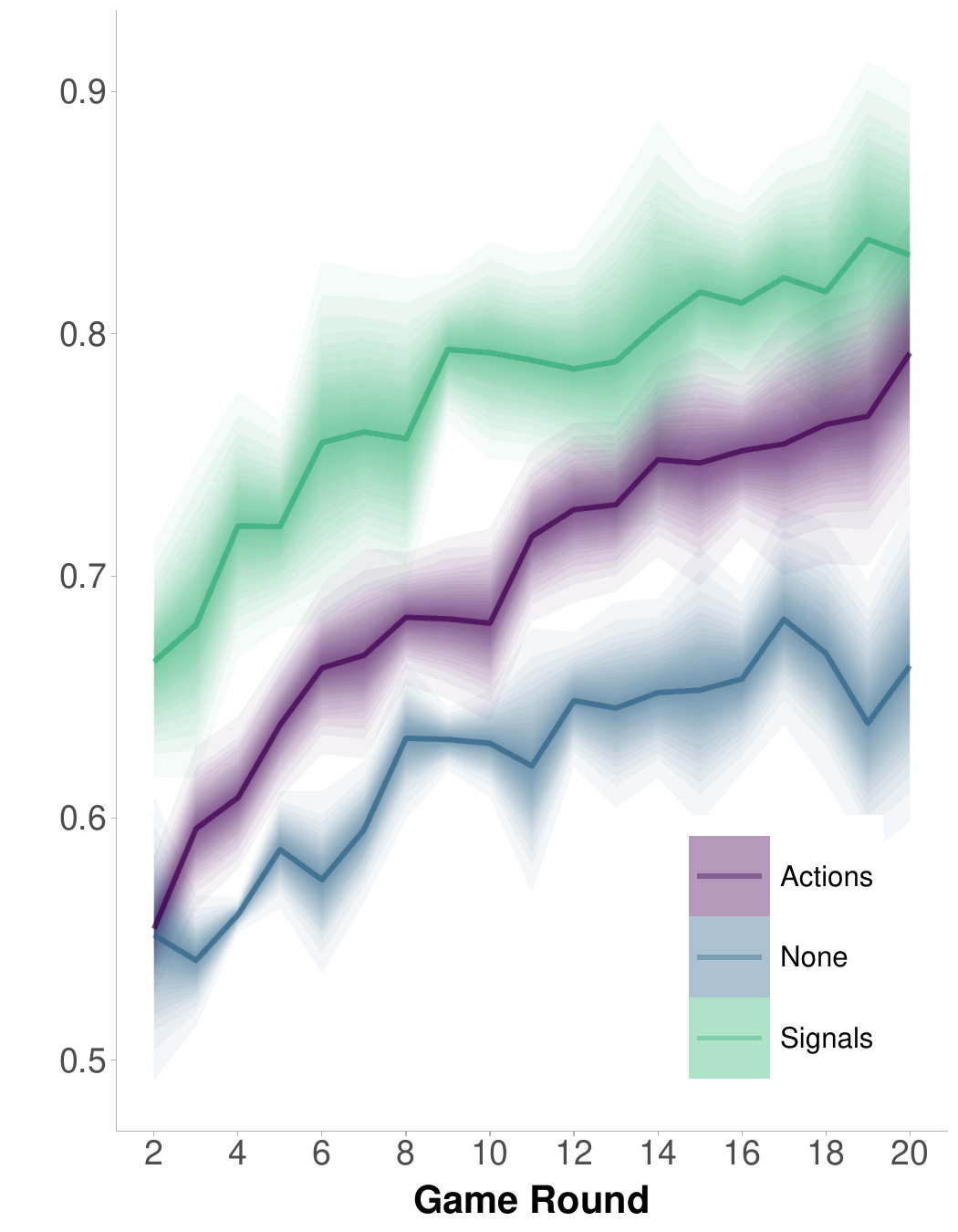} &
\includegraphics[scale=0.32]{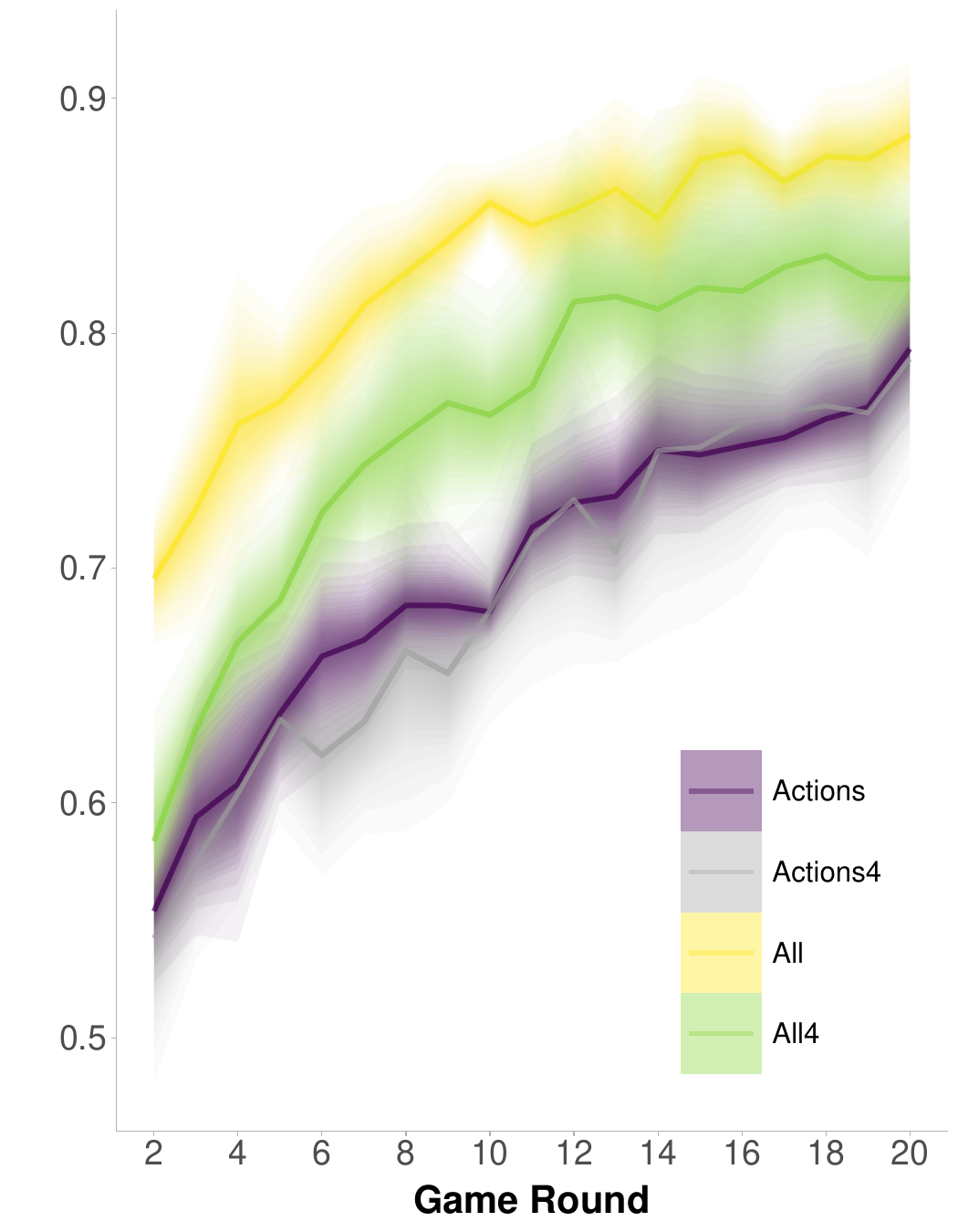} 
\end{tabular}
\end{center}
{\footnotesize \underline{Notes:} Both panels present the average frequency of correct actions in each treatment per each round, averaged across games. Shaded regions represent confidence intervals from 50\% (darkest) to 95\% (faintest) probability levels. Confidence intervals are constructed with a variance-covariance matrix clustered by session.}
\end{figure}

Panel (a) in Figure \ref{fig:CorrectGuesses} compares the performance in the {\sc actions} treatment with two benchmarks: the {\sc no info} and the {\sc signals} treatments. The former benchmark is the lower bound of learning rates, which is what one expects to happen in the {\sc actions} treatment if subjects ignore entirely each other's actions and base their decisions only on the sequence of private signals they receive. The latter benchmark is the upper bound, in which the signals of all members are public; this is what would happen if subjects could perfectly infer the signals of others from their actions. 

Panel (a) shows that, despite the complexity of the inference problem in the {\sc actions} treatment, people are able to extract useful information from others' actions and learn faster than they would without this information, as the comparison to the {\sc no info} treatment shows. For instance, notice that subjects chose the correct action with a higher probability in period 13 of the {\sc actions} treatment than in period 20 of the {\sc no info} treatment. Thus, the benefit of observing others' actions eventually exceeds the benefit of the additional 7 signals which are potentially revealed by the agents' second period actions. The same holds more generally if we compare period $t$ of the {\sc actions} treatment with period $t+7$ of the {\sc no info} treatment. However, if people observe each other's raw signals instead of actions then they learn even faster, as the comparison between {\sc actions} and {\sc signals} shows.\footnote{Table \ref{app_table_corr_none_actions_signals} and panel (a) of Figure \ref{app_fig_correctGuesses_diff} in Appendix A confirm that the difference in performance of {\sc actions} vs {\sc no info} treatment is significantly higher in later rounds ($9.6 \%$) compared with early ones ($5.3 \%$) with $p<0.1$. Table \ref{app_table_corr_none_actions_signals} in Appendix A compares performance of {\sc no info} vs {\sc actions} treatments as well as {\sc actions} vs {\sc signals} treatments separately for rounds 2 - 20 and rounds 11 - 20. In all pairwise comparisons we obtain significant difference between all treatments.} Both of these patterns are qualitatively consistent with theoretical predictions.

Panel (b) of Figure \ref{fig:CorrectGuesses} explores the effect of group size on learning rates in the two information structures: one with private signals, the {\sc actions} treatments, and another with public signals, the {\sc all} treatments. Interestingly, larger groups learn faster only when both signals and the actions of others are public ({\sc all4} versus {\sc all} treatments), but not when group members observe each other's actions only ({\sc actions4} versus {\sc actions} treatments).\footnote{Table \ref{app_table_corr_size} and panel (b) of Figure \ref{app_fig_correctGuesses_diff} in Appendix A confirm these results. Specifically, the probability of a correct action is $10 \%$ higher in the {\sc all} compared with {\sc all4} ($p<0.01$), whereas this difference is not statistically different from zero when comparing {\sc actions4} vs {\sc actions}.} The lack of a group size effect on learning in the {\sc actions} treatment is reminiscent of the theoretical result described in \cite{harel2020rational}, according to which the speed of learning does not change with the group size. This is, however, quite a loose interpretation of the theoretical result, given that the theory refers to the long-run effects at the limit as time tends to infinity. With regards to the level of consensus across group sizes, we find smaller groups to be more aligned than larger groups, especially in early rounds of the game.\footnote{See Figure \ref{app_fig_cons_size} and Table \ref{app_table_cons_size} in the Online Appendix for details.}

\medskip

\textit{\textbf{Observation 4:} Larger groups learn faster than small groups when people have access to both others' actions and others' signals. However, when people can only rely on others' actions, the group size does not affect the speed of learning.}

\section{Behavioral model}\label{sec:Model}

In this section we propose a simple behavioral alternative to the Bayesian model, and show that it reproduces the main empirical patterns observed in the experiment.

The information potentially available to agent $i$ at time $t$ includes her own previous-round signals $(s_{i,\tau})_{\tau < t}$, the signals of others $(s_{j,\tau})_{j \neq i, \tau < t}$ and the actions of others $(a_{j,\tau})_{j \neq i, \tau < t}$. The first is available in all treatments, the second is available in the {\sc signals} and {\sc all} treatments, and the last is available in the {\sc actions} and {\sc all} treatments.

Our data shows that in the {\sc all} treatment, subjects rely both on
the signals, as well as the actions of others. Accordingly, we assume
that agent $i$ relies on two quantities when making a decision in
round $t$: $A_{it}$, which depends on others' actions, and $S_{it}$,
that depends on the signals. Given $S_{it}$ and $A_{it}$, we assume
that agents choose the action $R_{it}$ with probability proportional
to $\exp(\beta S_{it} + \gamma A_{it})$ for some $\beta,\gamma \geq 0$. That is, agents use a logistic function to noisily respond to these stimuli. 

We now describe how $A_{it}$ and $S_{it}$ are calculated. When
deciding on an action at time $t$, agent $i$ chooses a random agent
$j$, and records that agent's action $A_{it} = a_{j,t-1}$ from the previous round; in this section, we identify the red actions and red signals with $+1$, and the green ones with $-1$, so that $a \in \{-1,+1\}$. Agent $i$ also calculates a normalized sum of signals
\begin{align*}
    S_{it} = \frac{\sum_{k,\tau < t}s_{k,\tau}}{(\sum_{k,\tau < t} 1)^{\psi}},
\end{align*}
where $\psi \geq 0$ is a third parameter of the model.

Since $s_{k,\tau} \in \{-1,+1\}$, $S_{it}$ is equal to the sum of the
signals (that is, the difference between the number of red and green
signals) normalized by a function of the number of signals. We assume
the general form $(\sum_{k,\tau \leq t} 1)^{\psi}$, with $\psi \geq
0$, which includes linear normalization ($\psi = 1$), square-root
normalization ($\psi = \frac{1}{2}$), and no normalization ($\psi = 0$) as special cases.

From a Bayesian perspective, the sum of the signals is a sufficient
statistic, regardless of the total number of signals; this would correspond to $\psi=0$. However,
empirical evidence shows that as the number of signals increases, a
given net signal difference becomes less persuasive: two red signals
alone are perceived as stronger evidence than one hundred and two red
signals accompanied by one hundred green signals
\citep{griffin1992weighing}. We observe the same pattern in our data.

Allowing the exponent $\psi$ to be estimated lets the data determine
how strongly the signal sum should be discounted as the number of
signals increases. A common approach in the literature \citep{Benjamin} is to use the sample proportion, which in our framework corresponds to the restriction $\psi=1$. Our experimental data allow us to relax this restriction and estimate the normalization parameter directly, thereby assessing which scaling of the signal history best fits observed behavior.


This model captures agents who (potentially) pay attention to the two
sources of information, even though the actions are redundant, and use
a noisy heuristic rather than optimizing subject to some cognitive or
attention constraint. This is not an unreasonable heuristic, since
imitating others is a potentially low-cognitive-cost alternative to
counting signals. As we show, this three-parameter model fits our data
remarkably well across all treatments. 

\subsection{Simulation of the Behavioral Model}

We first simulate the {\sc all} treatment for parameter values
\(\beta = 1/2\), \(\gamma = 1\) and \(\psi = \frac{1}{2}\). These values are chosen to
illustrate a potential mechanism underlying the patterns observed in
the data. We then use the experimental data to estimate the values of
these parameters that best fit observed betting behavior across
informational treatments.

In the {\sc signals} treatment, actions are not observable, and thus we
set \(\gamma = 0\). The same applies to the {\sc no info} treatment, where
the signal statistic \(S\) is constructed solely from an agent’s own
signals, as others’ signals are not available. An analogous restriction
applies in the {\sc actions} treatment.

\begin{figure}[h!]
\begin{center}
\caption{Simulation results: Evolution of the correct actions (frequency)}\label{fig:sim} 
\begin{tabular}{ccc}
\scriptsize{Panel (a): {\sc all} and {\sc signals}} &
                                                      \scriptsize{Panel (b): {\sc actions}, {\sc no info} and {\sc signals}} & \scriptsize{Panel (c): {\sc all} and {\sc actions}, by group size}\\
\includegraphics[scale=0.45]{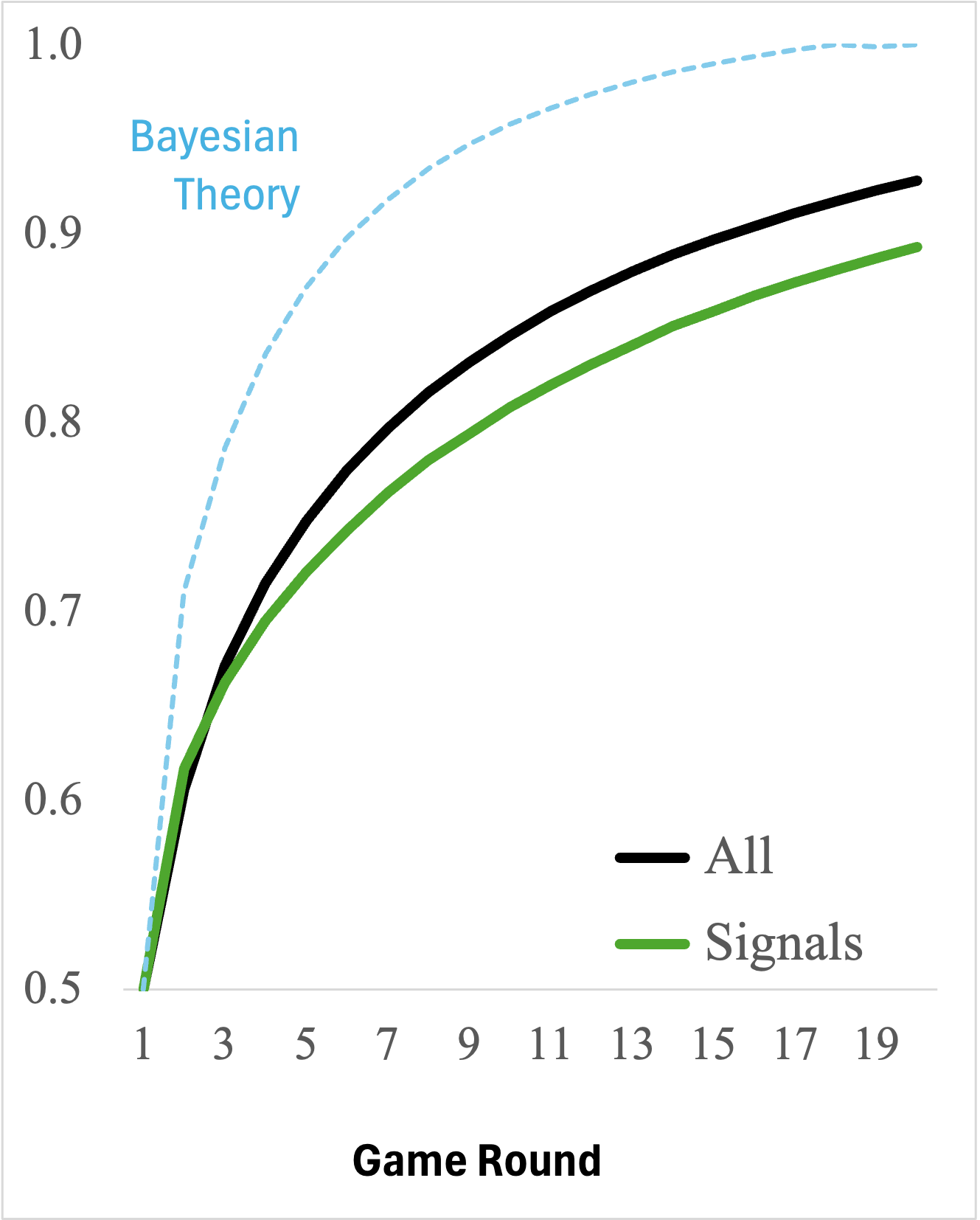} &
\includegraphics[scale=0.45]
{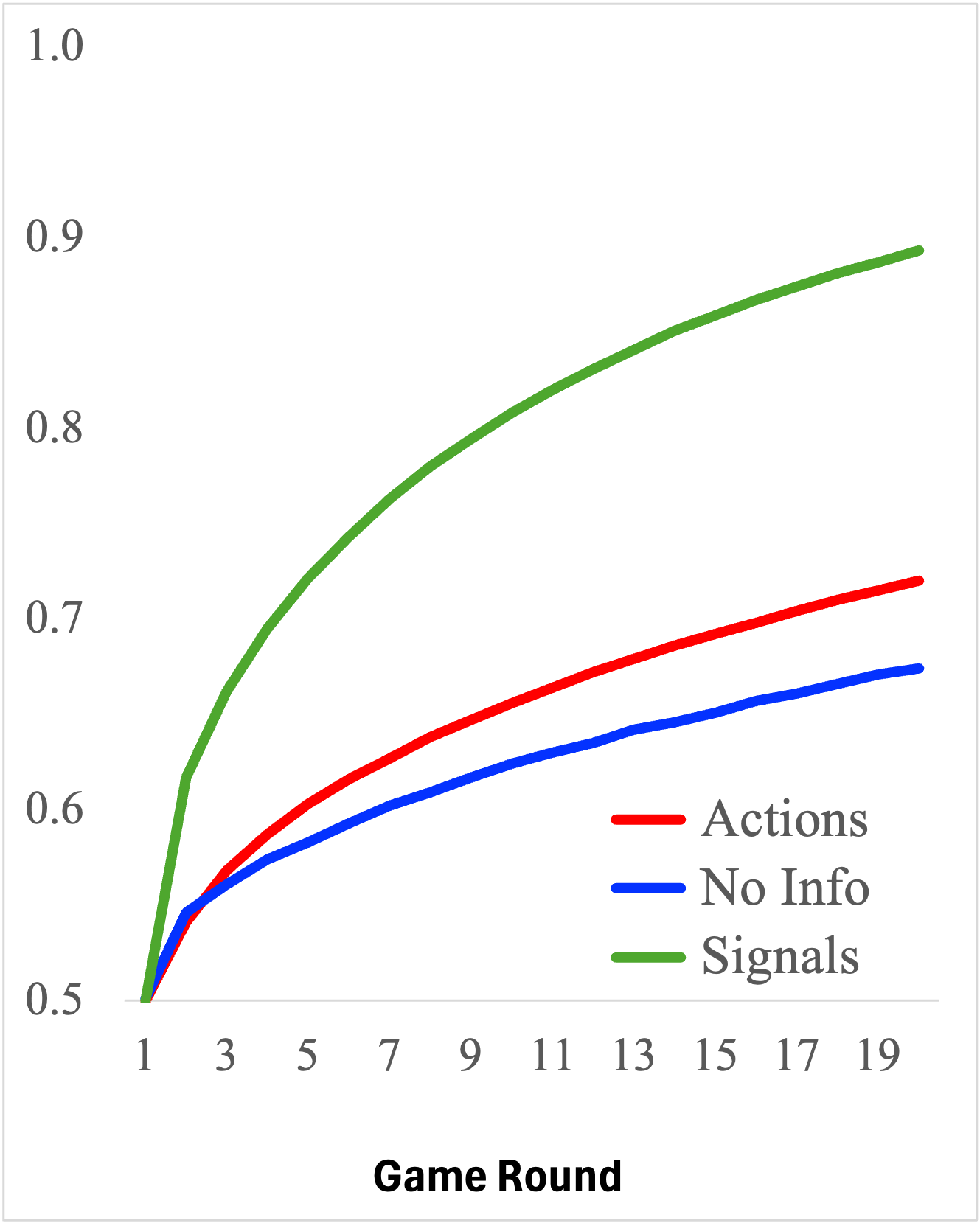} &
\includegraphics[scale=0.45]
{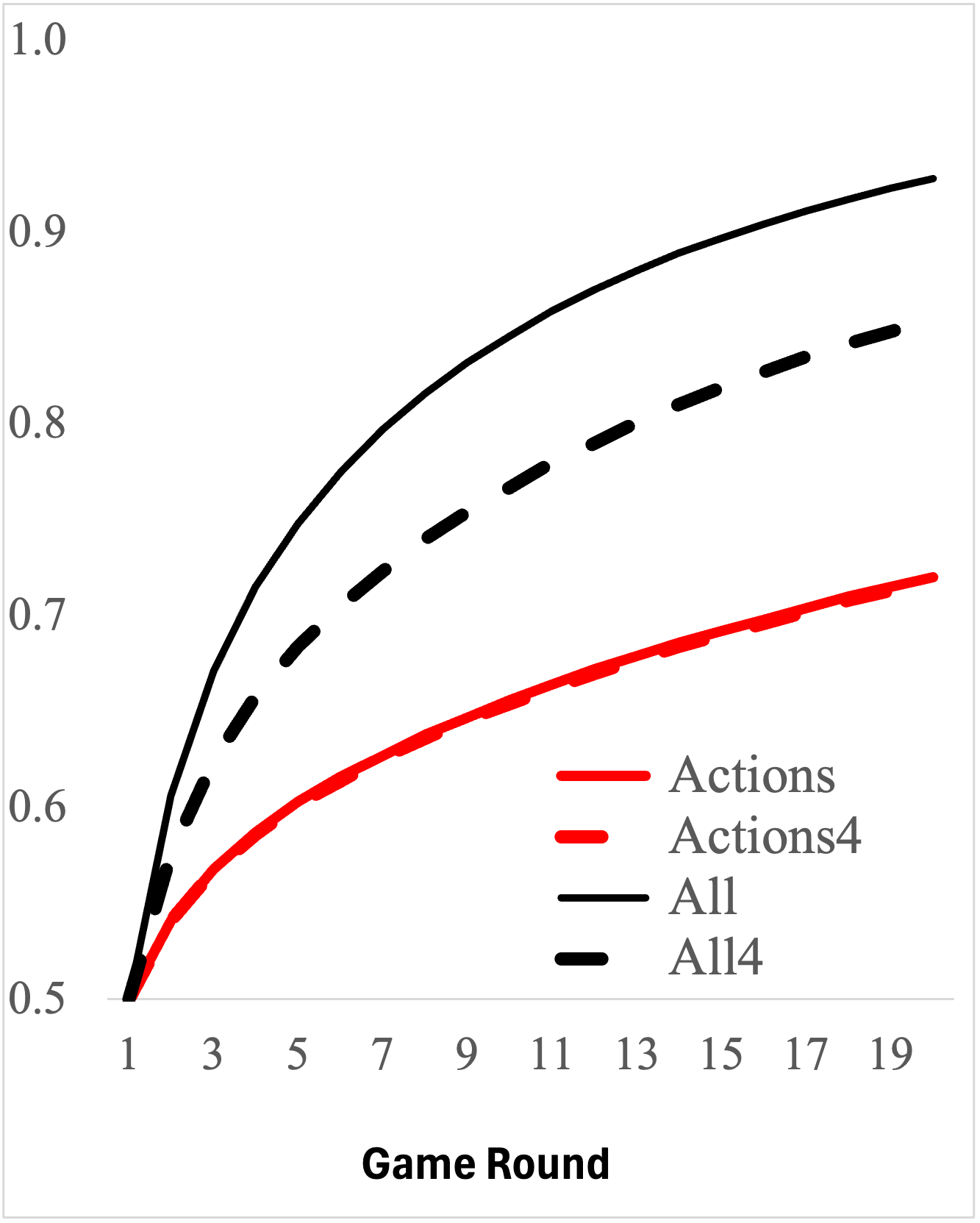} 

\end{tabular}
\end{center}
\end{figure}

Panel (a) in Figure~\ref{fig:sim} shows simulated predictions for the
probability of a correct guess in each round of the game under the
{\sc all} and {\sc signals} treatments. These closely match the
empirical patterns in panel (a) of Figure~\ref{fig:Correct}, showing
that participants are more likely to make correct choices in the {\sc
all} treatment, where they observe others’ actions in addition to
signals.

To build intuition for this result, note that the probability of a
mistake depends on both the sign and the magnitude of
\(\beta S_{it}+ \gamma A_{it}\). Assigning a strictly positive weight
\(\gamma > 0\) to \(A_{it}\) can reduce the overall probability of a mistake
because actions are correlated with signals. As a result, including
\(A_{it}\) often moves \(\beta S_{it} + \gamma A_{it}\) in the correct direction
(i.e., the sign of \(S_{it}\)), even though in some cases it may shift it in
the wrong direction.

Panel (b) shows simulation results for the {\sc actions}, {\sc no info},
and {\sc signals} treatments, which closely match the patterns in
panel (a) of Figure~\ref{fig:CorrectGuesses}. These results indicate
that participants are able to extract information from others’ actions
to some extent, although less effectively than from observing signals
directly.

Panel (c) reports simulated predictions for the effect of group size
on the probability of a correct choice. In the {\sc all} treatment,
larger groups are more likely to choose correctly, whereas in the {\sc
actions} treatment, groups of size 4 and 8 exhibit very similar
success probabilities. This pattern replicates the empirical findings
in panel (b) of Figure~\ref{fig:CorrectGuesses}.

\begin{figure}[h!]
  \centering
      \caption{Simulation results: Learning from others' actions in the {\sc all} treatment}\label{fig:sim-five-line} 
    \includegraphics[width=0.45\linewidth]{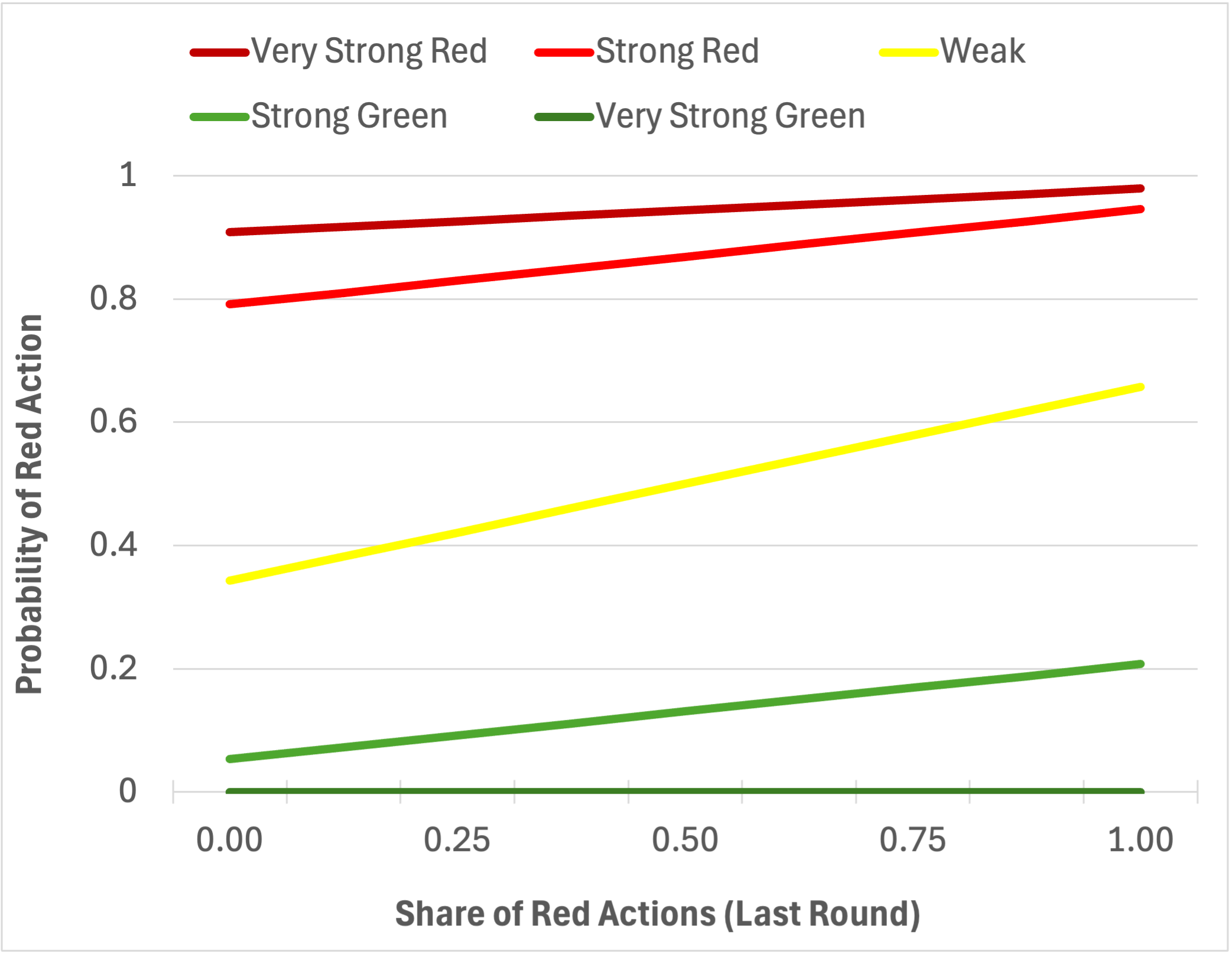}
\end{figure}

Next, we show that the model not only matches the aggregate results,
but also generates these predictions for the ``correct'' reasons. In
Figure~\ref{fig:sim-five-line}, we reproduce the empirical patterns in
Figure~\ref{fig:learning_from_guesses}, showing that participants are
more responsive to others’ actions when signals are weak. This is
intuitive, as the term \(\exp(\beta S_{it} + \gamma A_{it})\) implies that when
\(S_{it}\) is close to zero, variation in behavior is primarily driven by
\(A_{it}\), which captures the influence of others’ actions.

\subsection{Estimation of the Behavioral Model}

Leveraging our experimental data, we estimate the parameters
$\beta$, $\gamma$, and $\psi$ that govern our behavioral model for each
of the four information structures: {\sc no info}, {\sc actions},
{\sc signals}, and {\sc all}. We describe the data likelihood for the
{\sc all} treatment, where participants observe both the history of
signals and the past actions of others.

Let red signals and actions in the data be coded as $+1$ and green signals and
actions as $-1$. For participant $i$ in round $t$, define
the net signal imbalance and the number of observed signals as
$M_{it} \equiv \sum_{k \in N,\,\tau<t} s_{k\tau}$, $n_{it}\equiv \sum_{k \in N,\,\tau<t} 1$,
and let
\[
S_{it}
=
\frac{M_{it}}{(n_{it})^\psi}
\]
denote the signal index. In the model, participant $i$ also receives a
social signal $A_{it} \in \{-1,+1\}$, obtained by sampling one of the
other participants' actions from the previous round. Conditional on a
realization of $A_{it}$, the probability that participant $i$ chooses
red is
\[
\mathrm{P}_{\beta,\gamma,\psi}\left(a_{it} = R \mid S_{it}, A_{it}\right)
=
\Lambda\!\left(\beta S_{it}^g + \gamma A_{it}\right),
\]
where $\Lambda(x)=\mathrm{logit}^{-1}(x)$.

The realized social signal $A_{it}$ is latent from the econometrician's
perspective: the data reveal the set of previous-round actions from
which it is drawn, but not which action the participant samples in the
model. We therefore integrate over the empirical distribution of possible
realizations of $A_{it}$. Let
\[
p_{it}
=
\frac{\sum_{j \neq i} \mathbf{1}\{a_{j,t-1}=R\}}
     {\sum_{j \neq i} 1}
\]
denote the share of other participants who chose red in the previous
round. Then 
\begin{align*}
\Pr{A_{it}=+1}{(a_{j,t-1})_{j\neq i}} =
p_{it}
\end{align*}
and 
\begin{align*}
\Pr{A_{it}=-1}{(a_{j,t-1})_{j\neq i}} =
1-p_{it}.
\end{align*}
Thus, the model-implied conditional choice probability can be written as
\begin{equation}
\label{eq:behavioral_like}
q_{\beta,\gamma,\psi}(M_{it},n_{it},p_{it})
=
p_{it}\,\Lambda\!\left(\beta \frac{M_{it}}{n_{it}^\psi}+\gamma\right)
+
(1-p_{it})\,\Lambda\!\left(\beta \frac{M_{it}}{n_{it}^\psi}-\gamma\right),
\end{equation}
which is a two-component finite mixture of logit choice
probabilities, induced by uncertainty over the realization of the
latent social signal $A_{it}$.

Identification of the model parameters follows from
variation in both the signal history, summarized by
\((M_{it},n_{it})\), and the previous-round actions of others,
summarized by \(p_{it}\). Let $q_0(M_{it},n_{it},p_{it})$  denote the population
conditional choice probability. The behavioral
model imposes the restriction $q_0(M_{it},n_{it},p_{it})=q_{\beta,\gamma,\psi}(M_{it},n_{it},p_{it})$.

The parameter \(\gamma\) is identified from how choices vary with the share of
others choosing red, \(p_{it}\), holding signal information fixed. In
particular, consider observations with balanced signals,
\(M_{it}=0\). For any \(n_{it}>0\),
\[
q_0(0,n_{it},p_{it})
=
p_{it}\,\Lambda(\gamma)
+
(1-p_{it})\,\Lambda(-\gamma).
\]
Since \(\Lambda(-\gamma)=1-\Lambda(\gamma)\), this can be written as
\[
q_0(0,n_{it},p_{it})
=
\Lambda(-\gamma)
+
p_{it}\left[2\Lambda(\gamma)-1\right].
\]
Thus, for any two values \(p_1\neq p_2\) in the support of the data,
\[
\frac{q_0(0,n_{it},p_1)-q_0(0,n_{it},p_2)}{p_1-p_2}
=
2\Lambda(\gamma)-1.
\]
Because \(2\Lambda(\gamma)-1\) is strictly increasing in \(\gamma\),
variation in \(p_{it}\) when the signal imbalance is zero identifies
\(\gamma\).

 The parameter \(\beta\) is identified from the responsiveness of choices to the sign and
magnitude of the signal imbalance \(M_{it}\). Given \(\gamma\), define
\[
F_{\gamma,p}(z)
\equiv
p_{it}\,\Lambda(z+\gamma)
+
(1-p_{it})\,\Lambda(z-\gamma).
\]
For every \(p_{it}\in[0,1]\), this function is strictly increasing in \(z\).
Therefore, 
\[
z(M_{it},n_{it})
=
\beta \frac{M_{it}}{n_{it}^\psi}
=
F_{\gamma,p}^{-1}\!\left(q_0(M_{it},n_{it},p_{it})\right).
\]

Third, \(\beta\) and \(\psi\) are separately identified from variation in
\((M_{it},n_{it})\). Since
\[
z(M_{it},n_{it})=\beta \frac{M_{it}}{n_{it}^\psi},
\]
for any \(M_{it}\neq 0\) and two values \(n_1\neq n_2\) in the support of the
data,
\[
\frac{z(M_{it},n_1)}{z(M_{it},n_2)}
=
\left(\frac{n_2}{n_1}\right)^\psi.
\]
Hence,
\[
\psi
=
\frac{\log |z(M_{it},n_1)|-\log |z(M_{it},n_2)|}
     {\log n_2-\log n_1}.
\]
Once \(\psi\) is identified, \(\beta\) is identified from
\[
\beta
=
\frac{z(M_{it},n_{it})n_{it}^\psi}{M_{it}}.
\]

Equation~(\ref{eq:behavioral_like}) gives the model-implied conditional
choice probability for treatments in which subjects observe others'
actions, namely the {\sc actions} and {\sc all} treatments. In the
{\sc no info} and {\sc signals} treatments, subjects do not observe
others' actions, and we therefore impose $\gamma=0$. The conditional
choice probability then reduces to
\[
q_{\beta,\psi}(M_{it},n_{it})
=
\Lambda\!\left(\beta \frac{M_{it}}{(n_{it})^\psi}\right).
\]
Thus, the likelihood
contribution of an observed action
$y_{it}=\mathbf{1}\{a_{it}=R\}$ is Bernoulli with success probability
$q_{\beta,\psi}(M_{it},n_{it})$ in the {\sc no info} and
{\sc signals} treatments, and with success probability
$q_{\beta,\gamma,\psi}(M_{it},n_{it},p_{it})$ in the
{\sc actions} and {\sc all} treatments. We estimate the behavioral model
separately for the {\sc no info}, {\sc actions}, {\sc signals}, and
{\sc all} treatments using a Bayesian approach implemented via Markov
chain Monte Carlo (MCMC).\footnote{The parameter estimates are virtually
identical when the model is estimated by maximum likelihood (MLE).}

Table~\ref{table_behavioral_param} reports the estimated model
parameters. The estimates of signal sensitivity, $\beta$, are positive
in all treatments. They are also similar within pairs of treatments
that differ only in whether subjects observe others' signals but share
the same observability of actions: $\beta$ is close to one in the
{\sc none} and {\sc signals} treatments, and around 1.27 in the
{\sc actions} and {\sc all} treatments. Consistent with our previous
findings, participants respond to others' actions beyond the information
contained in signals, as indicated by $\gamma>0$. Sensitivity to actions
is larger in the {\sc actions} treatment, where others' actions convey
information about unobserved private signals. Nevertheless, participants
also place positive weight on others' actions in the {\sc all} treatment,
where this information is redundant because the underlying signals are
already observed.

The estimates of $\psi$ are closer to $1/2$ than to $1$ across
treatments, indicating that  participants appear to scale signal
imbalances less aggressively than the sample proportion would imply. Thus, the data favor a normalization closer to a
square-root scaling of the signal information rather than the sample
proportion normalization commonly used in the literature
\citep{Benjamin}.

Beyond the parameter values, we evaluate the effects of comparable changes in
\(S_{it}\) and \(A_{it}\) on choice probabilities. In particular, in the {\sc
all} treatment, we find that a one-standard-deviation increase in
\(S\) has an effect similar to that of a change in \(A\) from observing
a green to a red action, corresponding to an increase of 0.3 in the probability of a red action.

\begin{table}[ht]
  \captionsetup{skip=8pt}
  \centering
  \caption{Behavioral Model Parameters}
\label{table_behavioral_param}
\begin{tabular}{lcccc}
  \hline
Parameter & {\sc none} & {\sc signals} & {\sc actions} & {\sc all} \\ 
  \hline
  \hline
$\beta$ & 1.047 & 1.018 & 1.268 & 1.273 \\ 
   & [0.960, 1.137] & [0.880, 1.174] & [1.187, 1.356] & [1.118, 1.434] \\ 
  $\gamma$ & -- & -- & 0.968 & 0.648 \\ 
   & -- & -- & [0.910, 1.027] & [0.582, 0.710] \\ 
  $\psi$ & 0.524 & 0.570 & 0.516 & 0.628 \\ 
   & [0.487, 0.558] & [0.537, 0.602] & [0.488, 0.544] & [0.598, 0.656] \\ 
  $\Delta Pr(a^g_{it}=R)$ & 0.253 & 0.314 & 0.265 & 0.294 \\
   \vspace{2mm}
   ($1$ SD in $S$)  & [0.245, 0.261] & [0.307, 0.321] & [0.258, 0.272] & [0.283, 0.304] \\ 
  $\Delta Pr(a^g_{it}=R)$ & -- & -- & 0.449 & 0.313 \\
    \vspace{2mm}
    ($A=-1$ to $A=1$)       & -- & -- & [0.426, 0.473] & [0.283, 0.341] \\
  \hline
  $N$ & 15,200 & 15,580 & 25,840 & 28,880 \\ 
  Participants & 80 & 82 & 136 & 152 \\ 
   \hline
\end{tabular}
\vspace{1mm}
\begin{minipage}{\linewidth}
\footnotesize
\underline{Notes:} Point estimates represent the median across posterior
draws. 95\% credible intervals are reported in square brackets.
\end{minipage}
\end{table}

\begin{figure}[h!]
\begin{center}
  \caption{Estimation results: Correct actions and learning from others}
  \label{fig:behavioral_corr_learn} 
\begin{tabular}{ccc}
\scriptsize{Panel (a): {\sc all} and {\sc signals}} &
                                                      \scriptsize{Panel (b): {\sc actions}, {\sc no info} and {\sc signals}} & \scriptsize{Panel (c): {\sc all} and {\sc actions}, by group size}\\
\includegraphics[scale=0.25]{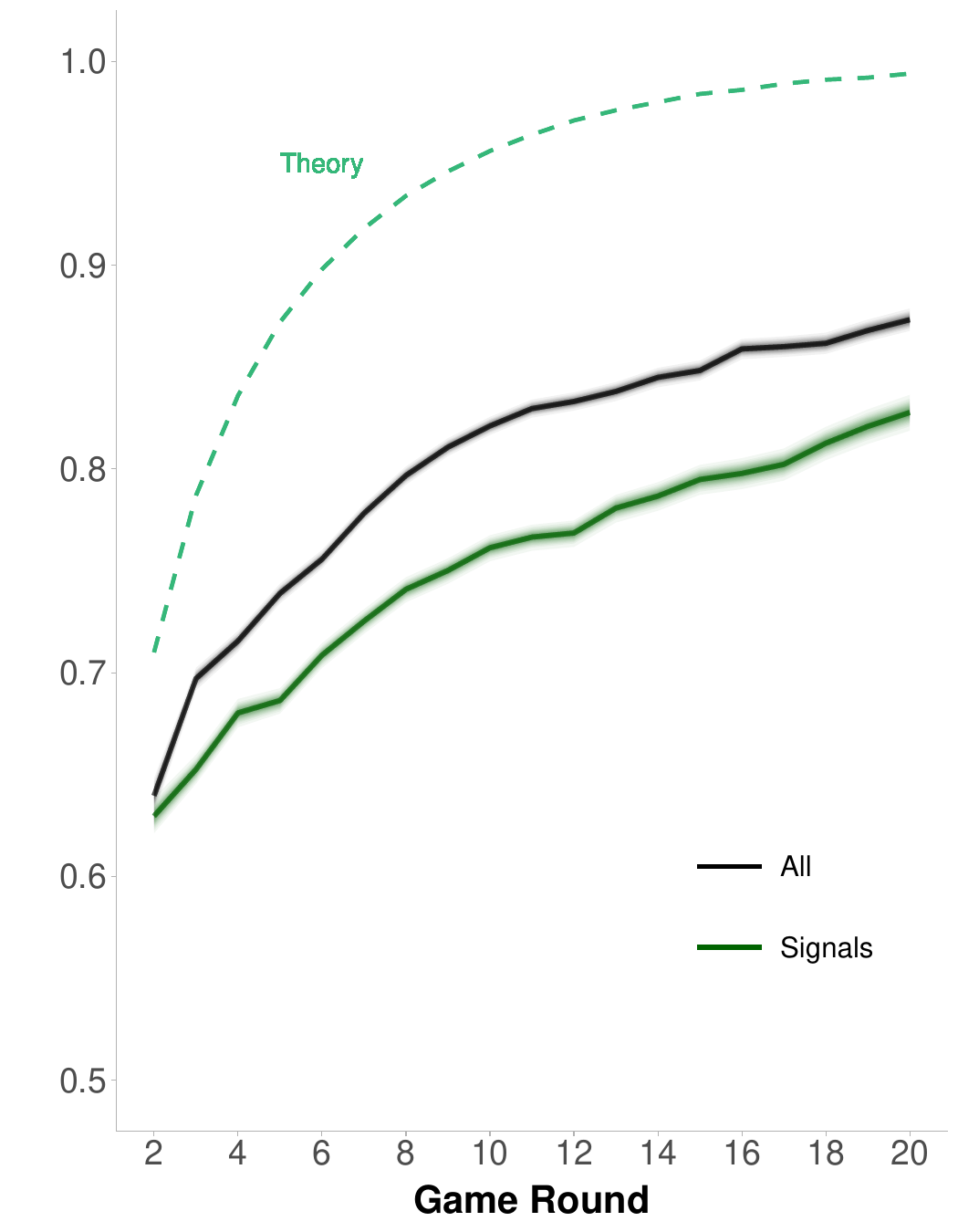} &
\includegraphics[scale=0.25]{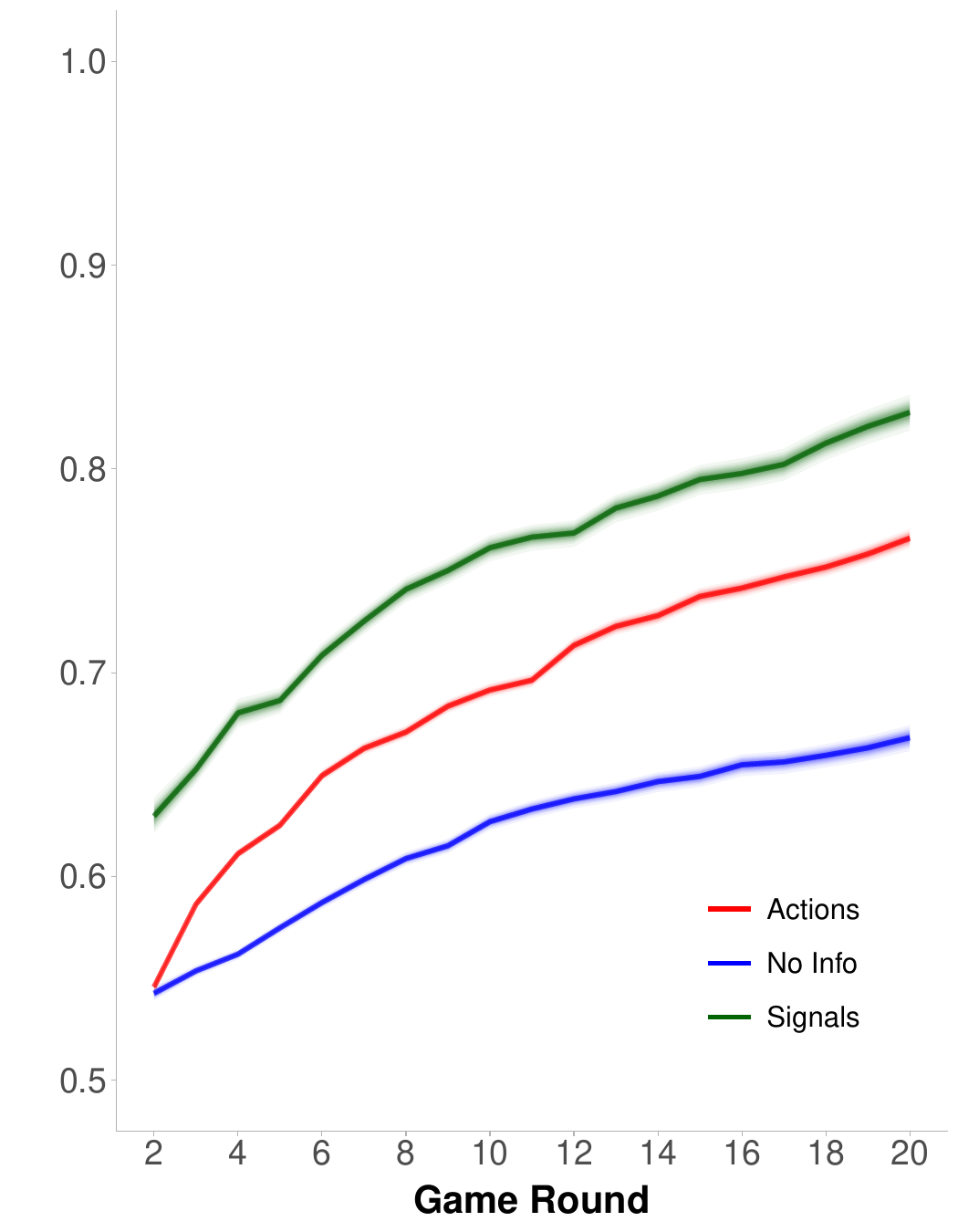}
                                                                            &
                                                                              \includegraphics[scale=0.25]{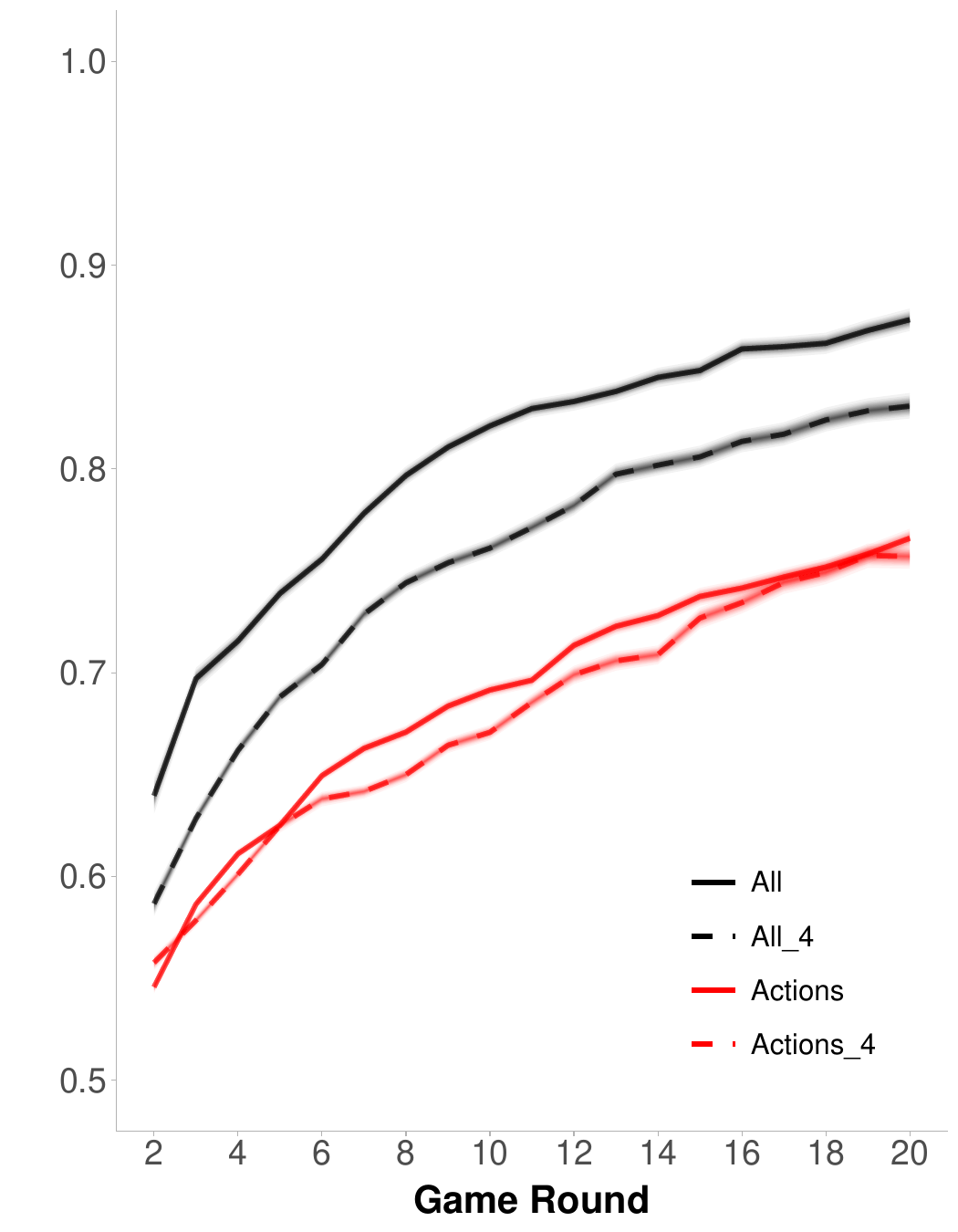}
\end{tabular}
\end{center}
{\footnotesize \underline{Notes:} Panel (a) presents the predicted probability of a correct action in
the {\sc signals} and {\sc all} treatments, computed from the
behavioral model and averaged by round across games and
participants. Panel (b) presents the predicted probability of a correct action in
the {\sc signals}, {\sc actions} and {\sc no info} treatments, computed from the
behavioral model and averaged by round across games and
participants. Panel (c) presents the predicted probability of a correct action in
the {\sc all}, {\sc all4}, {\sc actions} and {\sc actions4} treatments, computed from the
behavioral model and averaged by round across games and
participants. Shaded regions represent 95\% credible
posterior intervals from 50\% (darkest) to 95\% (faintest) probability
levels.}
\end{figure}

Panel (a) in Figure~\ref{fig:behavioral_corr_learn} presents the
predicted probability of a correct action in the {\sc signals} and
{\sc all} treatments, computed from the behavioral model and averaged
by round across games and participants. As in the simulation exercise
in Figure~\ref{fig:sim}, the behavioral model closely
matches the experimental data in both treatments, with the performance
gap driven by participants’ use of others’ actions in their
decision-making.

Panels (b) and (c) replicate the patterns in
Figure~\ref{fig:CorrectGuesses}. The behavioral model captures the
improvement in performance from incorporating others’ actions relative
to the {\sc none} treatment. In addition, at the estimated parameter
values, the model successfully reproduces the empirical finding that
larger groups perform better than smaller ones when both others’
actions and signals are observable.

\begin{figure}[h!]
\begin{center}
      \caption{Simulation results: Learning from others' actions in
        the {\sc all} treatment}
      \label{fig:behavioral_share_actions} 
\begin{tabular}{c}
\includegraphics[scale=0.4]{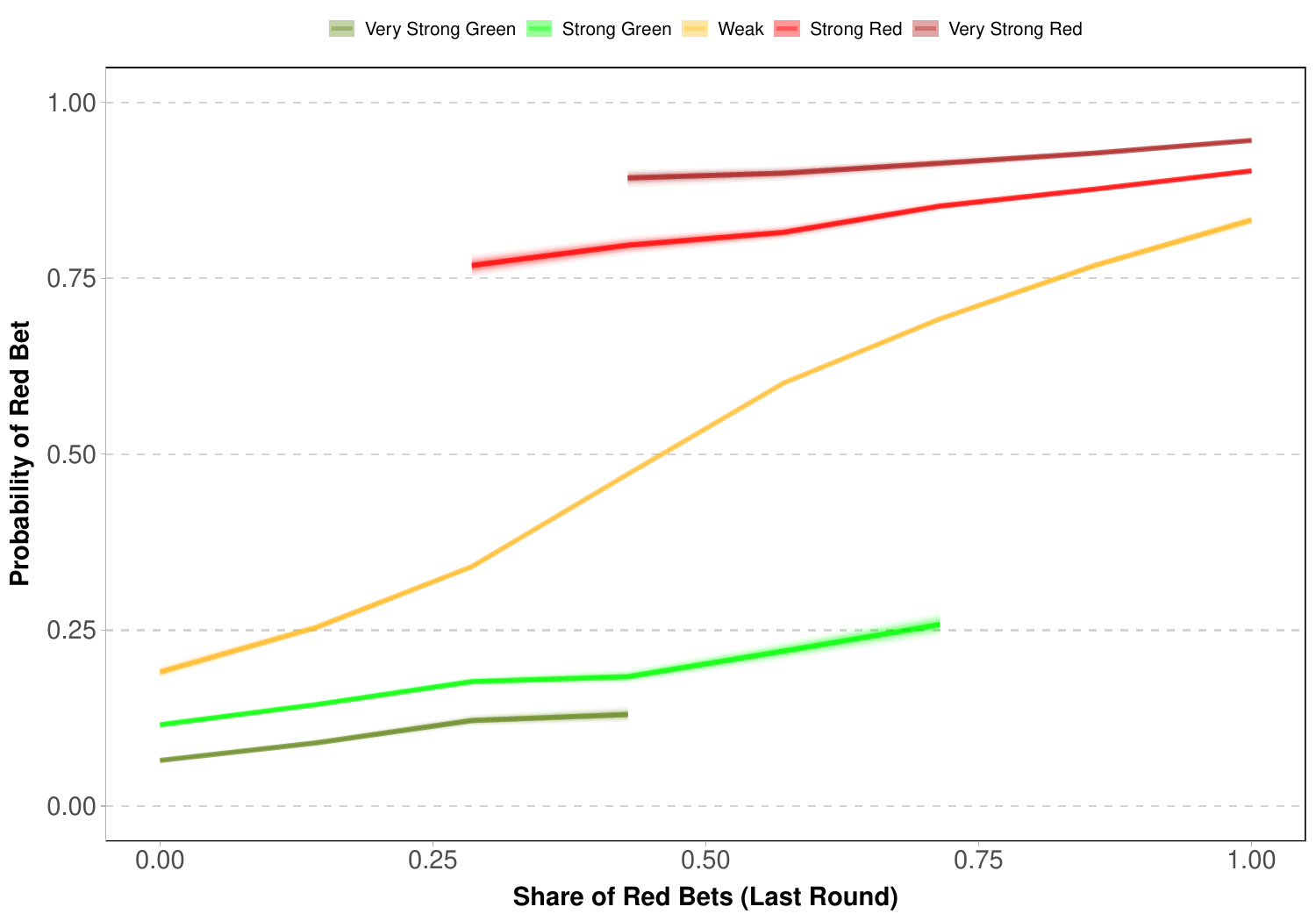}
\end{tabular}
\end{center}
{\footnotesize \underline{Notes:} This figure depicts the probability of a red action computed from the behavioral model and grouped by the share of red actions in the previous rounds as observed in the data within each signal strength category.}
\end{figure}

Figure \ref{fig:behavioral_share_actions} depicts the predicted probability of a red action from the
behavioral model grouped by the share of observed red actions in the previous round and
by signal strength categories, as defined in Section
\ref{sec:Results}. The model replicates the observed pattern in Figure
\ref{fig:learning_from_guesses}, whereby participants rely more
heavily on others’ actions when signals are \emph{weak}, consistent
with the increasing relative importance of \(\gamma\) as the
informativeness of \(S_{it}\) declines.

The behavioral model replicates participants’ behavior in the data
without relying on heterogeneity in sensitivity to signals or
actions. Nevertheless, the model is flexible enough to accommodate
heterogeneity across participants’ characteristics. Motivated by our analysis of IQ and the associated variation in
behavior, we allow \(\beta\) and \(\gamma\) to vary across
participants, capturing differences in their ability to extract
information from signals and others’ actions:\footnote{In the IQ
  heterogeneity specification, we hold $\psi$ fixed across
  participants since we do not find $\psi$ changes by IQ level.}
\[
\beta_{i}=\exp(\beta_{\text{common}}+\beta_{\text{low-IQ}}\,
\text{low-IQ}_i), \quad
\gamma_{i}=\exp(\gamma_{\text{common}}+\gamma_{\text{low-IQ}}\,
\text{low-IQ}_i),
\]
where \(\exp(\cdot)\) ensures that \(\beta_i,\gamma_i \geq 0\), as
required by the behavioral model.\footnote{The results do not rely on
this specific functional form or prior assumptions; the posterior
distributions of \(\beta\) and \(\gamma\) are concentrated on positive values.}

\begin{figure}[h!]
\begin{center}
\caption{Behavioral Model Parameters ({\sc all} treatment) as a
  function of IQ}\label{fig:behavioral_param_naive}
\begin{tabular}{c}
\includegraphics[scale=0.4]{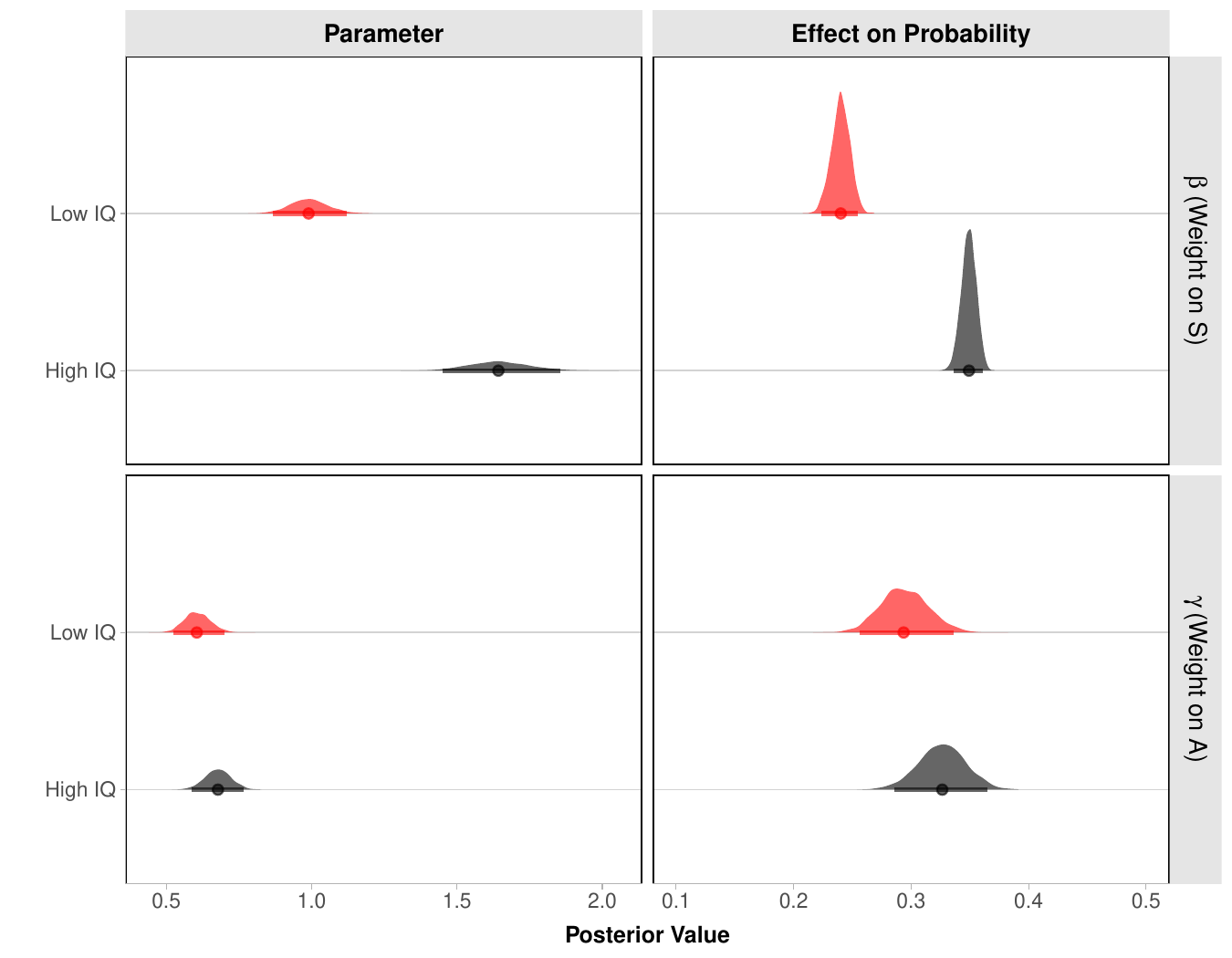}
\end{tabular}
\end{center}
{\footnotesize \underline{Notes:} The upper panels of this figure
  present posterior estimates of  $\beta$ and of the effect of
  $S_{it}$ on choice probability by participants’ IQ level. The lower panels present the posterior
distribution of $\gamma$ and of the effect of \(A_{it}\) on choice
probability by participants’ IQ level. Points represent posterior medians, and lines denote
95\% credible intervals.}
\end{figure}

Figure \ref{fig:behavioral_param_naive} reports the results from
estimating the heterogeneous behavioral model on data from the {\sc
all} treatment. Consistent with the analysis of individual responses
by IQ, we find that \emph{low-IQ} participants are substantially less
sensitive to signal information in \(S\) than \emph{high-IQ}
participants
(\(\beta_{\text{low-IQ}} = 0.99\) vs.
\(\beta_{\text{high-IQ}} = 1.64\)). This difference translates into a
markedly smaller effect on betting probabilities from a
one-standard-deviation increase in \(S\) for \emph{low-IQ}
participants (0.24) relative to \emph{high-IQ} participants (0.35),
with the 95\% credible interval for the difference excluding zero.

By contrast, \emph{low-IQ} and \emph{high-IQ} participants place
similar weight on others' actions
(\(\gamma_{\text{low-IQ}} = 0.61\) vs.
\(\gamma_{\text{high-IQ}} = 0.68\)), with the 95\% credible interval
for the difference including zero. Consistent with the individual
results summarized in Figure \ref{fig:RespSignals}, this implies that
the two groups respond similarly to a change in the observed action in
the previous round from green to red
(0.29 for \emph{low-IQ} vs. 0.33 for \emph{high-IQ}), again with the
95\% credible interval for the difference including zero.

The focus on this paper is experimental, and so we leave for future
work the in-depth theoretical analysis of this heuristic and its general
properties. Both the simulations and estimation on our data suggest some natural conjectures. 

First, in the presence of both signals and actions, imitation can be
beneficial. For a given value of $\beta$, which governs responses to
signals, some positive values of $\gamma$ improve outcomes relative to
the baseline case $\gamma = 0$, where others' actions are ignored. The
intuition is that, because individual responses are noisy, past
actions can contain useful additional information. One might expect this benefit to diminish once too much weight is placed on past actions, since all information ultimately originates in the signals. 

Second, the results from the {\sc actions} treatment indicate that group
size does not have a substantial effect on outcomes. This is
consistent with the theoretical results of \cite{harel2020rational},
although that setting assumes Bayesian agents with a high degree of
rationality. More generally, it is of interest to study the extent to
which imitation remains useful under cognitive limitations and other
departures from optimal behavior.

\section{Discussion}\label{sec:Discussion}

The imitation of the actions, opinions and values of others is a natural human tendency. It is an integral part of how infants and adults learn, and of how information is disseminated through society. It can also lead to inefficient outcomes, as has been highlighted by the social learning literature over the past few decades. In our setting, which features repeated interaction between participants and gradual accumulation of public raw data as well as others' opinions, we show that participants imitate each other, and that imitation improves their performance. 

This result was not a priori obvious. Optimal Bayesian behavior does not feature imitation, and behavioral economics does not offer a clear-cut prediction that imitation should improve efficiency, or indeed suggests the opposite. Clearly, it is highly inefficient for all agents to ignore the raw data and mimic each other, and so too much imitation is harmful. Hence the usefulness of imitation lies in moderation: both the signals and the actions have to play a role. We develop a simple behavioral model that reproduces these results in simulations. A fruitful avenue for future research is the further development and formal study of this model.


To study imitation we constructed a social learning environment with repeated actions. The experimental literature on such settings is small, and we believe that there is much room for further research. In particular, it would be interesting to understand the robustness of imitation to various interventions, such as the inclusion of a well-informed central information source, or of ideological agents whose actions are independent of the data, or a more general structure of heterogeneity in information quality of different agents. These questions are important in an age in which the dissemination of uncertain but critical information to the public is a major societal challenge.

\appendix 
\renewcommand\thefigure{\thesection.\arabic{figure}}
\renewcommand\thetable{\thesection.\arabic{table}}   

\clearpage

\section{Additional Aggregate Results\label{app_aggregate}}
\setcounter{figure}{0}
\setcounter{table}{0}   

\begin{figure}[h!]
\begin{center}
\begin{tabular}{cc}
\scriptsize{Panel (a): Correct Actions, by round} & \scriptsize{Panel (b): Consensus rates, by round} \\ 
\includegraphics[scale=0.3]{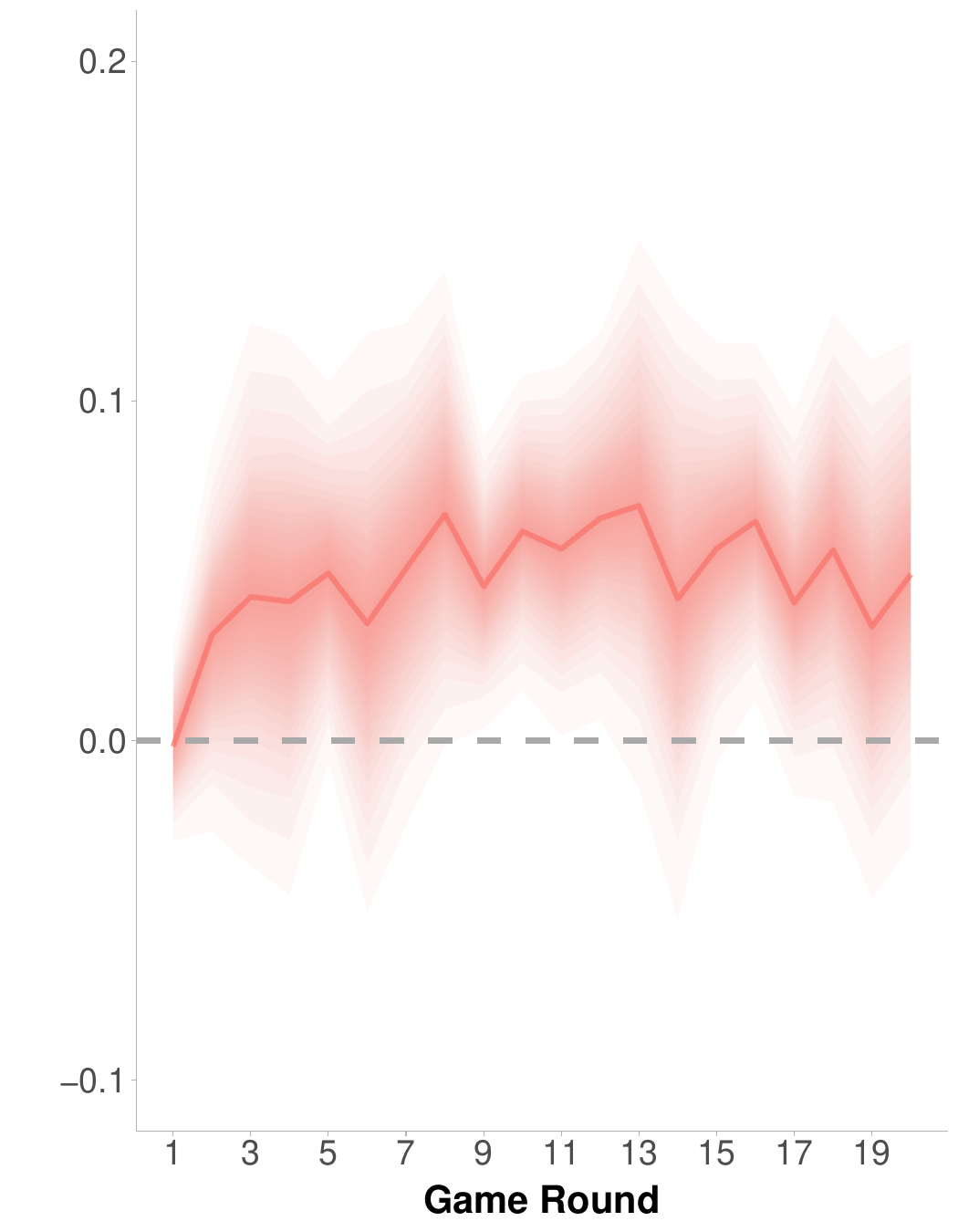}  & \includegraphics[scale=0.3]{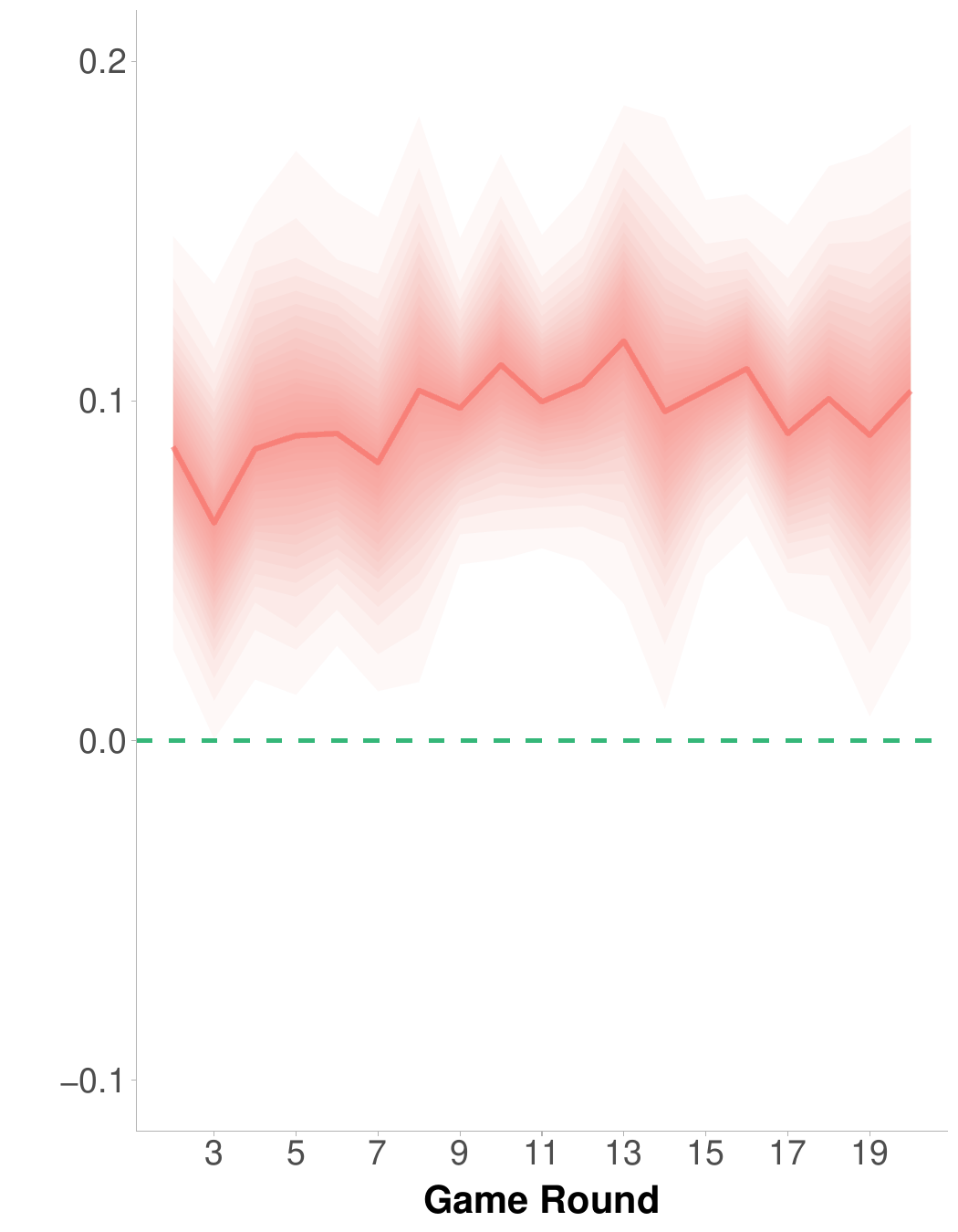} 
\end{tabular}
\caption{Differences between the {\sc all} and {\sc signals} treatments}\label{fig_app_corr_diff_signals_all} 
\end{center}
{\footnotesize \underline{Notes:} Panel (a) presents the average difference in the  frequency of correct actions in each treatment in each round, averaged across games. Panel (b) depicts the difference in consensus rates in each round. For panel (b) we exclude cases with equal number of green and red signals. Shaded regions represent confidence intervals from 50\% (darkest) to 95\% (faintest) probability levels. Confidence intervals are constructed with a variance-covariance matrix clustered by session.}
\end{figure}

\begin{figure}[h!]
\begin{center}
\begin{tabular}{cc}
 \scriptsize{Panel (a): Majority Correct Actions, by round} &  \scriptsize{Panel (b): Difference between {\sc all} and {\sc signals}, by round} \\ 
\includegraphics[scale=0.3]{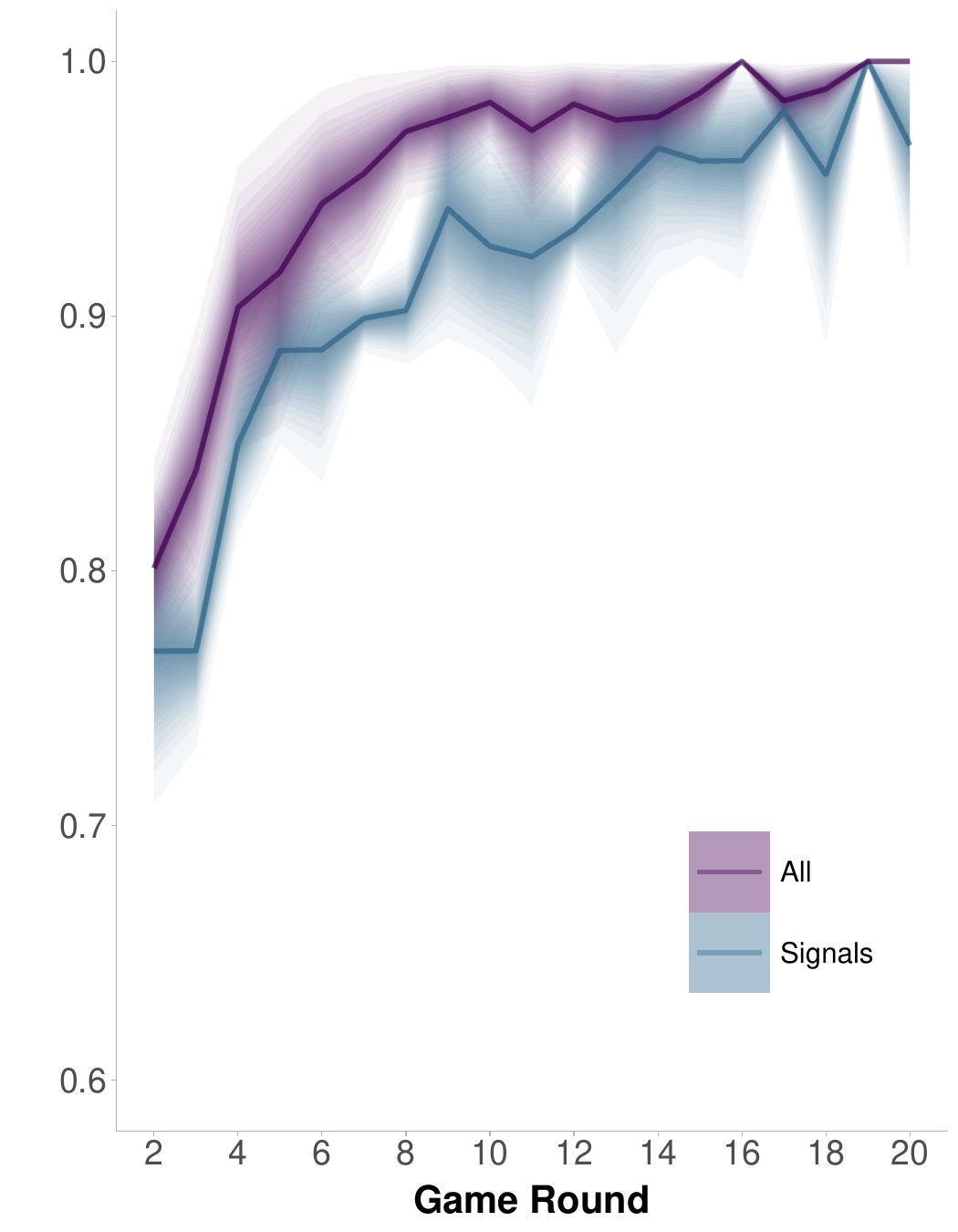}  & \includegraphics[scale=0.3]{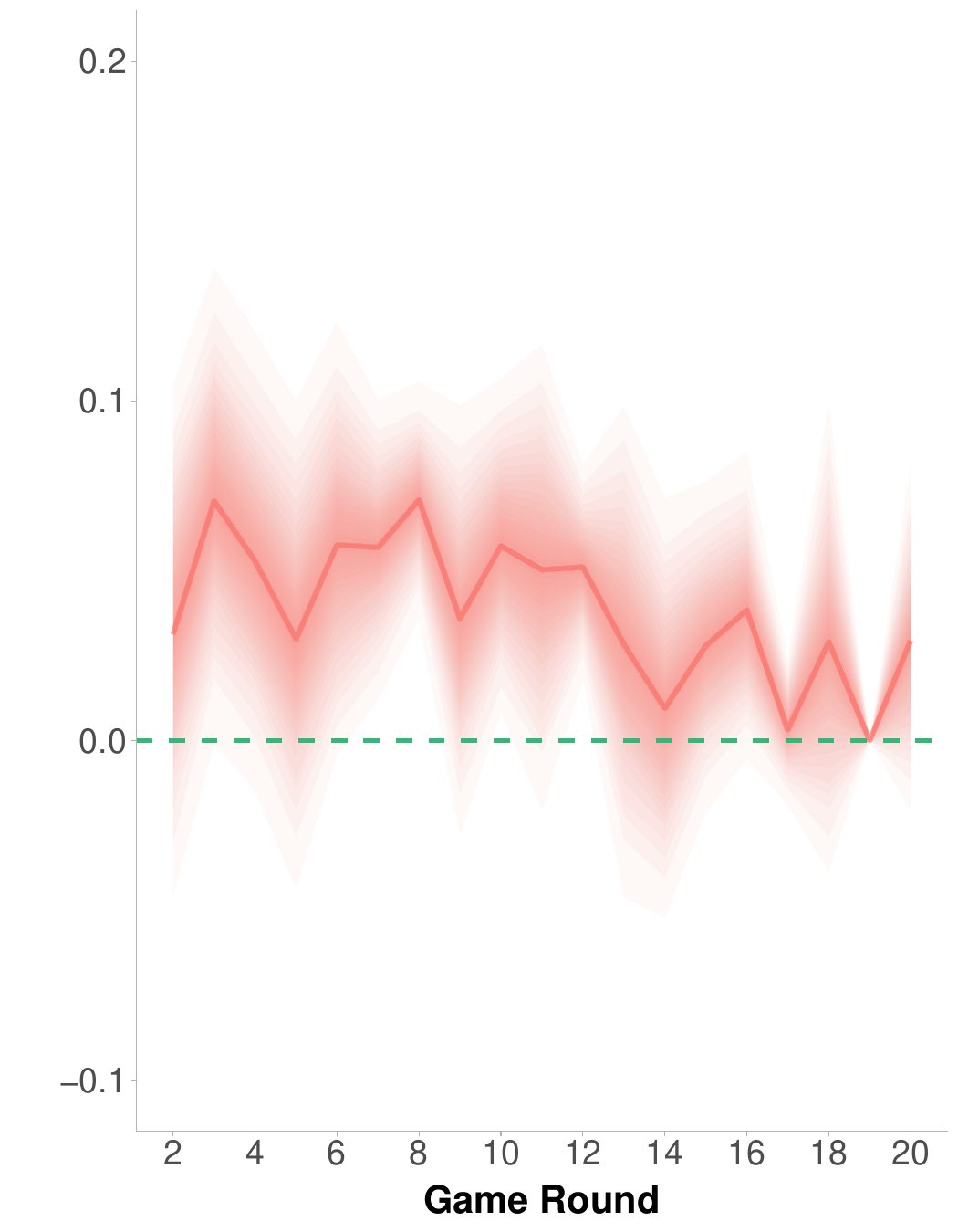}   
\end{tabular}
\caption{How often is the majority correct? {\sc all} versus {\sc signals} treatments}\label{fig_app_corr_maj_signals_all} 
\end{center}
{\footnotesize \underline{Notes:} Panel (a) presents the frequency of correct actions by the majority in each treatment per round, averaged across games. Panel (b) presents the average difference in the frequency of correct actions by the majority between the two treatments. We exclude cases with an equal number of green and red signals. Shaded regions represent confidence intervals from 50\% (darkest) to 95\% (faintest) probability levels. Confidence intervals are constructed with a variance-covariance matrix clustered by session.}
\end{figure}

\newpage

\begin{table}[!htbp] \centering 
\scriptsize
  \caption{Treatment Effects for {\sc signals} and {\sc all} treatments} 
  \label{table_app_signals_all} 
\begin{tabular}{@{\extracolsep{5pt}}lcccccc} 
\\[-1.8ex]\hline 
\hline \\[-1.8ex] 
 & \multicolumn{6}{c}{\textit{Dependent variable:}} \\ 
\cline{2-7} 
\\[-1.8ex] & \multicolumn{2}{c}{Correct Actions} & \multicolumn{2}{c}{Consensus Rate} & \multicolumn{2}{c}{Majority is Correct}\\ 
\\[-1.8ex] & (1) & (2) & (3) & (4) & (5) & (6)\\ 
\hline \\[-1.8ex] 
 {\sc all} (Baseline) & 0.702$^{***}$ & 0.701$^{***}$  & 0.812$^{***}$ & 0.809$^{***}$ & 0.803$^{***}$ & 0.808$^{***}$ \\ 
  & (0.012) & (0.013)  & (0.009) & (0.011)  & (0.013) & (0.014)  \\ 
  & & \\ 
 {\sc signals} (Effect) & $-$0.051$^{*}$ & $-$0.047$^{*}$ & $-$0.095$^{***}$ & $-$0.089$^{***}$ & $-$0.038$^{**}$ & $-$0.051$^{***}$  \\ 
  & (0.028) & (0.028)  & (0.028) & (0.029) & (0.016) & (0.016)\\ 
  & & \\ 
 {\sc signals} (Effect) $\times$ Late Rounds  &  & $-$0.007 &  & $-$0.011 &  & 0.026$^{*}$  \\ 
  &  & (0.017)  &  & (0.012)  &  & (0.015)  \\ 
  & & \\ 
\hline \\[-1.8ex] 
Game Round Fixed Effects & Yes & Yes & Yes & Yes & Yes & Yes \\ 
Observations & 44,460 & 44,460 & 5,499 & 5,499 & 5,499 & 5,499  \\ 
Adjusted R$^{2}$ & 0.021 & 0.021 & 0.135 & 0.135 & 0.062 & 0.062  \\ 
\hline 
\hline \\[-1.8ex] 
\end{tabular} 
\vspace{1mm}

{\footnotesize \underline{Notes:} $^{**}$p$<$0.05; $^{***}$p$<$0.01. Clustered standard errors by session in parentheses. Late rounds are 11 through 20.}
\end{table} 

\begin{figure}[h!]
\begin{center}
\caption{Frequency of correct actions, by information structure and group size}\label{app_fig_correctGuesses_diff} 
\begin{tabular}{cc}
\scriptsize{Panel (a): Information Structures} & \scriptsize{Panel (b): Group Size}\\
\includegraphics[scale=0.32]{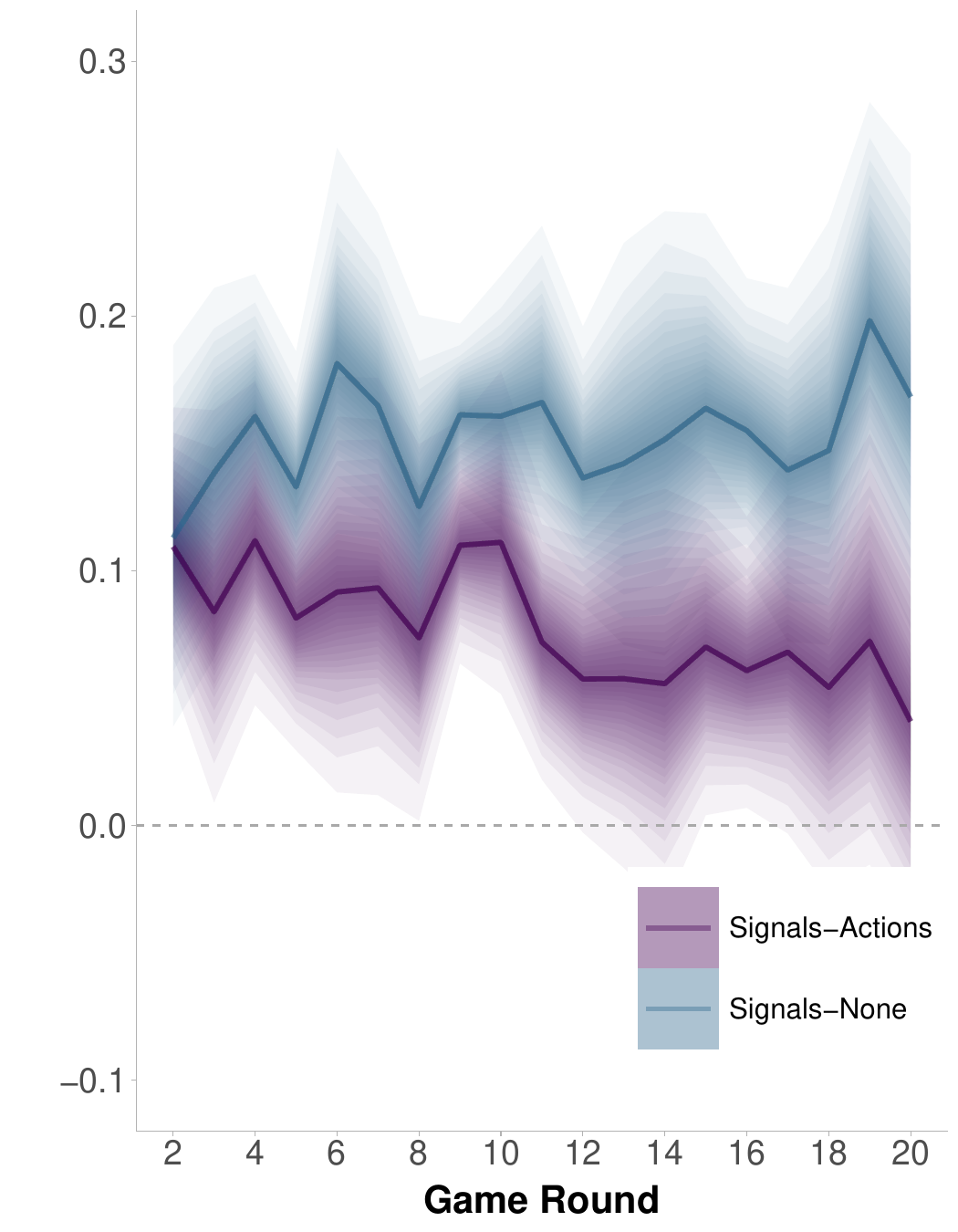} &
\includegraphics[scale=0.32]{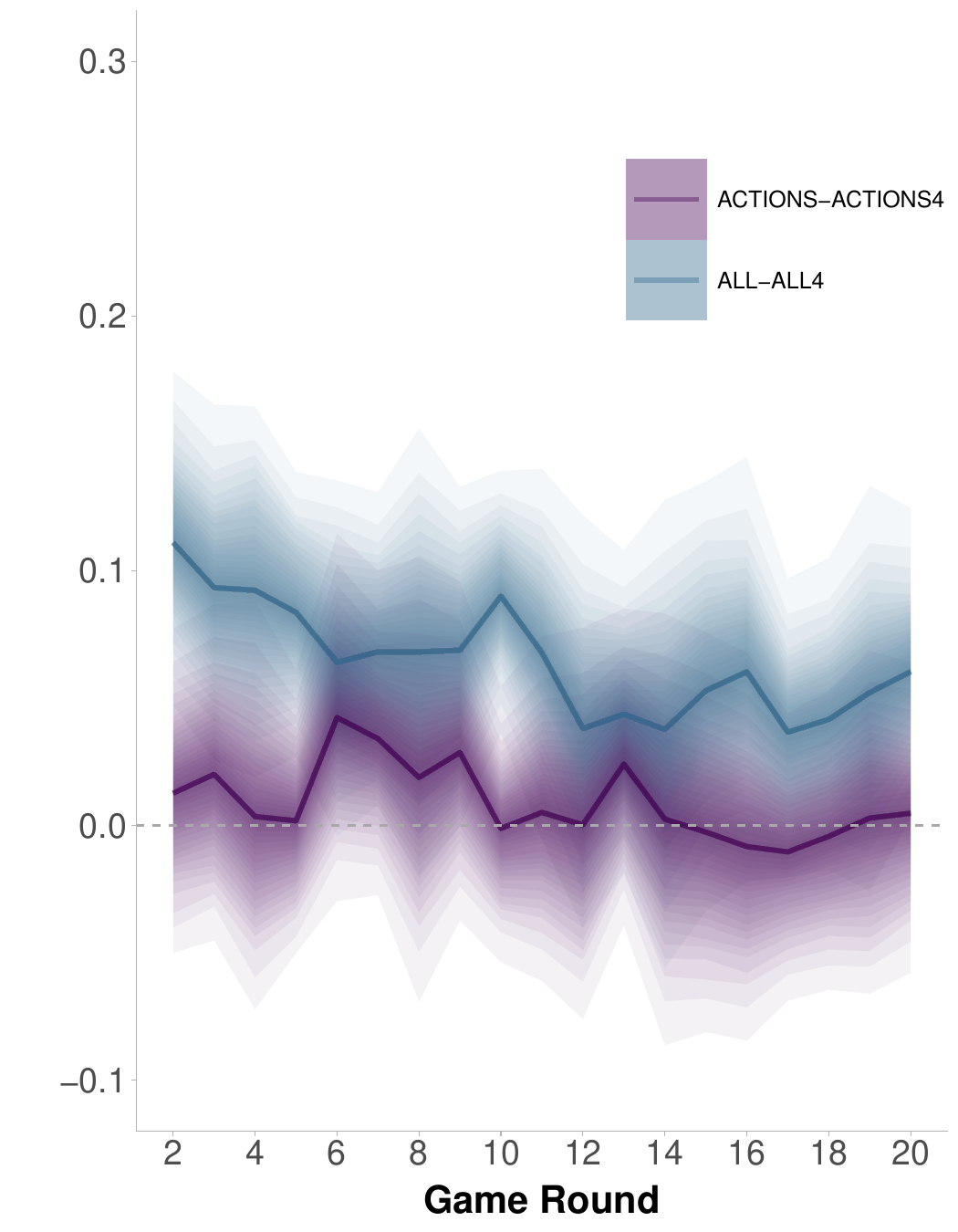} 
\end{tabular}
\end{center}
{\footnotesize \underline{Notes:} Both panels present the average frequency of correct actions in each treatment per each round, averaged across games. Shaded regions represent confidence intervals from 50\% (darkest) to 95\% (faintest) probability levels. Confidence intervals are constructed with a variance-covariance matrix clustered by session.}
\end{figure}

\begin{table}[!htbp] \centering 
\scriptsize
  \caption{Treatment Effects for {\sc no info}, {\sc actions}, and {\sc signals}} 
  \label{app_table_corr_none_actions_signals} 
\begin{tabular}{@{\extracolsep{5pt}}lcc} 
\\[-1.8ex]\hline 
\hline \\[-1.8ex] 
 & \multicolumn{2}{c}{\textit{Dependent variable:}} \\ 
\cline{2-3} 
\\[-1.8ex] & \multicolumn{2}{c}{Correct Actions} \\ 
\\[-1.8ex] & (1) & (2)\\ 
\hline \\[-1.8ex] 
 {\sc actions} (Baseline) & 0.583$^{***}$ & 0.572$^{***}$ \\ 
  & (0.017) & (0.015) \\ 
  & & \\ 
{\sc no info} (Effect) & $-$0.076$^{***}$ & $-$0.053$^{***}$ \\ 
  & (0.019) & (0.016) \\ 
  & & \\ 
{\sc no info} (Effect) $\times$ Late Rounds &  & $-$0.043$^{*}$ \\ 
  &  & (0.026) \\ 
  & & \\ 
 {\sc signals} (Treatment) & 0.078$^{***}$ & 0.096$^{***}$ \\ 
  & (0.029) & (0.028) \\ 
  & & \\ 
 {\sc signals} (Treatment) $\times$ Late Rounds &  & $-$0.034$^{**}$ \\ 
  &  & (0.013) \\ 
  & & \\ 
\hline \\[-1.8ex] 
Game Round Fixed Effects & Yes & Yes \\ 
Observations & 56,430 & 56,430 \\ 
Adjusted R$^{2}$ & 0.028 & 0.028 \\ 
\hline 
\hline \\[-1.8ex] 
\end{tabular} 
\vspace{1mm}

{\footnotesize \underline{Notes:} $^{**}$p$<$0.05; $^{***}$p$<$0.01. Clustered standard errors by session in parentheses. Late rounds are 11 through 20.}
\end{table} 

\begin{table}[!htbp] \centering 
\scriptsize
  \caption{Treatment Effects for Group Size} 
  \label{app_table_corr_size} 
\begin{tabular}{@{\extracolsep{5pt}}lcccc} 
\\[-1.8ex]\hline 
\hline \\[-1.8ex] 
 & \multicolumn{4}{c}{\textit{Dependent variable:}} \\ 
\cline{2-5} 
\\[-1.8ex] & \multicolumn{4}{c}{Correct Actions} \\ 
 & (1) & (2) & (3) & (4) \\ 
\hline \\[-1.8ex] 
 {\sc all4} (Baseline) & 0.614$^{***}$ & 0.603$^{***}$ &  &  \\ 
  & (0.017) & (0.016) &  & \\ 
   {\sc actions4} (Baseline) &  &  & 0.546$^{***}$ & 0.539$^{***}$ \\ 
  & & & (0.016) & (0.012) \\ 
  & & & & \\ 
{\sc all} (Effect) & 0.065$^{***}$ & 0.082$^{***}$ &  &  \\ 
  & (0.023) & (0.022) &  &  \\ 
  & & & & \\ 
 {\sc all} (Effect) $\times$ Late Rounds &  & $-$0.033 &  &  \\ 
  &  & (0.031) &  &  \\ 
  & & & & \\ 
{\sc actions} (Effect) &  &  & 0.008 & 0.019 \\ 
  &  &  & (0.022) & (0.020) \\ 
  & & & & \\ 
{\sc actions} (Effect) $\times$ Late Rounds &  &  &  & $-$0.021 \\ 
  &  &  &  & (0.024) \\ 
  & & & & \\ 
\hline \\[-1.8ex] 
Game Round Fixed Effects & Yes & Yes & Yes & Yes \\ 
Observations & 44,080 & 44,080 & 40,090 & 40,090 \\ 
Adjusted R$^{2}$ & 0.029 & 0.029 & 0.020 & 0.020 \\ 
\hline 
\hline \\[-1.8ex] 
\end{tabular} 
\vspace{1mm}

{\footnotesize \underline{Notes:} $^{**}$p$<$0.05; $^{***}$p$<$0.01. Clustered standard errors by session in parentheses. Late rounds are 11 through 20.}
\end{table}

\newpage

\section{Open-ended Questions \label{app:lda}}
\setcounter{figure}{0}
\setcounter{table}{0}   

We quantify the answers to the
open-ended questions designed to elicit participants' strategies throughout the game and analyze their relationship with participants' characteristics, specifically players' IQ.  

For each participant in the {\sc all4} and {\sc all} treatments, we examine the
three open-ended survey questions asked after all rounds have been
played:
\begin{enumerate}
    \item  What strategy did you use in the game (if any)? Please elaborate.
    \item  Did you look at the balls drawn for other players in your group? Did you
find them useful/not useful? Please elaborate. 
    \item Did you look at the bets made by other players in your group? Did you find them useful/not useful? Please elaborate.
\end{enumerate}

We implement a machine learning algorithm for a probabilistic topic model known as the structural topic model (STM) \citep{roberts2013structural}. Under this framework, a \emph{topic} is defined as a probability distribution over words and a participant's response in our data is modeled as a distribution over topics. Thus, each participant's response in the data can belong to multiple topics with a probability distribution estimated from the data. We are interested in the proportion of an answer spent covering each estimated topic.

Each participant's response to question $q \in \{(i),(ii),(iii)\}$, $r_q$,  has its own distribution over topics, $\mathbf{\theta}_{r_q}$. We label this parameter topic prevalence and interpret it as the proportion of each topic $k=1,\ldots,K$ in response $r_q$. We can think of each topic $k$ as drawn from a multinomial distribution with parameter $\mathbf{\theta}_{r_q}$. Conditional on the topic selected, word $w_{r_q,n}$ included in response $r_q$ is drawn from a multinomial distribution over the vocabulary $n=1,\ldots,N$ with parameter $\mathbf{\beta}_{k,n}$. This is a probability vector over the $N$ words in the vocabulary.

In probabilistic topic models such as the STM, the number of topics $K$ needs to be selected a priori. In doing so, there is a trade-off between interpretability (consistent with a lower $K$) and goodness-of-fit (consistent with a higher $K$). We favor the former and choose the low value of $K=4$ for the three open-ended questions. We find that $K=4$ gives us topics that are very straightforward to interpret and with a higher semantic coherence (i.e., the most probable words in a given topic co-occur together) than models with more topics.

The structural topic model allows for the inclusion of participants' characteristics to inform the topic prevalence. Specifically, $\mathbf{\theta}_{r_q} \sim
\mathrm{LogisticNormal}(X_{r_q} \gamma,\Sigma)$, where $X_d$ is a vector of participant characteristics. In addition to our binary measure of IQ,  we include several participant characteristics: \texttt{female}, which is an indicator variable that takes the value of one if the participant identifies as female. \texttt{stem}, which is an     indicator variable that takes the value of one if the participant's major is STEM.  \texttt{overconfidence} measures the extent of a participant's over-estimation of her IQ, which is given by the difference between the number of questions a participant believes she solved correctly and the actual number of correct answers. \texttt{risk} is  measured by the number of points invested in a risky asset as specified in a risky investment task solved at the end of the game.
     
The topic model for each open-ended question is estimated using a variational Expectation-Maximization (EM) algorithm, as implemented in the \texttt{stm} package in \texttt{R} \citep{roberts2013structural}. Prior to estimation, we pre-processed the raw responses using standard conventions: we stem words (i.e., reduce words to their root form), drop punctuation, as well as common
stop-words, and remove words that were used less than $0.5 \%$ over all responses.

\begin{figure}[h!]
\begin{center}
\includegraphics[scale=0.6]{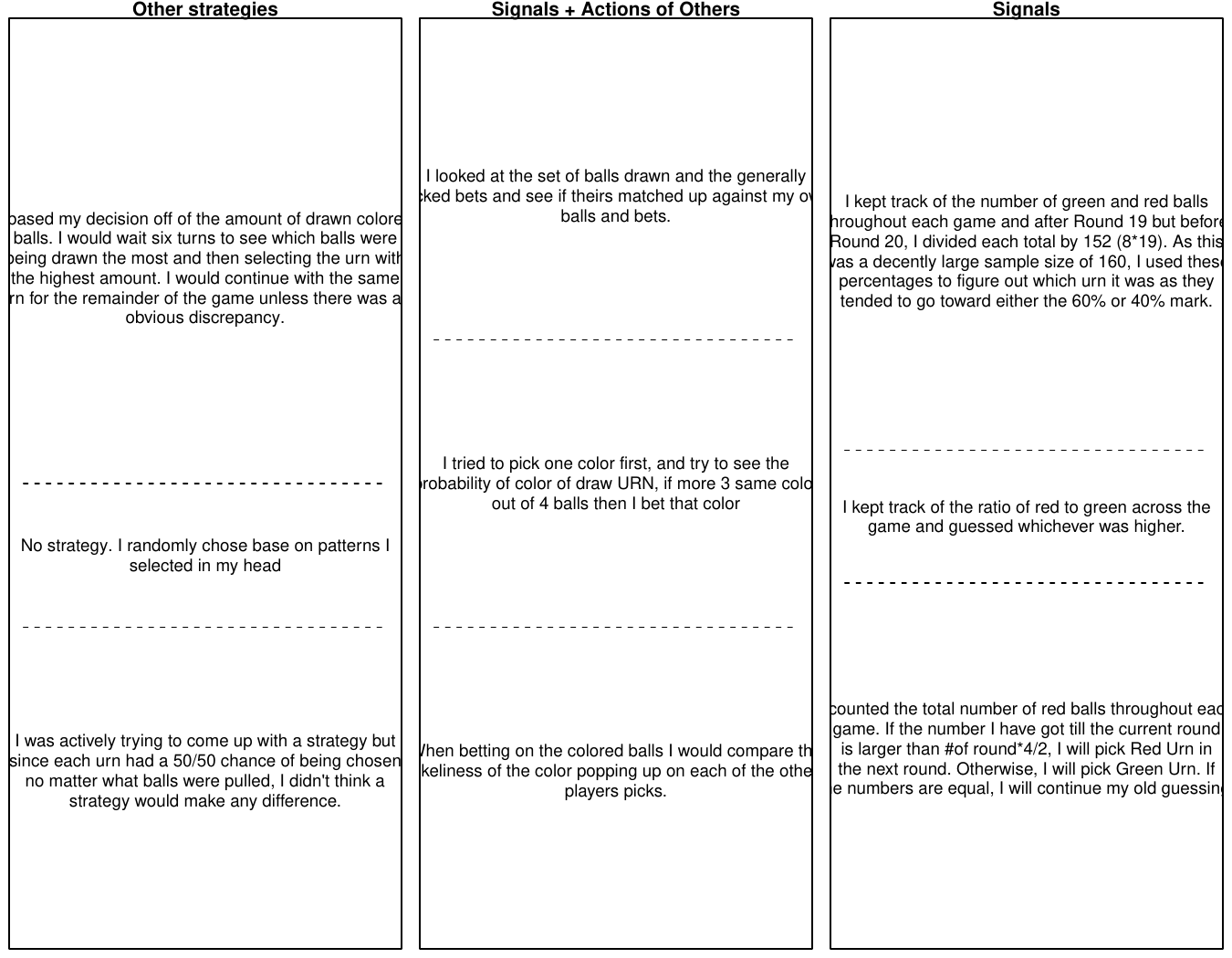}   
\caption{Example Responses to ``What strategy did you use in the game (if any)?'' by topic ({\sc all4} and {\sc all} treatments) \label{app_fig:ex_responses_strategy}}
\end{center}
{\footnotesize \underline{Notes:} Each panel presents the three most highly associated open-ended responses with each of the four estimated topics for the question ``What strategy did you used in the game (if any)?''}
\end{figure}

\begin{figure}[h!]
\begin{center}
\includegraphics[scale=0.6]{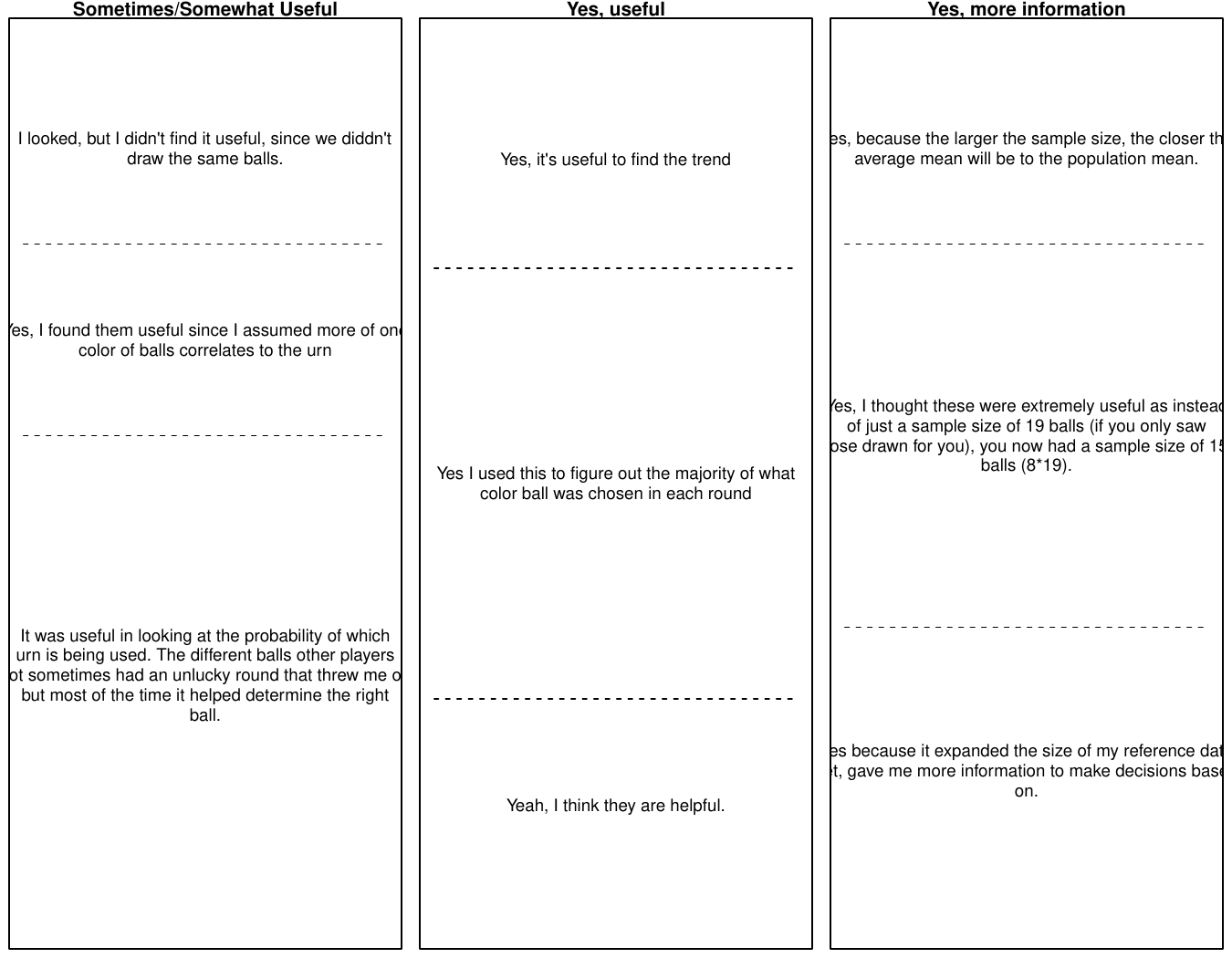}  
\caption{Example Responses to ``Did you look at the balls drawn for other players in your group?''  by topic ({\sc all4} and {\sc all} treatments) \label{app_fig:ex_responses_lookballs}}\label{app_fig:ex_lookballs} 
\end{center}
{\footnotesize \underline{Notes:} Each panel presents the three most highly associated open-ended responses with each of the four estimated topics for the question ``Did you look at the balls drawn for other players in your group? Did you
find them useful/not useful? Please elaborate.''}
\end{figure}

\begin{figure}[h!]
\begin{center}
\includegraphics[scale=0.6]{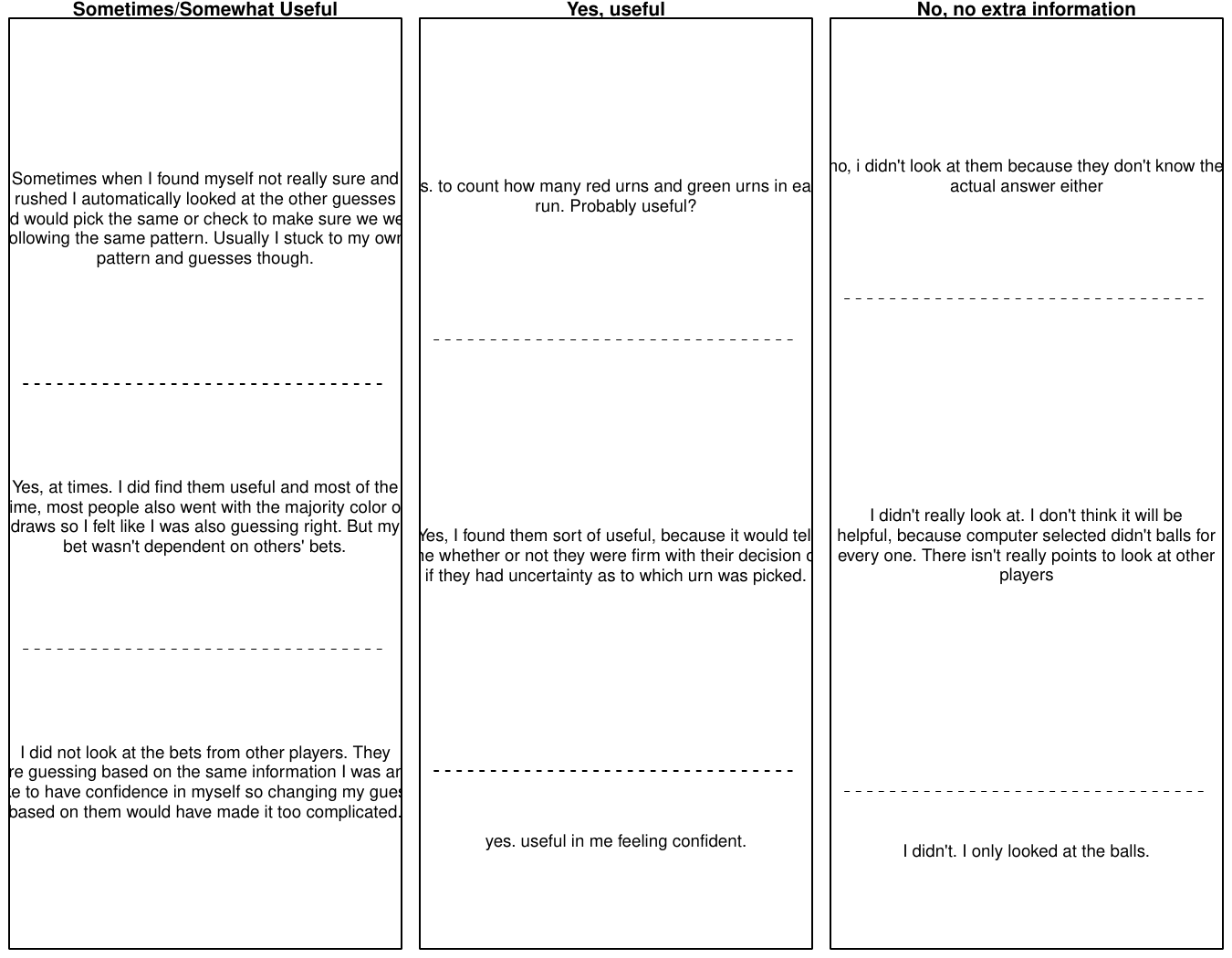}    
\caption{Example Responses to ```Did you look at the bets made by other players in your group?''  by topic ({\sc all4} and {\sc all} treatments) \label{app_fig:ex_responses_lookbets}}\label{app_fig:ex_lookbets} 
\end{center}
{\footnotesize \underline{Notes:} Each panel presents the three most highly associated open-ended responses with each of the four estimated topics for the question ``Did you look at the bets made by other players in your group? Did you
find them useful/not useful? Please elaborate.''}
\end{figure}

Figure \ref{app_fig:ex_responses_strategy} shows the labels of three estimated topics for the strategy question along with the actual responses that are estimated to be highly associated with each topic. The topics we labeled \emph{Other Strategies} encompass participants who either choose actions at random, ignoring signals and actions of others, or used heuristics that deviate from Bayesian updating. The topic we labeled \emph{Signals + Actions of Others} encompasses answers where participants emphasized using both signals and others' actions to make their choices. The topic labeled \emph{Signals} describes answers where participants said making their choices based on the aggregate number of green and red balls drawn throughout a game. Figures \ref{app_fig:ex_lookballs} and \ref{app_fig:ex_lookbets} show the labels and response examples for questions \emph{(ii)} and \emph{(iii)}, respectively. For these questions, the estimated topics capture the range from negative to positive reliance on signals and others' actions, respectively.  
    
\begin{figure}[h!]
\begin{center}
\begin{tabular}{ccc}
\scriptsize{Panel (a): Topic ``Signals''} & \scriptsize{Panel (b): Topic ``No Strategy''}   & \scriptsize{Panel (c): Topic ``Signals + Actions''}\\ 
\includegraphics[trim=6cm 6cm 6cm 6cm,clip = true,scale=0.9]{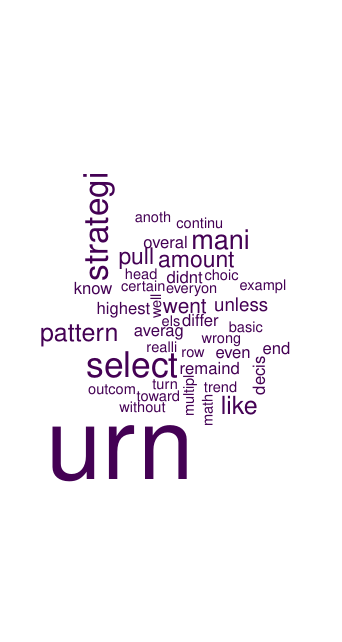}  &  \includegraphics[trim=4cm 4cm 4cm 4cm,clip = true,scale=0.58]{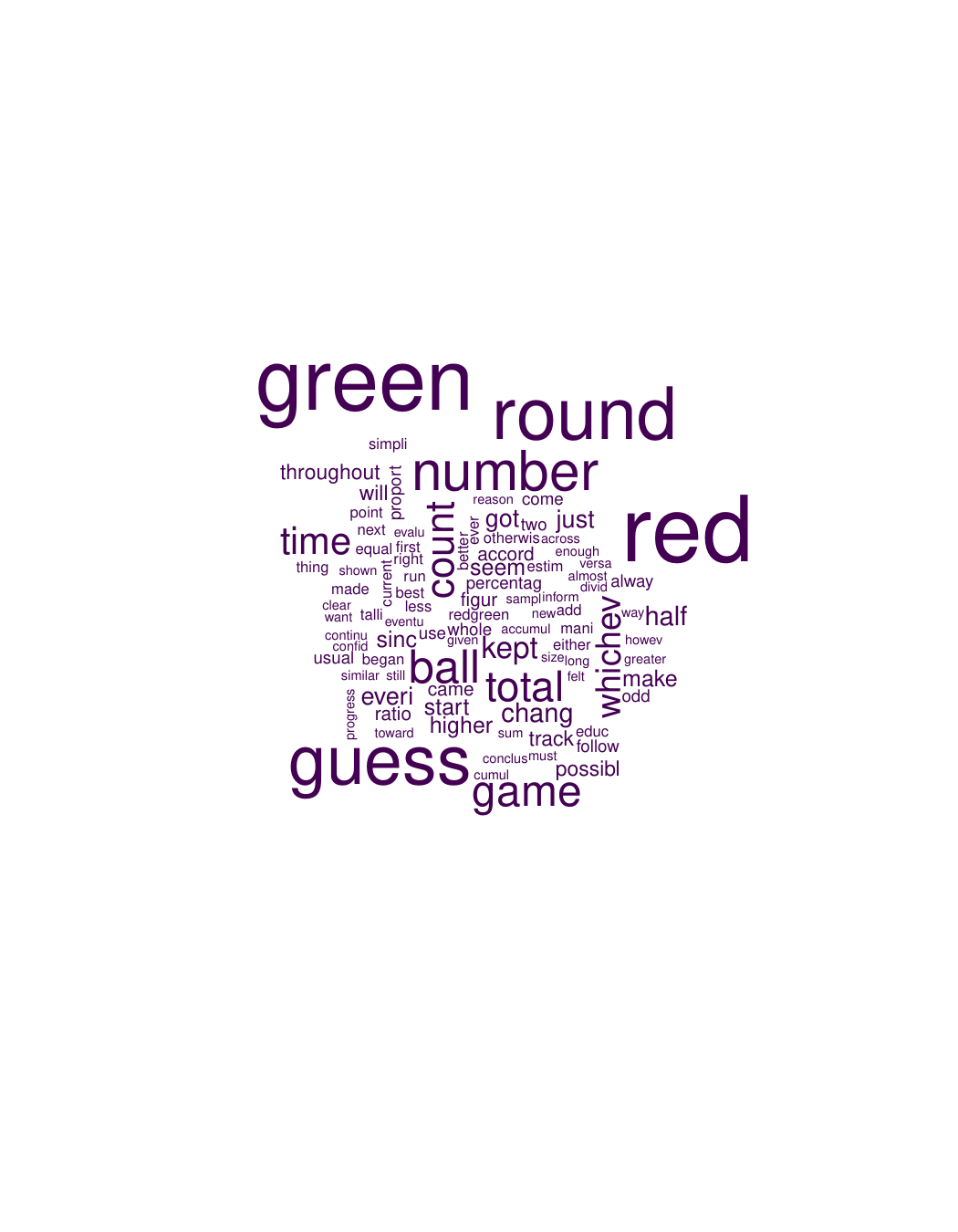} & \includegraphics[trim=5cm 6cm 4cm 4cm,clip = true,scale=0.72]{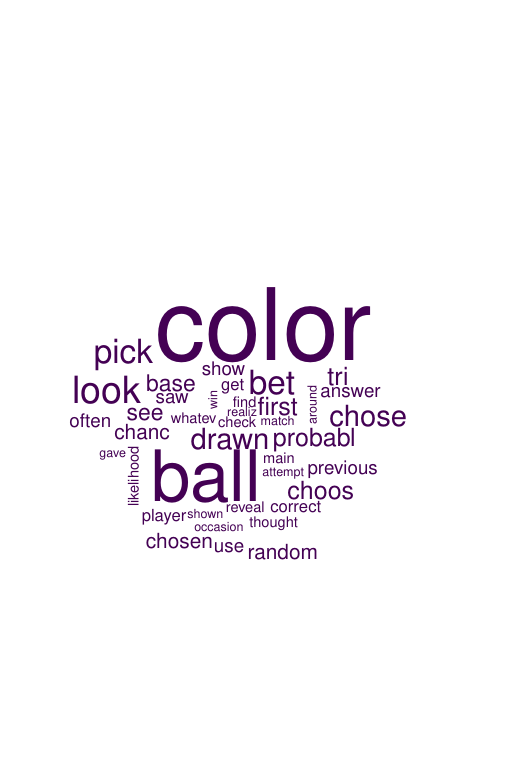}\\
\includegraphics[scale=0.31]{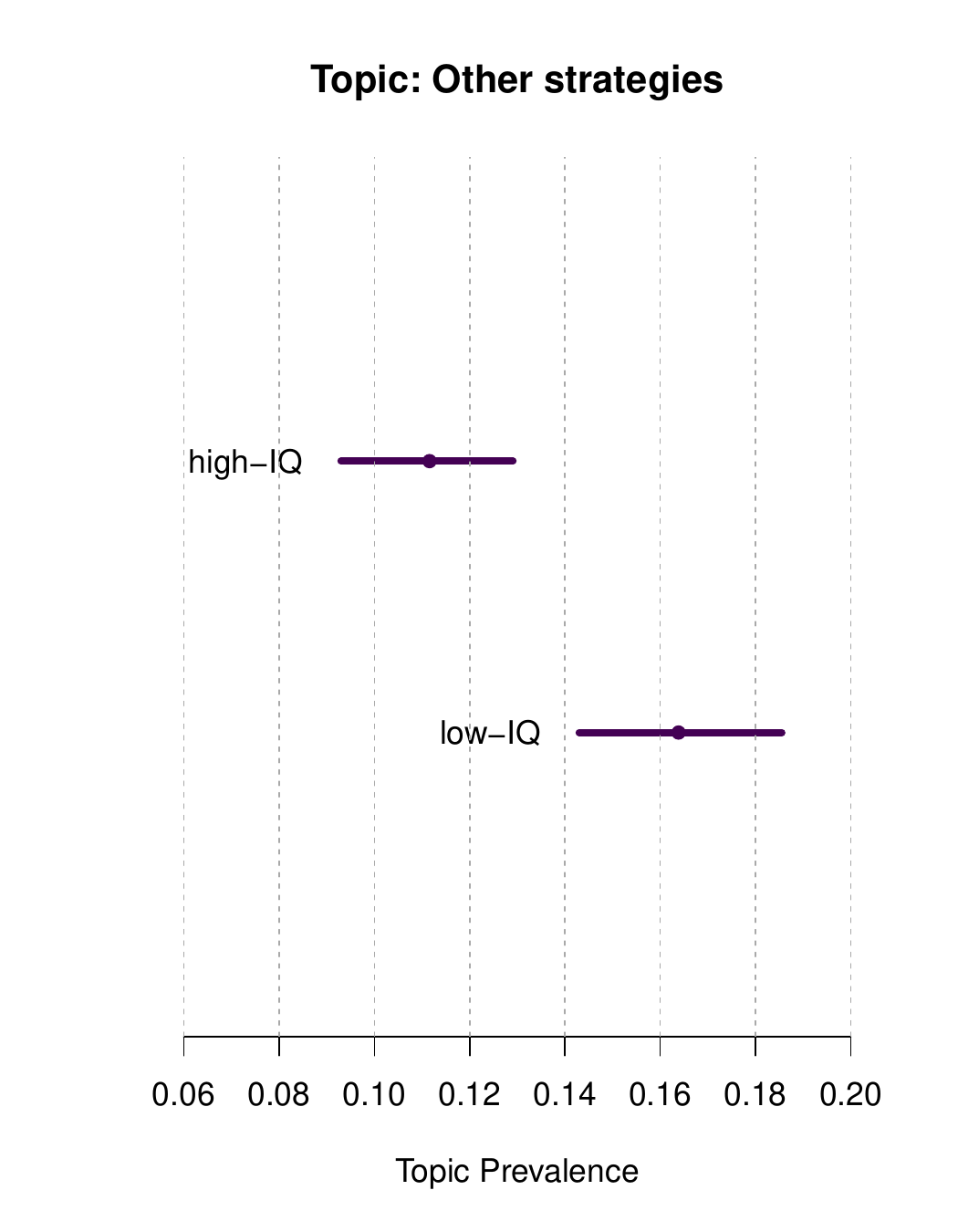}  &  \includegraphics[scale=0.31]{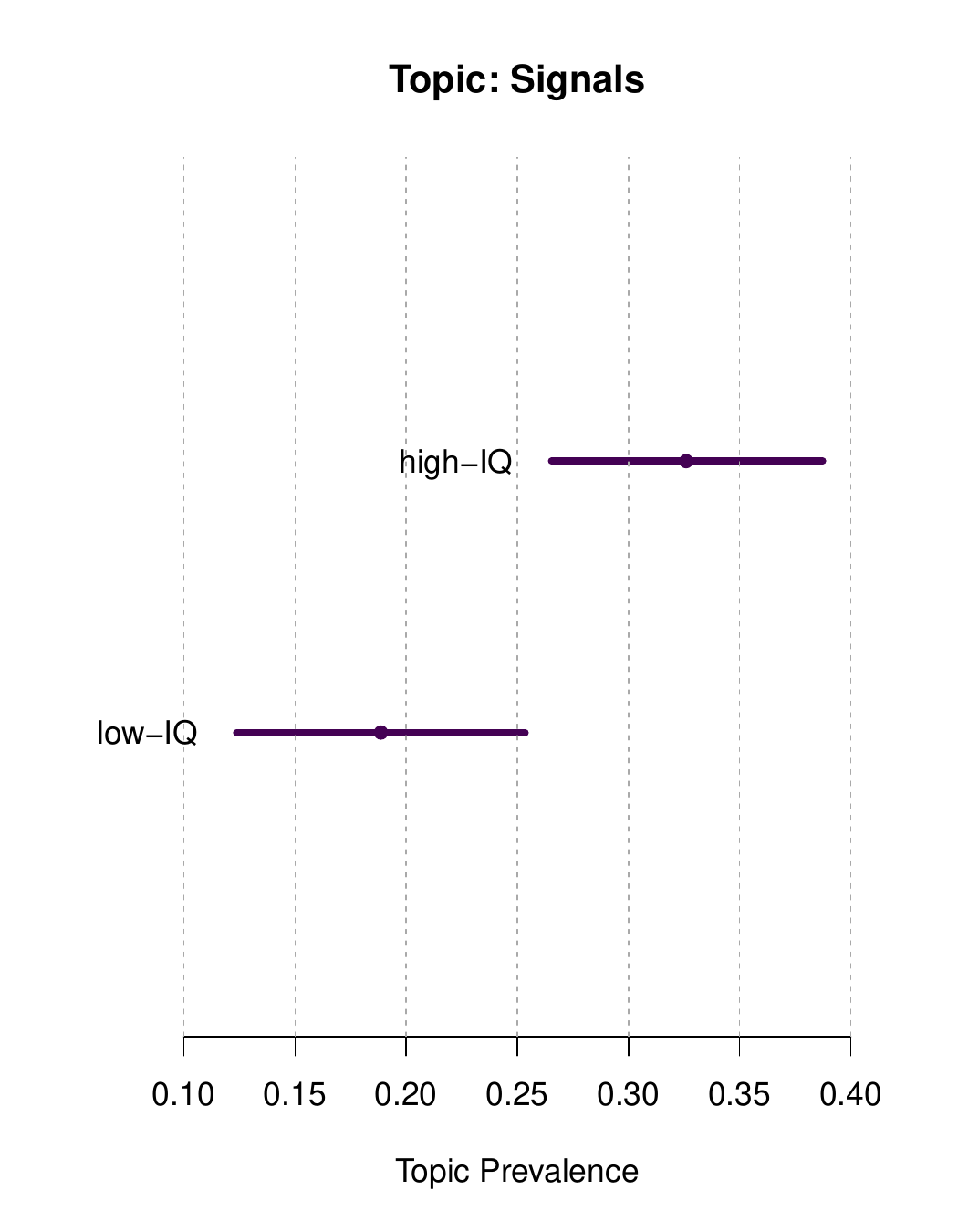} & \includegraphics[scale=0.31]{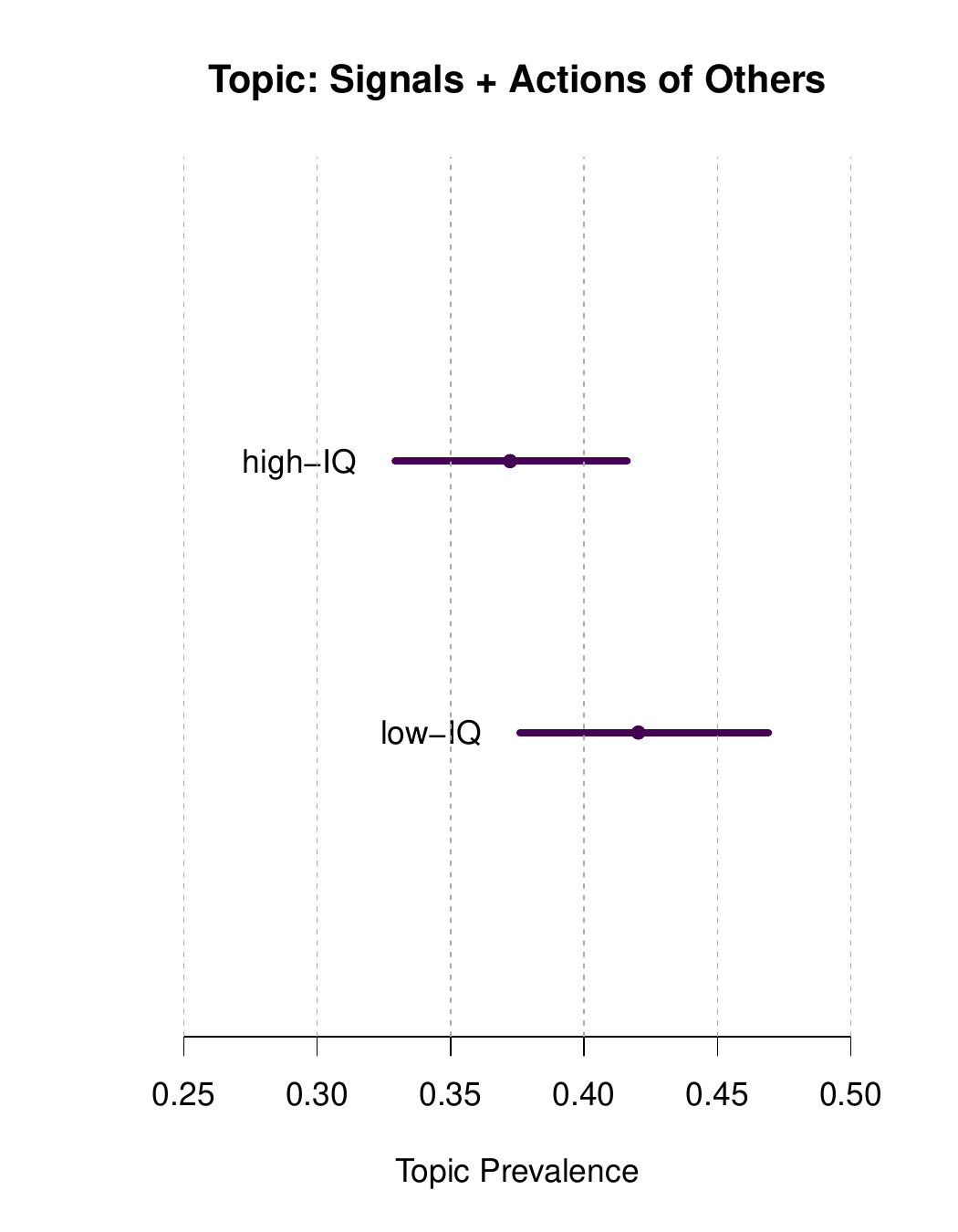}\\
\end{tabular}
\caption{Responses to ``What strategy did you use in the game (if any)?''  of low-IQ and high-IQ Participants ({\sc all4} and {\sc all} treatments)}\label{app_fig:prevalence_all} 
\end{center}
{\footnotesize \underline{Notes:} The upper row of each panel presents the word cloud of the top 100 words with the highest likelihood of being drawn from each topic, with the size of the word representing the magnitude of this likelihood. The lower row of each panel presents the estimated topic prevalence for the subgroups of low-IQ and high-IQ participants with $95 \%$ confidence intervals.}
\end{figure}

\begin{figure}[h!]
\begin{center}
\begin{tabular}{cc}
\scriptsize{Panel (a): Topic ``Sometimes/Somewhat Useful''} & \scriptsize{Panel (b): Yes, more information} \\ 
\includegraphics[trim=6cm 6cm 6cm 7cm,clip = true,scale=0.7]{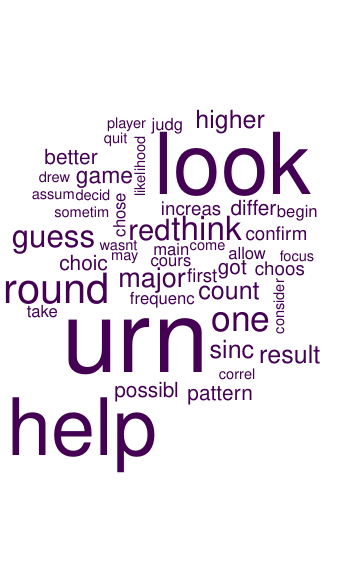}  &  \includegraphics[trim=5cm 7cm 5cm 7.3cm,clip = true,scale=0.8]{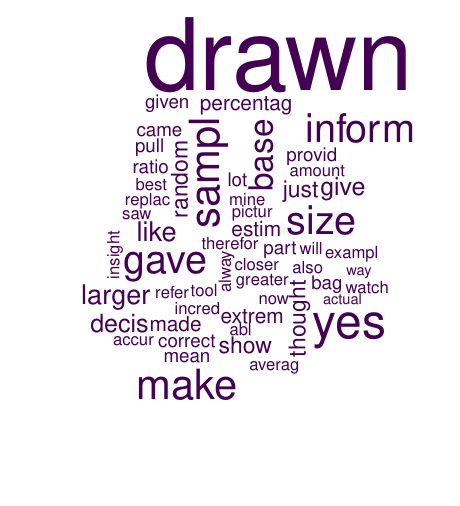}\\
\includegraphics[scale=0.35]{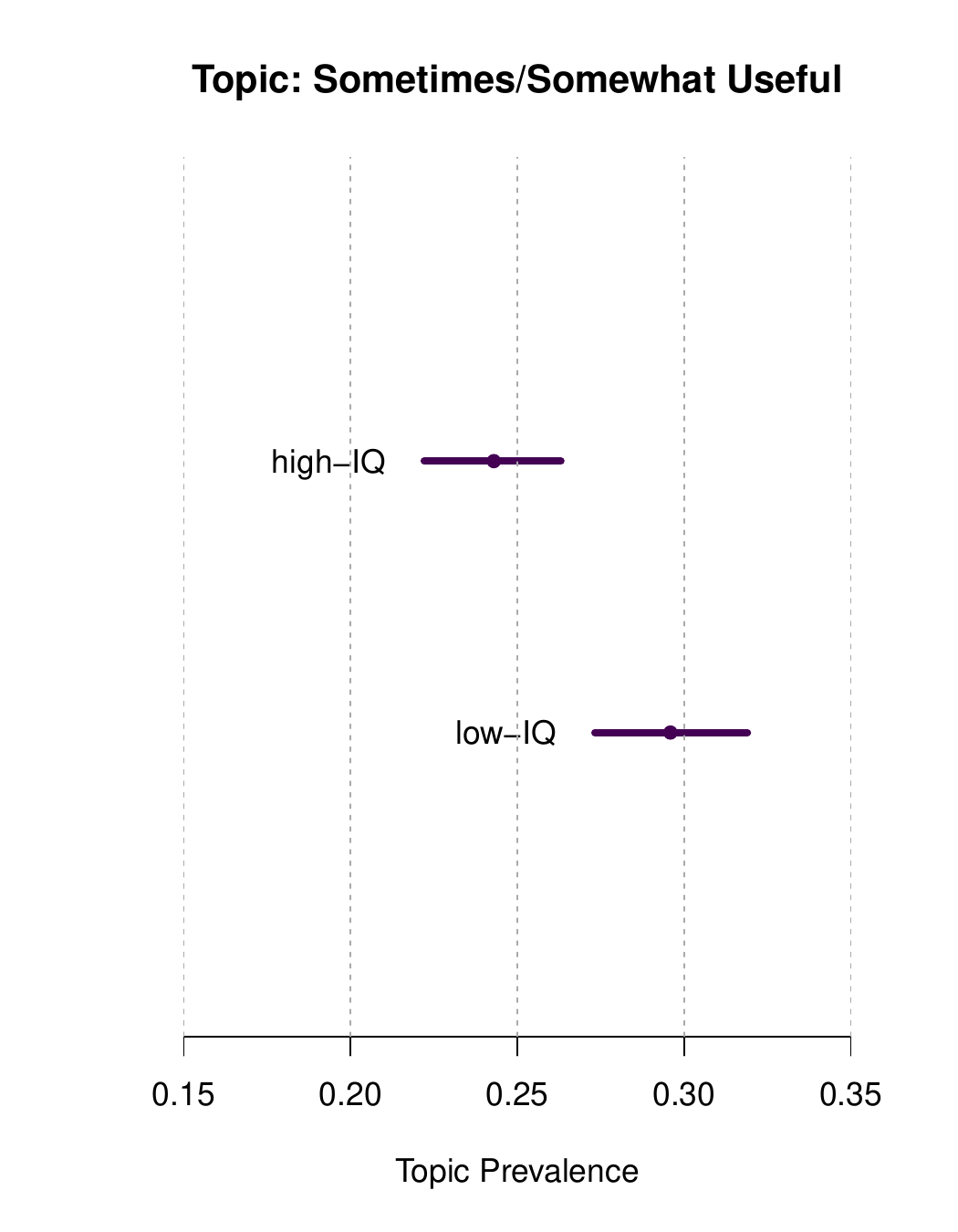}  &  \includegraphics[scale=0.35]{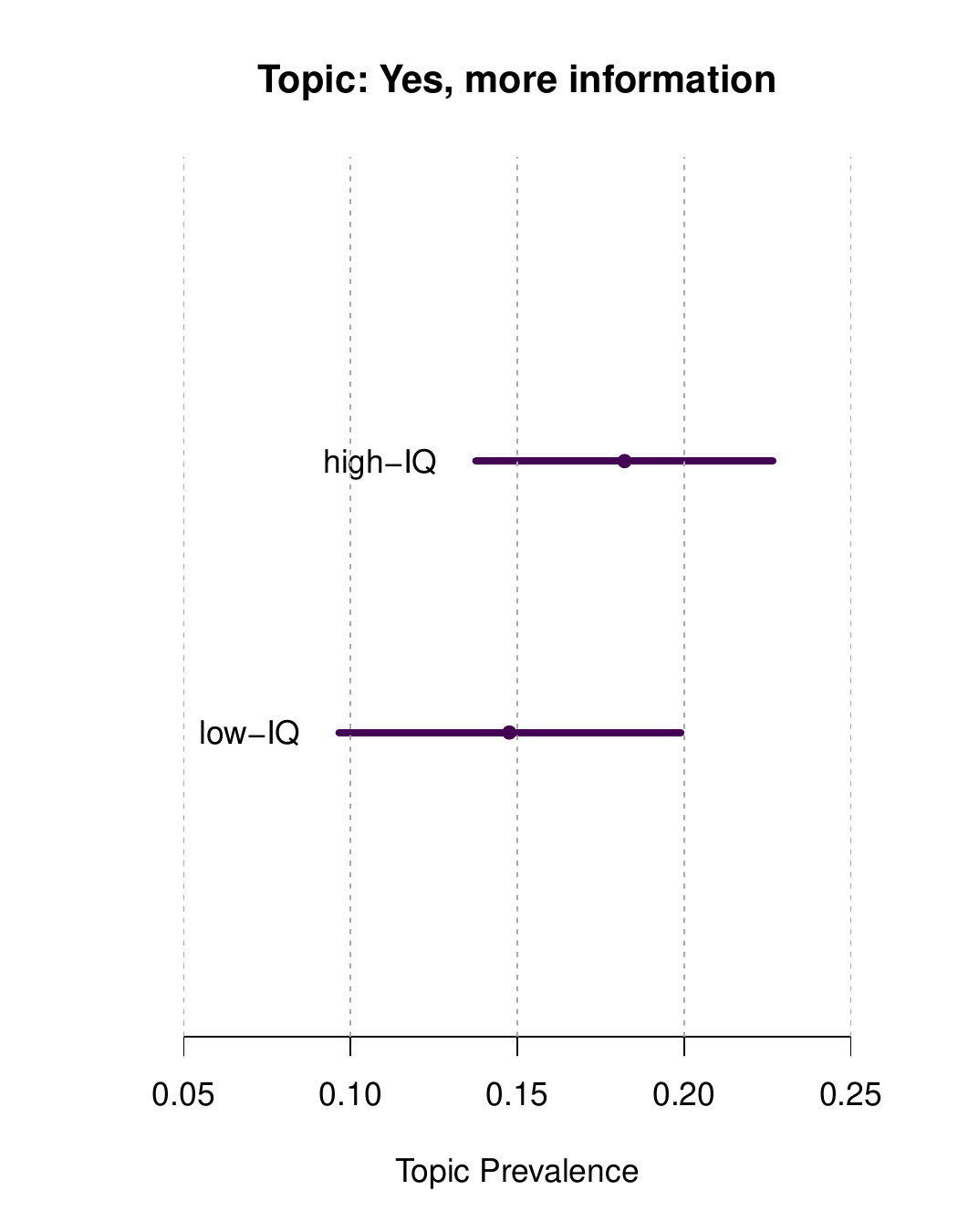}
\end{tabular}
\caption{Responses to ``Did you look at the balls drawn for other players in your group?''  of low-IQ and high-IQ Participants ({\sc all4} and {\sc all} treatments)\label{app_fig:prevalence_all_lookballs}} 
\end{center}
{\footnotesize \underline{Notes:} The upper row of each panel presents the word cloud of the top 100 words with the highest likelihood of being drawn from each topic, with the size of the word representing the magnitude of this likelihood. The lower row of each panel presents the estimated topic prevalence for the subgroups of low-IQ and high-IQ participants with $95 \%$ confidence intervals.}
\end{figure}

\begin{figure}[h!]
\begin{center}
\begin{tabular}{cc}
\scriptsize{Panel (a): Topic ``Yes, useful''} & \scriptsize{Panel (b): No, no extra information} \\ 
\includegraphics[trim=5.5cm 6cm 5cm 6cm,clip = true,scale=0.9]{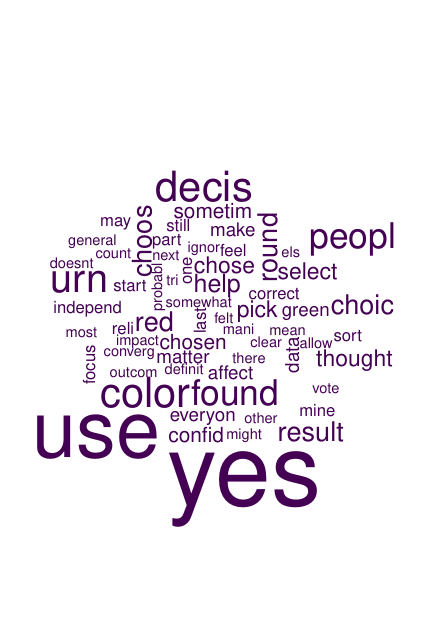}  &  \includegraphics[trim=5cm 6cm 5cm 8cm,clip = true,scale=0.9]{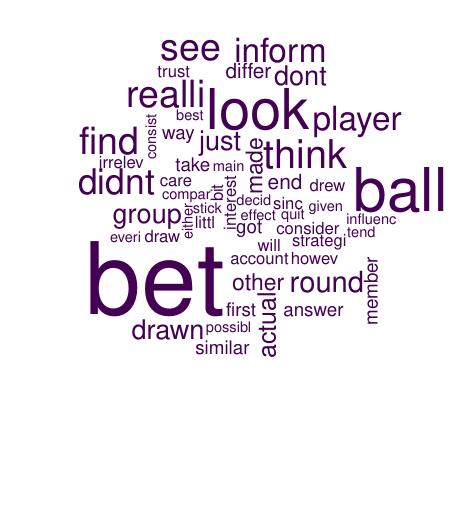}\\
\includegraphics[scale=0.35]{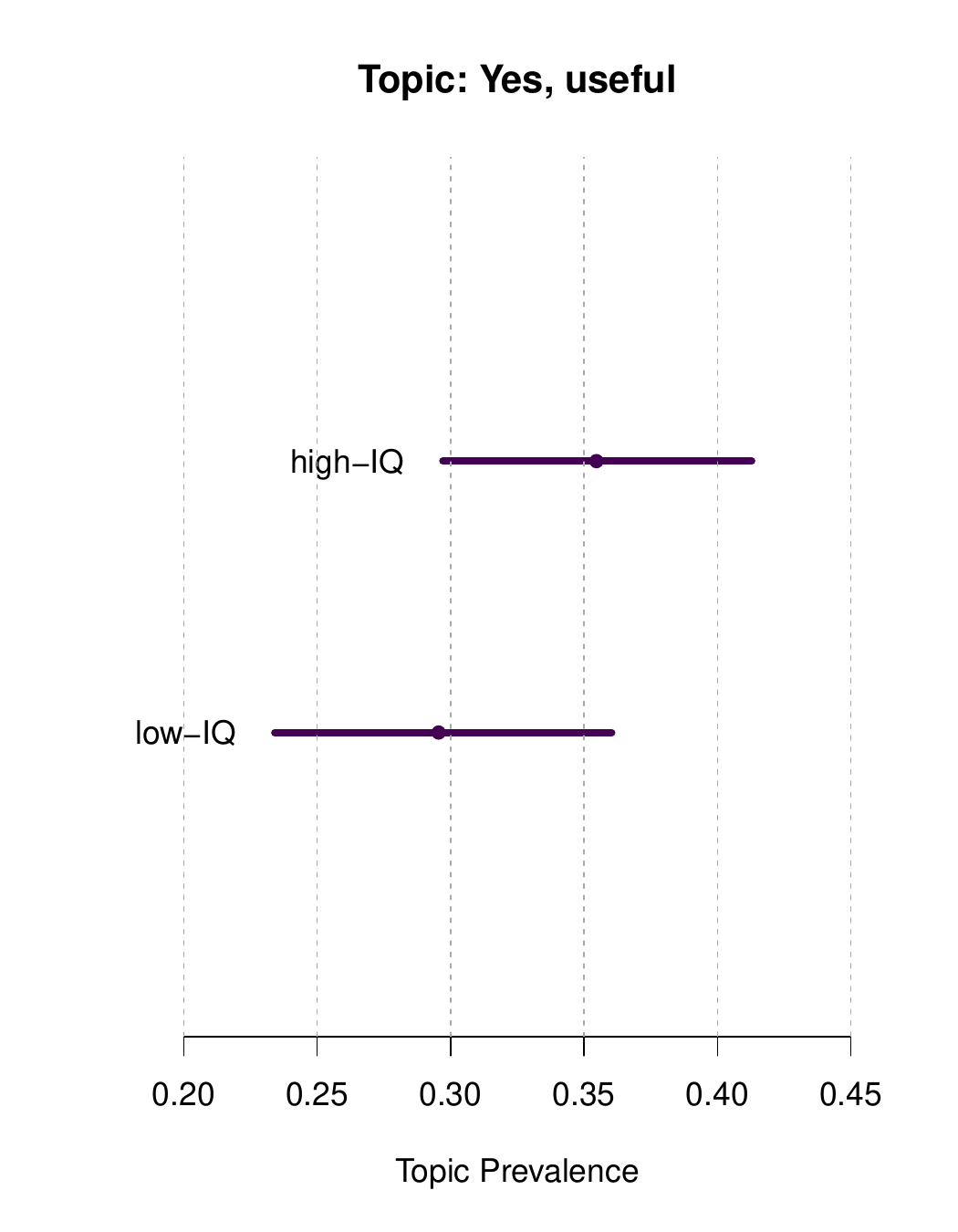}  &  \includegraphics[scale=0.35]{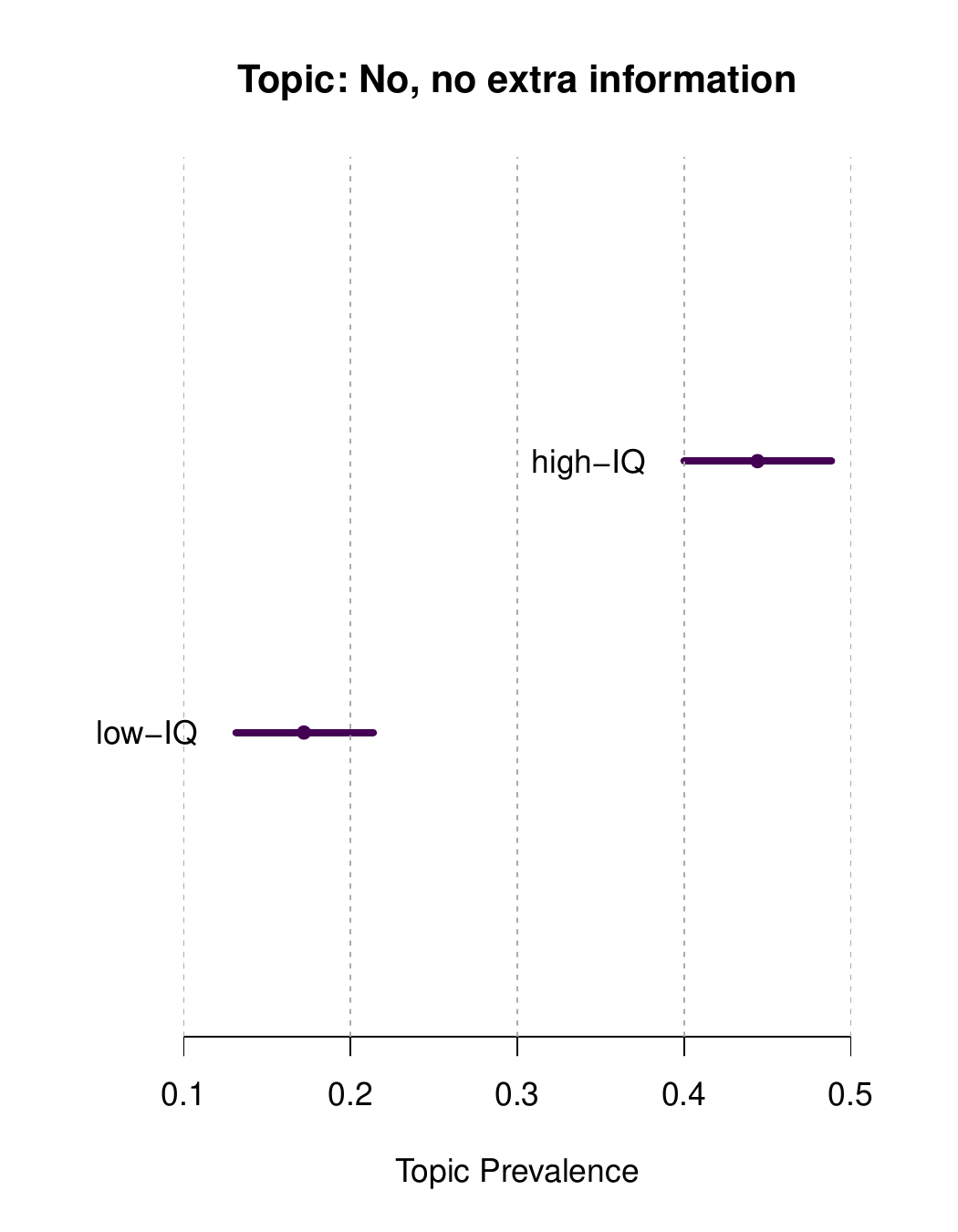}\\
\end{tabular}
\caption{Responses to ``Did you look at the bets made by other players in your group?''  of low-IQ and high-IQ Participants ({\sc all4} and {\sc all} treatments)}\label{app_fig:prevalence_all_lookbets} 
\end{center}
{\footnotesize \underline{Notes:} The upper row of each panel presents the word cloud of the top 100 words with the highest likelihood of being drawn from each topic, with the size of the word representing the magnitude of this likelihood. The lower row of each panel presents the estimated topic prevalence for the subgroups of low-IQ and high-IQ participants with 95\% confidence intervals.}
\end{figure}

Figure \ref{app_fig:prevalence_all} shows for each estimated topic in the strategy question \emph{(i)}, the relationship between our measure of IQ and the  estimates of topic prevalence, $\mathbf{\theta}_{r_q}$. We capture each topic with a word cloud of the top words associated with it. Figures \ref{app_fig:prevalence_all_lookballs} and \ref{app_fig:prevalence_all_lookbets} present these estimates for questions \emph{(ii)} and \emph{(iii)}, respectively.
    
We find that participants' description of their strategies is consistent with their actual behavior in the game according to their IQ type. In particular, high-IQ participants, who appear closer to Bayesian behavior in the game (see Figure \ref{fig:RespSignals}), described using strategies based on signals to a larger extent than low-IQ ones ($14 \%$ larger with $p<0.001$, see panel (a) in Figure \ref{app_fig:prevalence_all}) who, in turn, relied significantly more on heuristics which, in many cases, disregarded raw signals as well as actions of others (see panel (b) in Figure \ref{app_fig:prevalence_all}). Moreover, when directly asked ``Did you look at the balls drawn for other players in your group? Did you find them useful/not useful?'', low-IQ participants were more skeptical about the information provided by other players' signals, compared to high-IQ players ($5 \%$ more with $p<0.001$)

In terms of  explicitly looking at the actions of others, the topic prevalence from question \emph{(i)}, which we labeled  ``Signals + Actions of Others'' shows that high-IQ participants describe other players' actions as useful in a similar proportion to low-IQ ones ($37 \%$ for high-IQ participants versus $42 \%$ for low-IQ ones ($p=0.055$), see panel (c) in Figure \ref{app_fig:prevalence_all}). Moreover, the analysis of the answers to ```Did you look at the bets made by other players in your group? Did you find them useful/not useful?'' shows that high-IQ participants describe other players' actions as useful similarly to low-IQ ones ($36 \%$ \emph{vs} $29 \%$ with $p =0.07$, see panel (a) of Figure \ref{app_fig:prevalence_all_lookbets}), while low-IQ players described the social signal as sometimes/somewhat useful to a larger extent than high-IQ ones ($4 \%$ and $30 \%$ more ($p<0.05$) for topics ``Sometimes/Somewhat Useful'', respectively). Overall, these answers are consistent with a similar responsiveness to others' actions by IQ types found in the game data (see Figure \ref{fig:RespSignals}). At the same time, more than two thirds of those participants who find the information in others' actions redundant come from the group of high-IQ players (see panel (b) in Figure \ref{app_fig:prevalence_all_lookbets}), which is consistent with the behavior of a subset of high-IQ participants who take decisions consistent with Bayesian updating.

\section{Participants' Beliefs \label{app:beliefs}}
\setcounter{figure}{0}
\setcounter{table}{0}   
 
 We analyze the precision of subjects' incentivized beliefs, as
 elicited at the end of the experiment (see section \ref{app_beliefs}
 for details). For the analysis, we pool all participants' answers
 across treatments.\footnote{We do not have data for subjects in the
   first session of each treatment, because it was used to collect
   data and calibrate subjects’ payments in subsequent sessions.} In
 the beliefs section, each participant provides an estimate of the
 fraction of correct actions (in the last round of the game) in both
 her own treatment and other treatments. We combine these answers and
 define participant $i$'s precision as her mean squared error across
 all treatments: $MSE_i =
 \sum_{treat}(y_{treat}-\hat{y}_{i,treat})^2$, where $y_{treat}$
 denotes the actual fraction of correct actions in treatment $treat
 \in \{\text{{\sc no info,actions4,actions,signals,all4,all}\}}$ and $\hat{y}_{i,treat}$ denotes participant $i$'s estimate about the fraction of correct actions in treatment $treat$. Both actual values and beliefs are discretized in 10 equally spaced bins from $0 \%$ to $100 \%$, consistent with subjects' answers.
 
  To assess the difference between low-IQ and high-IQ subjects on their beliefs' precision, we estimate a regression of the form:
  \begin{equation*}
      MSE_i=\delta_{treat,i} + \beta_1 \cdot \text{\texttt{low-IQ}} + X_i^{'}\gamma +\epsilon_{i},
  \end{equation*}
 where \texttt{low-IQ} is a dummy variable that takes a value of one if a subject is low-IQ according to our IQ auxiliary measure and 0 if she is high-IQ. $\delta_{treat,i}$ is a fixed effect of the treatment assigned to subject $i$. $\mathbf{X}_i$ is a vector of participant $i$'s characteristics as defined in Appendix \ref{app:lda} including: \texttt{female}, \texttt{stem}, \texttt{overconfidence} and \texttt{risk}. $\epsilon_{i}$ denotes the error term with a variance-covariance matrix clustered by session. 
 
\begin{figure}[h!]
\begin{center}
\includegraphics[scale=0.41]{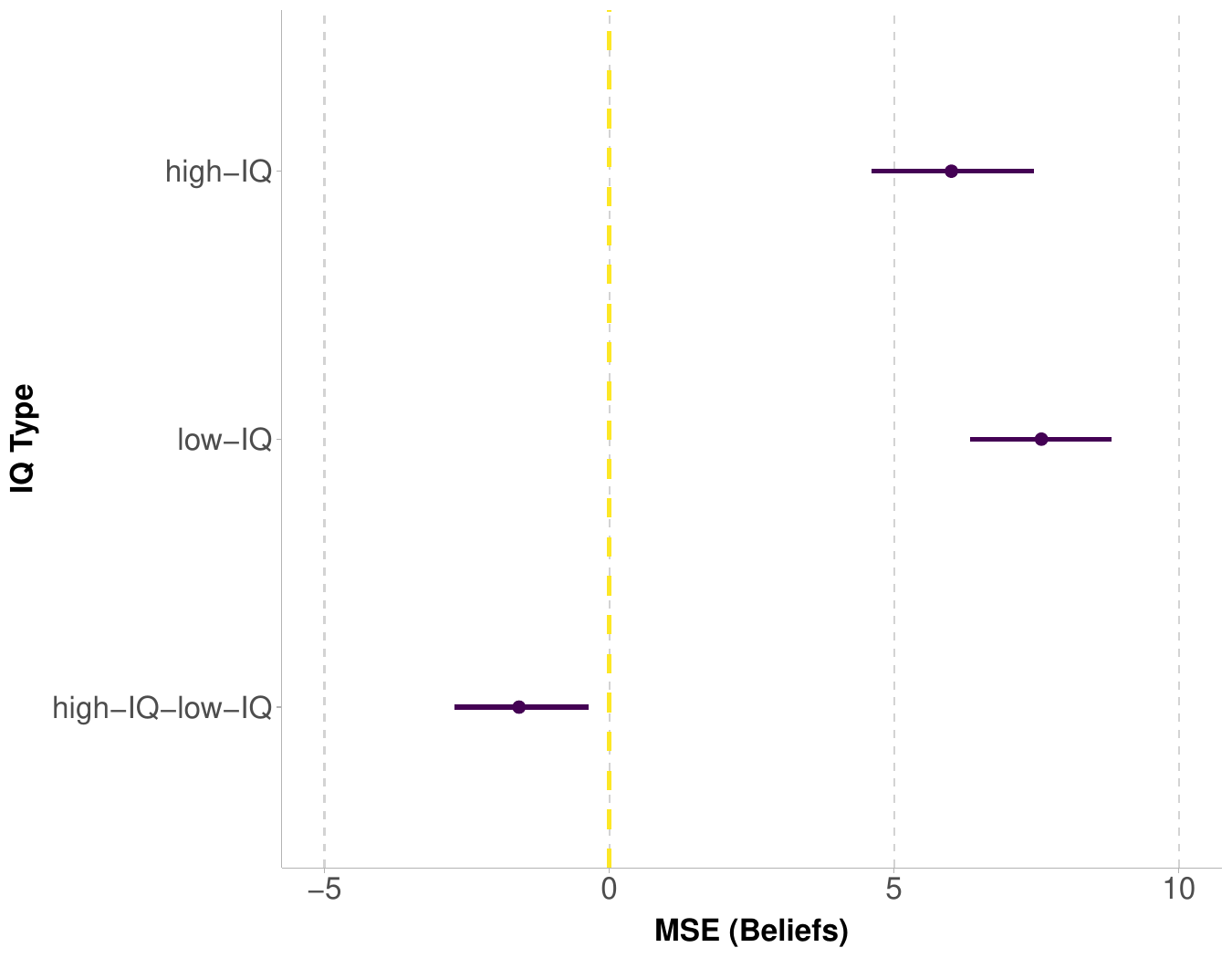}   
\caption{Beliefs' Accuracy and IQ Types \label{app_fig:beliefs_naive}}
\end{center}
{\footnotesize \underline{Notes:} The figure shows predicted mean squared errors for the beliefs of low-IQ and high-IQ participants along with $95 \%$ confidence intervals. For these estimated effects we use the estimated coefficients from column (3) in Table \ref{table_mse_iq}. We include as controls: \texttt{female}, \texttt{stem}, \texttt{overconfidence} and \texttt{risk} and set them to their median values in the data.}
\end{figure}

 The results of estimating the difference between low-IQ and high-IQ subjects on their beliefs' precision are presented in Table \ref{table_mse_iq} and visually in Figure \ref{app_fig:beliefs_naive}. Across all specifications, we find that low-IQ subjects are significantly less precise than their high-IQ counterparts. In particular, controlling for both treatment effects and participant characteristics (column (3) of Table \ref{table_mse_iq}), this effect translates into low-IQ participants miscalculating the actual fraction of correct actions by more than $10 \%$ (i.e., $\sqrt{1.58}=1.26$ bins of $10 \%$ increments each). 
 
\begin{table}[!htbp] \centering 
\footnotesize
  \caption{Precision of Beliefs (MSE) by IQ Type ({\sc all} treatments)} 
  \label{table_mse_iq} 
\begin{tabular}{@{\extracolsep{5pt}}lccc} 
\\[-1.8ex]\hline 
\hline \\[-1.8ex] 
 & \multicolumn{3}{c}{\textit{Dependent variable:}} \\ 
\cline{2-4} 
\\[-1.8ex] & \multicolumn{3}{c}{MSE} \\ 
\\[-1.8ex] & (1) & (2) & (3)\\ 
\hline \\[-1.8ex] 
 Intercept & 5.281$^{***}$ & 3.428$^{***}$ & 3.770$^{***}$ \\ 
  & (0.414) & (0.283) & (0.818) \\ 
 \texttt{low-IQ} & 1.097$^{**}$ & 1.088$^{**}$ & 1.580$^{***}$ \\ 
  & (0.537) & (0.545) & (0.602) \\ 
  & & & \\ 
\hline \\[-1.8ex] 
Treatment Fixed Effects & No & Yes & Yes \\ 
Participant Covariates & No & No & Yes \\ 
Observations & 476 & 476 & 476 \\ 
Adjusted R$^{2}$ & 0.005 & 0.012 & 0.013 \\ 
\hline 
\hline \\[-1.8ex] 
\end{tabular} 
\vspace{1mm}

{\footnotesize \underline{Notes:} $^{*}$p$<$0.1; $^{**}$p$<$0.05; $^{***}$p$<$0.01. Clustered standard errors by session in parentheses}
\end{table} 

 \clearpage
 
 \begin{center}
   {\huge Online Appendix}
 \end{center}

 \renewcommand\thefigure{\thesection.\arabic{figure}}
\renewcommand\thetable{\thesection.\arabic{table}}   

\section{Structure of the {\sc all} treatment and payments
\label{app_payments}}
\setcounter{figure}{0}
\setcounter{table}{0}   
Each participant earns \$10 for completing the session (participation fee). In addition, subjects earn money for other parts as described below.

\begin{itemize}
\item Part I: Main game 
\begin{itemize}
    \item 10 games with 20 rounds in each game
    \item random re-matching into groups of 8 subjects at the beginning of a game
    \item random round from a random game determines participants’ payments for this part: correct guess pays \$20, wrong guess pays \$5
\end{itemize} 
\item Part II: Beliefs 
\begin{itemize}
\item one questions is randomly selected for payment\footnote{For the {\sc no info}, {\sc actions}, and {\sc all} treatment we had 3 beliefs questions corresponding to these three information structures. In the {\sc signals} treatment, we had 4 beliefs questions each corresponding to one of the all four information structures.}, \$5 for a correct answer
\item questions appear in a random order for each subject
\item in the first session of each treatment, this part was not present because we were collecting the data used in the next sessions for subjects' payments. 
\end{itemize}
\item Part III: Risk Attitudes
\begin{itemize}
\item one of the two investment tasks is randomly selected for payment
\item points earned are converted into dollars using the rate 1 point $=$ 1 cent
\end{itemize}
\item Part IV: IQ and Overconfidence
\begin{itemize}
    \item six matrices for IQ measure, each correctly solved matrix earns 50 cents
    \item overconfidence is measured using two related dimensions: over-estimation and over-placement
    \item correct prediction for over-estimation task earns 50 cents
    \item in the first session of each treatment (those without beliefs questions) over-placement question is not incentivized, while in the remaining sessions it is and correct prediction is rewarded by 50 cents\footnote{The first session is used to collect data for comparing actual rank of students to that reported for the follow-up sessions.}
\end{itemize}
\item Part V: Open-ended Questionnaire for Strategies
\end{itemize}

\section{Instructions for {\sc all} treatment \label{app_instructions}}
\setcounter{figure}{0}
\setcounter{table}{0}   

\paragraph{Welcome.} You are about to participate in an experiment on decision-making. You will be paid for your participation in cash privately at the end of the session. Please turn off all electronic devices, especially phones. During the experiment you are not allowed to open or use any other applications on these laboratory computers, except for the interface of the experiment. 

The experiment consists of four parts: Part I of the experiment is the main and the longest part. The other parts (Part II, III and IV) are short. You will receive the instructions for each part of the experiment before that part begins. You have already earned \$7 for coming to the lab. In addition, you can earn money in each part of the experiment. The instructions for each part of the experiment will be very precise about that. 

\subsection*{Part I}

Part I of the experiment consists of 10 games. Each game consists of 20 rounds. Before the beginning of each game, you will be randomly assigned to a group of 8 players.  You will play all 20 rounds of the game with the same group of people. At the end of each game, you will again be randomly assigned to a new group of 8 players and will play 20 rounds with them, and so on. 

At the beginning of each game, the computer randomly selects one of the two Urns for your group for this game (independent of the urns selected in previous games): 

\begin{itemize}
    \item RED URN contains 6 RED balls and 4 GREEN balls
    \item GREEN URN contains 4 RED balls and 6 GREEN balls
\end{itemize}       

That is, there is 50\% chance that the RED URN is selected, in which case the urn has 6 RED balls and 4 GREEN balls and 50\% chance that the GREEN URN is selected, in which case the URN has 6 GREEN balls and 4 RED balls. We will refer to the selected urn as the URN.

The composition of the URN is determined once at the beginning of each game (before round 1) and stays the same throughout the game (in all 20 rounds). All players that were assigned to the same group share the same composition of the URN. At the beginning of a new game, after players are assigned to new groups but before round 1 begins, the computer again determines the composition of the URN for each group separately using the rule described above. Thus, the composition of the URN in your group stays the same in all 20 rounds of the same game but is not related to its composition in later games. 

In every round of a game, your task is to guess (bet on) the URN selected for your group at the beginning of this game. If you choose to bet on RED URN, it means that you are betting that the RED URN was selected for your group at the beginning of this game. If you choose to bet on GREEN URN, it means that you are betting that the GREEN URN was selected for your group at the beginning of this game. At the end of each round, one ball will be randomly drawn from the selected URN and its color revealed to you. The same thing happens with all other players in your group: each player bets on the URN selected for your group and then observes one randomly drawn ball from the selected URN. It is important that different players observe different draws, but that all drawn are made from the same URN.

\paragraph{Your payment in Part I.} To determine your payment in Part I, at the end of the experiment, the computer will select one game from the 10 games played. Each game is equally likely to be chosen for payment. Then the computer will select one of the 20 rounds in the selected game. If you guessed correctly the URN in the selected round of the selected game, then you will receive \$20. If your guess (bet) was wrong then you will receive \$5.

\paragraph{End-of-Round Information.} At the end of each round, you will be reminded of the URN you betted on and you will observe the color of the randomly drawn ball from the URN. Moreover, at the end of each round you will also observe the bets that other players in your group made in this round regarding the selected URN and the balls that were drawn from the selected URN for other players. This information will be summarized in the table on your screen. This table will keep track of all decision that you have made in this game (highlighted in yellow), as well as the decisions made by other players in your group. 

Note that the computer draws one ball for each member of your group in each round from the URN with replacement. That is, every ball that is drawn from the URN is placed back in the URN before the next draw. Therefore, if the URN contains, say, 6 RED balls and 4 GREEN balls then for each player in your group there is exactly 60\% chance that the drawn ball is RED and 40\% chance that it is GREEN. 

Are there any questions?

\subsection{Screenshots for {\sc all} treatment \label{app_screen}}

\begin{figure}[t]
\begin{center} 
\caption{Screenshot 1 from All treatment}\label{fig:screen1}
 \includegraphics[scale=0.35]{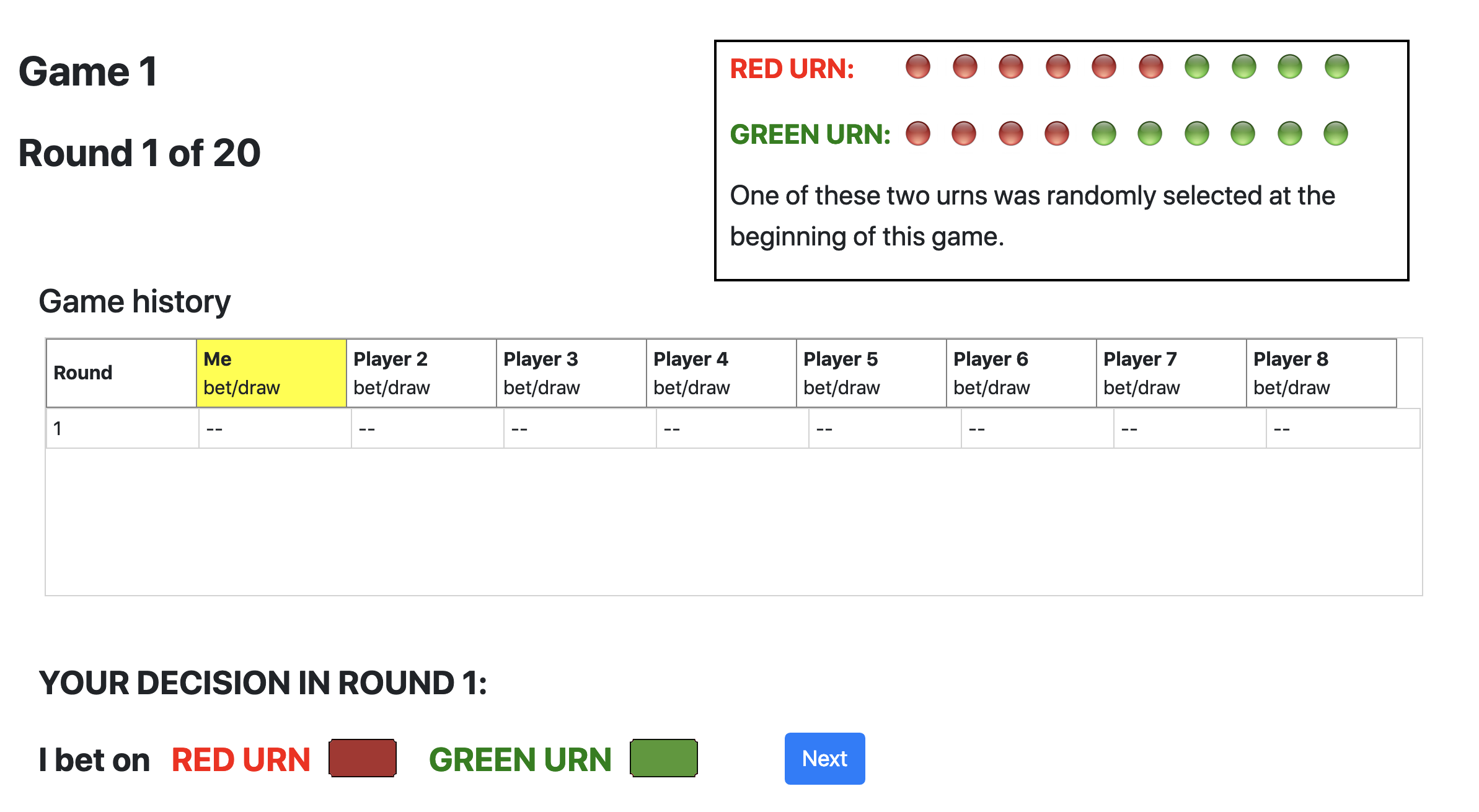}
\end{center} 
\end{figure}

This screenshot presented in Figure \ref{fig:screen1} shows round 1 of a game. On the top of the screen, subjects are reminded about the compositions of two urns. The game history table keeps track of all what has transpired in the current game, and the bottom of the screen is where subjects make their bets about which urn was selected for their group for this game. 

\begin{figure}[t]
\begin{center} 
\caption{Screenshot 2 from All treatment}\label{fig:screen2}
 \includegraphics[scale=0.35]{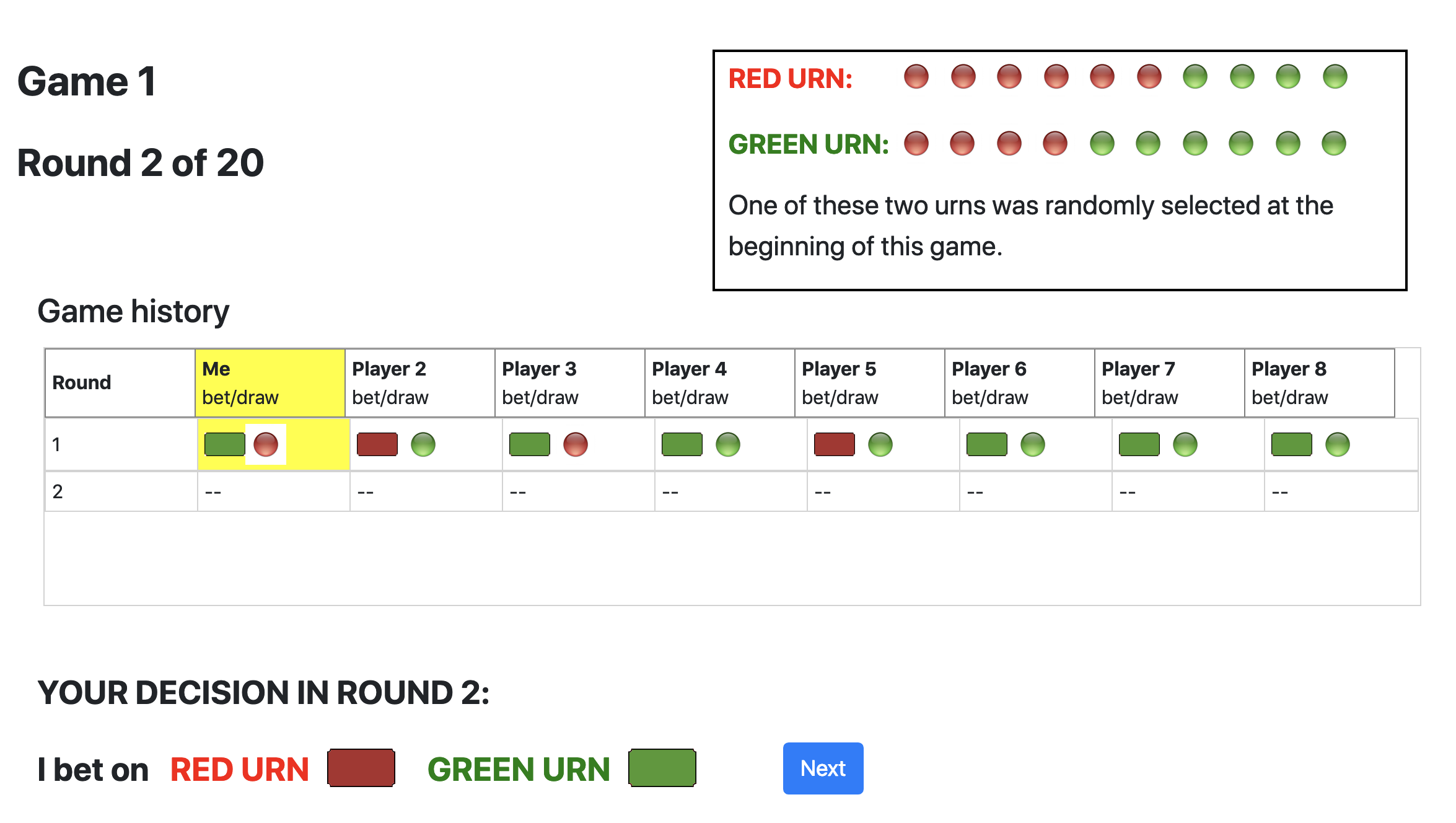}
\end{center} 
\end{figure}

The game history table starts filling up at the beginning of round 2 (as shown in Figure \ref{fig:screen2}), and it displays own bet and own draw as well as bets and draws of all group members (as seen on the second screenshot). Bets are always displayed as rectangles, i.e., lottery tickets, and signals are displayed as circles, i.e., balls drawn from the selected urn. 

\subsection{Strategy questions in {\sc all} treatment \label{app_strat}} 

Think about the main game in the experiment (betting on the urn selected for your group).
\begin{itemize}
\item What strategy did you use in the game (if any)? Please elaborate.
\item Did you look at the bets made by other players in your group? Did you find
them useful/not useful? Please elaborate.
\item Did you look at the balls drawn for other players in your group? Did you
find them useful/not useful? Please elaborate.
\item Was anything unclear about the game?
\item What is your gender?
\item What is your major?
\end{itemize}

\subsection{Beliefs questions in {\sc all} treatment \label{app_beliefs}} 
 
Below we present the three beliefs questions that subjects in the {\sc all} treatment were asked to answer. The formulation for the beliefs questions in the other treatments is adapted to the information structure subjects experienced in the treatment, but follow the same idea. The questions were presented in the random order across subjects. 

\textbf{Question 1.} In the past session, groups of 8 subjects played the same game that you just played. Namely, at the beginning of each game, the computer randomly selected one of the two urns: with probability 50\% the RED URN was selected, which contained 6 RED balls and 4 GREEN balls, and with probability 50\% the GREEN URN was selected, which contained 4 RED balls and 6 GREEN balls. Each game lasted for 20 rounds. In each round group members submitted their guesses (bets) about the URN selected for their group for this game. After making their choices, each group member observed the colors of the 8-randomly drawn balls from the selected urn (with replacement) and also guesses made by members of their group regarding the selected urn for this game. 

Please answer the following question:
What fraction of last round bets (bets submitted by subjects in round 20) were correct? The correct guess means that a subject guessed correctly which URN was selected for her group for this game.
\begin{center}
[[Radio buttons with ten options: 0\%-9\%, 10\%-19\%, ..., 90\% - 100\%]]
\end{center}
If this question is selected for payment, you will get paid \$5 if you chose correctly. 

\bigskip

\textbf{Question 2.} In the past session, groups of 8 subjects played game that was slightly different from the game you just played. Just like in your game, at the beginning of each game, the computer randomly selected one of the two urns: with probability 50\% the RED URN was selected, which contained 6 RED balls and 4 GREEN balls, and with probability 50\% the GREEN URN was selected, which contained 4 RED balls and 6 GREEN balls. Each game lasted for 20 rounds. In each round group members submitted their guesses (bets) about the URN selected for their group for this game. After making their choices, each group member observed the colors of the one randomly drawn ball from the selected urn and the guesses made by members of their group regarding the selected urn.

However, contrary to the game you just played, participants in these previous sessions did not observe the colors of the balls that were randomly drawn for other participants in their group. These past participants only observed the guesses that their group members made in each round and one randomly drawn ball from the selected urn.

Please answer the following question:
What fraction of last round bets (bets submitted by subjects in round 20) were correct? The correct guess means that a subject guessed correctly which URN was selected for her group for this game.

\begin{center}
[[Radio buttons with ten options: 0\%-9\%, 10\%-19\%, ..., 90\% - 100\%]]
\end{center}
If this question is selected for payment, you will get paid \$5 if you chose correctly. 

\bigskip

\textbf{Question 3.} In the past session, groups of n (parameter in the program) subjects played game that was slightly different from the game you just played. Just like in your game, at the beginning of each game, the computer randomly selected one of the two urns: with probability 50\% the RED URN was selected, which contained 6 RED balls and 4 GREEN balls, and with probability 50\% the GREEN URN was selected, which contained 4 RED balls and 6 GREEN balls. Each game lasted for 20 rounds. In each round group members submitted their guesses (bets) about the URN selected for their group for this game. After making their choices, each group member observed the colors of the one randomly drawn ball from the selected urn.

However, contrary to the game you just played, participants in this previous session were not provided with any additional information about the guesses made by other members of their group or the colors of the randomly drawn balls from the selected urn for other participants. The only information these past participants had was the color of one randomly drawn ball from the selected urn.

Please answer the following question:
What fraction of last round bets (bets submitted by subjects in round 20) were correct? The correct guess means that a subject guessed correctly which URN was selected for her group for this game.
\begin{center}
[[Radio buttons with ten options: 0\%-9\%, 10\%-19\%, ..., 90\% - 100\%]]
\end{center}
If this question is selected for payment, you will get paid \$5 if you chose correctly.

\subsection{Other Control Tasks \label{app_controls}}

\paragraph{Risk Attitudes.} Risk attitudes were measured using two investment tasks, in each of which subjects were endowed with 200 points (worth a total of \$2), any portion of which they could choose to invest in a risky project. In the first investment task, the risky project was successful 50\% of the time and had a return of 2.5 points for each point invested in it, while in the second investment task the risky project was successful 40\% of the times and returned 3 points for each point invested in it. Points not invested in the risky project had a return of 1 to 1 point. One of these two tasks was randomly selected for payment. This is one of the standard  methods used in the experimental literature to elicit subjects’ attitudes towards risk (see Gneezy and Potters (1997) and Charness, Gneezy, and Imas (2013)).  Administering this task twice with two sets of parameters allows to reduce measurement error (see ORIV technique developed by Gillen, Snowberg, and Yariv (2018)).

\paragraph{IQ and Overconfidence.} Subjects were asked to solve six matrices from the ICAR database (see ICAR, Condon and Revelle (2014)). Subjects earned 50 cents for each correctly solved matrix. So, the IQ of a subject is measured by the number of correctly solved matrices with the smallest number being 0 and the largest number being six. After solving these matrices, we asked subjects two questions:

\begin{enumerate}
    \item How many of the six puzzles do you think you correctly answered? You will receive 50 cents if you answer correctly.
    \item Now, think about 100 UCSD students. Where do you think you rank in terms of how many correct rotation cubes puzzles you got? For example, if you think you got the most correct, you should answer 1, while if you think you got the least correct, you should answer 100.  
\end{enumerate}

The last two questions are used to measure overconfidence of subjects. We chose to measure over-estimation and over-placement of subjects. Specifically, overestimation is the difference between how many ICAR questions a subject thinks she solved correctly minus how many she actually solved correctly, while the over-placement is the reported rank minus actual rank in a sample of the 100 randomly selected students. These measures are quite standard in the literature (see Chapman et al. 2019b).

\section{Physical \emph{vs} Virtual Lab \label{sec:AppOnlineSessions}}
\setcounter{figure}{0}
\setcounter{table}{0}   

We compare outcomes and behavior observed in the {\sc all} treatment across sessions conducted in the physical lab at UCSD and in the virtual lab at OSU. For each location we have four sessions with 64 subjects all together at UCSD and 88 subjects at OSU. In both locations, we used the standard subjects' pool of undergraduate students. 

\begin{figure}[h!]
\begin{center}
\caption{Aggregate statistics in the {\sc all} treatment}\label{fig:CorrectOnline} 
\begin{tabular}{cc}
\scriptsize{Correct Guesses, by round} & \scriptsize{Consensus, by round}
\\ 
\includegraphics[scale=0.3]{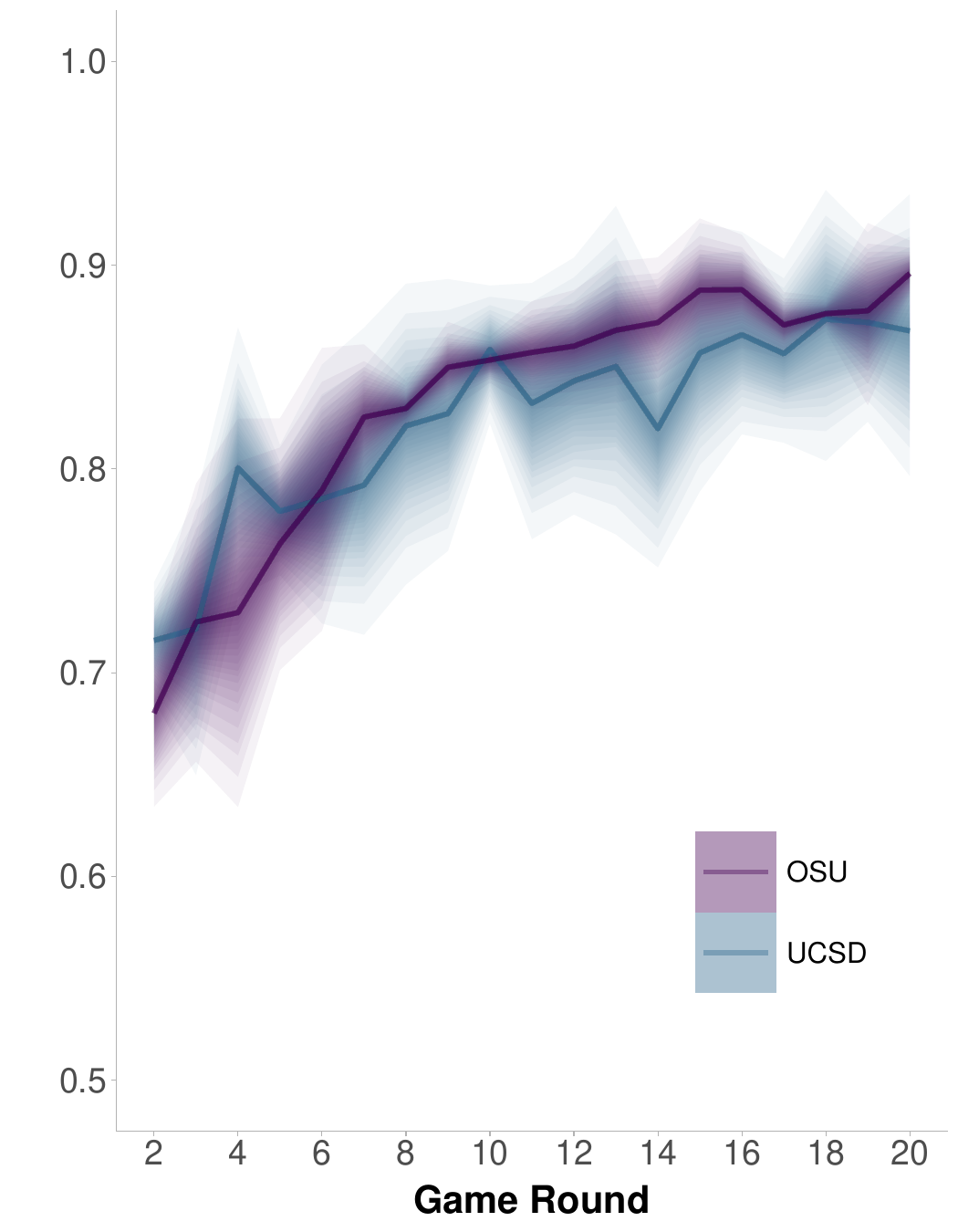} & \includegraphics[scale=0.3]{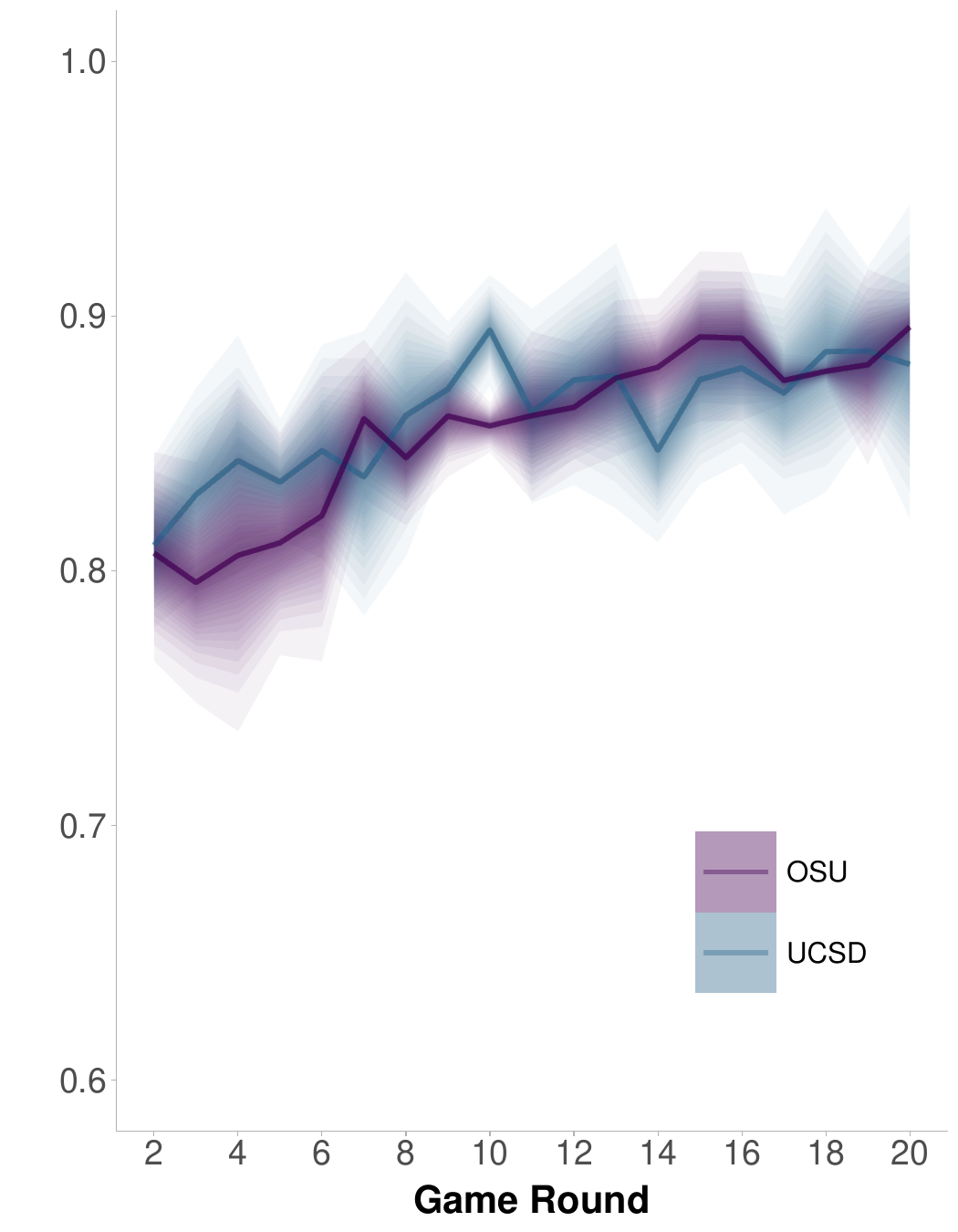} 
\end{tabular}
\end{center}
{\footnotesize \underline{Notes:} Panel (a) presents the average frequency of correct guesses in the {\sc all} treatment in each round, averaged across games. Panel (b) depicts the evolution of consensus in each round, i.e., the relative size of the majority, averaged across games.}
\end{figure}

\begin{figure}[h!]
\begin{center}
\caption{Individual responsiveness to signals and others' actions in {\sc all} treatment}\label{fig:IndOnline} 
\begin{tabular}{cc}
\scriptsize{Responsiveness to signals} & \scriptsize{Responsiveness to others' actions}\\ 
\includegraphics[scale=0.3]{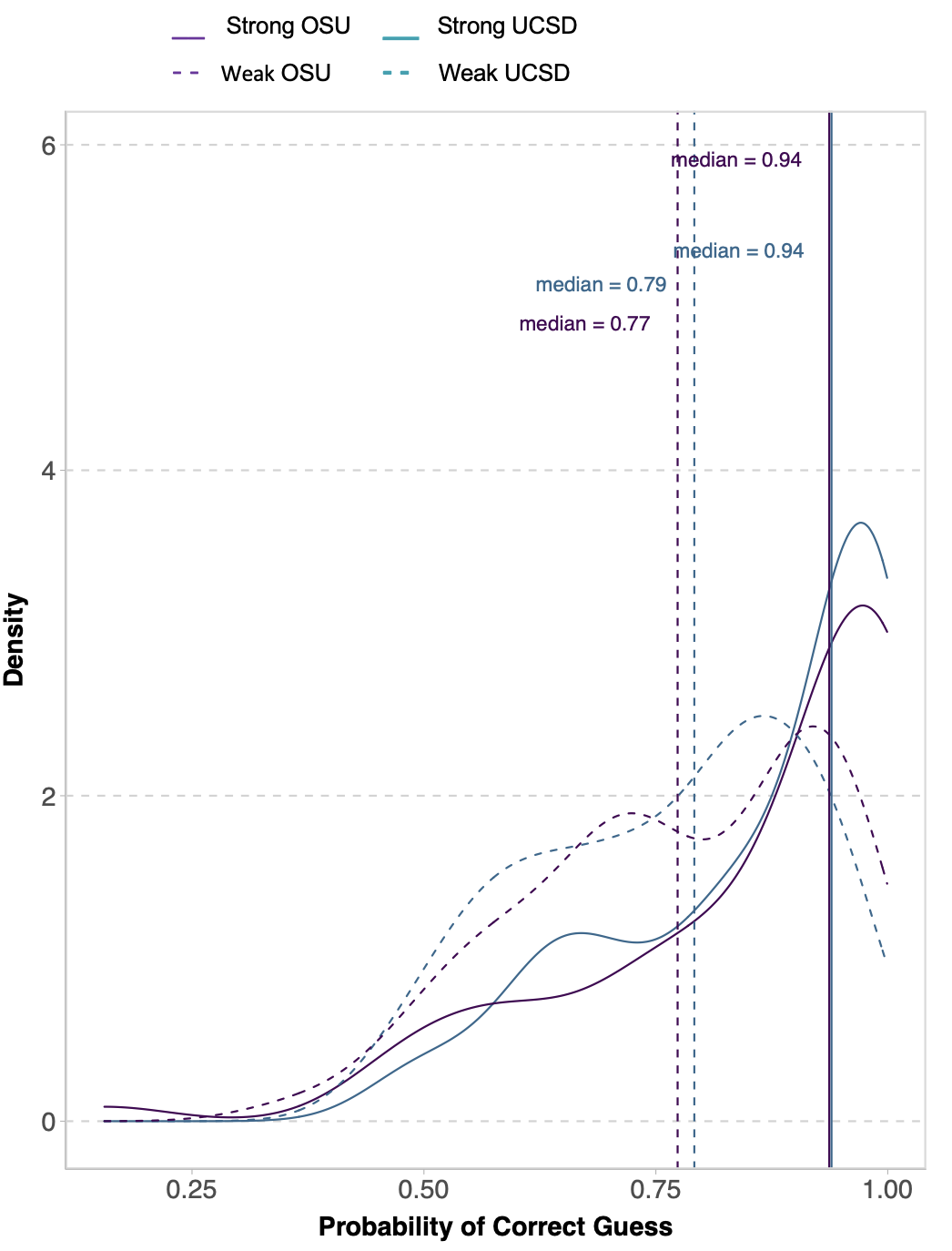} & \includegraphics[scale=0.31]{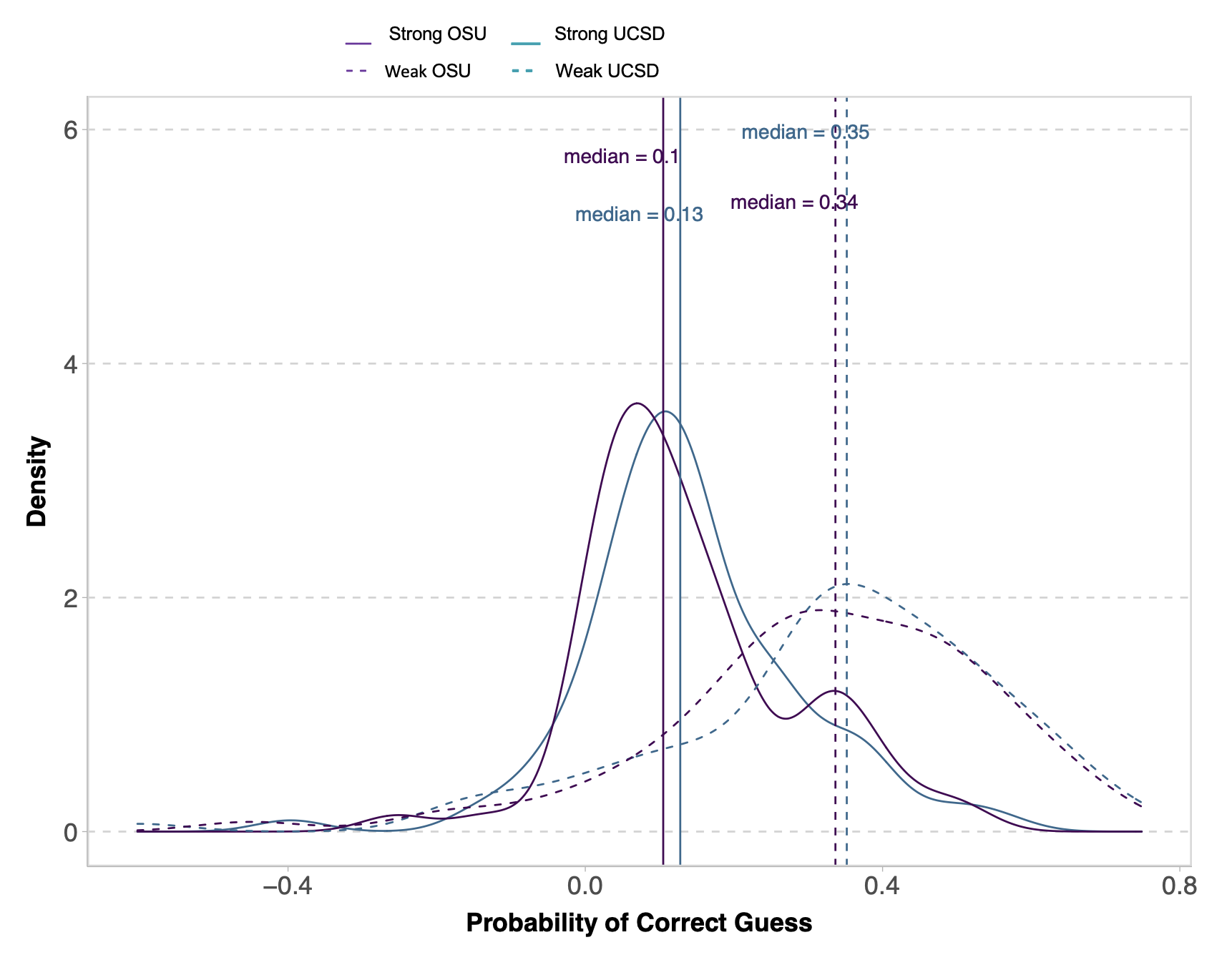} 
\end{tabular}
\end{center}
{\footnotesize \underline{Notes:} Panel (a) presents Kernel distributions of participants' responsiveness to signals separately for weak and strong signals. The vertical lines depict median responsiveness for each group. Responsiveness to signals is given by the probability that a participant's action matches signal majority, when actions of others group members in the previous round are split equally between green and red. Panel (b) presents Kernel distributions of participants' responsiveness to others' actions separately for weak and strong signals. The vertical lines depict median responsiveness for each group. Responsiveness to others' actions is measured by the change in the probability of choosing the action of the majority of signals when all versus none of the other group members choose the majority of signals in the last round.}
\end{figure}

\clearpage

Figure \ref{fig:CorrectOnline} presents two main outcomes of interest in each location: fraction of correct guesses in Panel (a) and consensus rates in Panel (b). The evolution and the levels of both outcomes are extremely similar in the two locations, which is confirmed by the statistical analysis. Regression analysis detects no significant differences between two locations with $p>0.10$ in all comparisons. Figure \ref{fig:IndOnline} complements aggregate results by depicting individual responsiveness to signals and to others' actions in the two locations. We find no significant differences between the distribution  of individual behavior of subjects in the two locations, as measured by these two statistics.

We conclude by noting that we detect no significant differences in neither aggregate results nor in the individual level results between sessions conducted in person at UCSD and sessions conducted online at OSU, which is why the analysis in the main text pools together these sessions.

\section{Learning Across Games\label{app_learn_games}}
\setcounter{figure}{0}
\setcounter{table}{0}   

\begin{figure}[h!]
\begin{center}
\begin{tabular}{cc}
\scriptsize{Panel (a): Correct Actions, by round} & \scriptsize{Panel (b): Consensus, by round}\\ 
\includegraphics[scale=0.32]{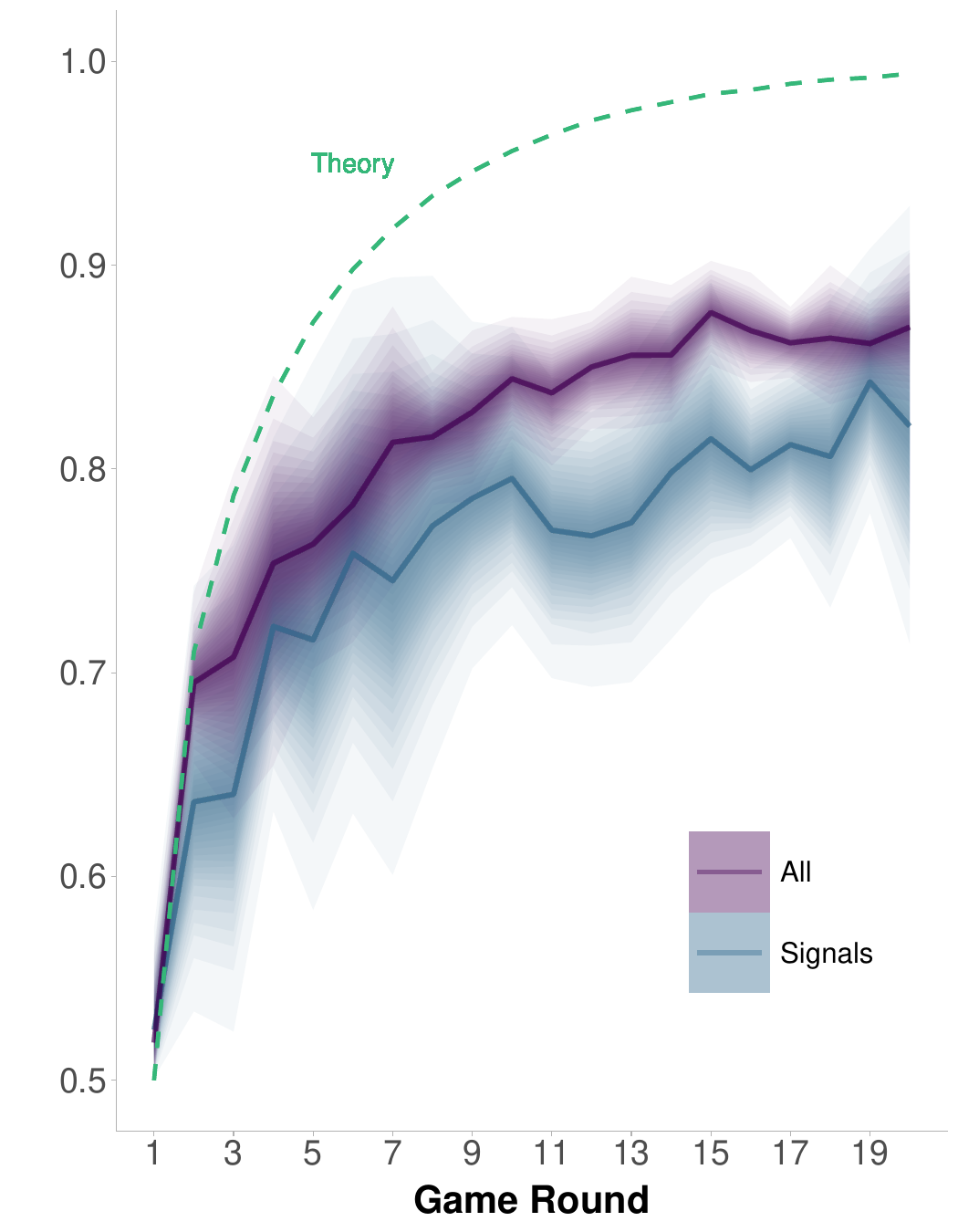}  & \includegraphics[scale=0.32]{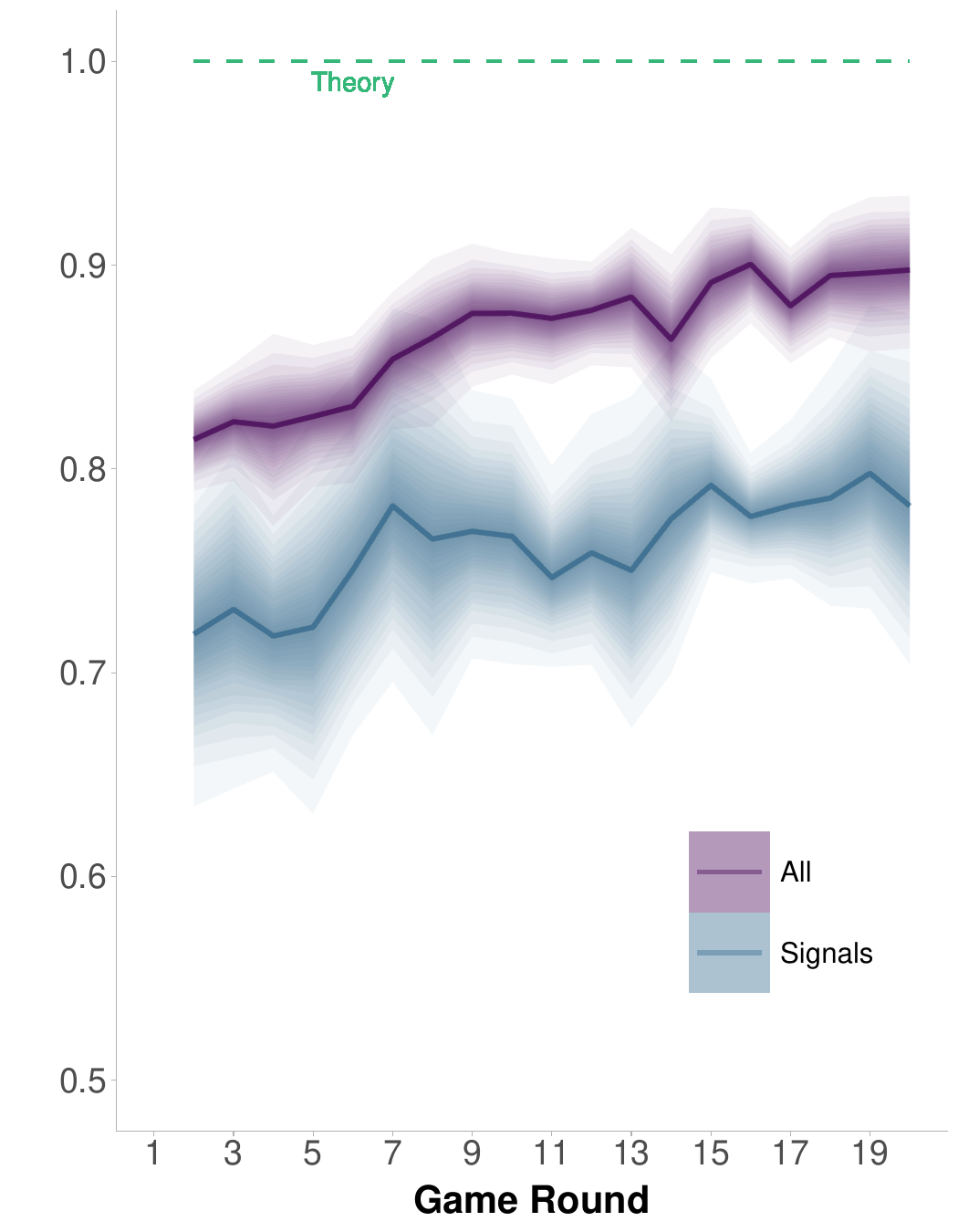}  \\
\multicolumn{2}{c}{\footnotesize{First 5 Games}}\\
\includegraphics[scale=0.32]{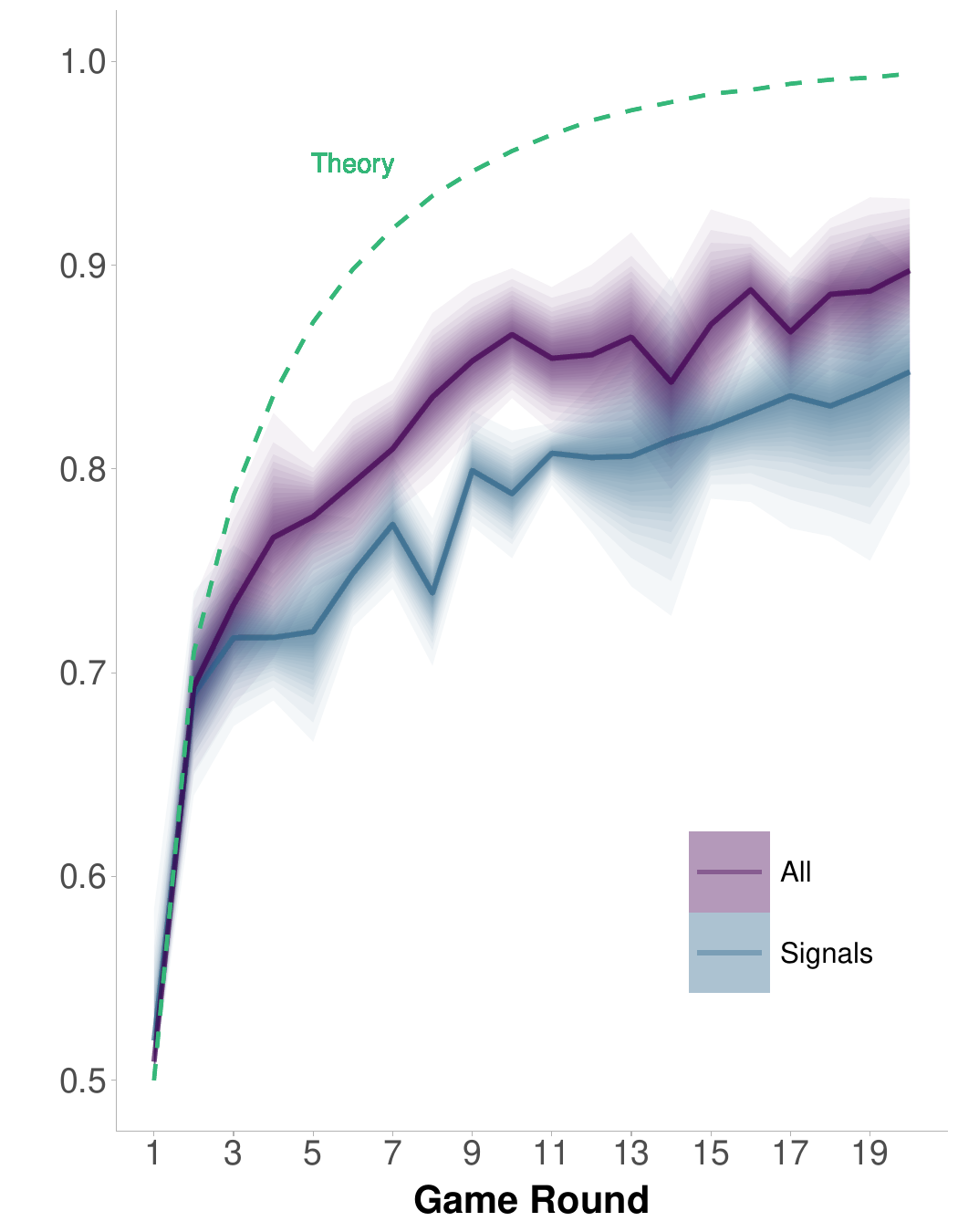}  & \includegraphics[scale=0.32]{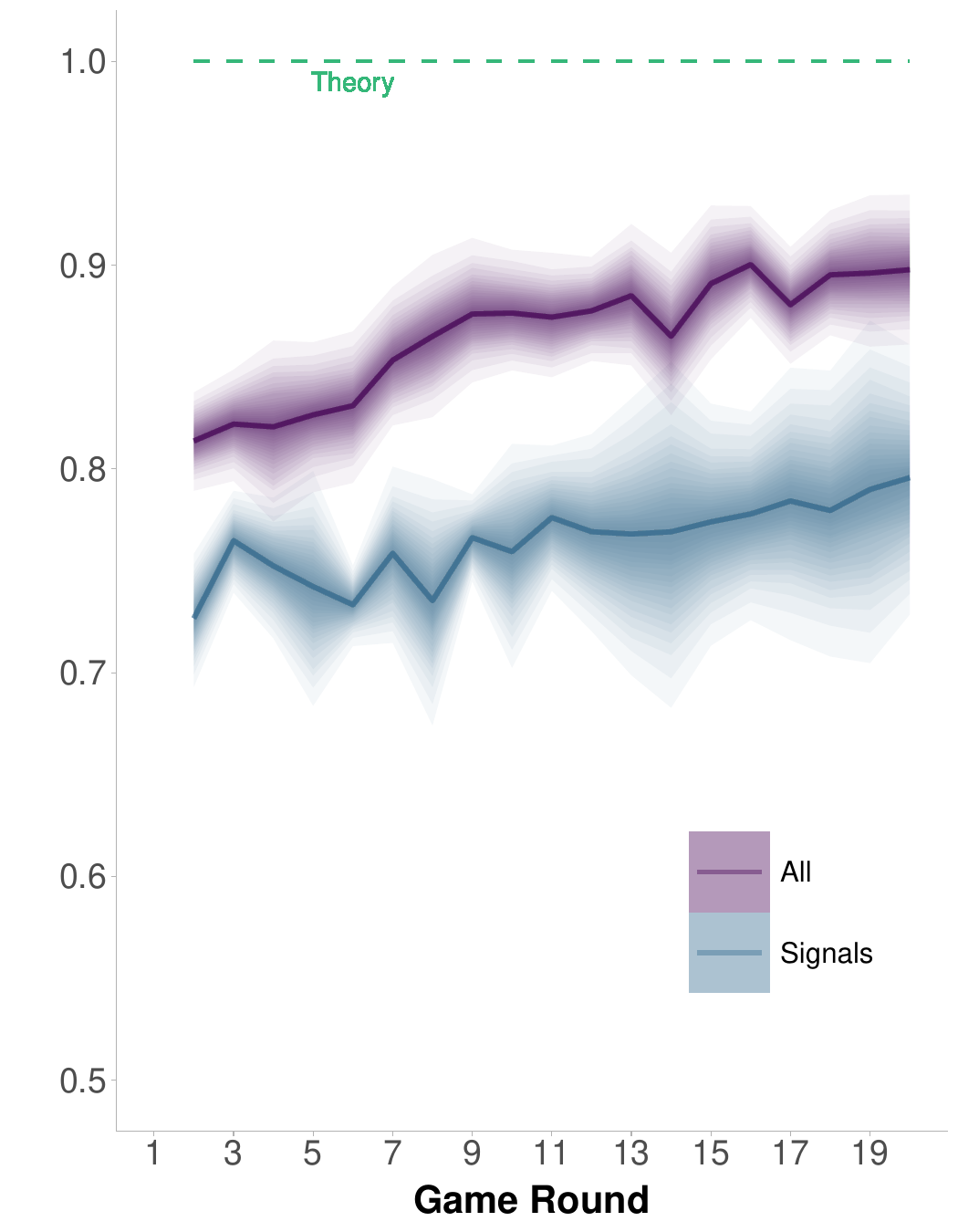} \\
\multicolumn{2}{c}{\footnotesize{Last 5 Games}}
\end{tabular}
\caption{{\sc all} and {\sc signals} treatments (early versus late games)}\label{app_fig:Correct_learn} 
\end{center}
{\footnotesize \underline{Notes:} Panel (a) presents the average frequency of correct actions in each treatment in each round, averaged across the first 5 games (top figure) and the last 5 games (bottom figure). Panel (b) depicts the evolution of consensus in each round, i.e., the relative size of the majority, averaged across the first 5 games (top figure) and the last 5 games (bottom figure). For Panel (b) we exclude cases with equal number of green and red signals. Shaded regions represent confidence intervals from 50\% (darkest) to 95\% (faintest) probability levels. Confidence intervals are constructed with a variance-covariance matrix clustered by session.}
\end{figure}

\begin{figure}[h!]
\begin{center}
\begin{tabular}{cc}
\scriptsize{Panel (a): Information Structures} & \scriptsize{Panel (b): Group Size}\\
\includegraphics[scale=0.32]{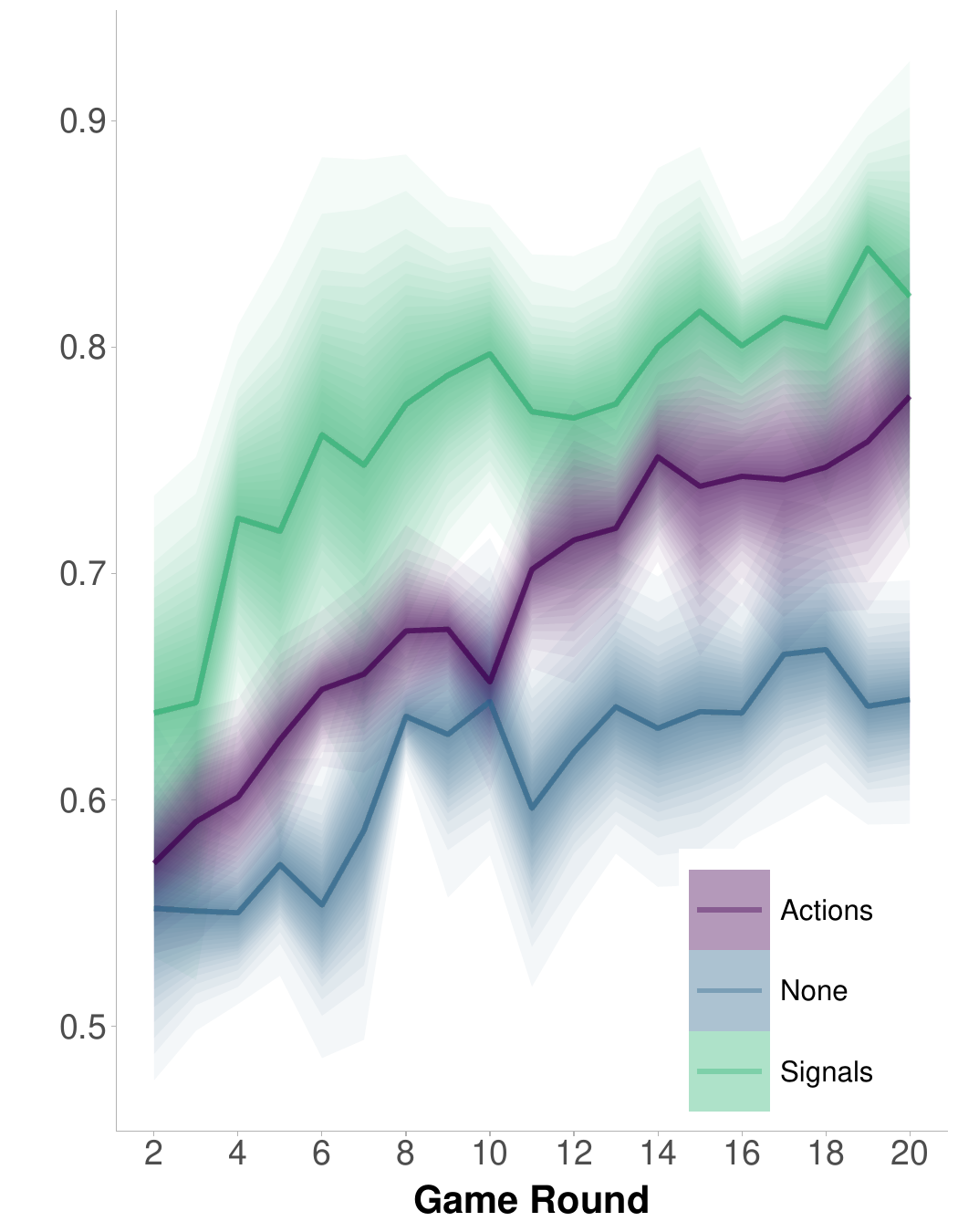} &
\includegraphics[scale=0.32]{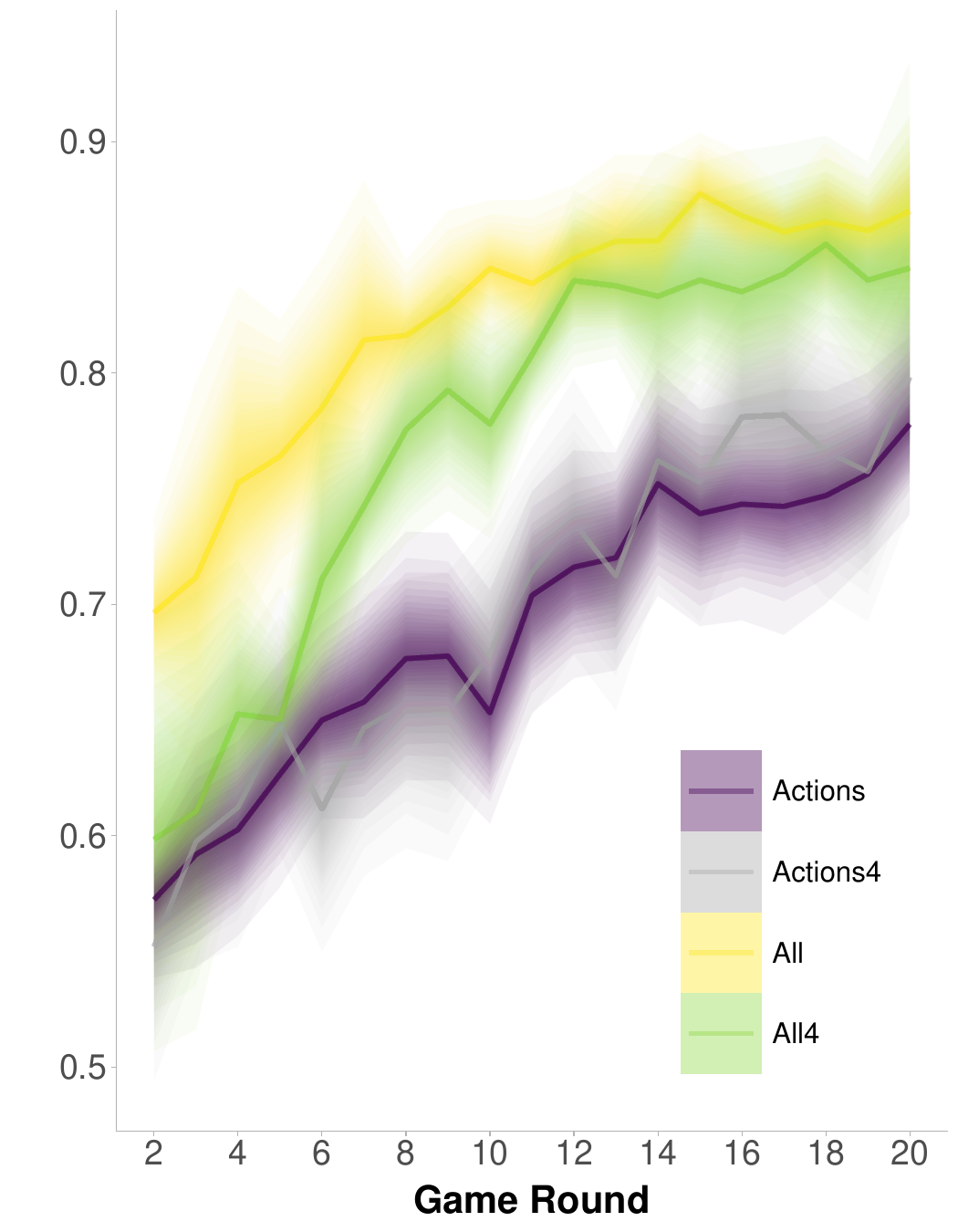} \\
\multicolumn{2}{c}{\footnotesize{First 5 Games}}\\
\includegraphics[scale=0.32]{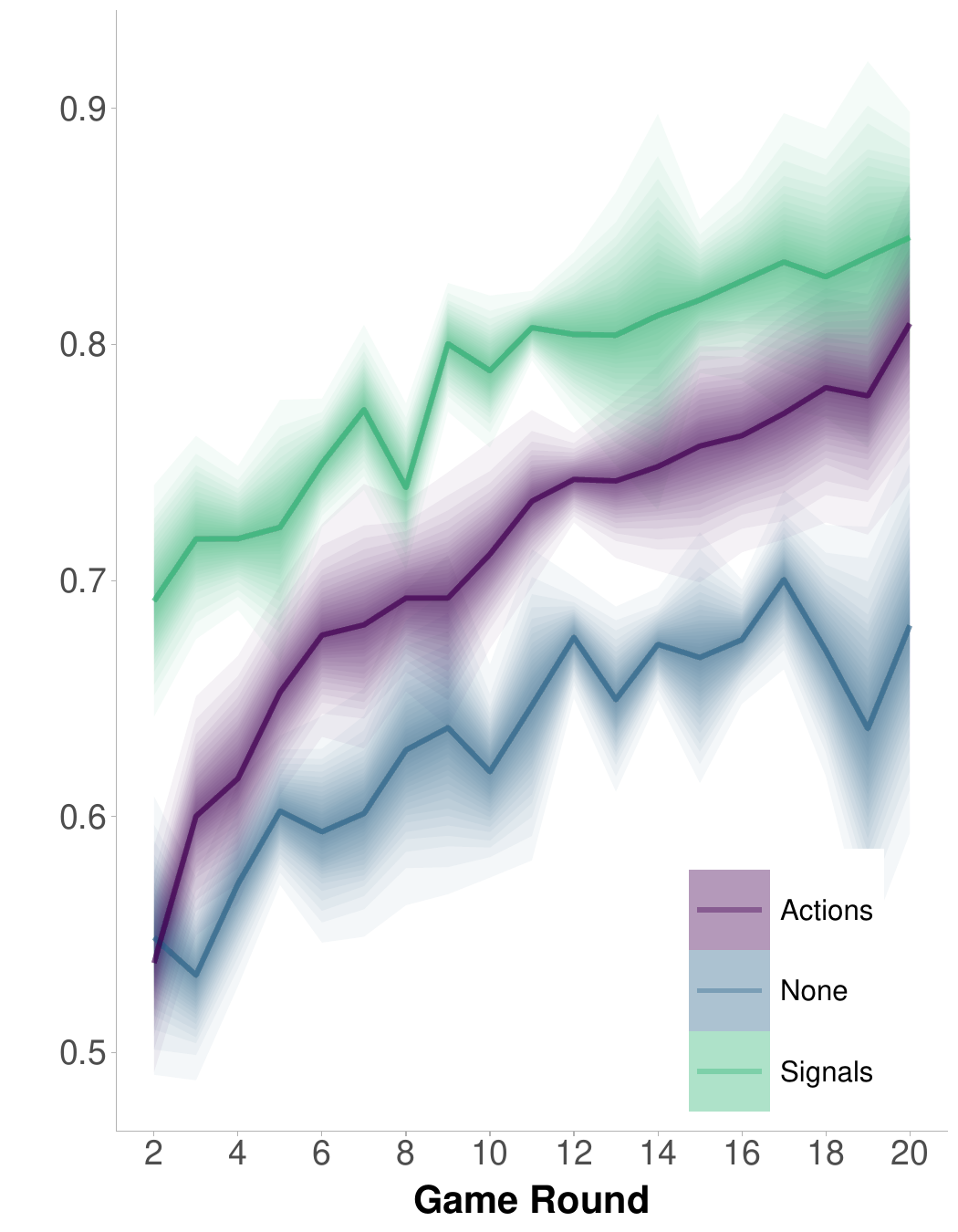} &
\includegraphics[scale=0.32]{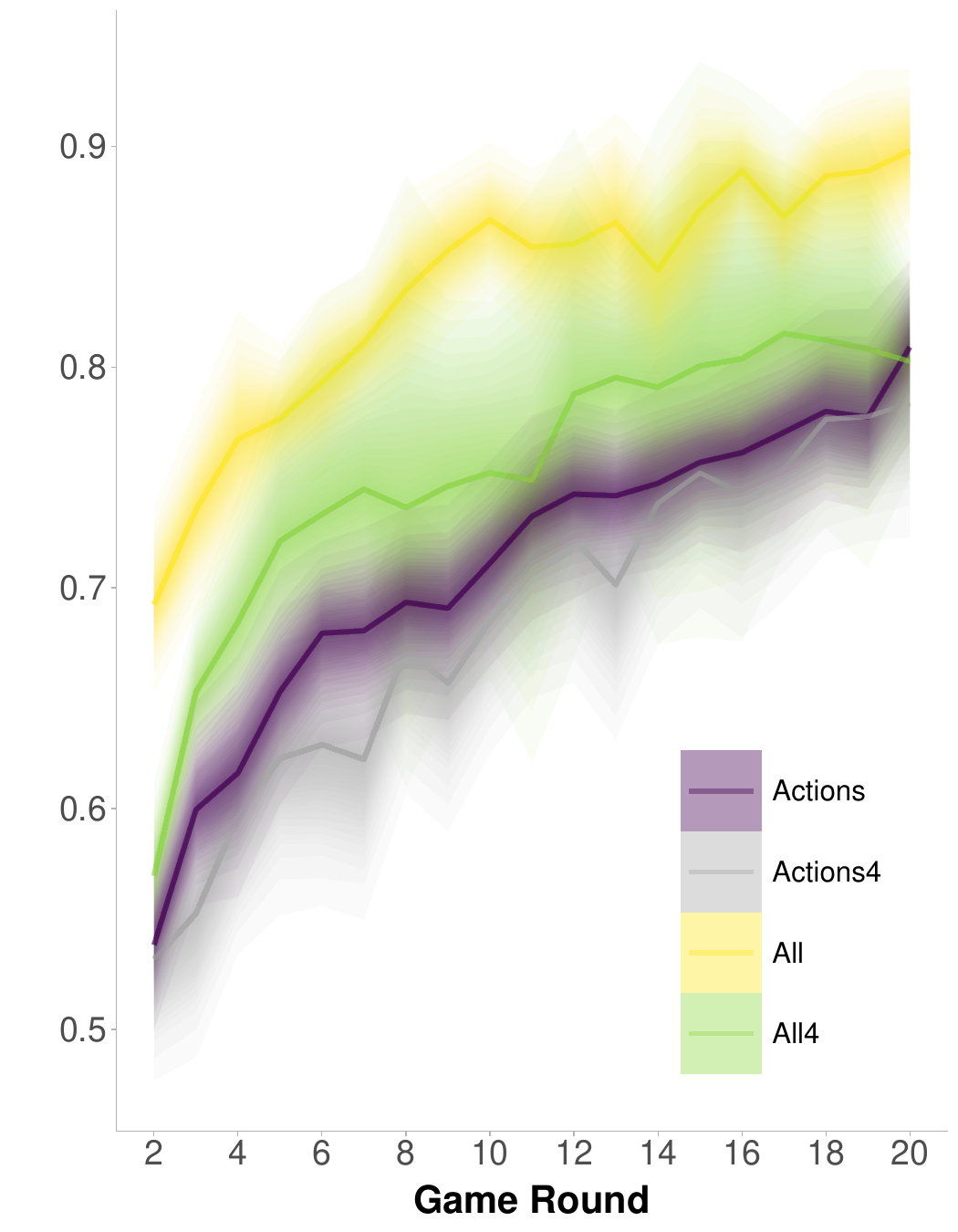} \\
\multicolumn{2}{c}{\footnotesize{Last 5 Games}}
\end{tabular} 
\caption{Frequency of correct actions (early versus late games), by information structure and group size}\label{app_fig:CorrectGuesses_learn} 
\end{center}
{\footnotesize \underline{Notes:} Both panels present the average frequency of correct actions in each treatment per each round, averaged across the first 5 games (top figure) and the last 5 games (bottom figure). Shaded regions represent confidence intervals from 50\% (darkest) to 95\% (faintest) probability levels. Confidence intervals are constructed with a variance-covariance matrix clustered by session.}
\end{figure}

\section{Learning From Others' Actions\label{app_learn_others}}
\setcounter{figure}{0}
\setcounter{table}{0}   

\begin{figure}[!h]
    \begin{center}
    \begin{tabular}{cc}
    \scriptsize{Panel (a): Weak Signals $=$ 75$^{\text{th}}$-25$^{\text{th}}$ percentiles} &  \scriptsize{Panel (b): Weak Signals $=$ 70$^{\text{th}}$-30$^{\text{th}}$ percentiles} \\
    \includegraphics[scale=0.26]{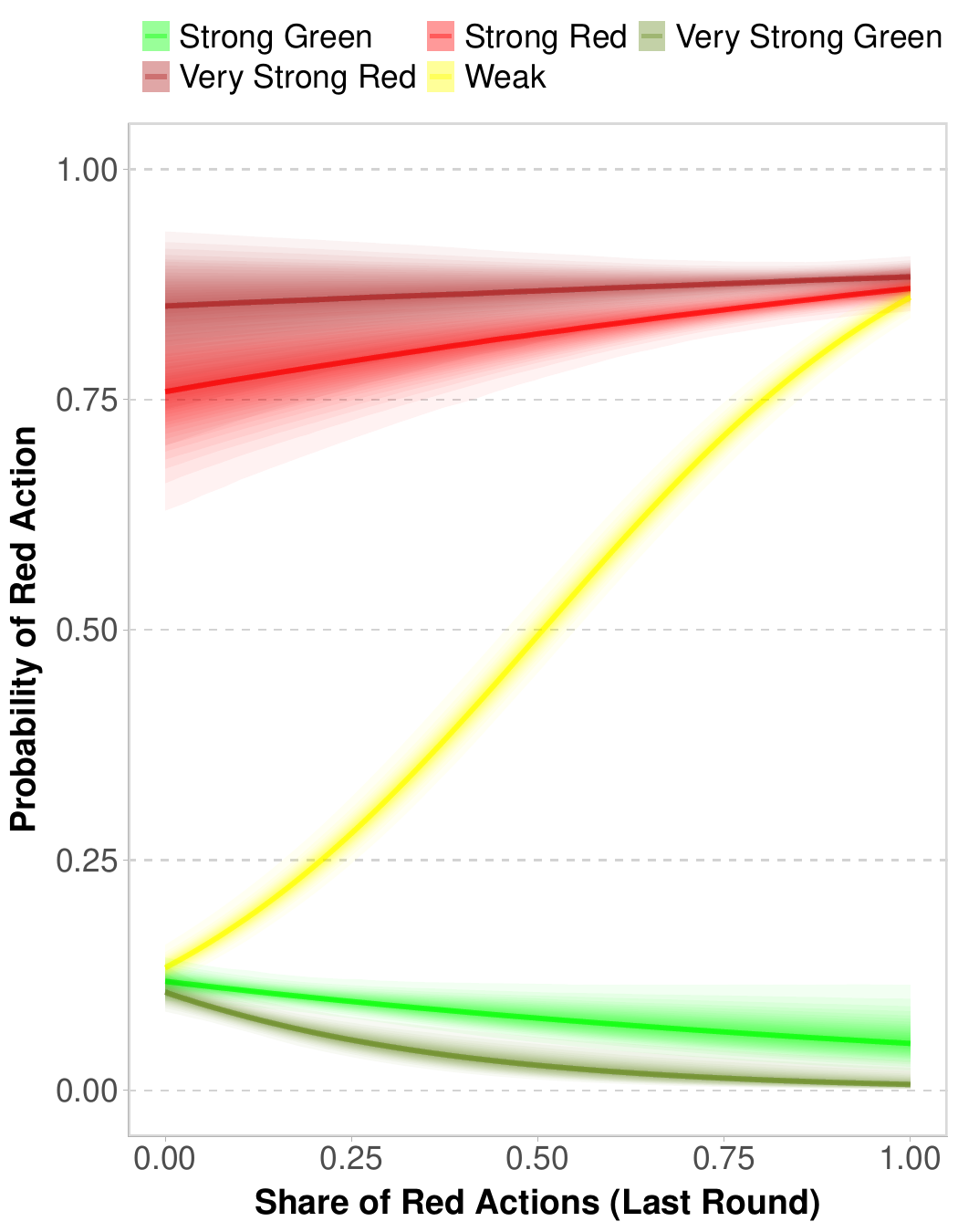} &   \includegraphics[scale=0.26]{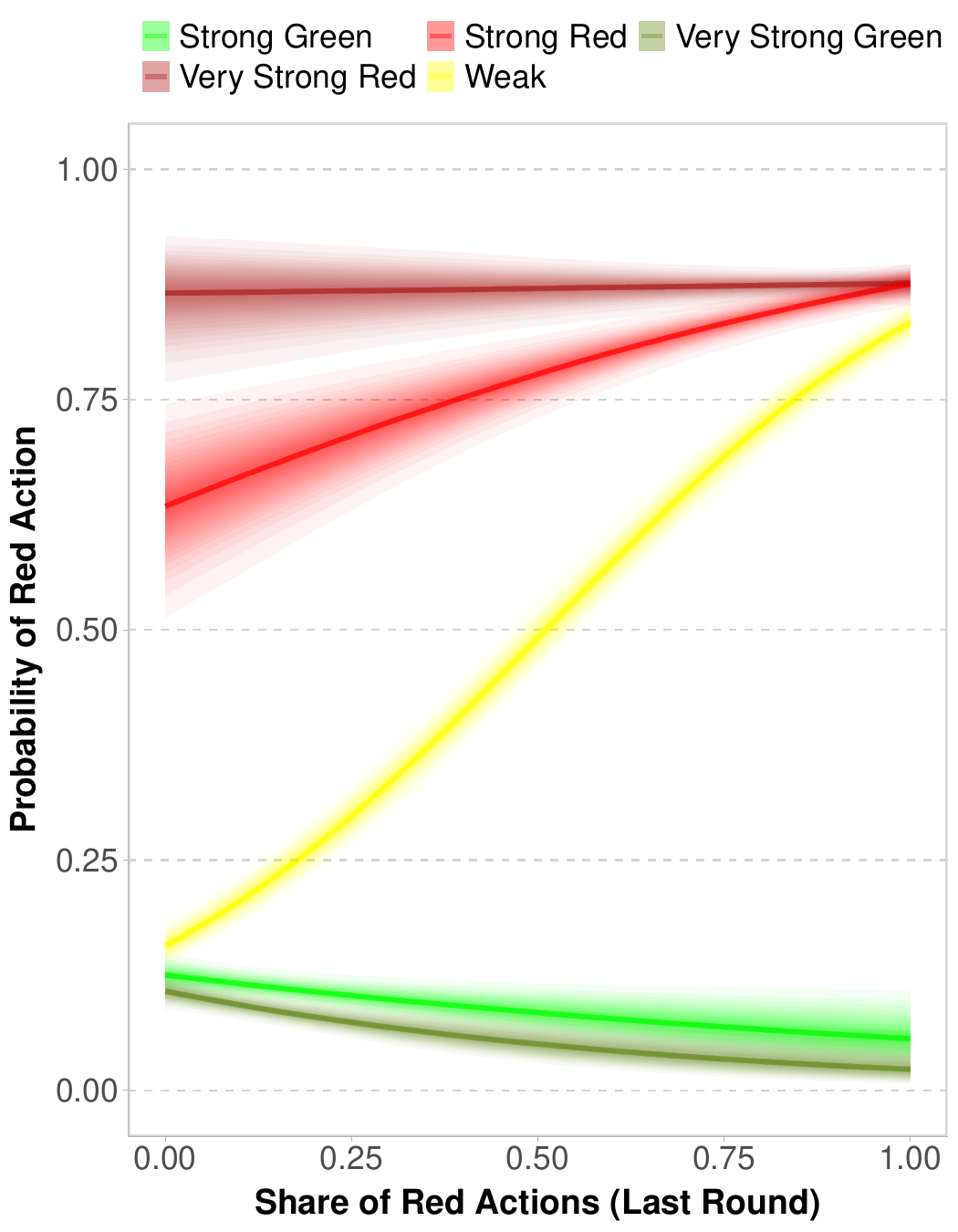}\\
        \scriptsize{Panel (c): Weak Signals $=$ 65$^{\text{th}}$-35$^{\text{th}}$ percentiles} &  \scriptsize{Panel (d): Weak Signals $=$ 60$^{\text{th}}$-40$^{\text{th}}$ percentiles} \\
      \includegraphics[scale=0.26]{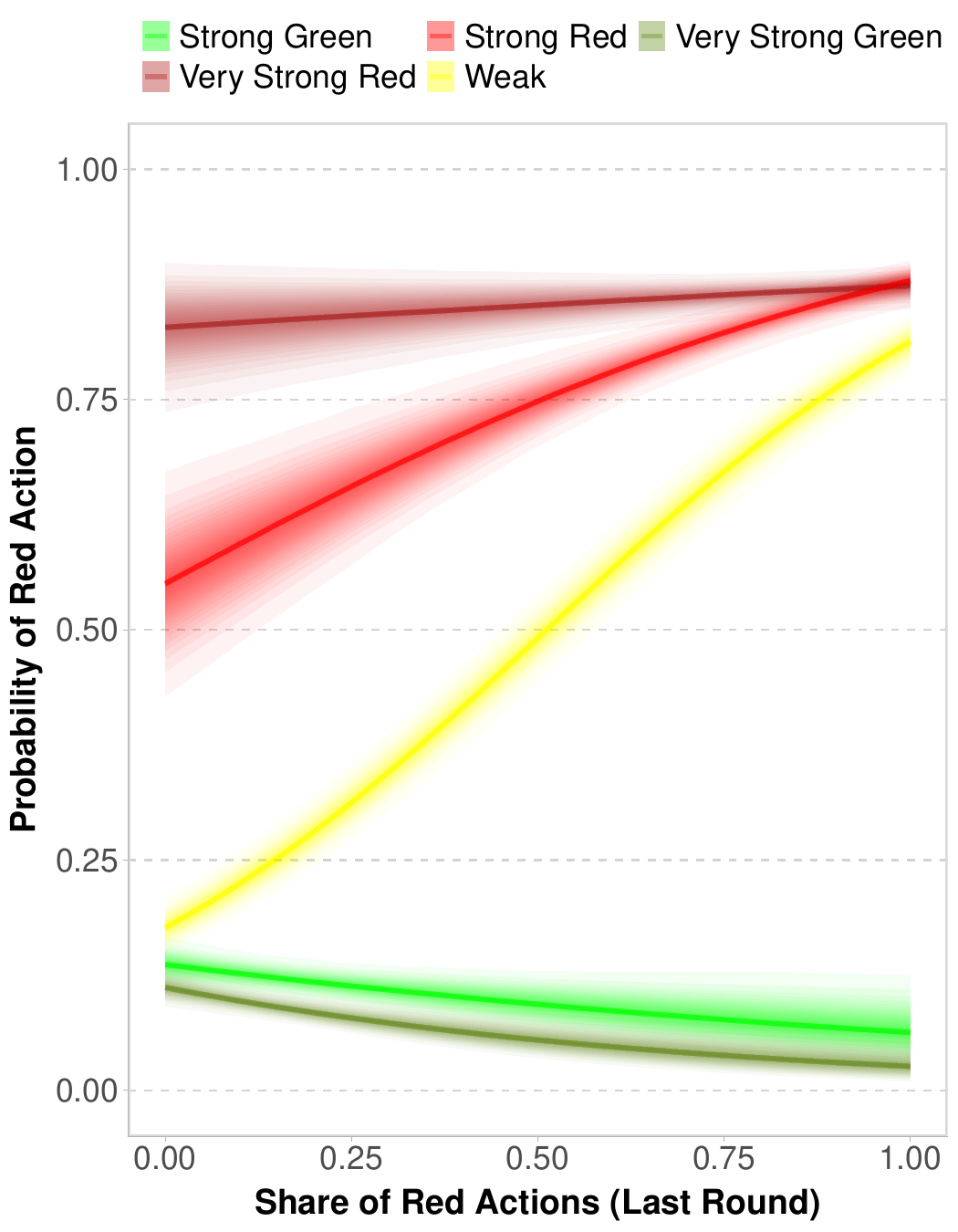} &
        \includegraphics[scale=0.26]{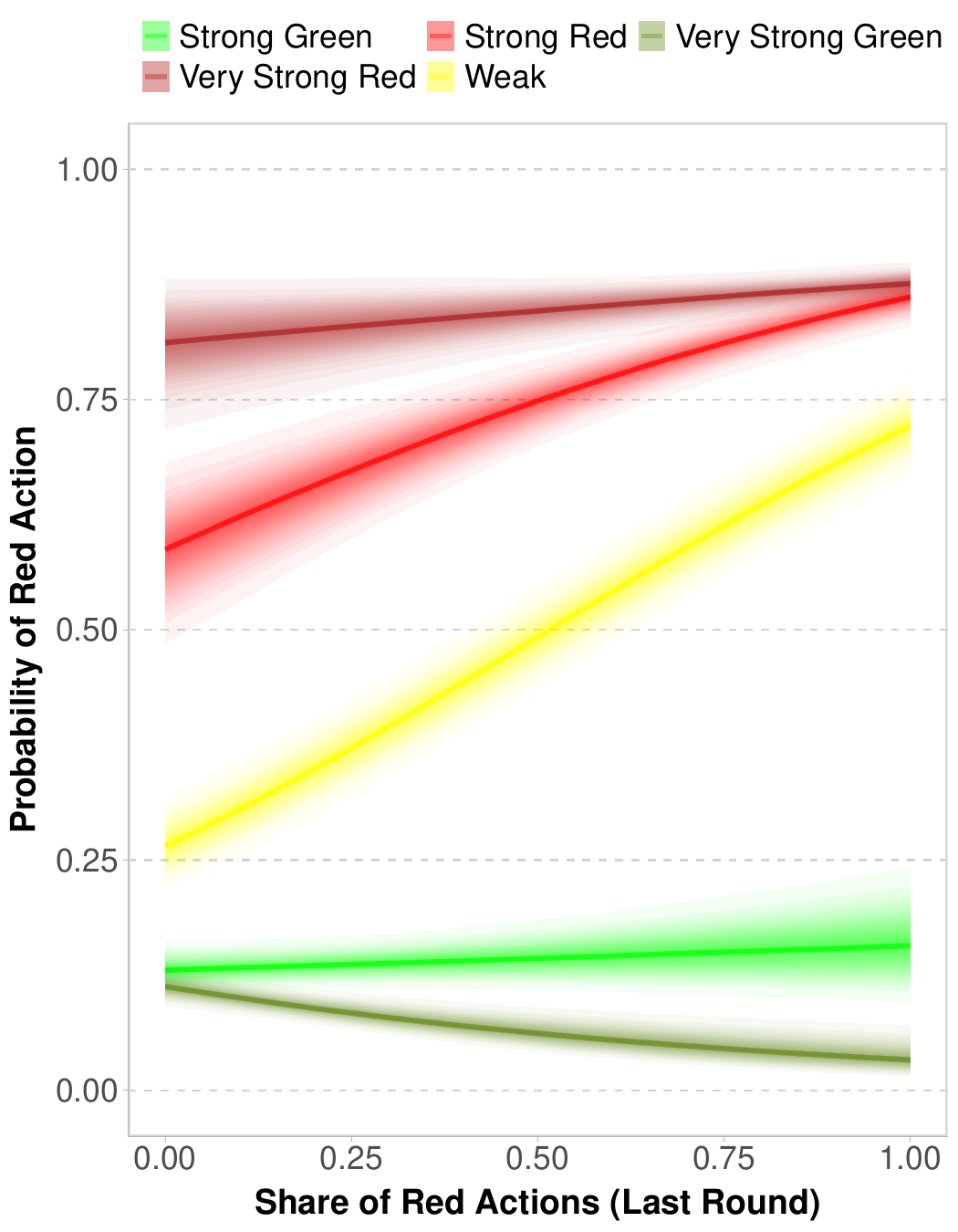} \\
        \end{tabular}
    \caption{Different Cutoffs for Signal Strength in {\sc all} treatment}
    \label{app_fig_learning_from_guesses_rob} 
    \end{center}
    
    {\footnotesize \underline{Notes:} Each panel depicts the probability of guessing red as a function of the share of red actions of other group members in the previous round. The estimates are obtained from a Bayesian logistic regression of subjects' actions on the share of others' actions in the previous round conditional on signal strength. For panel (a) we use the classification in the main text: \emph{Very Strong Green} with $percentile=(0,0.1]$, \emph{Strong Green} with $percentile=(0.1,0.25]$, \emph{Weak} with $percentile=(0.25,0.75]$,  \emph{Strong Red} with $percentile=(0.75,0.9]$ and \emph{Very Strong Red} with $percentile=(0.9,1)$. For the cutoffs of panels (b), (c) and (d) we reduce the \emph{Weak} signals category and increase the ``strong" categories proportionally: \emph{Very Strong Green} with $percentile=(0,0.1+x]$, \emph{Strong Green} with $percentile=(0.1+x,0.25+x]$, \emph{Weak} with $percentile=(0.25+x,0.75-x]$,  \emph{Strong Red} with $percentile=(0.75-x,0.9-x]$ and \emph{Very Strong Red} with $percentile=(0.9-x,1)$, where $x \in \{0.05,0.1,0.15\}$.}
\end{figure}

\begin{figure}[h!]
    \begin{center}
     \begin{tabular}{cc}
        \scriptsize{Panel (a): Effect of others' actions} &  \scriptsize{Panel (b): In sample fit} \\
    \includegraphics[scale=0.35]{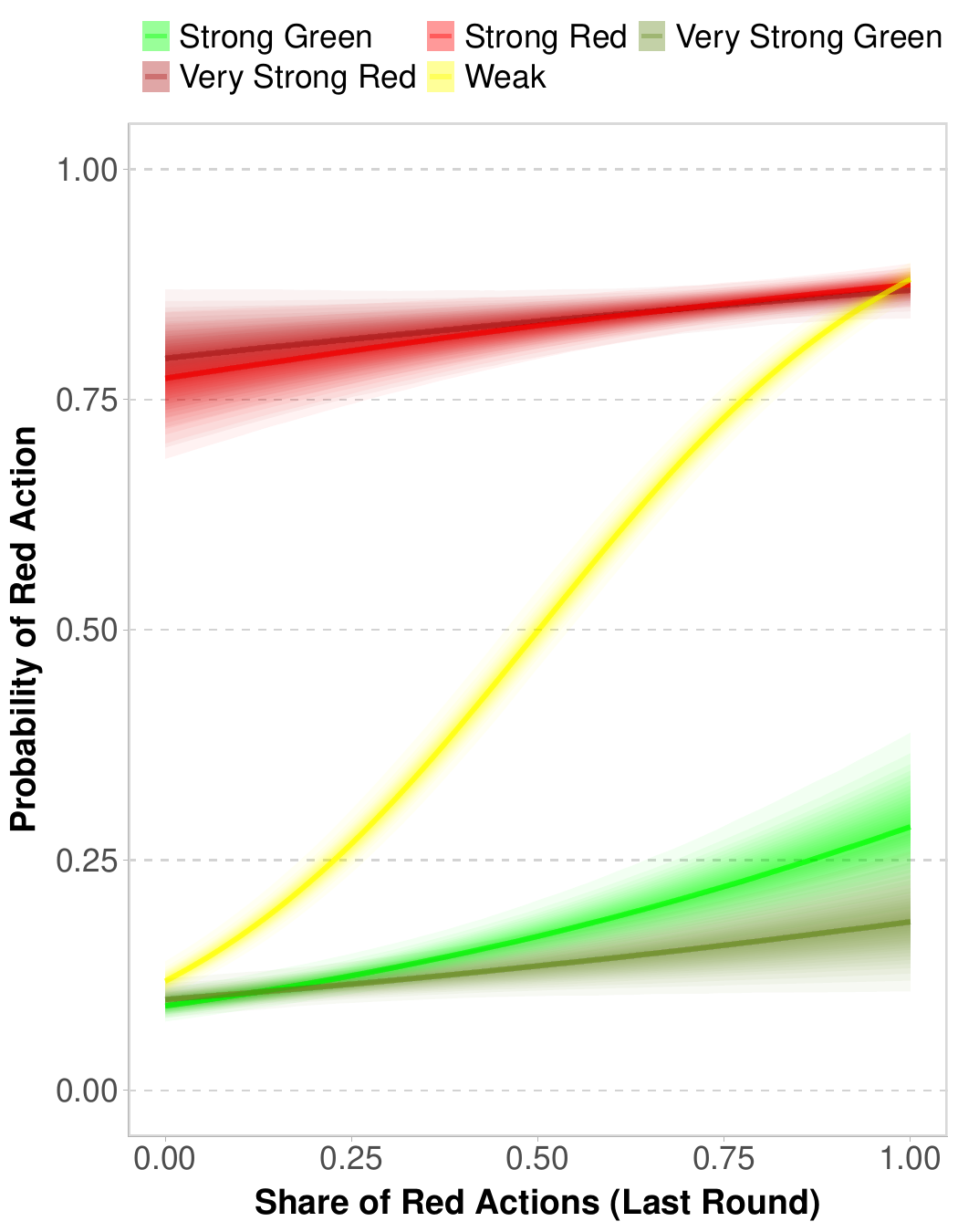} & 
    \includegraphics[scale=0.35]{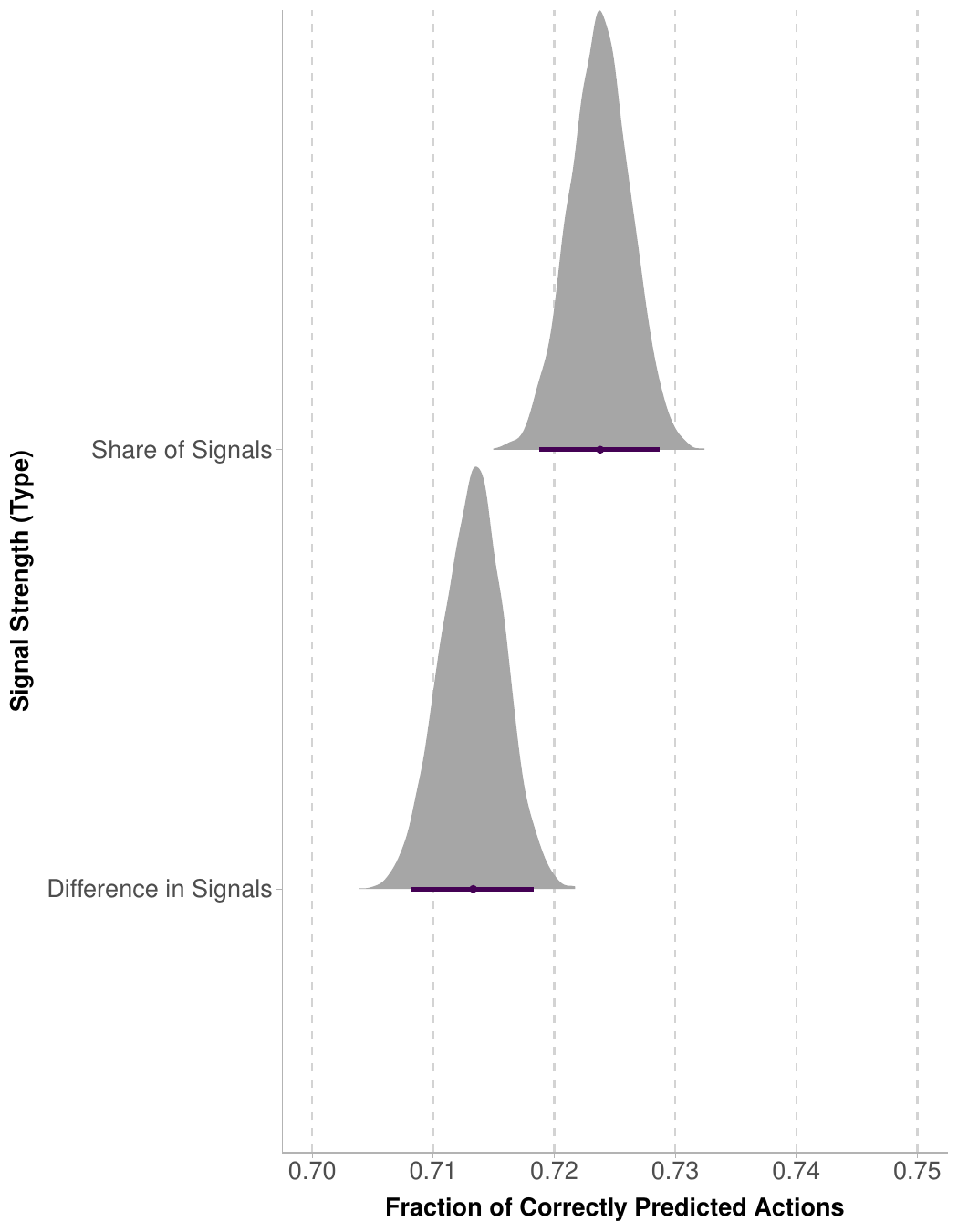}
    \end{tabular}
    \caption{Alternative Signal Strength Measures (Difference in Signals \emph{vs} Share of Signals)}
    \label{app_fig_pred_diff_share} 
    \end{center}
    
    {\footnotesize \underline{Notes:} Panel (a) depicts the probability of choosing red as a function of the share of red actions of other group members conditional on signal strength as constructed from the share of red signals. Shaded regions represent $95$\% credible intervals from 50\% (darkest) to 95\% (faintest) probability levels. Panel (b) depicts the posterior distributions, along with 95\% credible intervals, of the fraction of correctly predicted guesses using either the difference in red and green signals or the share of red signals to construct signal strength. For both measures of signal strength, we estimate a Bayesian logistic regression of subjects' actions on the share of others' actions in the previous round conditional on signal strength and session random effects.}
\end{figure}

\section{Additional Aggregate Results \label{app_agg_additional}}
\setcounter{figure}{0}
\setcounter{table}{0}   

\begin{table}[!hbp] \centering 
\scriptsize
  \caption{Treatment Effects by Group Size} 
  \label{app_table_cons_size} 
\begin{tabular}{@{\extracolsep{5pt}}lcccc} 
\\[-1.8ex]\hline 
\hline \\[-1.8ex] 
 & \multicolumn{4}{c}{\textit{Dependent variable:}} \\ 
\cline{2-5} 
\\[-1.8ex] & \multicolumn{4}{c}{Consensus Rate} \\ 
\\[-1.8ex] & (1) & (2)& (3) & (4)\\ 
\hline \\[-1.8ex] 
{\sc actions} (Baseline) & 0.686$^{***}$ & 0.681$^{***}$ &  & \\ 
  & (0.011) & (0.010) &  & \\ 
  {\sc all} (Baseline)  &  &  & 0.820$^{***}$ & 0.812$^{***}$\\ 
  & & & (0.010) & (0.011)\\ 
  & & & & \\ 
{\sc actions4} (Effect) & 0.090$^{***}$ & 0.101$^{***}$ \\ 
  & (0.010) & (0.009) \\ 
  & & \\ 
{\sc actions4} (Effect) $\times$ Late Rounds &  & $-$0.020$^{*}$ \\ 
  &  & (0.010) \\ 
  & & \\ 
{\sc all4}(Effect) & & & 0.049$^{***}$ & 0.064$^{***}$ \\ 
 & & & (0.009) & (0.011) \\ 
 & & & & \\ 
{\sc all4} (Effect) $\times$ Late Rounds & & &  & $-$0.028$^{**}$ \\ 
  & & &  & (0.014) \\ 
\hline \\[-1.8ex] 
Game Round Fixed Effects & Yes & Yes & Yes & Yes \\ 
Observations & 5,550 & 5,550 & 6,841 & 6,841 \\ 
Adjusted R$^{2}$ & 0.160 & 0.161 & 0.058 & 0.061\\ 
\hline 
\hline \\[-1.8ex] 
\end{tabular} 

\vspace{1mm}
{\footnotesize \underline{Notes:} $^{*}$p$<$0.1; $^{**}$p$<$0.05; $^{***}$p$<$0.01. Clustered standard errors by session in parentheses. Late rounds are 11-20.}
\end{table}

\begin{figure}[h!]
\begin{center}
\caption{Consensus Rates, by Group Size}\label{app_fig_cons_size} 
\begin{tabular}{cc}
\scriptsize{Panel (a): {\sc actions4} \emph{vs} {\sc actions}} & \scriptsize{Panel (b): {\sc all4} \emph{vs} {\sc all}}\\
\includegraphics[scale=0.32]{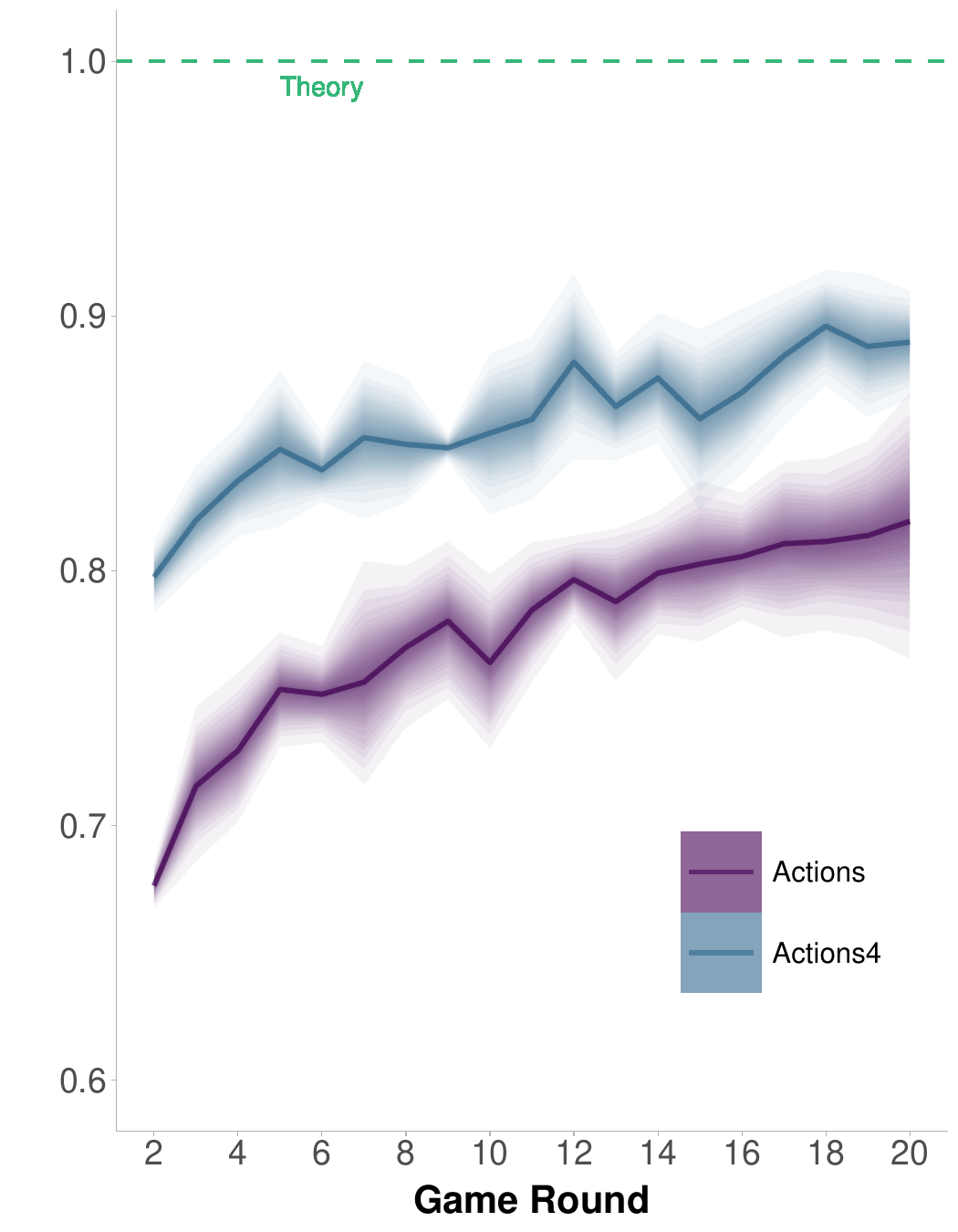} &
\includegraphics[scale=0.32]{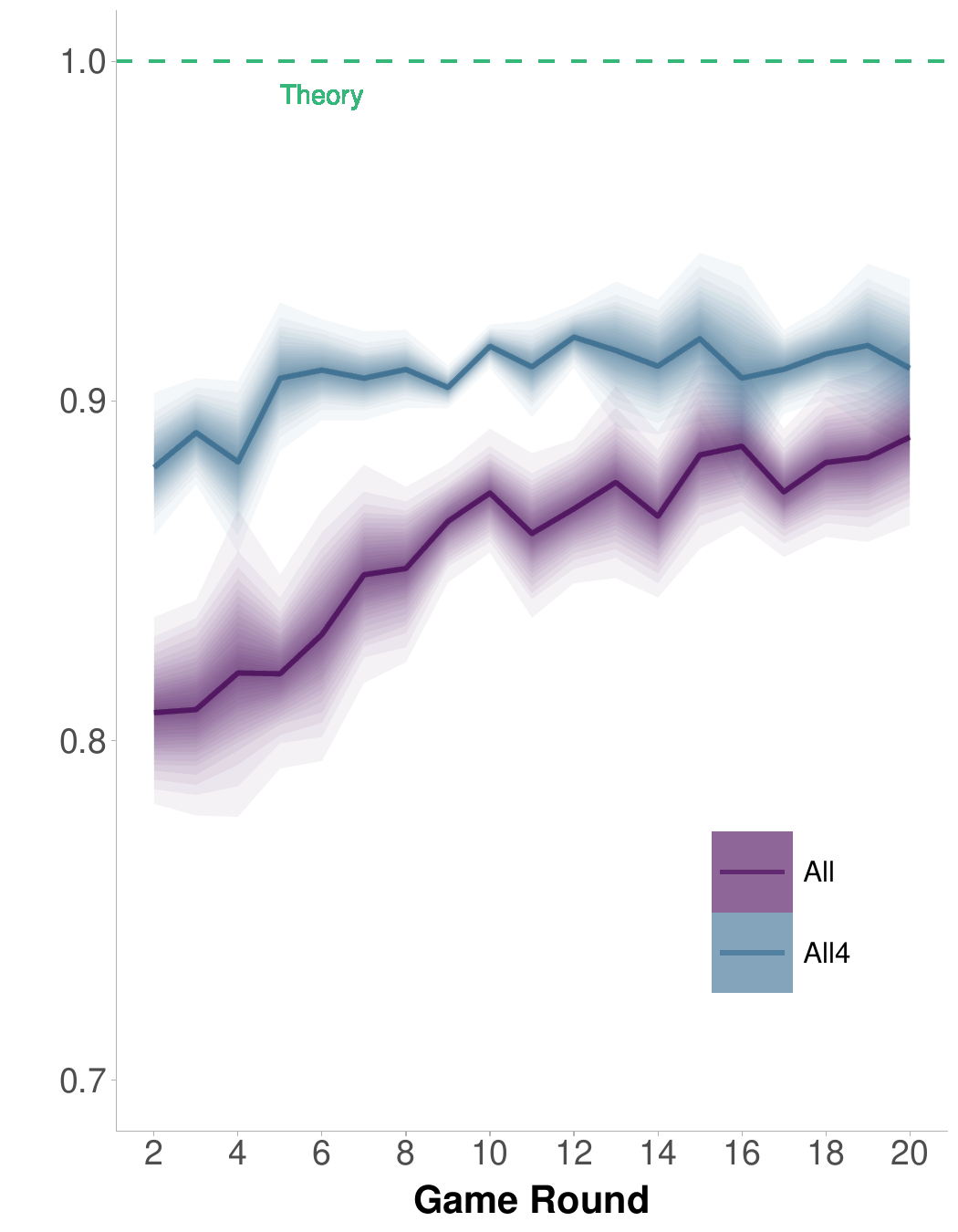} 
\end{tabular}
\end{center}
{\footnotesize \underline{Notes:} Both panels present the consensus rates in each treatment per each round, averaged across games within a session. Shaded regions represent confidence intervals from 50\% (darkest) to 95\% (faintest) probability levels. Confidence intervals are constructed with a variance-covariance matrix clustered by session.}
\end{figure}

\begin{table}[!hbp] \centering 
\footnotesize
  \caption{Probability of Red Bet by Signal Strength ({\sc all} versus {\sc signals})} 
  \label{app_table_red_bet} 
\begin{tabular}{@{\extracolsep{5pt}}lcccc} 
\\[-1.8ex]\hline 
\hline \\[-1.8ex] 
 & \multicolumn{4}{c}{\textit{Dependent variable:}} \\ 
\cline{2-5} 
\\[-1.8ex] & \multicolumn{4}{c}{Red Bet} \\ 
\\[-1.8ex] & (1) & (2) & (3) & (4)\\ 
\hline \\[-1.8ex] 
 Constant & 0.088$^{***}$ & 0.199$^{***}$ & 0.174$^{***}$ & $-$0.078 \\ 
  & (0.011) & (0.023) & (0.056) & (0.060) \\ 
  & & & & \\ 
{\sc signals} & 0.032 & 0.032 & 0.046$^{*}$ & 0.300$^{***}$ \\ 
  & (0.032) & (0.032) & (0.026) & (0.026) \\ 
  & & & & \\ 
 \emph{Strong Green} & 0.020$^{**}$ & 0.015$^{*}$ & 0.032$^{**}$ & 0.031$^{*}$ \\ 
  & (0.009) & (0.009) & (0.016) & (0.016) \\ 
  & & & & \\ 
 \emph{Weak} & 0.369$^{***}$ & 0.315$^{***}$ & 0.321$^{***}$ & 0.306$^{***}$ \\ 
  & (0.019) & (0.026) & (0.024) & (0.022) \\ 
  & & & & \\ 
 \emph{Strong Red} & 0.765$^{***}$ & 0.753$^{***}$ & 0.747$^{***}$ & 0.740$^{***}$ \\ 
  & (0.021) & (0.022) & (0.020) & (0.017) \\ 
  & & & & \\ 
 \emph{Very Strong Red} & 0.790$^{***}$ & 0.790$^{***}$ & 0.786$^{***}$ & 0.780$^{***}$ \\ 
  & (0.022) & (0.022) & (0.024) & (0.021) \\ 
  & & & & \\ 
 \emph{Strong Green} $\times$ {\sc signals} & 0.028$^{***}$ & 0.025$^{**}$ & 0.006 & 0.003 \\ 
  & (0.011) & (0.011) & (0.024) & (0.022) \\ 
  & & & & \\ 
 \emph{Weak} $\times$ {\sc signals} & $-$0.111$^{***}$ & $-$0.108$^{***}$ & $-$0.120$^{***}$ & $-$0.123$^{***}$ \\ 
  & (0.030) & (0.030) & (0.031) & (0.030) \\ 
  & & & & \\ 
 \emph{Strong Red} $\times$ {\sc signals}& $-$0.089 & $-$0.090 & $-$0.116$^{*}$ & $-$0.122$^{*}$ \\ 
  & (0.076) & (0.074) & (0.067) & (0.063) \\ 
  & & & & \\ 
 \emph{Very Strong Red} $\times$ {\sc signals} & $-$0.066 & $-$0.065 & $-$0.099 & $-$0.114$^{*}$ \\ 
  & (0.068) & (0.068) & (0.069) & (0.067) \\ 
  & & & & \\ 
\hline \\[-1.8ex] 
Game Round Fixed Effects & No & Yes & Yes & Yes \\ 
Game Fixed Effects & No & No & Yes & Yes \\ 
Participant Fixed Effects & No & No & No & Yes \\ 
Observations & 44,460 & 44,460 & 44,460 & 44,460 \\ 
Adjusted R$^{2}$ & 0.271 & 0.274 & 0.285 & 0.321 \\ 
\hline 
\hline \\[-1.8ex] 
\end{tabular} 
\vspace{1mm}

{\footnotesize \underline{Notes:} $^{*}$p$<$0.1; $^{**}$p$<$0.05; $^{***}$p$<$0.01. Clustered standard errors by session in parentheses. }
\end{table} 

\section{Additional Individual Results \label{app_ind_additional}}
\setcounter{figure}{0}
\setcounter{table}{0}   

For participant covariates we include: \texttt{female}, which is an indicator variable that takes the value of one if the participant identifies as female and zero, otherwise. \texttt{stem}, which is an indicator variable that takes the value of one if the participant's major is STEM and zero, otherwise.  \texttt{overconfidence} measures the extent of a participant's over-estimation of her IQ, which is given by the difference between the number of questions a participant believes she solved correctly and the actual number of correct answers. \texttt{risk} is  measured by the number of points invested in a risky asset as specified in a risky investment task solved at the end of the game.

\begin{table}[!hbp] \centering 
\footnotesize
  \caption{Probability of Correct Guess by IQ Level ({\sc all} versus {\sc signals})} 
  \label{app:table_corr_iq} 
\begin{tabular}{@{\extracolsep{5pt}}lcccc} 
\\[-1.8ex]\hline 
\hline \\[-1.8ex] 
 & \multicolumn{4}{c}{\textit{Dependent variable:}} \\ 
\cline{2-5} 
\\[-1.8ex] & \multicolumn{4}{c}{Correct Guess} \\ 
\\[-1.8ex] & (1) & (2) & (3) & (4)\\ 
\hline \\[-1.8ex] 
 Constant & 0.740$^{***}$ & 0.614$^{***}$ & 0.577$^{***}$ & 0.506$^{***}$ \\ 
  & (0.021) & (0.019) & (0.022) & (0.038) \\ 
  & & & & \\ 
{\sc signals} & $-$0.055$^{**}$ & $-$0.055$^{**}$ & $-$0.055$^{**}$ & $-$0.041 \\ 
  & (0.026) & (0.026) & (0.026) & (0.029) \\ 
  & & & & \\ 
 \texttt{IQ} & 0.024$^{***}$ & 0.024$^{***}$ & 0.024$^{***}$ & 0.031$^{***}$ \\ 
  & (0.004) & (0.004) & (0.004) & (0.006) \\ 
  & & & & \\ 
 \texttt{female} &  &  &  & 0.023 \\ 
  &  &  &  & (0.022) \\ 
  & & & & \\ 
 \texttt{stem} &  &  &  & 0.018 \\ 
  &  &  &  & (0.021) \\ 
  & & & & \\ 
 \texttt{overconfidence} &  &  &  & 0.015$^{**}$ \\ 
  &  &  &  & (0.007) \\ 
  & & & & \\ 
 \texttt{risk} &  &  &  & 0.0002 \\ 
  &  &  &  & (0.0001) \\ 
  & & & & \\ 
 \texttt{IQ} $\times$ {\sc signals} & 0.003 & 0.003 & 0.003 & 0.001 \\ 
  & (0.011) & (0.011) & (0.011) & (0.010) \\ 
  & & & & \\ 
\hline \\[-1.8ex] 
Game Round Fixed Effects & No & Yes & Yes & Yes \\ 
Game Fixed Effects & No & No & Yes & Yes \\ 
Participant Covariates & No & No & No & Yes \\ 
Observations & 44,460 & 44,460 & 44,460 & 44,460 \\ 
Adjusted R$^{2}$ & 0.014 & 0.032 & 0.034 & 0.037 \\ 
\hline 
\hline \\[-1.8ex]  
\end{tabular} 

\vspace{1mm}
{\footnotesize \underline{Notes:} $^{*}$p$<$0.1; $^{**}$p$<$0.05; $^{***}$p$<$0.01. Clustered standard errors by session in parentheses.}
\end{table} 

\begin{table}[!hbp] \centering 
\footnotesize
  \caption{Probability of Correct Guess by low-IQ/high-IQ Subjects ({\sc all} versus {\sc signals})} 
  \label{app:table_corr_naive} 
\begin{tabular}{@{\extracolsep{5pt}}lcccc} 
\\[-1.8ex]\hline 
\hline \\[-1.8ex] 
 & \multicolumn{4}{c}{\textit{Dependent variable:}} \\ 
\cline{2-5} 
\\[-1.8ex] & \multicolumn{4}{c}{Correct Guess} \\ 
\\[-1.8ex] & (1) & (2) & (3) & (4)\\ 
\hline \\[-1.8ex] 
 Constant & 0.863$^{***}$ & 0.737$^{***}$ & 0.701$^{***}$ & 0.653$^{***}$ \\ 
  & (0.015) & (0.014) & (0.015) & (0.033) \\ 
  & & & & \\ 
{\sc signals} & $-$0.047 & $-$0.047 & $-$0.047 & $-$0.044 \\ 
  & (0.045) & (0.045) & (0.045) & (0.042) \\ 
  & & & & \\ 
 \emph{low-IQ} & $-$0.074$^{***}$ & $-$0.074$^{***}$ & $-$0.074$^{***}$ & $-$0.088$^{***}$ \\ 
  & (0.016) & (0.016) & (0.016) & (0.026) \\ 
  & & & & \\ 
 \texttt{female} &  &  &  & 0.023 \\ 
  &  &  &  & (0.024) \\ 
  & & & & \\ 
 \texttt{stem} &  &  &  & 0.023 \\ 
  &  &  &  & (0.020) \\ 
  & & & & \\ 
 \texttt{overconfidence} &  &  &  & 0.009 \\ 
  &  &  &  & (0.007) \\ 
  & & & & \\ 
 \texttt{risk} &  &  &  & 0.0003$^{*}$ \\ 
  &  &  &  & (0.0001) \\ 
  & & & & \\ 
 \emph{low-IQ} $\times$ {\sc signals} & 0.010 & 0.010 & 0.010 & 0.017 \\ 
  & (0.031) & (0.031) & (0.031) & (0.031) \\ 
  & & & & \\ 
\hline \\[-1.8ex]
Game Round Fixed Effects & No & Yes & Yes & Yes \\ 
Game Fixed Effects & No & No & Yes & Yes \\ 
Participant Covariates & No & No & No & Yes \\ 
Observations & 44,460 & 44,460 & 44,460 & 44,460 \\ 
Adjusted R$^{2}$ & 0.012 & 0.029 & 0.031 & 0.035 \\ 
\hline 
\hline \\[-1.8ex]  
\end{tabular} 
\vspace{1mm}

{\footnotesize \underline{Notes:} $^{*}$p$<$0.1; $^{**}$p$<$0.05; $^{***}$p$<$0.01. Clustered standard errors by session in parentheses.  }
\end{table} 

\begin{figure}[h!]
    \begin{center}
    \begin{tabular}{ccc}
    \includegraphics[scale=0.3]{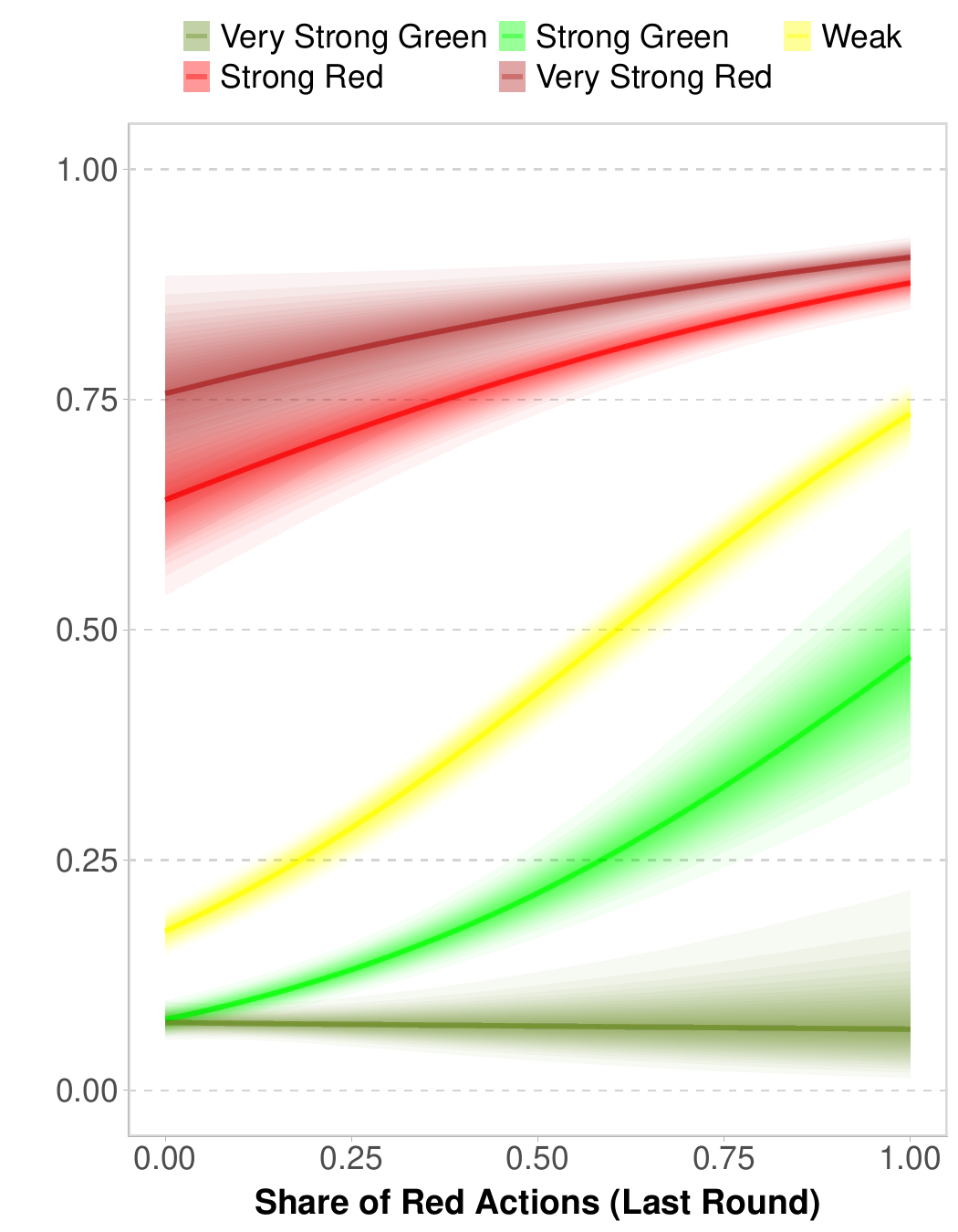} &  \includegraphics[scale=0.3]{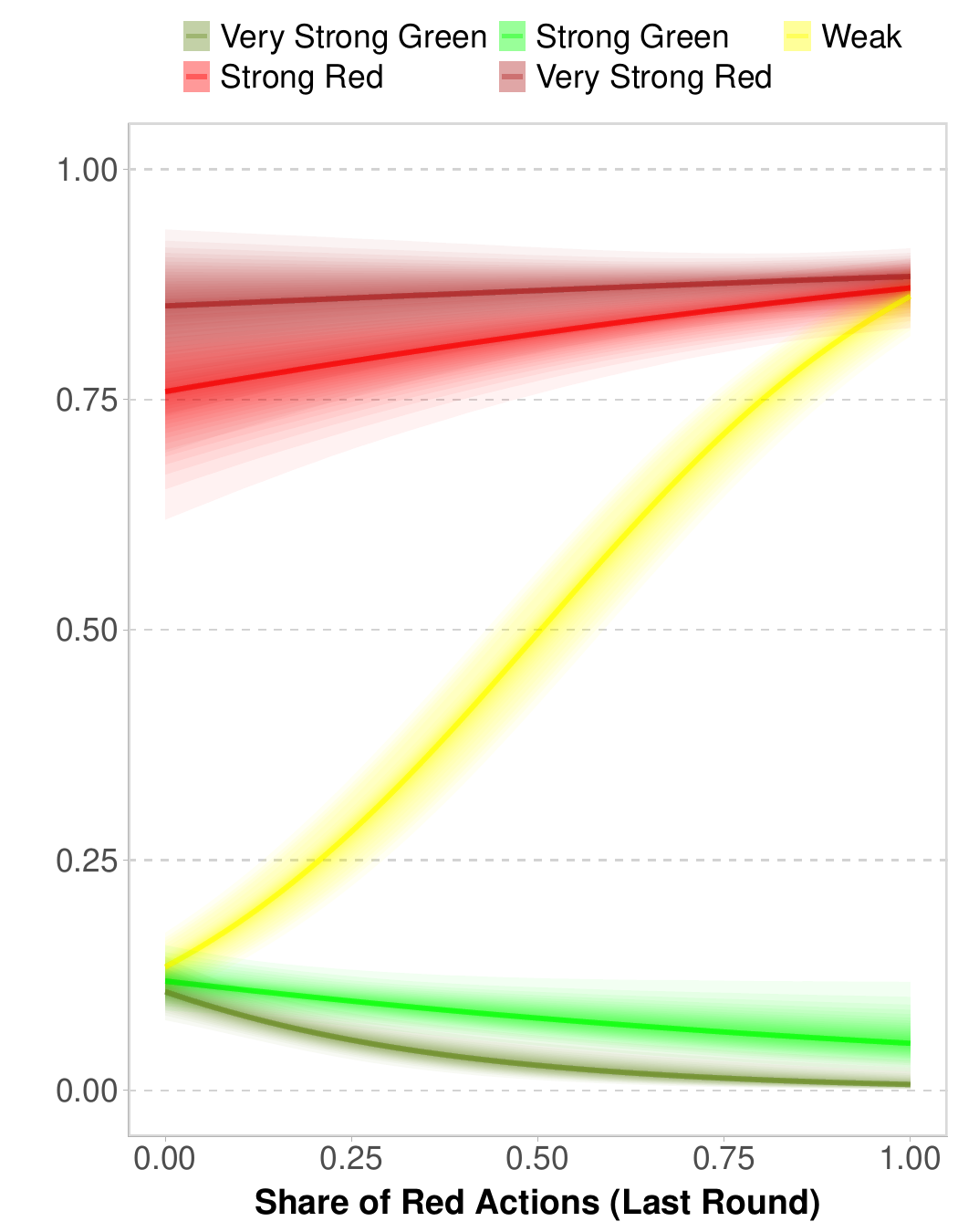}  & \includegraphics[scale=0.3]{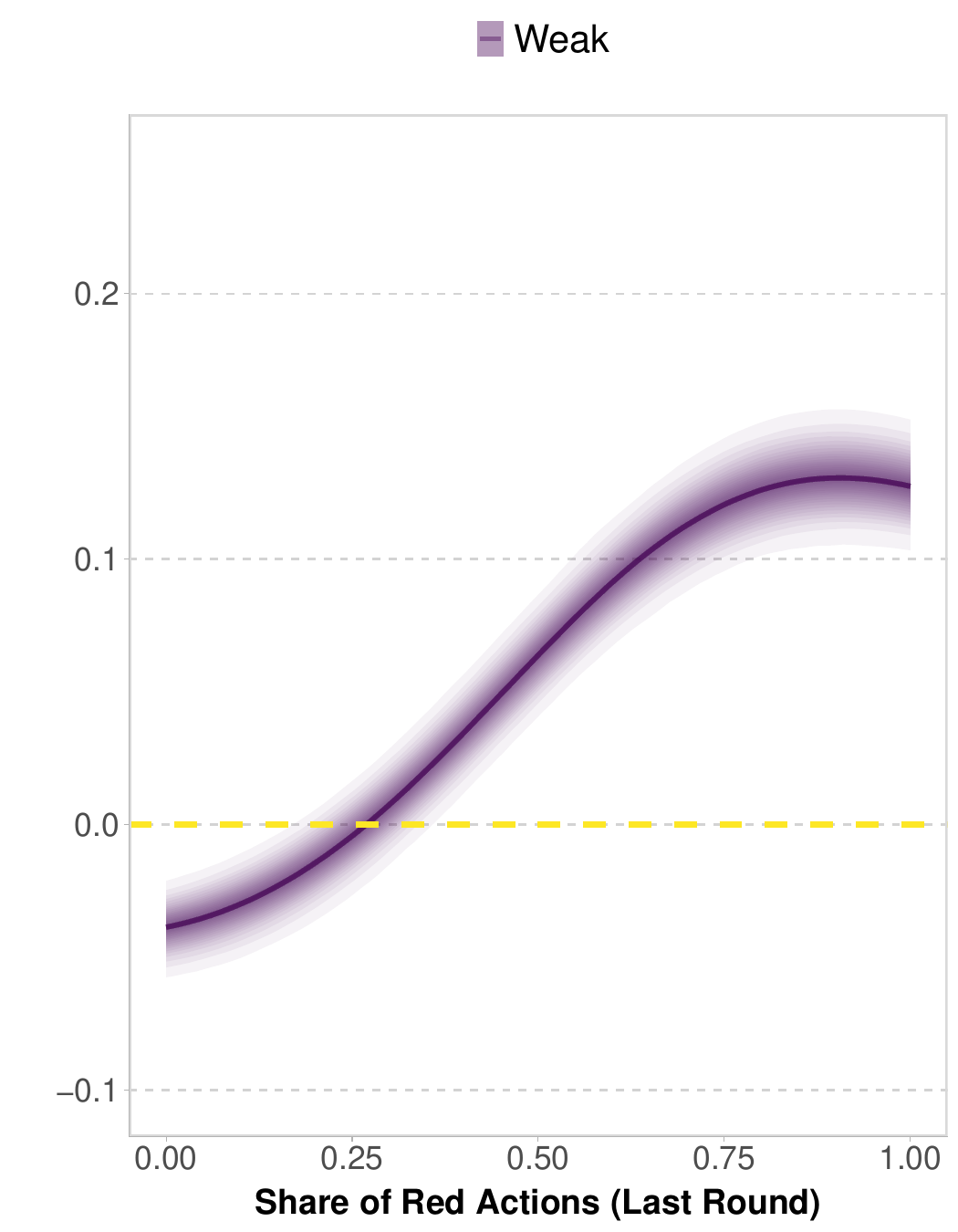} \\
    \scriptsize{Panel (a): Small Groups} & \scriptsize{Panel (b): Large Groups} & \scriptsize{Panel (c): Large - Small Groups}
     \end{tabular}
    \caption{Learning from others' actions in {\sc all} treatments}
    \label{fig:Actions_others_ALL}
    \end{center}

    {\footnotesize \underline{Notes:} Panel (a) depicts the probability of choosing red as a function of the share of red actions of other group members in the {\sc all4} treatment, obtained from a Bayesian logistic regression of the subject's action on the share of others' actions in the previous round conditional on the difference between red and green signals. Solid lines depict the median of the posterior distribution and dashed lines depict $95 \%$ confidence intervals. Panel (b) presents the same exercise for the {\sc all} treatment. Panel (c) shows the difference in the probability of a red bet between the {\sc all} and {\sc all4} treatments for a \emph{Weak} signal strength.} 
\end{figure}

\begin{figure}[h!]
    \begin{center}
    \begin{tabular}{cc}
    \includegraphics[scale=0.35]{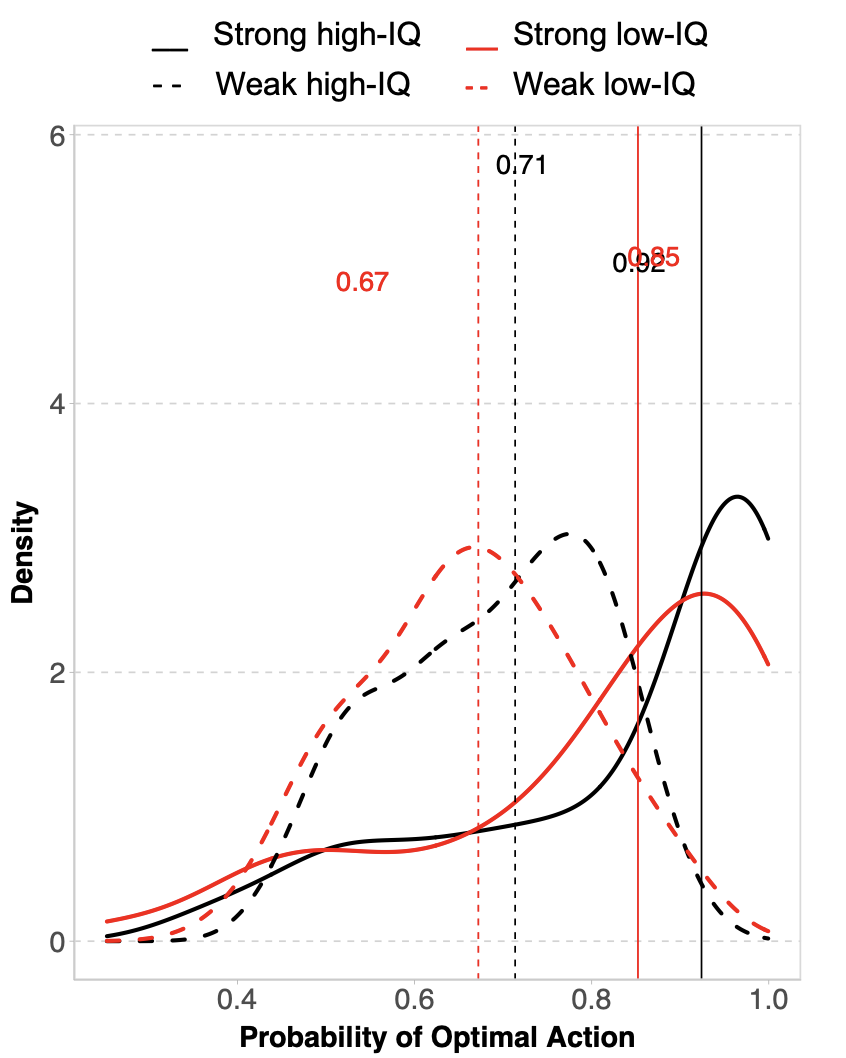} &  \includegraphics[scale=0.35]{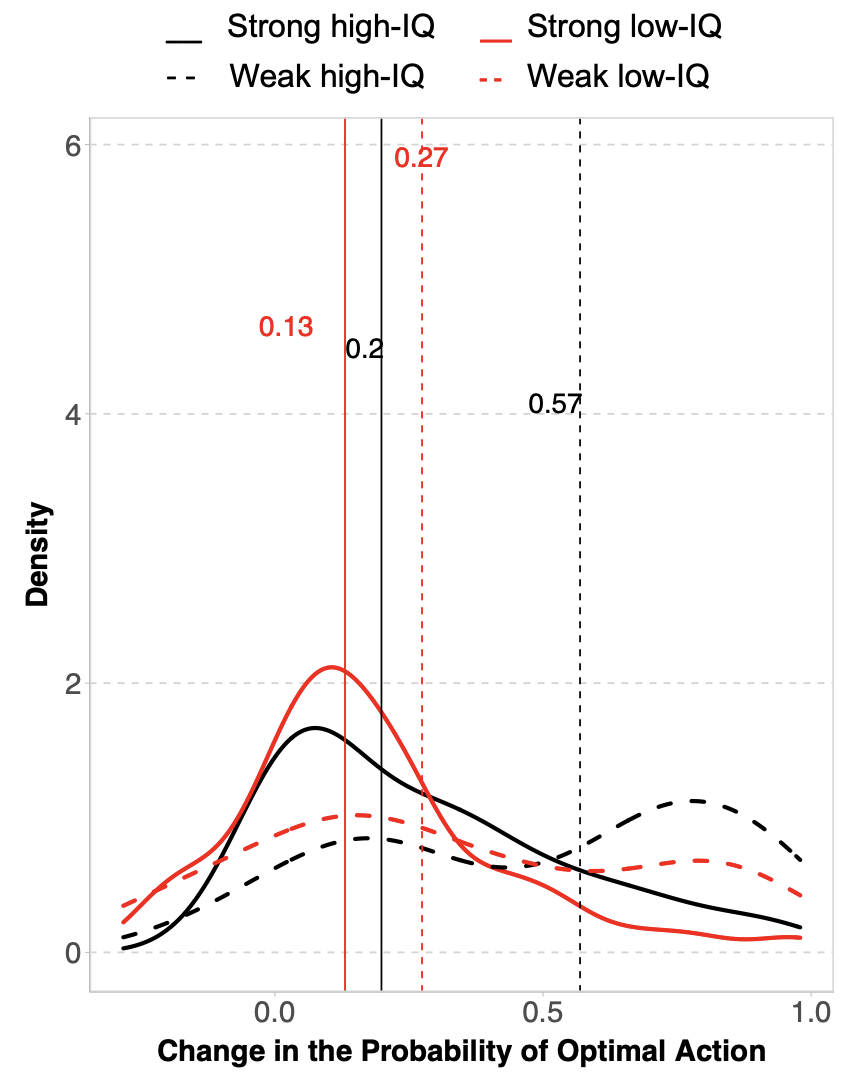}  \\
    \scriptsize{Panel (a): Responsiveness to signals} & \scriptsize{Panel (b): Responsiveness to actions} 
     \end{tabular}
    \caption{Responsiveness to signals and actions, individual level data ({\sc actions})}
    \label{fig:resp_actions}
    \end{center}

    {\footnotesize \underline{Notes:} Panel (a) shows the kernel distributions of participants’ responsiveness to signals for both weak and strong signals. The vertical lines and the numbers next to them depict median responsiveness for each group. Responsiveness to signals is calculated based on equation (3) in the main text for {\sc actions}. Responsiveness is given by the probability that a participant’s action matches signal majority, i.e., the probability of an optimal action. Panel (b) shows the kernel distributions of participants’ responsiveness to others’ actions for weak and strong signals. The vertical lines and the numbers next to them depict median responsiveness for each group. Responsiveness to others’ actions is measured by the change in the probability of choosing the action of the majority of signals when all versus none of the other group members choose the majority of signals in the last round.}
\end{figure}

\section{Informational Value of Others' Actions \label{app:rmse}}
One of the main findings from comparing the {\sc signals} and {\sc all}
treatments is that participants in the {\sc all} information treatment
outperform those in the {\sc signals} treatment. In this Appendix we
decompose this improvement in participants’ performance across treatments into the contributions of
private information from signals and social information from others’
actions. We quantify the contribution of each source relative to the
performance implied by Bayesian updating. Specifically, we compute the
root mean squared error (RMSE) between the Bayesian benchmark
probabilities of correct actions (as shown in Table \ref{tab:simulation}) and model-predicted probabilities
under two specifications: one in which participants’ behavior depends
only on signals, and another in which it depends on both signals and
others’ actions.

As before, let \(v_{it}^g\) denote the correct action of participant
\(i\) in round \(t\) of game \(g\), where
\(v_{it}^g \sim \mathrm{Bernoulli}(p_{it}^g)\). We model the
probability of a correct action as a flexible function of the
information contained in signals and in others’ actions, while
controlling for treatment differences and round effects.
\begin{equation}
  \label{eq:decomposition}
  p_{it}^g
  =
  \mathrm{logit}^{-1}\!\left(
  \alpha_{ig}
  + \lambda_t
  + \delta_{treat[i]}
  + \beta_1 \,\text{log-odds}^{\text{correct}}_{it}
  + \beta_2 \, share_{v_{-i,t-1}}
  \right),
\end{equation}
where, \(\alpha_{ig}\) is a participant--game-specific random intercept,
\(\lambda_t\) is a round-specific fixed effect, and
\(\delta_{treat[i]}\) is a fixed effect for the treatment assigned to
participant \(i\).

We capture the information contained in signals using the posterior
log-odds in favor of the true state \(\omega\) relative to the
alternative state \(\omega^{c}\):
\[
\text{log-odds}^{\text{correct}}_{it}
=
\log\left(
\frac{\mathbb{P}(\omega \mid I_{i,t})}
     {\mathbb{P}(\omega^{c} \mid I_{i,t})}
\right)
=
\begin{cases}
\text{log-odds}_{it}, & \text{if } \omega = R,\\
-\text{log-odds}_{it}, & \text{if } \omega = G,
\end{cases}
\]
where \(\text{log-odds}_{it}\), defined in
equation~(\ref{eq:log_odds}), measures the evidence in favor of
\(\omega = R\).

The information conveyed by others’ actions is summarized by the share
of others’ correct choices in the previous round,
\(share_{v_{-i,t-1}}\).

We estimate the parameters in equation~(\ref{eq:decomposition}) using a
Bayesian logistic regression that pools participants’ choices from the
{\sc signals} and {\sc all} treatments.

\begin{figure}[h!]
  \begin{center}
    \begin{tabular}{ccc}
      \scriptsize{Panel (a): Treatment Effect} &
                                                          \scriptsize{Panel
                                                          (b):
                                                          Signals' Effect} &
      \scriptsize{Panel
                                                          (b):
                                                          Actions' Effect}\\ 
\includegraphics[scale=0.25]{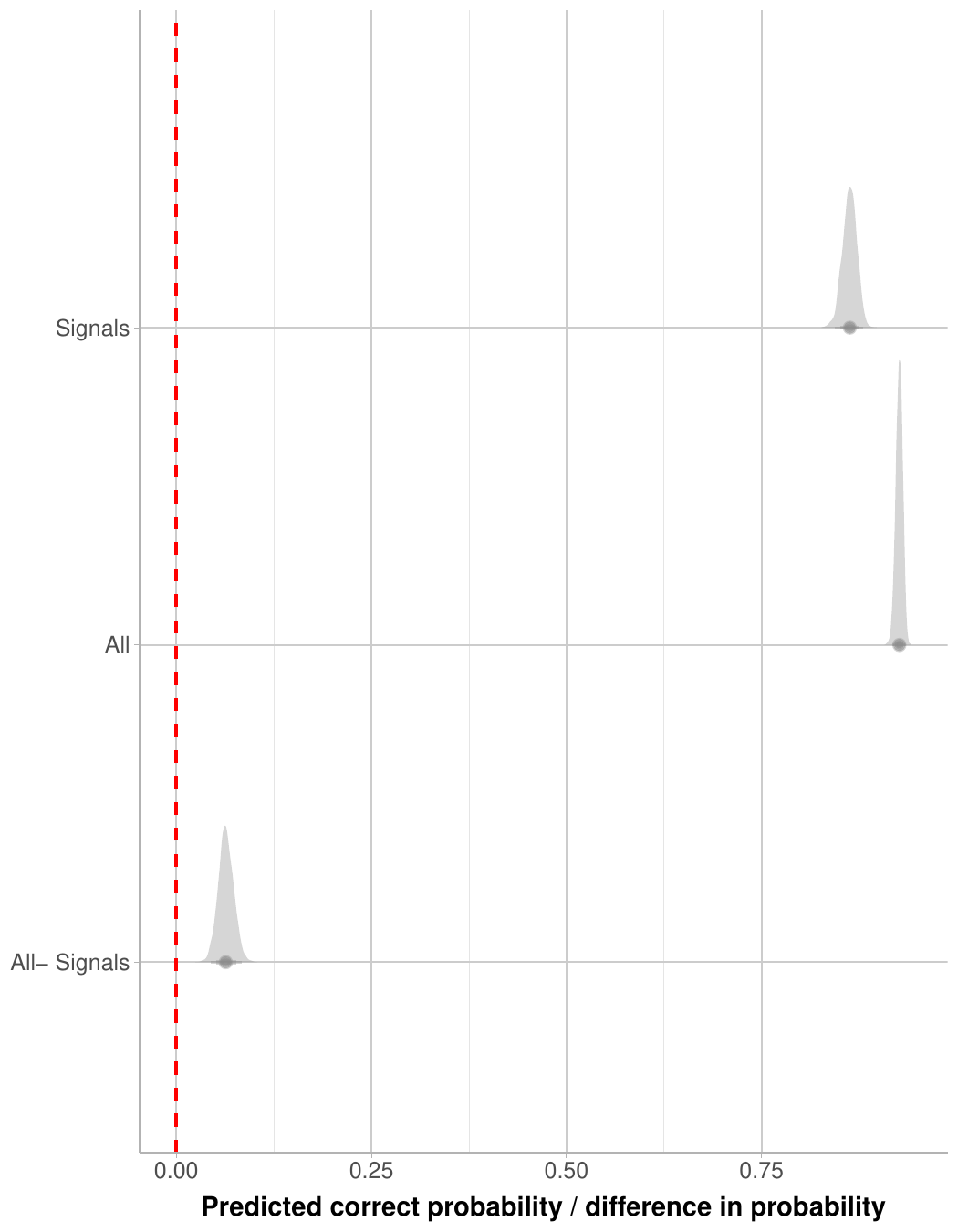}
      &
        \includegraphics[scale=0.25]{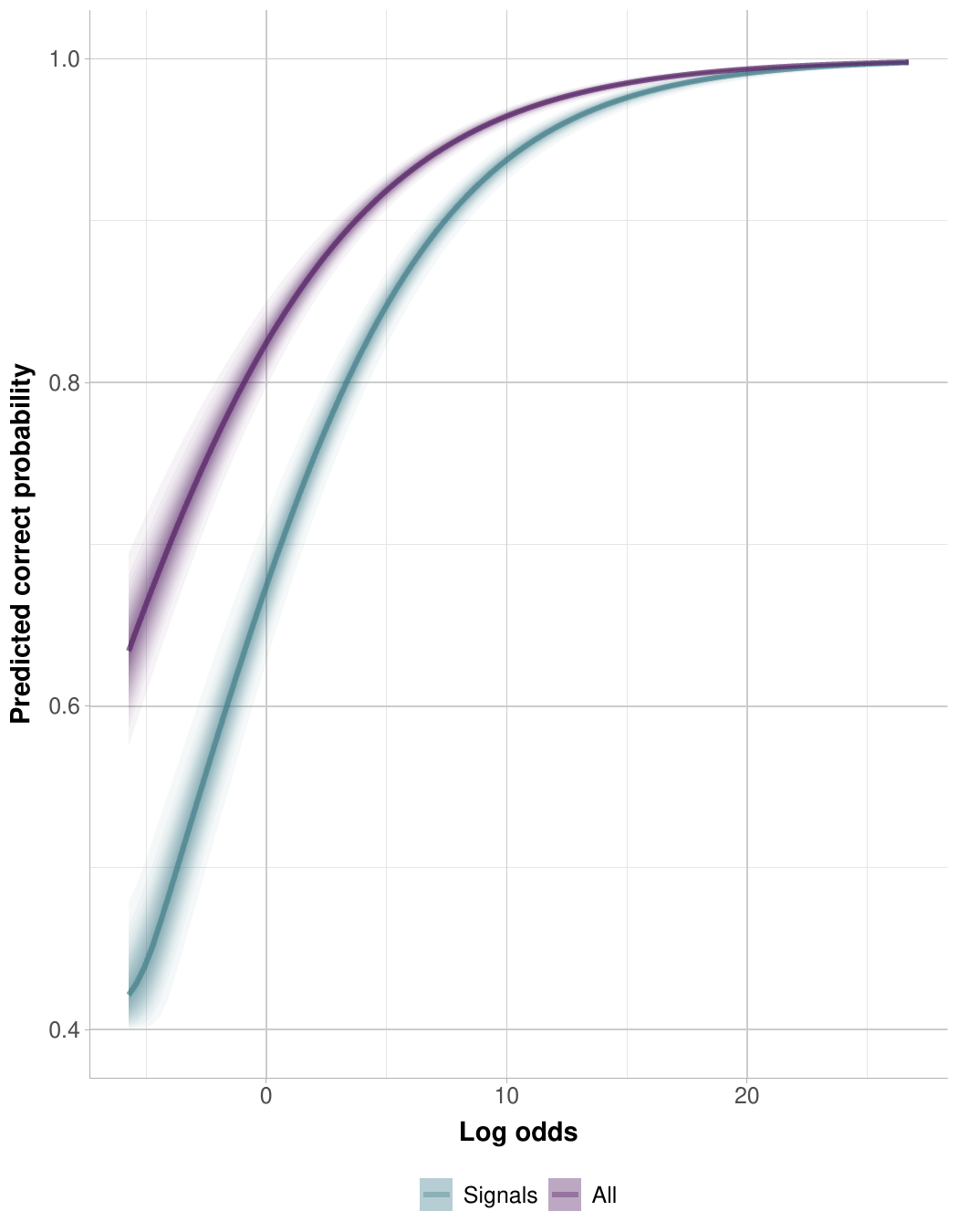}
                                                                             &
                                                                               \includegraphics[scale=0.25]{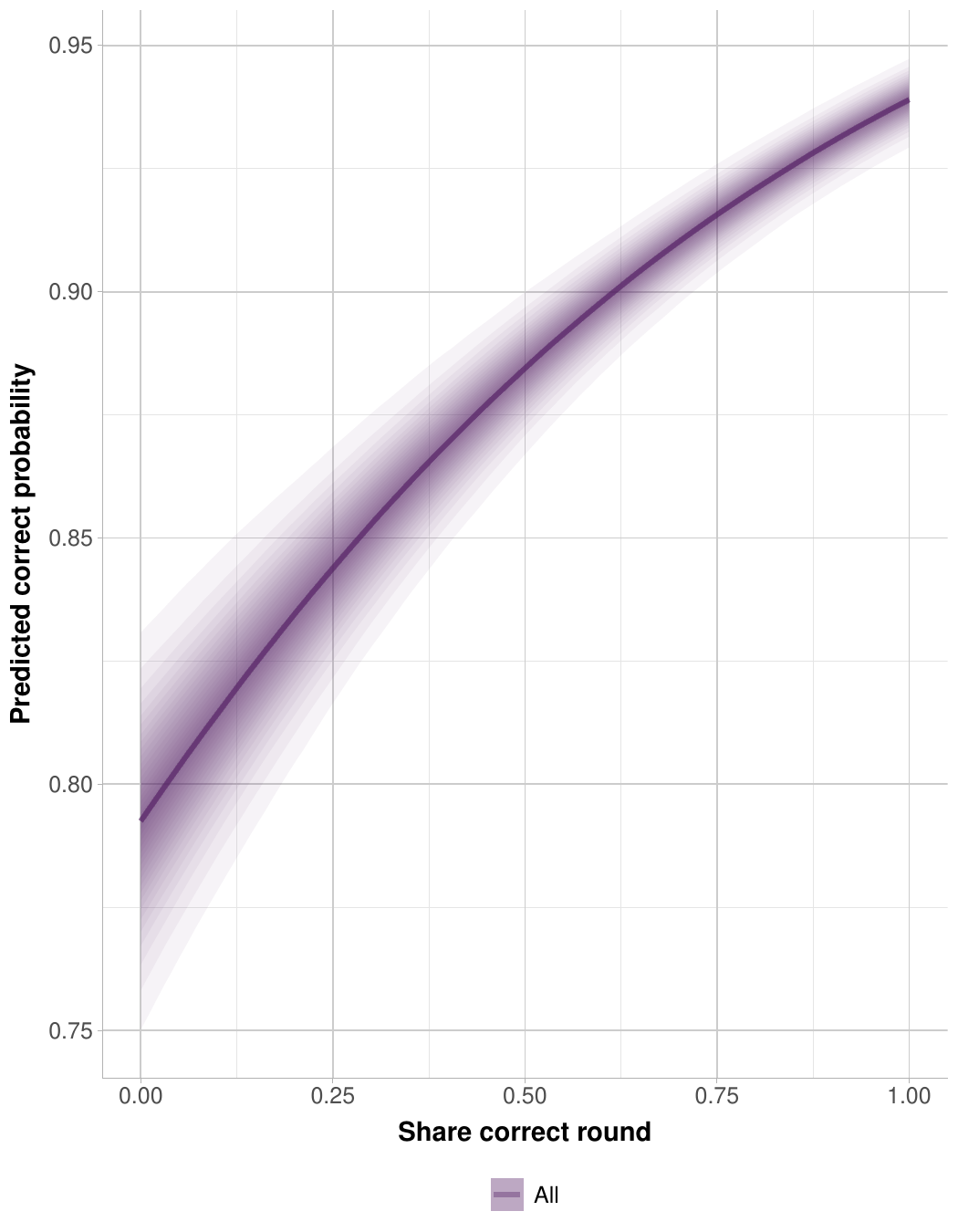}

    \end{tabular} 
\caption{Effect of Information on Signals and Actions}\label{fig:decomposition} 
\end{center}
{\footnotesize \underline{Notes:} Panel (a) presents the posterior probability of a correct choice for the {\sc signals} and {\sc all} treatments, holding fixed both the signal information and the information from others’ actions. Panel (b) presents the posterior probability of a correct choice as log-odds vary from their minimum to maximum values in the data, for both the {\sc signals} and {\sc all} treatments, while holding the information from others’ actions fixed. Panel (c) presents the posterior probability of a correct choice as the share of others’ correct choices varies from its minimum to maximum values in the data for the {\sc all} treatment, while holding the signal information fixed.}
\end{figure}

Figure \ref{fig:decomposition} summarizes the main results from the
estimation of equation~(\ref{eq:decomposition}). The left panel
reports the posterior distribution of the probability of a correct
action under the {\sc signals} and {\sc all} treatments, averaging
across participants, games, and rounds, while holding fixed the
information from signals (\(\text{log-odds}_{it}\)) and others’
actions (\(share_{v_{-i,t-1}}\)) at their median values in the data.
We find a significant increase in performance of 6
percentage points when participants are assigned to the {\sc all}
treatment relative to the {\sc signals} treatment, that cannot be
directly attributed to the informational value of signals or others' actions.

The middle panel reports the effect of signals on performance by
treatment. We show the probability of a correct action as
\(\text{log-odds}^{\text{correct}}_{it}\) varies from its minimum to its
maximum in the data, while holding others’ actions
(\(share_{v_{-i,t-1}}\)) fixed at their median values. For participants
in both the {\sc signals} and {\sc all} treatments, performance
improves substantially as signals become more informative about the
true state. This effect is stronger for participants in the {\sc all}
treatment, particularly when signals are less informative.

The right panel isolates the effect of others’ actions on performance
by reporting the probability of a correct action as
\(share_{v_{-i,t-1}}\) varies from zero to one, while holding the
informativeness of signals
(\(\text{log-odds}^{\text{correct}}_{it}\)) fixed at its median value.
Consistent with the results in the main text (Observation~2),
participants in the {\sc all} treatment improve their performance by
relying on the redundant information conveyed by others’ actions. This
improvement remains substantial even after accounting for the
informativeness of observed signals, increasing the probability of a
correct action by more than 10 percentage points as the share of
others’ correct actions in the previous round rises from zero to one.

Using the estimated parameters, we construct a \emph{baseline}
predicted probability of a correct action driven only by variation
across participants and games (\(\alpha_{ig}\)), rounds
(\(\lambda_t\)), and treatments (\(\delta_{treat[i]}\)), setting
\(\beta_1 = \beta_2 = 0\) and thereby shutting down the contribution
of both signals and others’ actions.

We then construct predicted probabilities driven by signals by adding
to the \emph{baseline} specification the information provided by
signals, \(\beta_1 \,\text{log-odds}^{\text{correct}}_{it}\), while
keeping \(\beta_2 = 0\).

Finally, we construct predicted probabilities driven by
\emph{signals + actions} by augmenting the \emph{signals}
specification with the effect of others’ actions,
\(\beta_2 \, share_{v_{-i,t-1}}\).

\begin{figure}[h!]
  \begin{center}
\includegraphics[scale=0.45]{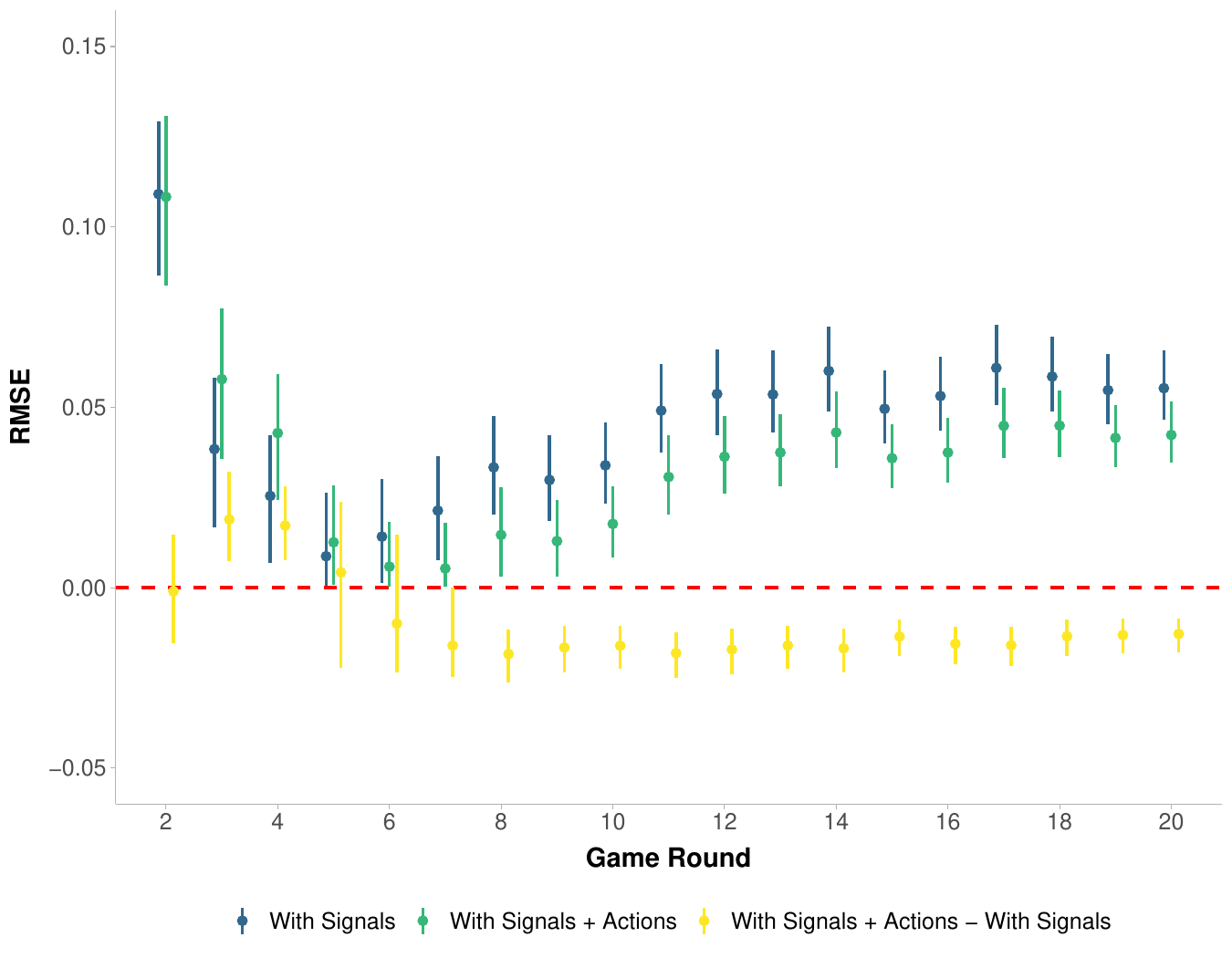}
\caption{RMSE relative to Optimal Behavior}\label{fig:rmse} 
\end{center}
{\footnotesize \underline{Notes:} This figure reports the posterior median root mean squared error (RMSE)
between the Bayesian benchmark probabilities of correct actions and the
model-predicted probabilities under two specifications. Predicted
probabilities based on \emph{signals} include treatment fixed effects,
game-round fixed effects, and signal information measured using
posterior log-odds. Predicted probabilities based on \emph{signals +
actions} additionally incorporate social information from others'
actions, measured as the share of others’ correct choices in the
previous round.}
\end{figure}

Figure \ref{fig:rmse} reports, for each game round, the additional
informational value of others’ actions beyond signals by comparing the
posterior RMSE of predicted probabilities relative to the Bayesian
benchmark under the \emph{signals} and \emph{signals + actions}
specifications. Averaging across rounds, incorporating others’ actions
reduces RMSE by approximately 1 percentage point. This reduction is
driven by improved accuracy from round seven onward, where the RMSE
decreases by 1.5 percentage points.

Table \ref{tab:rmse} reports the median RMSE estimates across
scenarios. Overall, the largest improvement in accuracy comes from
incorporating signals, which reduces RMSE by approximately 3
percentage points relative to the \emph{baseline} scenario, with a
particularly large effect in later rounds (up to 6 percentage points).
Beyond the information provided by signals, incorporating the
redundant information contained in others’ actions further improves
participants’ performance relative to the Bayesian benchmark, with an
additional reduction in RMSE of 1 percentage point, equivalent to one-third of the
informational value of signals.

\begin{table}[h!]
  \centering
  \caption{The Effect of Signals and information on RMSE.}
\label{tab:rmse}
\begin{tabular}{lccccc}
\hline
Round & Baseline &  Signals & S + A & (S-B) & (S+A - Signals) \\
  \hline
  \hline
Round 2 & 0.216 & 0.109 & 0.108 & -0.106 & -0.001  \\
Round 3 & 0.135 & 0.038 & 0.058 & -0.096 & 0.019  \\
Round 4 & 0.098 & 0.025 & 0.043 & -0.072 & 0.017  \\
Round 5 & 0.057 & 0.009 & 0.013 & -0.049 & 0.004  \\
Round 6 & 0.036 & 0.014 & 0.006 & -0.022 & -0.010 \\
Round 7 & 0.017 & 0.021 & 0.005 & 0.004 & -0.016 \\
Round 8 & 0.004 & 0.033 & 0.015 & 0.029 & -0.018 \\
Round 9 & 0.010 & 0.030 & 0.013 & 0.020 & -0.017 \\
Round 10 & 0.022 & 0.034 & 0.018 & 0.012 & -0.016  \\
Round 11 & 0.043 & 0.049 & 0.031 & 0.006 & -0.018 \\
Round 12 & 0.054 & 0.054 & 0.036 & -0.000 & -0.017  \\
Round 13 & 0.062 & 0.054 & 0.037 & -0.008 & -0.016  \\
Round 14 & 0.077 & 0.060 & 0.043 & -0.017 & -0.017  \\
Round 15 & 0.071 & 0.050 & 0.036 & -0.022 & -0.014  \\
Round 16 & 0.083 & 0.053 & 0.037 & -0.030 & -0.016  \\
Round 17 & 0.101 & 0.061 & 0.045 & -0.040 & -0.016  \\
Round 18 & 0.108 & 0.058 & 0.045 & -0.049 & -0.014 \\
Round 19 & 0.113 & 0.055 & 0.042 & -0.058 & -0.013  \\
Round 20 & 0.124 & 0.055 & 0.042 & -0.069 & -0.013 \\
\hline
Average & 0.075 & 0.045 & 0.035 & -0.030 & -0.010 \\
\hline
\end{tabular}
\parbox{\linewidth}{\footnotesize\raggedright
\underline{Notes:} This table reports posterior median RMSE between
the probabilities of correct actions under the Bayesian Benchmark and
predicted probabilities for three different scenarios: (i)
\emph{baseline} includes a treatment fixed effect and game-round fixed
effects. (ii) \emph{signals} additionally incorporate signal
information measured using log-odds. (iii) \emph{signals + actions} additionally incorporate others' actions, measured by the share of others' correct choices in the previous round.}
\end{table}

\section{Heterogeneity in the Effect of Redundant Information \label{app:hetero}}
\setcounter{figure}{0}
\setcounter{table}{0}
The main finding that participants perform better in the {\sc all}
treatment relative to the {\sc signals} treatment, as shown in
Figure~\ref{fig:Correct}, reflects an average treatment effect that
masks heterogeneity across participants in their likelihood of making
a correct bet. In this Appendix, we show that the performance advantage
of the {\sc all} treatment over the {\sc signals} treatment is robust
to conditioning on the distribution of prediction quality across
participants.

For this purpose, we model the probability of making a correct bet at
the individual level using a flexible binary-response specification
with individual-specific random effects.

For a given treatment ({\sc all} or {\sc signals}, the data generating process for the correct action of participant $i$ in round
$t$ of game $g$, $v_{it}^g$, is given by $v_{it}^g \sim
\text{Bernoulli}(p_{it}^g)$, where
\begin{equation*}
  p_{it}^g=\text{logit}^{-1}(\alpha_{i} + \beta_{it} + \upsilon_{g}),
\end{equation*}
where $\alpha_{i}$ is a participant-specific random intercept,
$\beta_{it}$ is a participant-specific random effect for round $t$. We allow a flexible
correlation pattern across individual effects:
\[
\begin{pmatrix}
\alpha_{i}\\
\beta_{it}
\end{pmatrix}
\sim
\mathcal N\!\left(
\begin{pmatrix}
0\\
0
\end{pmatrix},
\Sigma_u
\right).
\]

Finally, $\upsilon_{g} \sim \mathcal N(0,\sigma_g^2)$ denotes the
game-specific random intercept.

We leverage variation across rounds and games within participants to
recover subject-level response probabilities via Bayesian logistic
regression. Using the posterior estimates of individual probabilities,
we replicate the analysis in Section~\ref{sec:RedundantInfo}, comparing
how often subjects correctly guess the state in each game round across
treatments by quantiles of these individual probabilities.

\begin{figure}[h!]
\begin{center}
\includegraphics[scale=0.55]{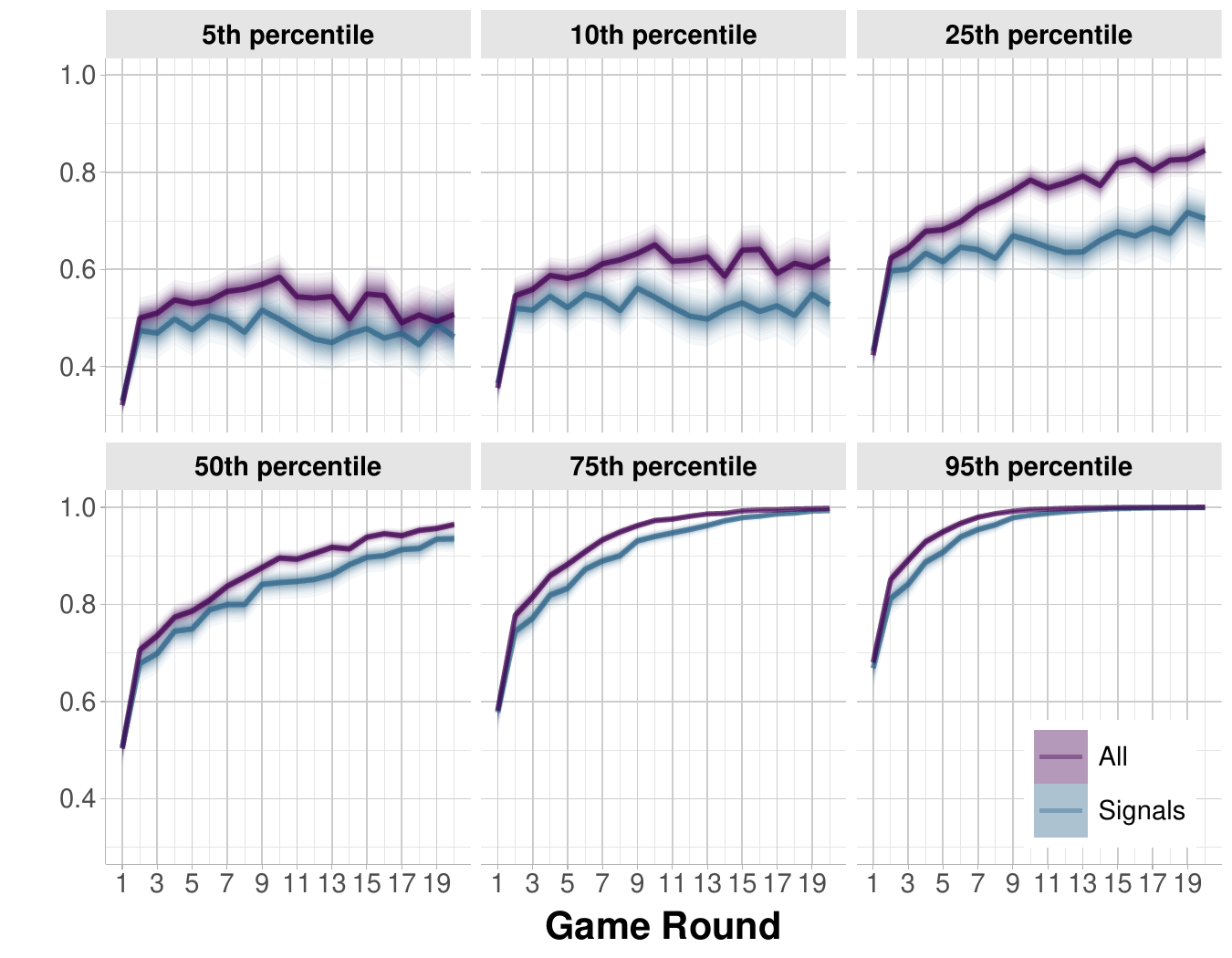}   
\caption{Statistics in the {\sc all} and {\sc signals}  treatments by
  participant heterogeneity \label{app_fig:effect_hetero}}
\end{center}
{\footnotesize \underline{Notes:} The figure presents the average
  probability of correct actions in each round, averaged across
  games. Shaded regions represent credible intervals from 50\%
  (darkest) to 95\% (faintest) levels.}
\end{figure}

Figure~\ref{app_fig:effect_hetero} summarizes these results. Consistent
with the aggregate findings, subjects across quantiles perform better
in the {\sc all} treatment relative to {\sc signals}. Moreover, the
effect of redundant information is significantly larger for
participants in the lower quantiles of the distribution (i.e., the 5th,
10th, and 25th percentiles). Although the gap between treatments is
smaller at higher quantiles, it remains statistically significant and
economically meaningful (i.e., around $5$ p.p.), particularly in earlier rounds.

\begin{figure}[h!]
\begin{center}
\includegraphics[scale=0.55]{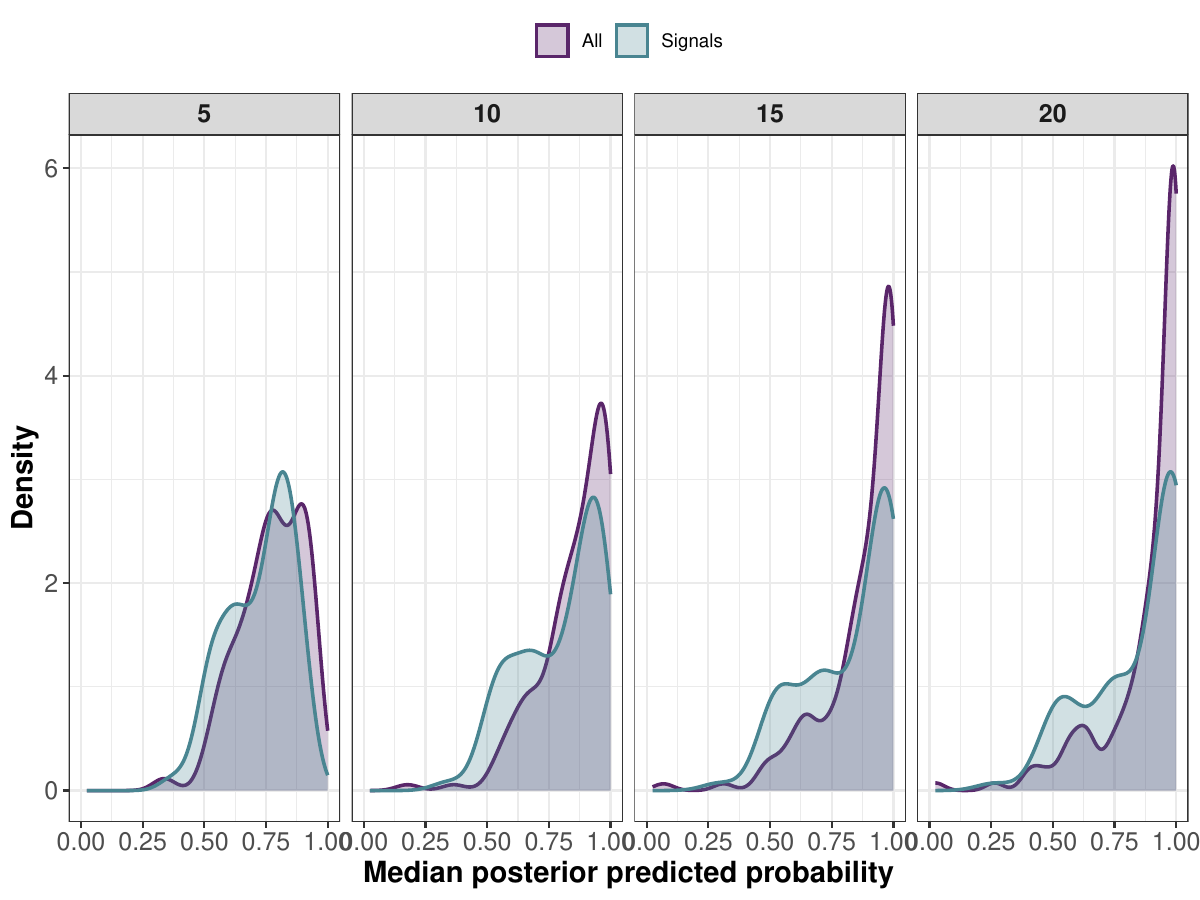}   
\caption{Statistics in the {\sc all} and {\sc signals}  treatments by
  participant heterogeneity \label{app_fig:effect_hetero_dens}}
\end{center}
{\footnotesize \underline{Notes:} The figure presents the average
  probability of correct actions in each round, averaged across
  games. Shaded regions represent credible intervals from 50\%
  (darkest) to 95\% (faintest) levels.}
\end{figure}

Figure~\ref{app_fig:effect_hetero_dens} shows the full distribution of
individual-level correct probabilities across treatments and game
rounds. Consistent with the aggregate findings, as the game progresses,
individuals in both {\sc signals} and {\sc all} treatments improve
their performance. However, we find substantial heterogeneity across
participants, with individual probabilities ranging from below 0.25 to
above 0.95. Comparing treatments, we find that in later rounds (i.e.,
rounds 10, 15, and 20) a larger
share of individuals performs better in the {\sc all} treatment than
in the {\sc signals} treatment.

\section{Covariate Balance in Participant Characteristics\label{app:balance}}
\setcounter{figure}{0}
\setcounter{table}{0}

In Appendix~\ref{sec:AppOnlineSessions}, we show that aggregate
outcomes and individual-level behavior are very similar and
statistically indistinguishable across experimental sessions conducted
in the physical lab at UCSD and in the virtual lab at OSU. However,
because half of the experimental sessions in the {\sc all} and {\sc
signals} treatments were conducted at different locations, with
distinct pools of college students (UCSD and OSU), we assess in this
Appendix potential imbalances in participants’ characteristics across
treatments and their implications for inference on individual
behavior.

\begin{figure}[h!]
\begin{center}
\includegraphics[scale=0.55]{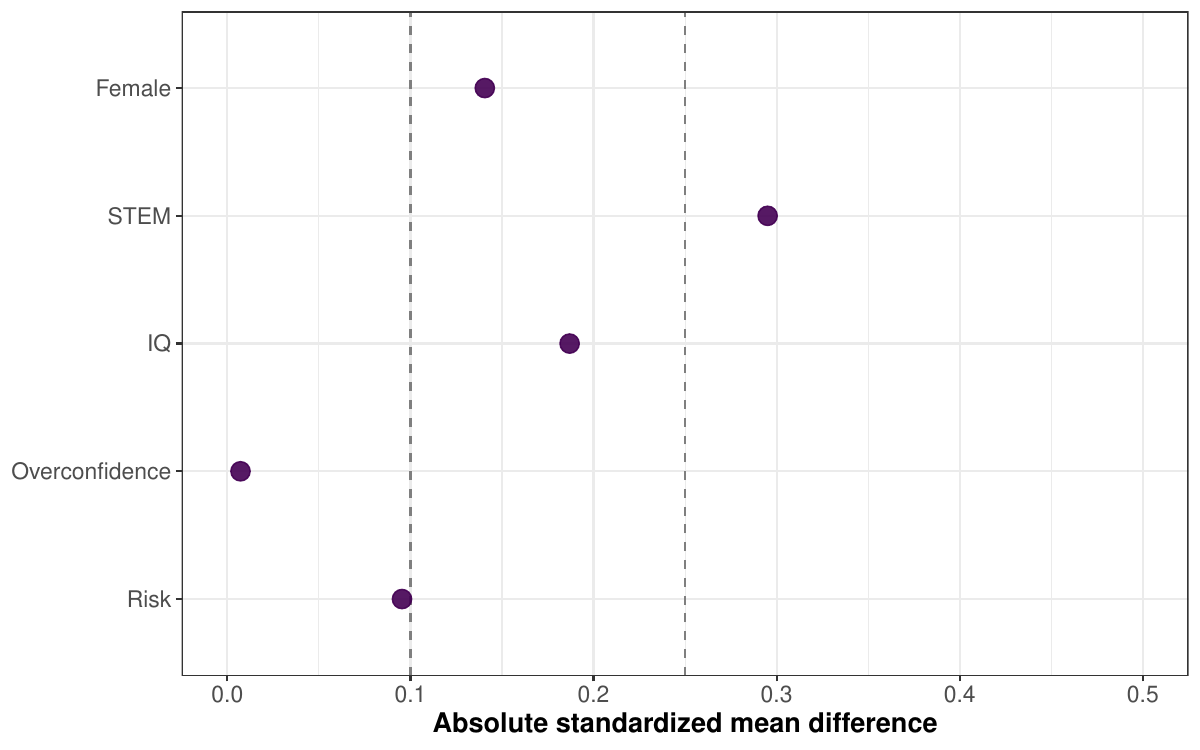}   
\caption{Covariate Balance  \label{app_fig:cov_balance}}
\end{center}
{\footnotesize \underline{Notes:} The figure presents absolute
  standardized mean differences  in covariate value between {\sc all} and {\sc signals}
treatments.}
\end{figure}

As a measure of covariate balance, Figure~\ref{app_fig:cov_balance}
reports absolute standardized mean differences between the {\sc all}
and {\sc signals} treatments. We find almost no imbalance for the
constructed covariates \texttt{overconfidence} and \texttt{risk}
(less than 0.1), moderate imbalance for \texttt{female} and the
constructed \texttt{iq} (less than 0.25), and substantial imbalance for \texttt{STEM},
with 31\% of participants in the {\sc all} treatment having a STEM
major, compared to 21\% in the {\sc signals} treatment.

We summarize the extent of imbalance across covariates by estimating
the propensity score, defined in our setting as
\[
\Pr(T_i = \text{{\sc all}} \mid X_i),
\]
where \(X_i\) is the vector of participant covariates
(\texttt{overconfidence}, \texttt{risk}, \texttt{female},
\texttt{iq}, \texttt{STEM}). We estimate the propensity score using a
logistic regression of \(T_i\) on \(X_i\).

\begin{figure}[h!]
\begin{center}
\includegraphics[scale=0.55]{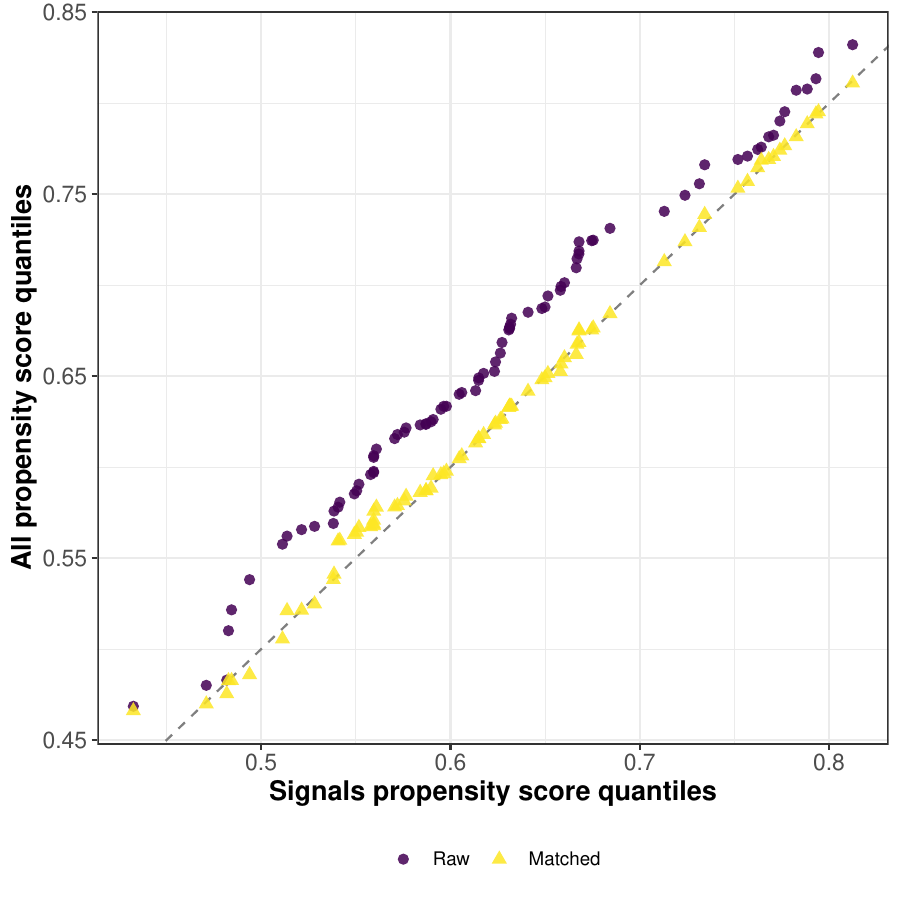}   
\caption{QQ Plot of Propensity Score for Treatment Assignment ({\sc all} and {\sc signals}) \label{app_fig:qq_plot}}
\end{center}
{\footnotesize \underline{Notes:} The purple dots represent empirical
  QQ estimates for the raw data. The yellow dots represent QQ
  estimates for the matched data. The 45-degree line indicates
  identical distributions. The propensity score is estimated with a
  logistic regression of treatment ({\sc all} and {\sc signals}) on
  \texttt{female}, \texttt{stem}, \texttt{iq}, \texttt{overconfidence}
and \texttt{risk}.}
\end{figure}

Figure~\ref{app_fig:qq_plot} compares the distribution of the estimated
propensity score between the {\sc signals} and {\sc all} treatments by
contrasting their quantiles across groups. The purple dots lie
consistently above the 45-degree line, indicating that participants in
the {\sc all} treatment differ systematically from those in the {\sc
signals} treatment in their observable characteristics.

Based on the estimated propensity score, we implement one-to-one
optimal matching. Each of the 82 participants in the {\sc signals}
treatment is matched to one of the 152 participants in the {\sc all}
treatment, yielding a total of 164 matched observations (82 matched
pairs), with 70 unmatched participants in the {\sc all} treatment.

The yellow triangles in Figure~\ref{app_fig:qq_plot}, which lie close
to the 45-degree line, indicate that matching yields nearly identical
propensity score distributions across the {\sc all} and {\sc signals}
treatments at most quantiles.

\begin{figure}[h!]
  \begin{center}
    \begin{tabular}{cc}
      \scriptsize{Panel (a): Correct Actions, by round} &
                                                          \scriptsize{Panel
                                                          (b):
                                                          Treatment
                                                          Differences
                                                          in Correct Actions, by round}\\ 
\includegraphics[scale=0.35]{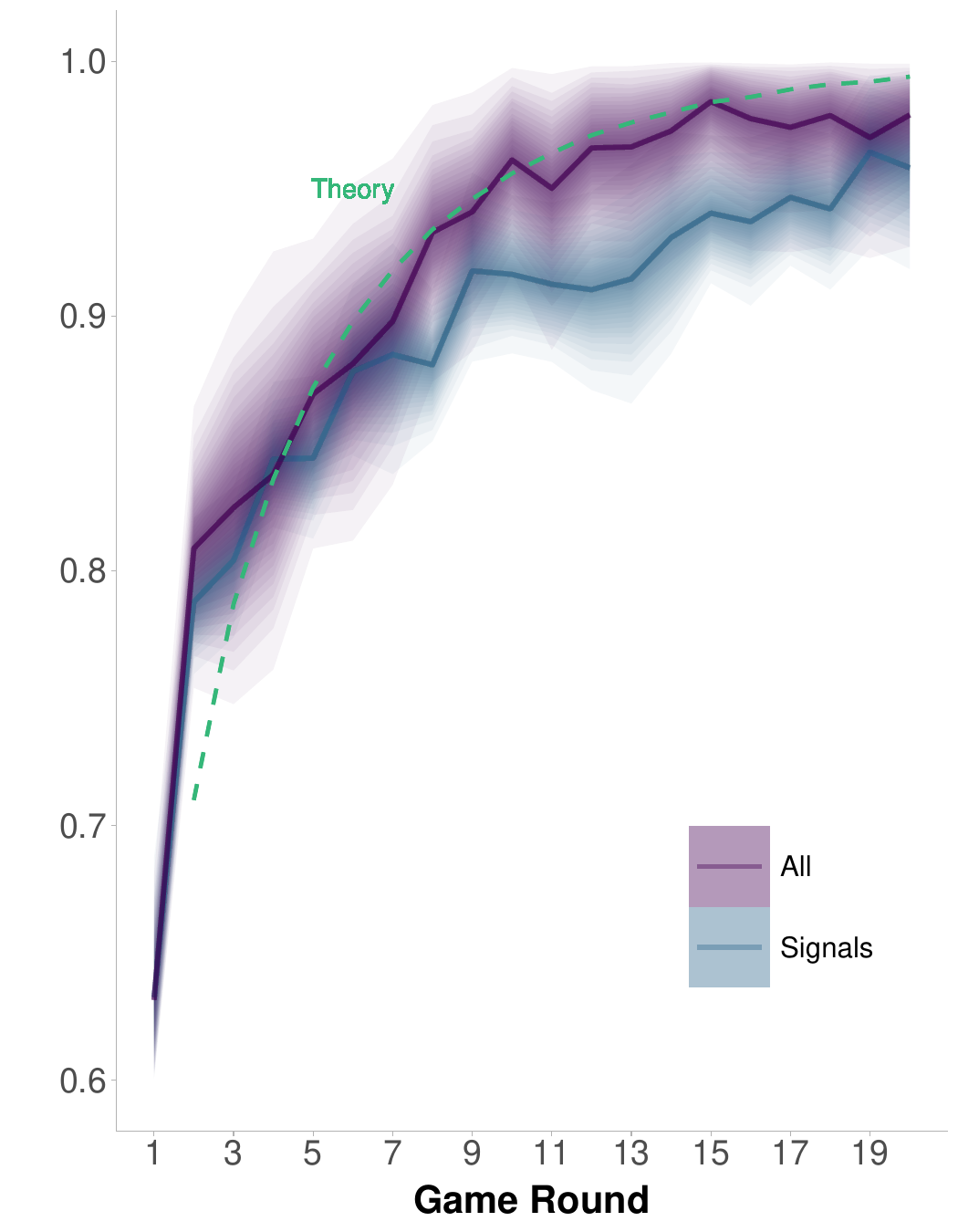}
      & \includegraphics[scale=0.35]{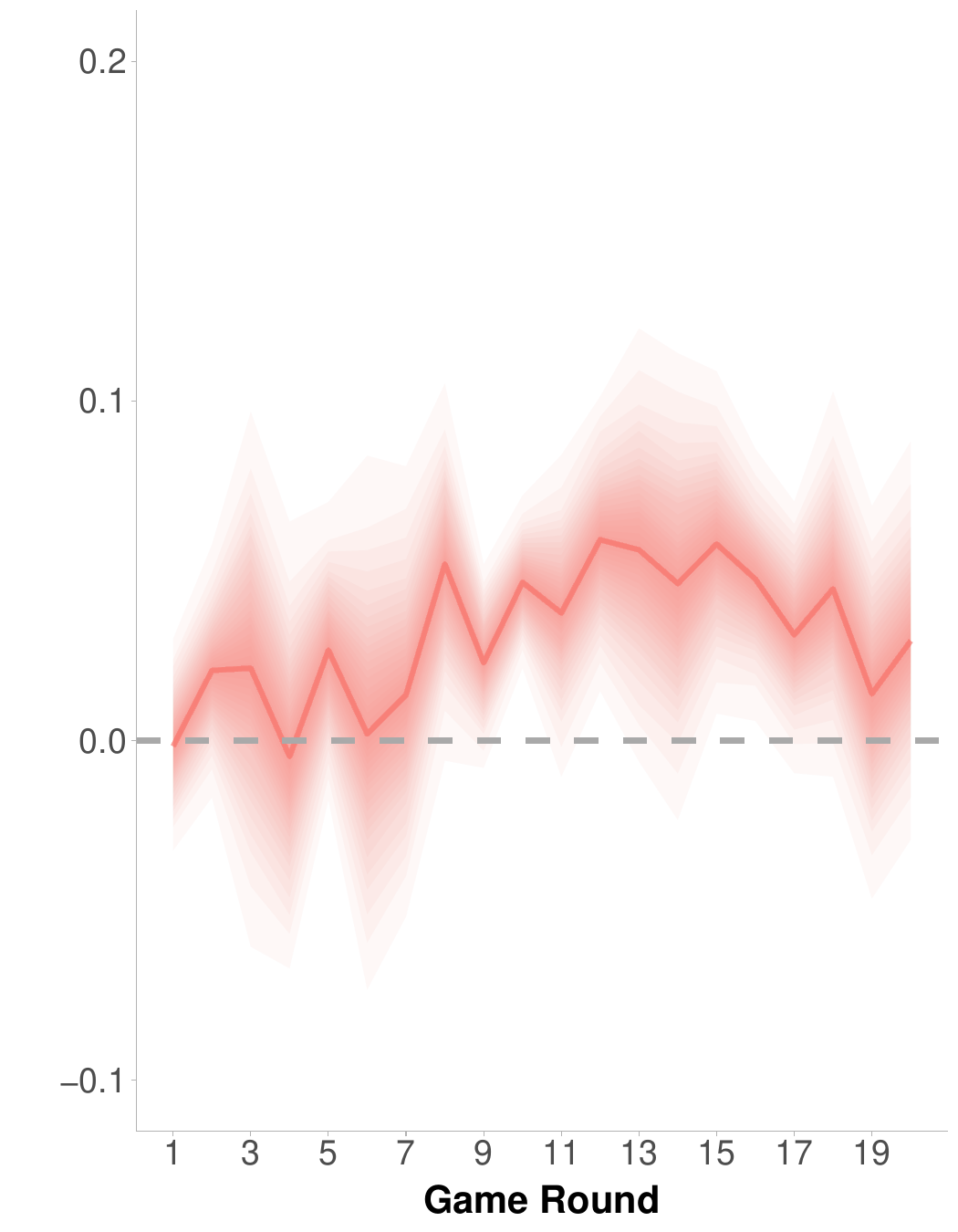}
    \end{tabular} 
\caption{Aggregate statistics in the {\sc all} and {\sc signals}
  treatments for the Exact Matched Sample}\label{fig:Correct_bal} 
\end{center}
{\footnotesize \underline{Notes:} Panel (a) presents the average
  frequency of correct actions in each round, averaged across
  games. Panel (b) presents the treatment difference in the  average
  frequency of correct actions in each round, averaged across
  games. Shaded regions represent $95\%$ confidence intervals from
  50\% (darkest) to 95\% (faintest) probability levels. Confidence
  intervals are constructed with a variance-covariance matrix
  clustered by session.}
\end{figure}

Using the matched sample, we replicate the main analysis comparing the
fraction of correct actions between the {\sc signals} and {\sc all}
treatments. Figure~\ref{fig:Correct_bal} reports the results of this
exercise. 

First, in the balanced subsample, the main aggregate finding remains
unchanged: participants perform better in the {\sc all} treatment than
in the {\sc signals} treatment. Moreover, relative to the full sample,
participants in the matched sample perform significantly better in
both treatments, with outcomes approaching the theoretical
probabilities of correct actions under Bayesian updating. Moreover, compared to the raw data, individuals in the
matched sample perform significantly better in both treatments, almost
matching the theoretical probabilities of correct actions under
Bayesian updating and common knowledge of rationality.  

\section{Random Actions in the Data \label{app:random}}
Our main analysis examines how individuals learn from others’ signals
and actions, and the extent to which their behavior deviates from the
Bayesian optimum. In this Appendix, we assess the extent to which such
deviations are driven by participants who behave as if random, ignoring
available information. We focus on the {\sc none} and {\sc signals}
treatments, where relevant information is directly provided through
observed signals.

To quantify random behavior, we estimate the effect of
information \(I_{i,t}\) on individual choices. We measure the
information from signals with the posterior
log-odds,
\begin{equation}
  \label{eq:log_odds}
 \text{log-odds}_{it}
 =
 \log\left(
 \frac{Pr(\omega = R \mid I_{i,t})}
      {Pr(\omega = G \mid I_{i,t})}
 \right),
\end{equation}
which captures the relative evidence in favor of \(\omega = R\)
relative to \(\omega = G\).\footnote{Similar results obtain when using
the difference between the number of red and green signals.} In
particular, under the {\sc none} treatment,
\(I_{i,t}=(s_{i,\tau})_{\tau<t}\), whereas under the {\sc signals}
treatment,
\(I_{i,t}=(s_{j,\tau})_{j\in N,\tau<t}\). 

Given this measure of signals’ informativeness, we model participant
\(i\)’s choice in round \(t\) of game \(g\), denoted \(a^g_{it}\), as
\[
a^g_{it} \sim \mathrm{Bernoulli}(p^g_{it}),
\]
where
\begin{equation*}
p^g_{it}
=
\mathrm{logit}^{-1}\!\left(
\alpha_i + \upsilon_g + \lambda_t + \beta_i\,\text{log-odds}_{it}
\right).
\end{equation*}
The parameter of interest is \(\beta_i\), which captures participant
\(i\)’s sensitivity to the posterior log-odds. The term \(\alpha_i\) is
a participant-specific random intercept, \(\upsilon_g\) is a
game-specific random intercept, and \(\lambda_t\) is a round-specific
random intercept. We assume
\[
\alpha_i \sim \mathcal N(0,\sigma_\alpha^2),
\qquad
\beta_i \sim \mathcal N(0,\sigma_\beta^2),
\qquad
\upsilon_g \sim \mathcal N(0,\sigma_\upsilon^2),
\qquad
\lambda_t \sim \mathcal N(0,\sigma_\lambda^2).
\]

We estimate responsiveness to information using a Bayesian logistic
regression. Within this framework, we evaluate the posterior
distribution of \(\beta_i\), focusing on whether it places substantial
mass near zero, consistent with random behavior  or instead indicates learning, i.e., \(\beta_i > 0\).

\begin{figure}[h!]
  \begin{center}
    \begin{tabular}{cc}
      \scriptsize{Panel (a): Aggregate Effect of Information} &
                                                          \scriptsize{Panel
                                                          (b):
                                                          Individual Effect of Information}\\ 
\includegraphics[scale=0.35]{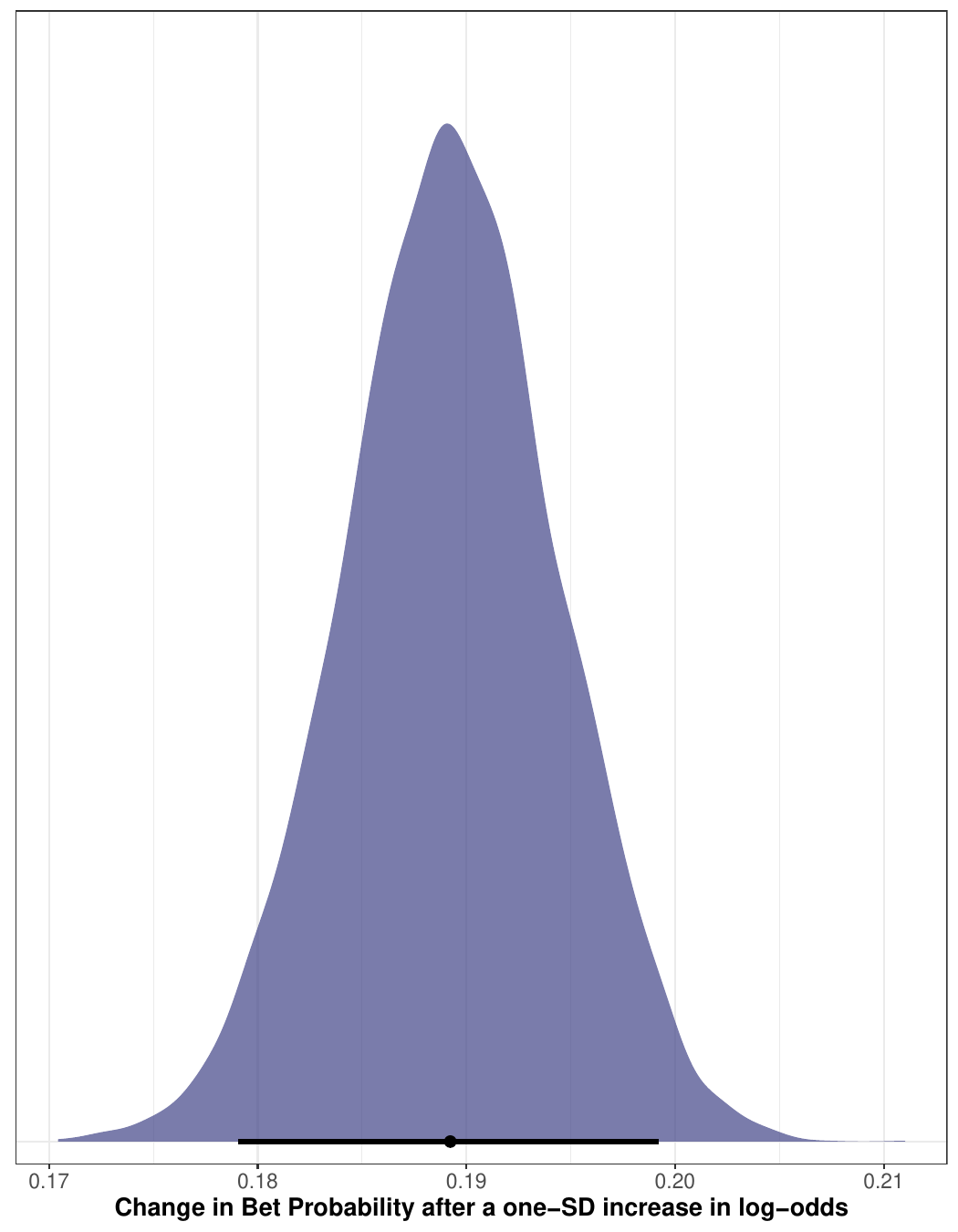}
      & \includegraphics[scale=0.35]{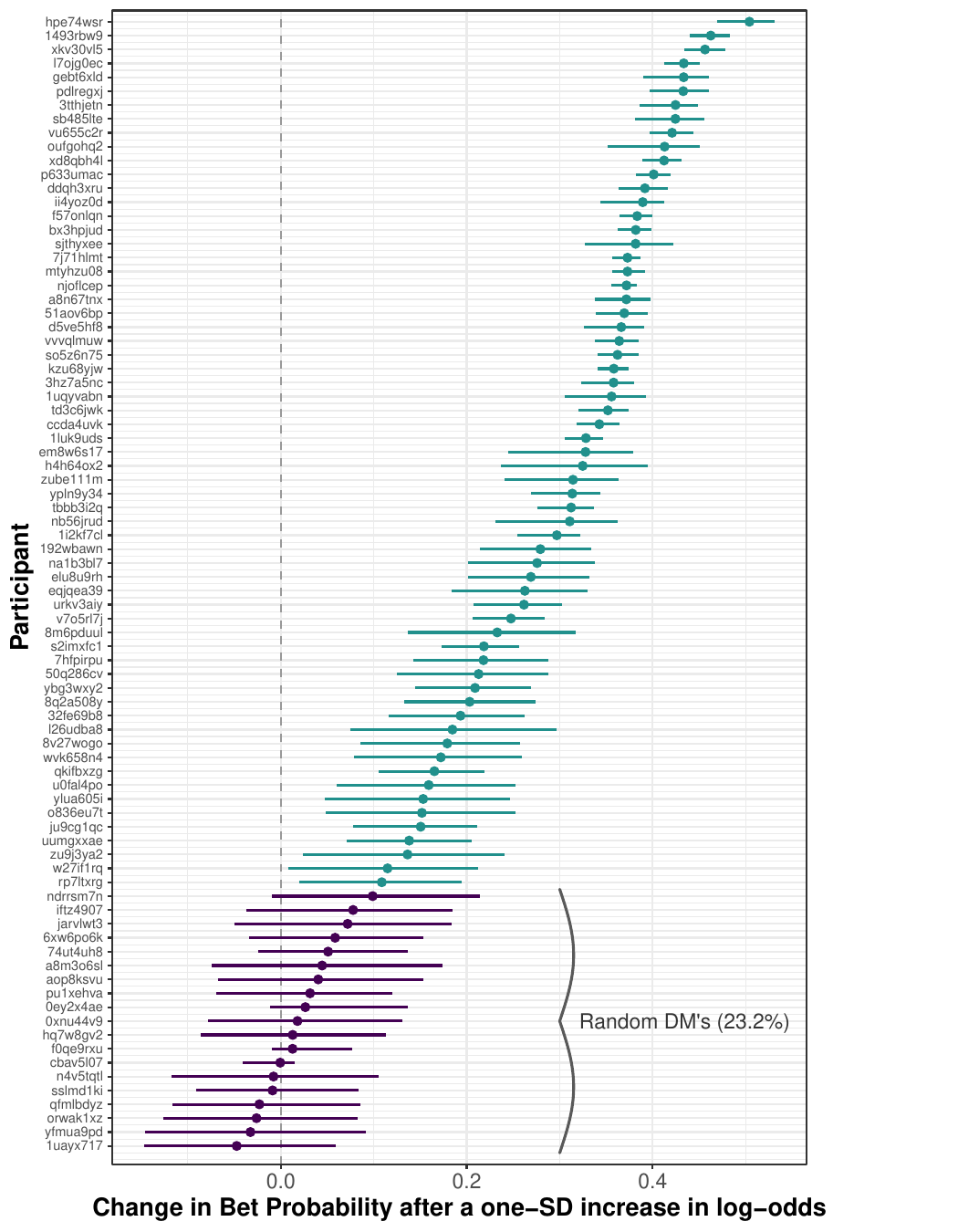}
    \end{tabular} 
\caption{Effect of Information on Actions ({\sc all} Treatment)}\label{fig:random_signals} 
\end{center}
{\footnotesize \underline{Notes:} Panel (a) presents the posterior
  change in bet probability after a one-SD increase in the log-oddds
  ratio, where the slope parameter is common across participants
  ($\beta_i=\beta$, for all $i$). Panel (b) presents the posterior
  change in bet probability after a one-SD increase in the log-odds
  ratio, where the slope parameter, $\beta_i$ is individual-specific}
\end{figure}

\begin{figure}[h!]
  \begin{center}
    \begin{tabular}{cc}
      \scriptsize{Panel (a): Aggregate Effect of Information} &
                                                          \scriptsize{Panel
                                                          (b):
                                                          Individual Effect of Information}\\ 
\includegraphics[scale=0.35]{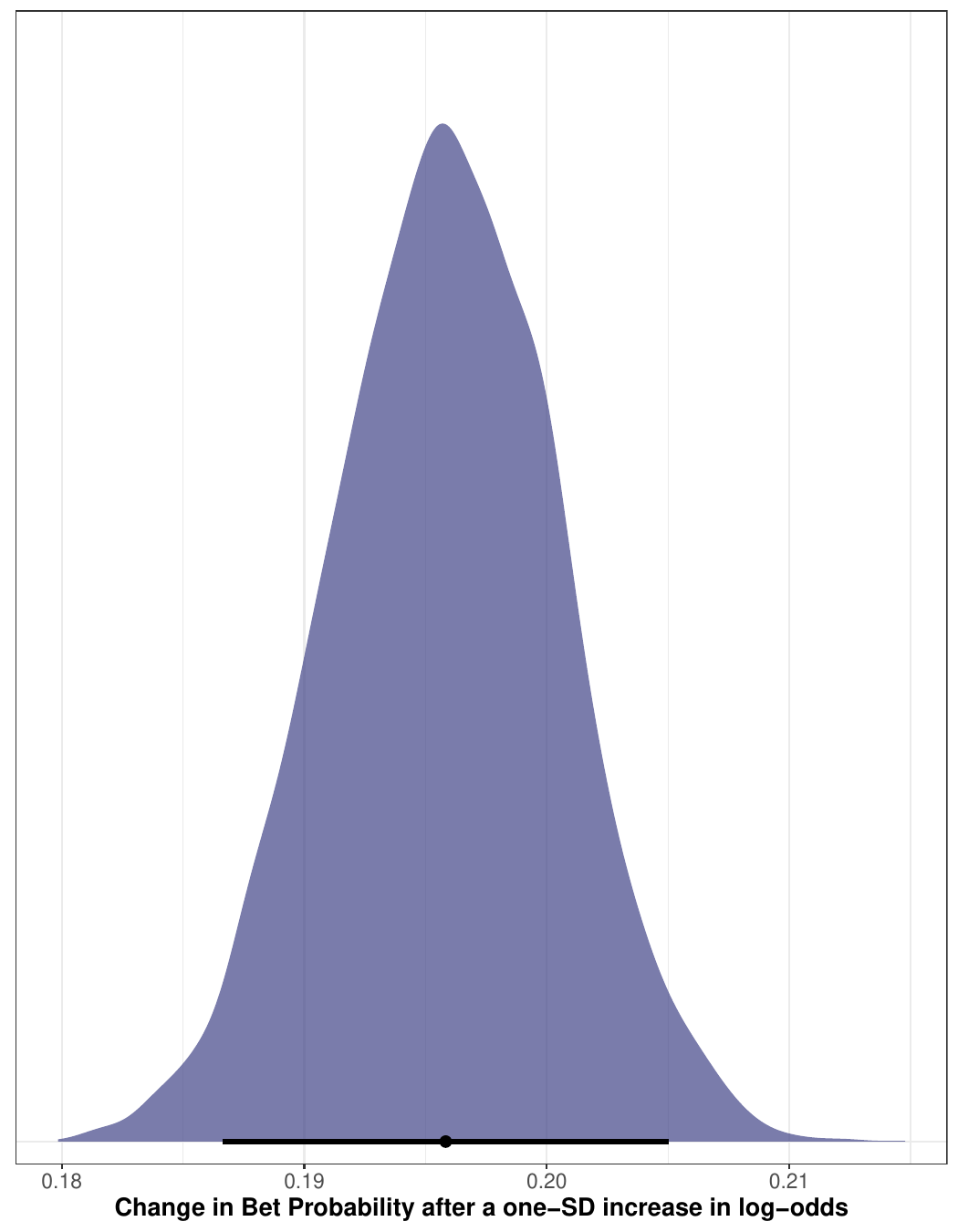}
      & \includegraphics[scale=0.35]{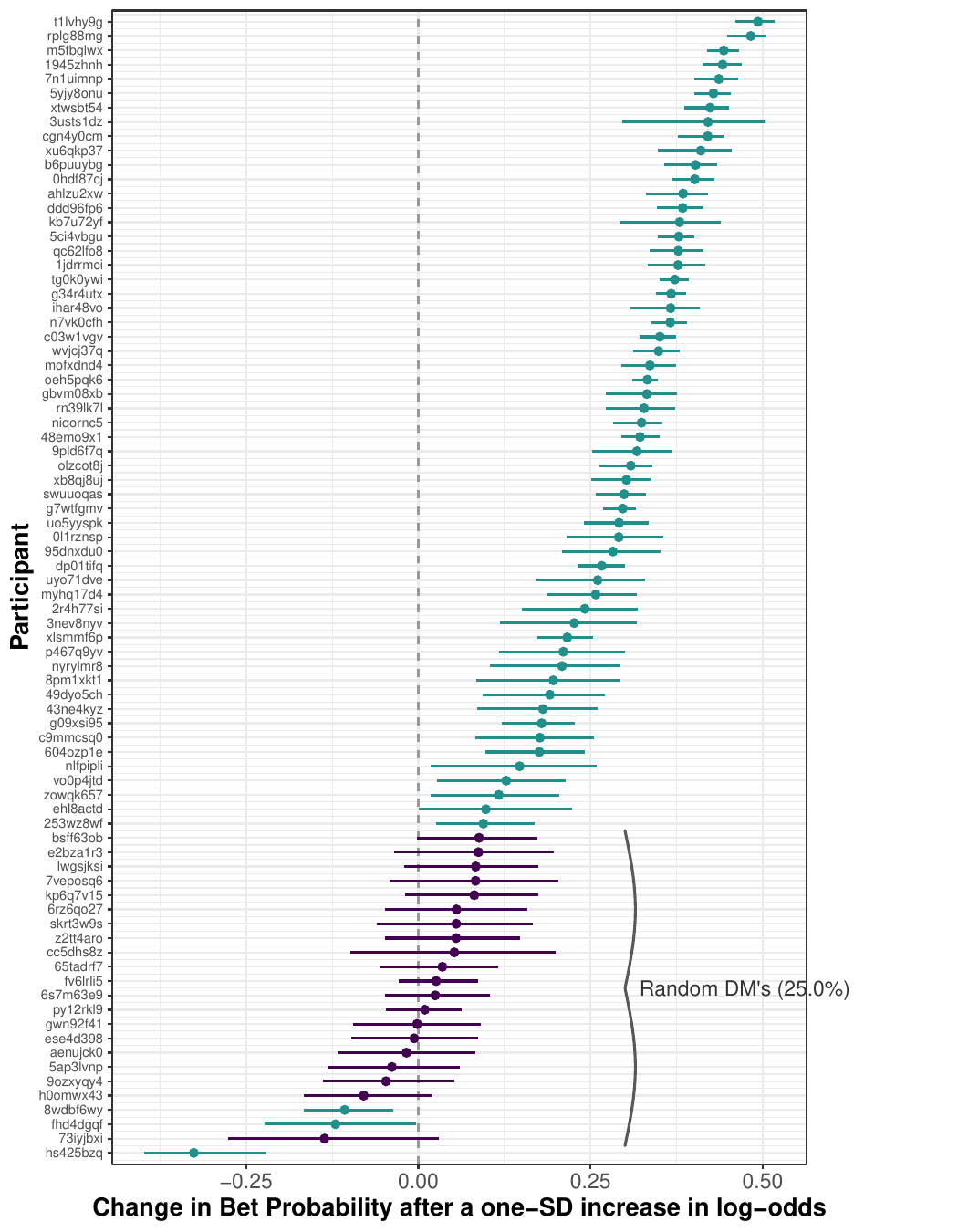}
    \end{tabular} 
\caption{Effect of Information on Actions ({\sc none} Treatment)}\label{fig:random_none} 
\end{center}
{\footnotesize \underline{Notes:} Panel (a) presents the posterior
  change in bet probability after a one-SD increase in the log-oddds
  ratio, where the slope parameter is common across participants
  ($\beta_i=\beta$, for all $i$). Panel (b) presents the posterior
  change in bet probability after a one-SD increase in the log-odds
  ratio, where the slope parameter, $\beta_i$ is individual-specific}
\end{figure}

Figures \ref{fig:random_signals} and \ref{fig:random_none} report the
results of this estimation for the {\sc signals} and {\sc none}
treatments, respectively. Panel (a) in these figures reports the
posterior change in the probability of betting Red following a one-standard-deviation increase in the log-odds (in favor of $\omega = R$) across participants, corresponding to a homogeneous effect of information. 

For both the {\sc signals} and {\sc none} treatments, we find a
sizable effect of signals on behavior, with an increase of
approximately 19 percentage points in the probability of betting
red. However, this average effect masks substantial heterogeneity
across participants. For more than half of participants in both
treatments, the effect of a one-standard-deviation increase in
log-odds favoring \(\omega = R\) exceeds 25 percentage points. 

On the other hand, around a quarter of participants in both treatments have posterior 95\% credible intervals
for \(\beta_i\) that include zero, suggesting weak or no
responsiveness to information, consistent with random choices.

\bibliography{refs}

\end{document}